\documentclass[%
 superscriptaddress,
 reprint,
 showpacs,preprintnumbers,
 nofootinbib,
 amsmath,amssymb,
 aps,
 prd,
 noeprint,
 floatfix,
]{revtex4-1}
\usepackage{graphicx}
\usepackage{xcolor}
\usepackage{appendix}
\usepackage[caption=false]{subfig}
\usepackage{hyperref}
\usepackage{comment}
\hypersetup{
    colorlinks=true,
    linkcolor=blue,
    filecolor=blue,      
    urlcolor=blue,
    pdftitle={ND-NIWG},
    }

\newcommand{\aref}[1]{\hyperref[#1]{App.~\ref*{#1}}}
\usepackage{lipsum}
\usepackage{amssymb,amsmath,amstext,amsthm,amsfonts}
\usepackage[utf8]{inputenc}
\usepackage{url}
\usepackage[colorinlistoftodos]{todonotes}
\usepackage{soul}
\usepackage{esvect}
\usepackage{multirow}
\usepackage{enumitem}
\usepackage{array}
\usepackage{lineno}
\usepackage{booktabs}  
\usepackage{multirow}  
\usepackage[capitalise]{cleveref}
\usepackage[normalem]{ulem}

\definecolor{LightCyan}{rgb}{0.88,1,1}
\usepackage{xcolor,colortbl}

\input{commands}
\makeatletter
\renewcommand*{\@fnsymbol}[1]{\ensuremath{\ifcase#1\or
  *\or \dagger\or \ddagger\or \mathsection\or \mathparagraph\or
  \|\or \#\or \star\or \circ\or \bullet\or
  **\or \dagger\dagger\or \ddagger\ddagger\or \mathsection\mathsection\or
  \mathparagraph\mathparagraph\else\@ctrerr\fi}}
\makeatother

\begin{document}

\title{Constraining Neutrino Interaction Uncertainties for Neutrino Oscillation Measurements at the T2K Experiment}

\newcommand{\INSTHD}{\affiliation{University Autonoma Madrid, Department of Theoretical Physics, 28049 Madrid, Spain}}
\newcommand{\INSTFE}{\affiliation{Boston University, Department of Physics, Boston, Massachusetts, U.S.A.}}
\newcommand{\INSTD}{\affiliation{University of British Columbia, Department of Physics and Astronomy, Vancouver, British Columbia, Canada}}
\newcommand{\INSTGA}{\affiliation{University of California, Irvine, Department of Physics and Astronomy, Irvine, California, U.S.A.}}
\newcommand{\INSTI}{\affiliation{IRFU, CEA, Universit\'e Paris-Saclay, F-91191 Gif-sur-Yvette, France}}
\newcommand{\INSTGB}{\affiliation{University of Colorado at Boulder, Department of Physics, Boulder, Colorado, U.S.A.}}
\newcommand{\INSTFH}{\affiliation{Duke University, Department of Physics, Durham, North Carolina, U.S.A.}}
\newcommand{\INSTEF}{\affiliation{ETH Zurich, Institute for Particle Physics and Astrophysics, Zurich, Switzerland}}
\newcommand{\INSTIG}{\affiliation{VNU University of Science, Vietnam National University, Hanoi, Vietnam}}
\newcommand{\INSTIE}{\affiliation{CERN European Organization for Nuclear Research, CH-1211 Gen\'eve 23, Switzerland}}
\newcommand{\INSTEG}{\affiliation{University of Geneva, Section de Physique, DPNC, Geneva, Switzerland}}
\newcommand{\INSTHJ}{\affiliation{University of Glasgow, School of Physics and Astronomy, Glasgow, United Kingdom}}
\newcommand{\INSTJG}{\affiliation{Ghent University, Department of Physics and Astronomy, Proeftuinstraat 86, B-9000 Gent, Belgium}}
\newcommand{\INSTDG}{\affiliation{H. Niewodniczanski Institute of Nuclear Physics PAN, Cracow, Poland}}
\newcommand{\INSTCB}{\affiliation{High Energy Accelerator Research Organization (KEK), Tsukuba, Ibaraki, Japan}}
\newcommand{\INSTIB}{\affiliation{University of Houston, Department of Physics, Houston, Texas, U.S.A.}}
\newcommand{\INSTED}{\affiliation{Institut de Fisica d'Altes Energies (IFAE) - The Barcelona Institute of Science and Technology, Campus UAB, Bellaterra (Barcelona) Spain}}
\newcommand{\INSTJC}{\affiliation{Institut f\"ur Physik, Johannes Gutenberg-Universit\"at Mainz, Staudingerweg 7, 55128 Mainz, Germany}}
\newcommand{\INSTHH}{\affiliation{Institute For Interdisciplinary Research in Science and Education (IFIRSE), ICISE, Quy Nhon, Vietnam}}
\newcommand{\INSTEI}{\affiliation{Imperial College London, Department of Physics, London, United Kingdom}}
\newcommand{\INSTGF}{\affiliation{INFN Sezione di Bari and Universit\`a e Politecnico di Bari, Dipartimento Interuniversitario di Fisica, Bari, Italy}}
\newcommand{\INSTBE}{\affiliation{INFN Sezione di Napoli and Universit\`a di Napoli, Dipartimento di Fisica, Napoli, Italy}}
\newcommand{\INSTBF}{\affiliation{INFN Sezione di Padova and Universit\`a di Padova, Dipartimento di Fisica, Padova, Italy}}
\newcommand{\INSTBD}{\affiliation{INFN Sezione di Roma and Universit\`a di Roma ``La Sapienza'', Roma, Italy}}
\newcommand{\INSTEB}{\affiliation{Institute for Nuclear Research of the Russian Academy of Sciences, Moscow, Russia}}
\newcommand{\INSTHI}{\affiliation{International Centre of Physics, Institute of Physics (IOP), Vietnam Academy of Science and Technology (VAST), 10 Dao Tan, Ba Dinh, Hanoi, Vietnam}}
\newcommand{\INSTJD}{\affiliation{ILANCE, CNRS – University of Tokyo International Research Laboratory, Kashiwa, Chiba 277-8582, Japan}}
\newcommand{\INSTHA}{\affiliation{Kavli Institute for the Physics and Mathematics of the Universe (WPI), The University of Tokyo Institutes for Advanced Study, University of Tokyo, Kashiwa, Chiba, Japan}}
\newcommand{\INSTID}{\affiliation{Keio University, Department of Physics, Kanagawa, Japan}}
\newcommand{\INSTIF}{\affiliation{King's College London, Department of Physics, Strand, London WC2R 2LS, United Kingdom}}
\newcommand{\INSTCC}{\affiliation{Kobe University, Kobe, Japan}}
\newcommand{\INSTCD}{\affiliation{Kyoto University, Department of Physics, Kyoto, Japan}}
\newcommand{\INSTEJ}{\affiliation{Lancaster University, Physics Department, Lancaster, United Kingdom}}
\newcommand{\INSTII}{\affiliation{Lawrence Berkeley National Laboratory, Berkeley, California, U.S.A.}}
\newcommand{\INSTBA}{\affiliation{Ecole Polytechnique, IN2P3-CNRS, Laboratoire Leprince-Ringuet, Palaiseau, France}}
\newcommand{\INSTFC}{\affiliation{University of Liverpool, Department of Physics, Liverpool, United Kingdom}}
\newcommand{\INSTFI}{\affiliation{Louisiana State University, Department of Physics and Astronomy, Baton Rouge, Louisiana, U.S.A.}}
\newcommand{\INSTIH}{\affiliation{Joint Institute for Nuclear Research, Dubna, Moscow Region, Russia}}
\newcommand{\INSTHB}{\affiliation{Michigan State University, Department of Physics and Astronomy,  East Lansing, Michigan, U.S.A.}}
\newcommand{\INSTCE}{\affiliation{Miyagi University of Education, Department of Physics, Sendai, Japan}}
\newcommand{\INSTDF}{\affiliation{National Centre for Nuclear Research, Warsaw, Poland}}
\newcommand{\INSTFJ}{\affiliation{State University of New York at Stony Brook, Department of Physics and Astronomy, Stony Brook, New York, U.S.A.}}
\newcommand{\INSTEH}{\affiliation{STFC, Rutherford Appleton Laboratory, Harwell Oxford,  and  Daresbury Laboratory, Warrington, United Kingdom}}
\newcommand{\INSTGJ}{\affiliation{Okayama University, Department of Physics, Okayama, Japan}}
\newcommand{\INSTCF}{\affiliation{Osaka Metropolitan University, Department of Physics, Osaka, Japan}}
\newcommand{\INSTGG}{\affiliation{Oxford University, Department of Physics, Oxford, United Kingdom}}
\newcommand{\INSTIC}{\affiliation{University of Pennsylvania, Department of Physics and Astronomy,  Philadelphia, Pennsylvania, U.S.A.}}
\newcommand{\INSTGC}{\affiliation{University of Pittsburgh, Department of Physics and Astronomy, Pittsburgh, Pennsylvania, U.S.A.}}
\newcommand{\INSTGD}{\affiliation{University of Rochester, Department of Physics and Astronomy, Rochester, New York, U.S.A.}}
\newcommand{\INSTHC}{\affiliation{Royal Holloway University of London, Department of Physics, Egham, Surrey, United Kingdom}}
\newcommand{\INSTBC}{\affiliation{RWTH Aachen University, III. Physikalisches Institut, Aachen, Germany}}
\newcommand{\INSTJF}{\affiliation{School of Physics and Astronomy, University of Minnesota, Minneapolis, Minnesota, U.S.A.}}
\newcommand{\INSTJB}{\affiliation{Departamento de F\'isica At\'omica, Molecular y Nuclear, Universidad de Sevilla, 41080 Sevilla, Spain}}
\newcommand{\INSTFB}{\affiliation{University of Sheffield, School of Mathematical and Physical Sciences, Sheffield, United Kingdom}}
\newcommand{\INSTDI}{\affiliation{University of Silesia, Institute of Physics, Katowice, Poland}}
\newcommand{\INSTIA}{\affiliation{SLAC National Accelerator Laboratory, Stanford University, Menlo Park, California, U.S.A.}}
\newcommand{\INSTBB}{\affiliation{Sorbonne Universit\'e, CNRS/IN2P3, Laboratoire de Physique Nucl\'eaire et de Hautes Energies (LPNHE), Paris, France}}
\newcommand{\INSTJE}{\affiliation{South Dakota School of Mines and Technology, 501 East Saint Joseph Street, Rapid City, SD 57701, United States}}
\newcommand{\INSTCH}{\affiliation{University of Tokyo, Department of Physics, Tokyo, Japan}}
\newcommand{\INSTBJ}{\affiliation{University of Tokyo, Institute for Cosmic Ray Research, Kamioka Observatory, Kamioka, Japan}}
\newcommand{\INSTCG}{\affiliation{University of Tokyo, Institute for Cosmic Ray Research, Research Center for Cosmic Neutrinos, Kashiwa, Japan}}
\newcommand{\INSTHF}{\affiliation{Institute of Science Tokyo, Department of Physics, Tokyo}}
\newcommand{\INSTGI}{\affiliation{Tokyo Metropolitan University, Department of Physics, Tokyo, Japan}}
\newcommand{\INSTHG}{\affiliation{Tokyo University of Science, Faculty of Science and Technology, Department of Physics, Noda, Chiba, Japan}}
\newcommand{\INSTB}{\affiliation{TRIUMF, Vancouver, British Columbia, Canada}}
\newcommand{\INSTJH}{\affiliation{University of Toyama, Faculty of Science, Toyama, Japan}}
\newcommand{\INSTDJ}{\affiliation{University of Warsaw, Faculty of Physics, Warsaw, Poland}}
\newcommand{\INSTDH}{\affiliation{Warsaw University of Technology, Institute of Radioelectronics and Multimedia Technology, Warsaw, Poland}}
\newcommand{\INSTIJ}{\affiliation{Tohoku University, Faculty of Science, Department of Physics, Miyagi, Japan}}
\newcommand{\INSTFD}{\affiliation{University of Warwick, Department of Physics, Coventry, United Kingdom}}
\newcommand{\INSTEA}{\affiliation{Wroclaw University, Faculty of Physics and Astronomy, Wroclaw, Poland}}
\newcommand{\INSTHE}{\affiliation{Yokohama National University, Department of Physics, Yokohama, Japan}}
\newcommand{\INSTH}{\affiliation{York University, Department of Physics and Astronomy, Toronto, Ontario, Canada}}

\INSTHD
\INSTFE
\INSTD
\INSTGA
\INSTI
\INSTGB
\INSTFH
\INSTEF
\INSTIG
\INSTIE
\INSTEG
\INSTHJ
\INSTJG
\INSTDG
\INSTCB
\INSTIB
\INSTED
\INSTJC
\INSTHH
\INSTEI
\INSTGF
\INSTBE
\INSTBF
\INSTBD
\INSTEB
\INSTHI
\INSTJD
\INSTHA
\INSTID
\INSTIF
\INSTCC
\INSTCD
\INSTEJ
\INSTII
\INSTBA
\INSTFC
\INSTFI
\INSTIH
\INSTHB
\INSTCE
\INSTDF
\INSTFJ
\INSTEH
\INSTGJ
\INSTCF
\INSTGG
\INSTIC
\INSTGC
\INSTGD
\INSTHC
\INSTBC
\INSTJF
\INSTJB
\INSTFB
\INSTDI
\INSTIA
\INSTBB
\INSTJE
\INSTCH
\INSTBJ
\INSTCG
\INSTHF
\INSTGI
\INSTHG
\INSTB
\INSTJH
\INSTDJ
\INSTDH
\INSTIJ
\INSTFD
\INSTEA
\INSTHE
\INSTH

\author{K.\,Abe}\INSTBJ
\author{S.\,Abe}\INSTCH
\author{H.\,Adhikary}\INSTDJ
\author{R.\,Akutsu}\INSTCB
\author{H.\,Alarakia-Charles}\INSTEJ
\author{Y.I.\,Alj Hakim}\INSTFB
\author{S.\,Alonso Monsalve}\INSTEF
\author{L.\,Anthony}\INSTEI
\author{S.\,Aoki}\INSTCC
\author{K.A.\,Apte}\INSTEI
% ...

\collaboration{The T2K Collaboration}\noaffiliation
	
\begin{abstract}

\noindent
In the context of neutrino oscillation measurements from the T2K experiment, the off-axis near detector ND280 plays a crucial role in constraining the incoming neutrino flux and neutrino-nucleus interaction cross sections. The result is a robust control over systematic uncertainties in the fit of neutrino oscillation parameters to the data at the T2K far detector, Super-Kamiokande. This paper details the methodology and results of these constraints in the context of the latest neutrino oscillation analysis from T2K. It describes how a new neutrino cross-section model and refined flux prediction are parameterized and fit to data in new ND280 event selections. Additionally, this work reports the results of extensive robustness studies, including fits with alternative interaction models, consistency checks against publicly available cross-section measurements, and \textit{p}-value evaluations, to demonstrate the reliability and robustness of our methodology. Finally, we present a sensitivity study demonstrating that the upgraded ND280, with improved acceptance and a lower hadron threshold, may enhance future constraints and further reduce systematic uncertainties in oscillation measurements.

\end{abstract}

\maketitle

\section{Introduction}
\label{sec:introduction}
Neutrino oscillations have been observed in a variety of neutrino experiments~\cite{Fukuda:1998mi,Ahmad:2002jz,Eguchi:2002dm,Ahn:2006zza,Adamson:2014vgd,Super-Kamiokande:2002ujc}. In the Pontecorvo-Maki-Nakagawa-Sakata (PMNS) scenario~\cite{Pontecorvo:1967fh,Maki:1962mu}, neutrino oscillations are governed by the PMNS matrix that describes the three neutrino flavor eigenstates (\nue, \num, $\nu_\tau$) as linear combinations of the mass eigenstates ($\nu_1$, $\nu_2$, $\nu_3$). This unitary matrix is parameterized by four parameters: three mixing angles ($\theta_{12}$, $\theta_{23}$, and $\theta_{13}$) and a phase ($\delta_{CP}$), which allows for the possibility of CP violation in the lepton sector. To describe the oscillation probabilities, the neutrino squared mass splittings ($\Delta m^2_{12}$, $\Delta m^2_{13}$, $\Delta m^2_{32}$) are also needed.

While the mixing angles and mass splittings are being measured with increasing precision~\cite{NOvA:2021nfi,T2K:2023mcm}, current and planned long-baseline (LBL) neutrino experiments are also focusing on the determination of still unknown quantities, namely the neutrino mass ordering (MO) and the value of $\delta_{CP}$, which can be measured precisely by comparing \nue and \nueb appearance rates in \num and \numb beams. These experiments will also have the statistical sensitivity to look for new physics by searching for differences between inferred oscillation probabilities and those allowed within the PMNS neutrino mixing paradigm~\cite{DUNE:2020fgq}. 

The oscillation probabilities and, therefore, appearance/disappearance probabilities are a function of the neutrino energy, a quantity which cannot be predicted {\em a priori} since neutrino beams are not monochromatic. In current experiments, incoming neutrino energies can be inferred only from the kinematics of the final-state particles produced in neutrino-nucleus interactions. To predict how a given appearance probability will manifest in observed event rates, some assumptions about the nature of neutrino interactions are needed. This is done by relying on theoretical models that, while useful, introduce uncertainties and need to be compared and tuned against existing data. 

In LBL experiments, the near detector measures the neutrino flux before oscillations occur, whereas the far detector measures the flux after propagation over the long baseline, where oscillations have taken place. A comparison between near- and far-detector event rates does not, in general, lead to perfect cancellation of all the systematic uncertainties. This is because there are important differences between the neutrinos observed in the near- and far-detectors, coming from 1) oscillations being present at the far detector but not the near detector, leading to different neutrino flux spectra, 2) detector acceptances, and 3) interaction target materials, all of which prevent using direct extrapolation methods. Instead, the analysis relies on using near-detector data to constrain systematic parameters in the simulation of the experiment---for example, neutrino interaction, flux, and detection uncertainties---and propagate these constraints to make predictions at the far detector. Thus, the analysis is sensitive to the choice of model and the parametrization of the associated uncertainties.
    
The Tokai-to-Kamioka experiment (T2K) is a long-baseline neutrino oscillation experiment located in Japan~\cite{Abe:2011ks}. At the J-PARC accelerator complex, a beam of mainly muon (anti)neutrinos is produced and detected by a series of near detectors located 280\,m from the neutrino production target. In particular, data collected by the off-axis near detector (ND280) are used to characterize the neutrino beam before it oscillates, providing constraints on the product of the T2K flux with the neutrino interaction cross section. The neutrino beam is detected again by the far detector Super-Kamiokande, located in the Kamioka mine, 295\,km away from J-PARC, where the effect of neutrino oscillations is appreciable. 
     
T2K studies the disappearance probabilities of muon neutrinos (\num) and antineutrinos (\numb) as well as the appearance probabilities of electron neutrinos (\nue) and antineutrinos (\nueb), providing useful information on $\theta_{13}$ and enabling precise measurements of $\theta_{23}$ and $\Delta m^2_{32}$~\cite{T2K:2023mcm}, while giving the first hints towards CP violation in the lepton sector~\cite{T2K:2023smv}. Although dominated by statistical uncertainties, T2K measurements also have non-negligible systematic uncertainties. In particular, the uncertainties related to the neutrino interaction model are a significant contributor to the total systematic uncertainty on the calculated event rates at the far detector. These uncertainties are expected to become the dominant source of errors in future measurements by the Hyper-Kamiokande~\cite{Hyper-KamiokandeProto-:2015xww} or DUNE~\cite{DUNE:2020lwj}. To overcome this issue, together with the construction of more capable near detectors, it is essential to use more accurate neutrino interaction models~\cite{Katori:2016yel}. In parallel, this requires the development of an adequate and comprehensive set of uncertainties that describe plausible model variation.

In this paper, we describe the T2K near-detector analysis that is used for T2K's oscillation measurements reported in Ref.~\cite{T2K:2025yoy}. The core of the analysis is a fit of a parameterized flux, neutrino interaction, and ND280 detector model to binned data from a variety of event selections. With respect to the previous analysis~\cite{T2K:2023smv}, this work includes the use of a more refined parameterization of the uncertainties related to the neutrino interactions, the use of a new incoming neutrino flux tuning, and the deployment of updated event selections leveraging information about outgoing protons and photons. \\
The paper is organized as follows: after describing the T2K experiment in \autoref{sec:ND}, details about the novelties with respect to the previous analysis are given in \autoref{sec:updates}. The flux model is then described in \autoref{sec:fluxpred}, while all the details about the neutrino interaction models and the corresponding parameterization are given in \autoref{sec:intModel}. \autoref{sec:NDfit} reports the details and the results of the tuning of these models against T2K data, while \autoref{sec:FDS} describes the testing and the validation of the new parameterization against a series of simulated data samples. Data results are first compared with respect to existing cross-section measurements in \autoref{subsec:benchmark} and then further discussed in \autoref{sec:discussion}. In \autoref{sec:sensitivity}, projected future sensitivities to constrain the neutrino interaction and flux models, including the constraints possible with the \ndup~\cite{T2K:2019bbb}, are reported before the paper concludes with \autoref{sec:summary-conclusions}.

\section{The T2K Experiment}
\label{sec:ND}
T2K's (anti)neutrino beam is produced by the Japan Proton Accelerator Research Complex (J-PARC) in Tokai, Ibaraki prefecture. Protons are accelerated to 30 GeV kinetic energy and are focused on a 90-cm-long graphite target. Outgoing charged hadrons are selected from these interactions by three magnetic horns that can switch polarity. The first T2K runs were performed with a current of $\pm$250 kA. The hadrons travel through a decay volume, where many of the mesons decay predominantly into muons and muon (anti)neutrinos, with small components of electron (anti)neutrinos, forming the neutrino beam.

The near-detector complex consists of three detector systems: one located on the beam axis (INGRID), one at 1.5$^{\circ}$ (WAGASCI-BabyMind), and one at 2.5$^\circ$ off-axis (ND280). INGRID primarily monitors the spatial profile and direction of the neutrino beam~\cite{Abe:2011xv}. WAGASCI-BabyMind, which is not used in the analysis described in this paper, is used to measure the neutrino cross section on carbon and water~\cite{T2K:2020txr,T2K:2025kda}. ND280 measures the product of the neutrino flux and interaction cross section at the same off-axis angle as the far detector, constraining relevant systematics for the oscillation analysis~\cite{Abe:2011ks}.

\begin{figure}[htbp]
    \centering
    \includegraphics[width=0.3\textwidth]{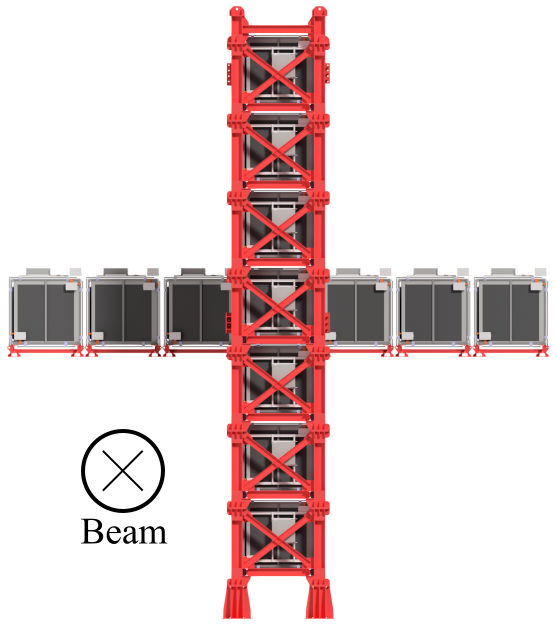}
    \caption{Drawing of the on-axis near detector INGRID. The neutrino beam is centered on the intersection of the arms. The beam direction is included in the figure.}
    \label{fig:ingrid}
\end{figure}

INGRID comprises 14 identical modules of iron and scintillator layers, which are arranged as a cross that covers an area of $10\times10~\text{m}^2$ and centered on the beam axis as shown in \autoref{fig:ingrid}. Each module alternates 6.5 cm-thick iron plates with 1 cm-thick scintillator planes, comprising a total of 9 iron and 11 scintillator planes. The module is surrounded by additional scintillator planes that act as veto detectors. Each module exposes a $1.24\times1.24~\text{m}^2$ area to the beam and has a target mass of 7.1~t. INGRID measures the beam direction with an accuracy better than 0.4~mrad, well within the required precision of $\pm1~\text{mrad}$ for the oscillation analysis~\cite{Abe:2011xv}.

\begin{figure}[htbp]
    \centering
    \includegraphics[width=0.4\textwidth]{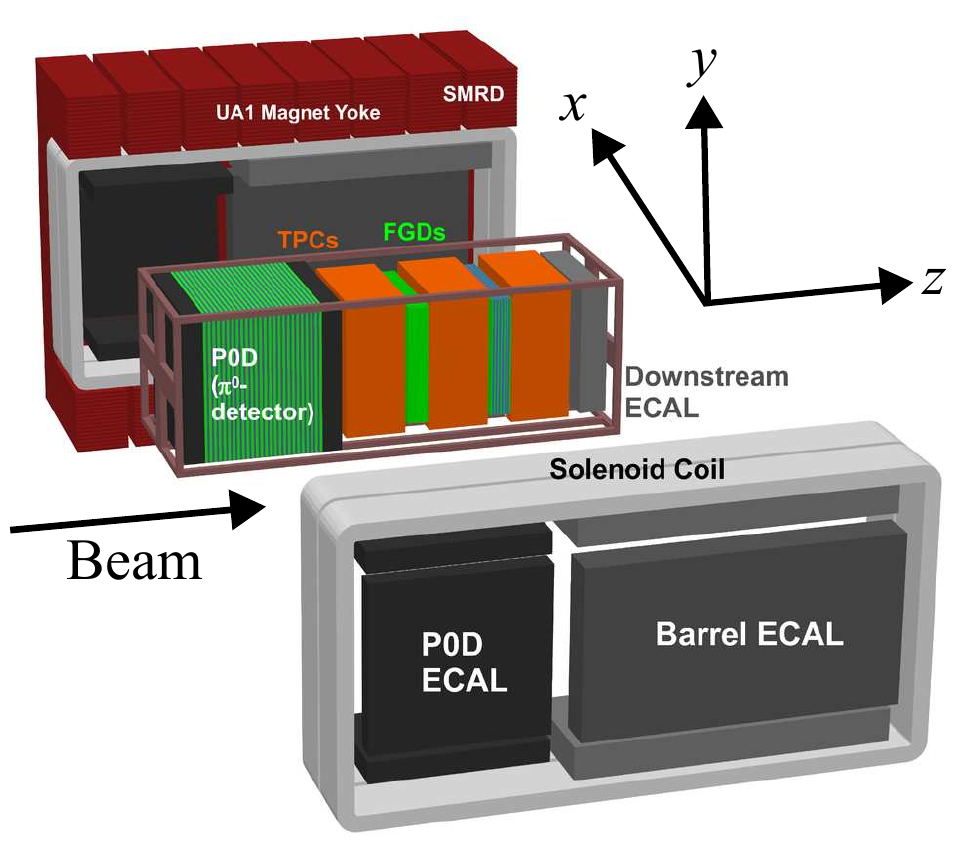}
    \caption{Drawing of the off-axis near detector ND280 in its original configuration. Coordinates, beam direction, and sub-detector labels are included.}
    \label{fig:nd280}
\end{figure}

ND280 is shown in \autoref{fig:nd280}. It measures $5.6~\text{m} \times 6.1~\text{m} \times 7.6~\text{m}$ (width $\times$ height $\times$ length) using a right-handed coordinate system with $z$ pointing along the nominal neutrino beam axis. It consists of multiple sub-detectors enclosed in the refurbished UA1~\cite{Astbury:1978uia,UA1:1980ndt} and NOMAD~\cite{NOMAD:1997pcg} magnet, which provides a magnetic field of 0.2~T. The magnet is instrumented with slabs of plastic scintillator that serve as a Side Muon Range Detector (SMRD)~\cite{Aoki:2013swe}, used both as a veto for the cosmic rays and as a detector for highly penetrating particles produced in neutrino interactions in the inner detectors. A detector optimized to measure neutrino interactions on water that produce $\pi^0$ (P$\emptyset$D)~\cite{Assylbekov:2011sh} is made of scintillator layers alternated with high-density polyethylene bags filled with water. Two fine-grained detectors (FGDs)~\cite{T2KND280FGD:2012umz} are used as the main targets for neutrino interactions. The upstream FGD (FGD1) consists of 15 polystyrene scintillator modules each measuring $186.4~\text{cm}\times186.4~\text{cm}\times2.02~\text{cm}$. Each module is composed of alternating layers of scintillator bars oriented along the $x$- and $y$-directions. Each layer contains 192 9.6 mm-wide square bars, approximately 2 m long. The downstream FGD (FGD2) contains six passive water modules, each sandwiched by polystyrene scintillator modules identical to those in FGD1. The FGDs are located between three Time Projection Chambers (TPCs)~\cite{T2KND280TPC:2010nnd}. The TPCs use an $\text{Ar}:\text{CF}_4:i\text{C}_4\text{H}_{10}$ gas mixture in a 95:3:2 concentration and have a space point resolution of $\sim$1 mm. An electromagnetic calorimeter (ECal)~\cite{T2KUK:2013wkh} surrounds the P$\emptyset$D, the TPCs, and the FGDs. The ECal consists of alternating layers of lead and plastic scintillator and provides additional particle identification capabilities. The FGDs and TPCs form the central tracking region of ND280 and are referred to simply as the tracker throughout the paper. This analysis selects events occurring in either FGD and uses the FGDs and TPCs for particle identification. The FGDs track charged particles and perform particle identification with a proton-to-pion misidentification probability below 10\%. The momentum for particles contained in the FGDs is measured by range. The TPCs are three-dimensional trackers that measure momentum through the curvature of the tracks in the magnetic field, with a resolution of $\sim$10\% for particles with momentum transverse to the magnetic field of around 1 GeV$/c$. 
The TPCs also provide excellent particle identification with a muon-to-electron misidentification of 0.2\% for tracks below 1 GeV$/c$. 

The Far Detector (FD), also placed 2.5$^\circ$ off-axis, is Super-Kamiokande ~\cite{Super-Kamiokande:2002weg}, a 50~kton water Cherenkov detector. It is located in the Kamioka mine, 295 km from the proton beam target. The inner detector is instrumented with 11,146 20-inch PMTs, and the outer detector is instrumented with around 1,885 8-inch PMTs. Commonly referred to as SK or Super-K, Super-Kamiokande will be strictly referred to as FD to emphasize the use of beam data in the far detector.

T2K started data-taking in 2010 and collected data with $\nu_\mu$- and $\bar{\nu}_\mu$-dominated beams by reversing the polarity of the beamline horn current. The results presented here are from an exposure of 1.97$\times$10$^{21}$ protons on target (POT) taken in $\nu$-mode and 1.63$\times$10$^{21}$ POT in \nub-mode.

\section{Updates from the previous analysis}
\label{sec:updates}
In this section, we summarize the improvements to the analysis presented in Ref.~\cite{T2K:2023smv}, which are detailed in the following sections.    

\begin{itemize}
    \item \textbf{ND280 selections:} 
    The selection of meson-less charged-current muon neutrino interactions  ($\nu_\mu\text{CC}0\pi$) at ND280 is split depending on whether or not a proton or photon is observed in the final state, increasing the number of samples per FGD in the $\nu$-mode beam from three to five. Photon tagging, using information from the ECal and TPCs, separates events with multiple charged pions and single neutral pions that were previously combined into a single sample. Consequently, new proton secondary interaction (SI) and ECal-related detector-response uncertainties were studied and employed in this analysis. The new ND280 samples are detailed in \autoref{subsec:ndsel}.

    \item \textbf{FD selections:} 
    A new selection was developed at FD, which targets single charged-pion production events from charged-current muon neutrino interactions ($\nu_\mu$CC$1\pi^+$-like). The selection includes both events with two prompt visible muon and pion rings, and events with a prompt visible muon ring and a delayed Michel electron, in which the charged pion is above and below the Cherenkov threshold, respectively. Along with this sample, there are five other samples: a one-ring muon- (1R$\mu$) and electron-like ring (1Re) in both neutrino and antineutrino beam mode, and a 1Re with one decay electron (1Re1de) coming from the decay of a charged pion. These samples will not be discussed in detail in this paper, but more information is provided in Refs.~\cite{T2K:2023smv,T2K:2025yoy}.
    
    \item \textbf{Neutrino flux model:}
    The neutrino flux calculation now includes hadron-scattering data from a replica of the T2K target, taken in 2010 by the NA61/SHINE collaboration. This improves on the 2009 data by measuring the yields of charged kaons and protons and collecting three times as much statistics on pions. The modeling of the cooling water flow in the horns and the non-hadronic uncertainties were also improved, sometimes leading to a marginal increase in the uncertainty. 
    The net effect is to almost halve the flux uncertainties between 2-7 GeV, slightly decrease the uncertainties below 0.5 GeV, while staying almost stable between 0.5-2 GeV. The flux simulation and uncertainties are discussed in \autoref{sec:fluxpred}.
    
    \item \textbf{Neutrino interaction model:}
    The uncertainties related to the neutrino interaction model were significantly expanded to reflect the new selections at the near and far detectors, and to account for improved characterization of neutrino interactions from nuclear theory and neutrino cross-section measurements. Theory-driven uncertainties related to the spectral-function-based charged current quasi-elastic (CCQE) model were developed, targeting effects such as Pauli blocking, distortion and/or absorption of the outgoing nucleon, and the uncertainties on the nuclear shell structure. An uncertainty on the strength of the proton final state interactions (FSI) was also added. The previous two-particle-two-hole (2p2h) shape uncertainties were split according to the initial state pair type ($np$ and $nn$ uncertainties). Uncertainties on single-pion production (SPP) now include the effect of the nuclear binding energy in resonant pion production, and uncertainties that target variations in the nucleon and pion kinematics. Two separate normalization uncertainties for \num and \numb charged-current single $\pi^0$ production were also introduced to reflect the photon-tagged sample at the near detector. An updated set of uncertainties for multi-pion production via shallow inelastic and deep inelastic scattering was also developed. An in-depth discussion of these uncertainties is given in \autoref{subsec:xsecuncert}.
    
    \item \textbf{New simulated data studies:} Given the updated event selections and neutrino interaction model, new simulated data studies to assess out-of-model effects were performed to test it. For instance, given the new flexibility added in the low-energy and low-momentum transfer region, a simulated data set with a prediction completely different than the default model was performed. The complete set of simulated data studies is discussed in \autoref{sec:FDS}.
    
\end{itemize}

\section{Flux prediction and uncertainties}
\label{sec:fluxpred}
The method for evaluating the neutrino flux prediction and the propagation of the associated uncertainties are the same as in previous results~\cite{T2K:2021xwb,T2K:2019bcf,Abe:2012av,T2K:2023smv}. Interactions inside the target are simulated by FLUKA 2011.2x~\cite{Bohlen:2014buj,Ferrari:2005zk}, while the particles exiting the target are tracked through the horn field using the GEANT3-based JNUBEAM package~\cite{T2K:2012bge}. The predictions for particles exiting the target's surface are tuned to yields measured by the NA61/SHINE experiment. Particles that leave the target are given a weight computed as follows:
\begin{equation}
w(p,\theta,z,i) = \frac{dn^\mathrm{NA61}(p,\theta,z,i)}{dn^\mathrm{MC}(p,\theta,z,i)}
\end{equation}
where $\mathrm dn$ is the POT-normalized differential yield for data ($\mathrm dn^\mathrm{NA61}$), and simulation Monte Carlo ($\mathrm dn^\mathrm{MC}$), with exiting momentum $p$, polar angle $\theta$, and longitudinal position $z$ along the target for an exiting particle of type $i$. The flux prediction and associated uncertainties are given for bins of true neutrino energy and neutrino flavor. The total flux systematic uncertainty is encoded in a covariance matrix constructed from plausible variations of hadron production and beamline modeling on the flux prediction.

Compared with the previous T2K analysis~\cite{T2K:2023smv}, where NA61/SHINE data taken in 2009 with a replica of the T2K target were included for the first time~\cite{NA61SHINE:2016nlf}, 2010 data~\cite{NA61SHINE:2018rhe} have been added in this analysis. 

\begin{figure}[htb]
    \centering 
    \includegraphics[width=0.9\columnwidth]{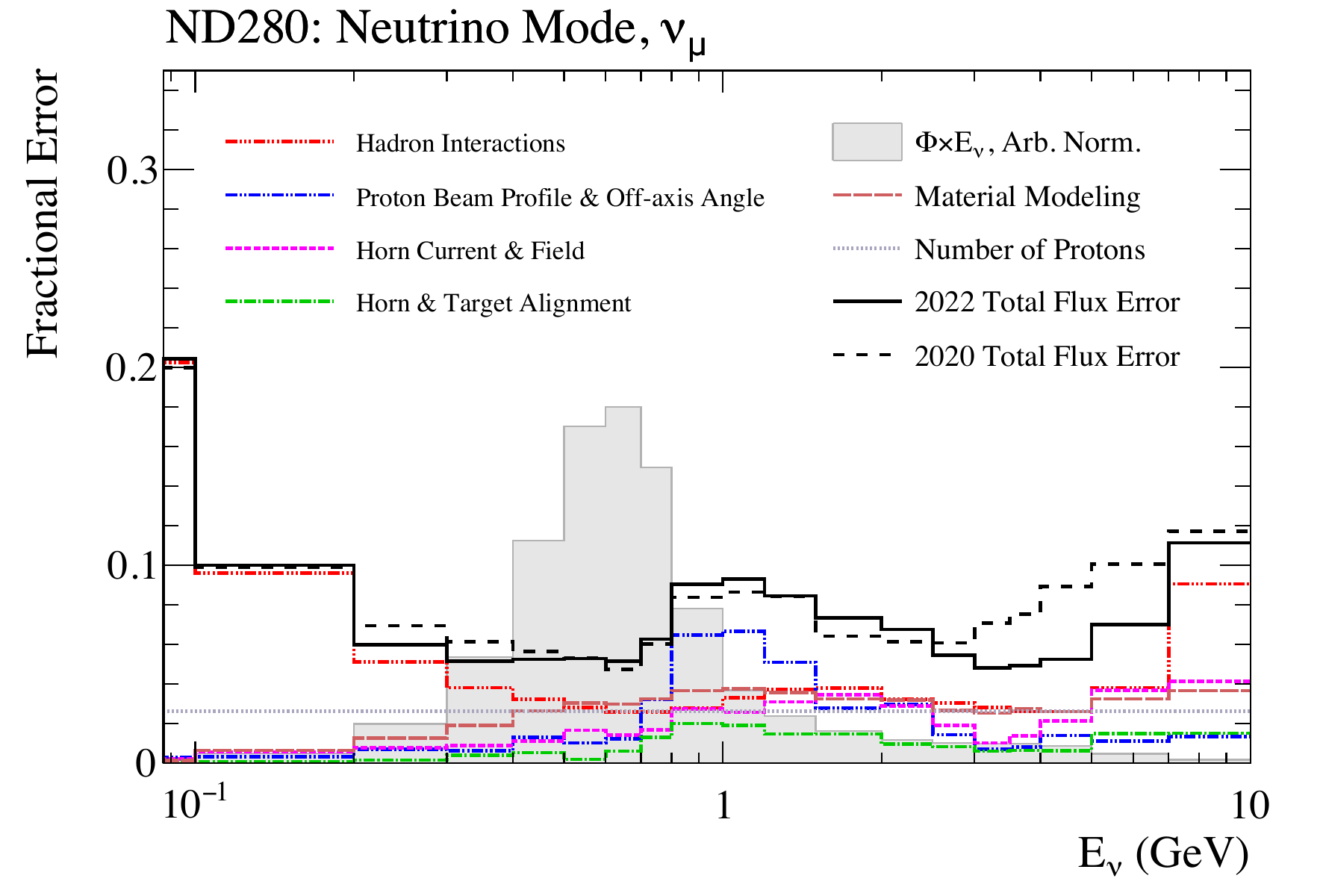}
    \includegraphics[width=0.9\columnwidth]{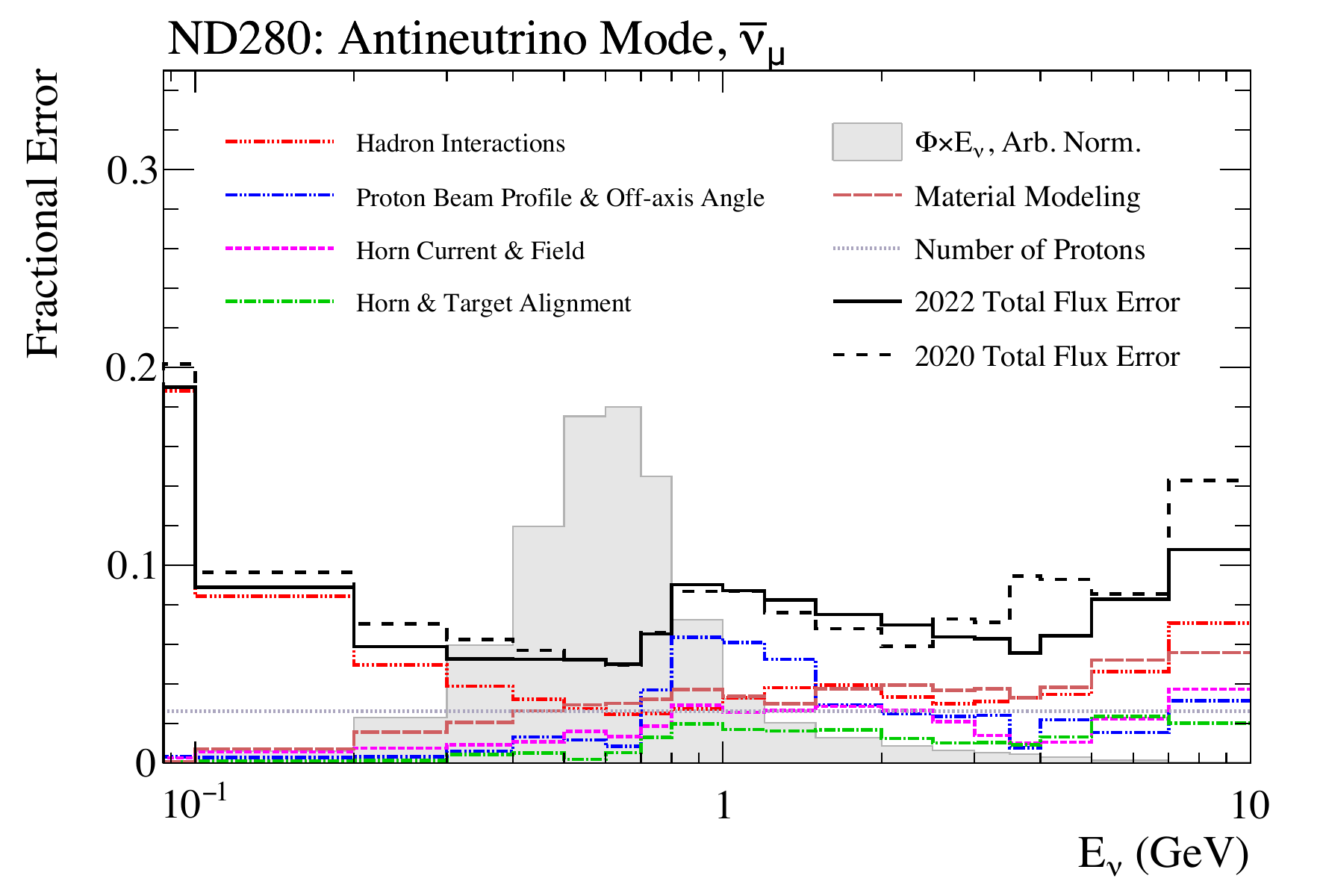}
    \includegraphics[width=0.9\columnwidth]{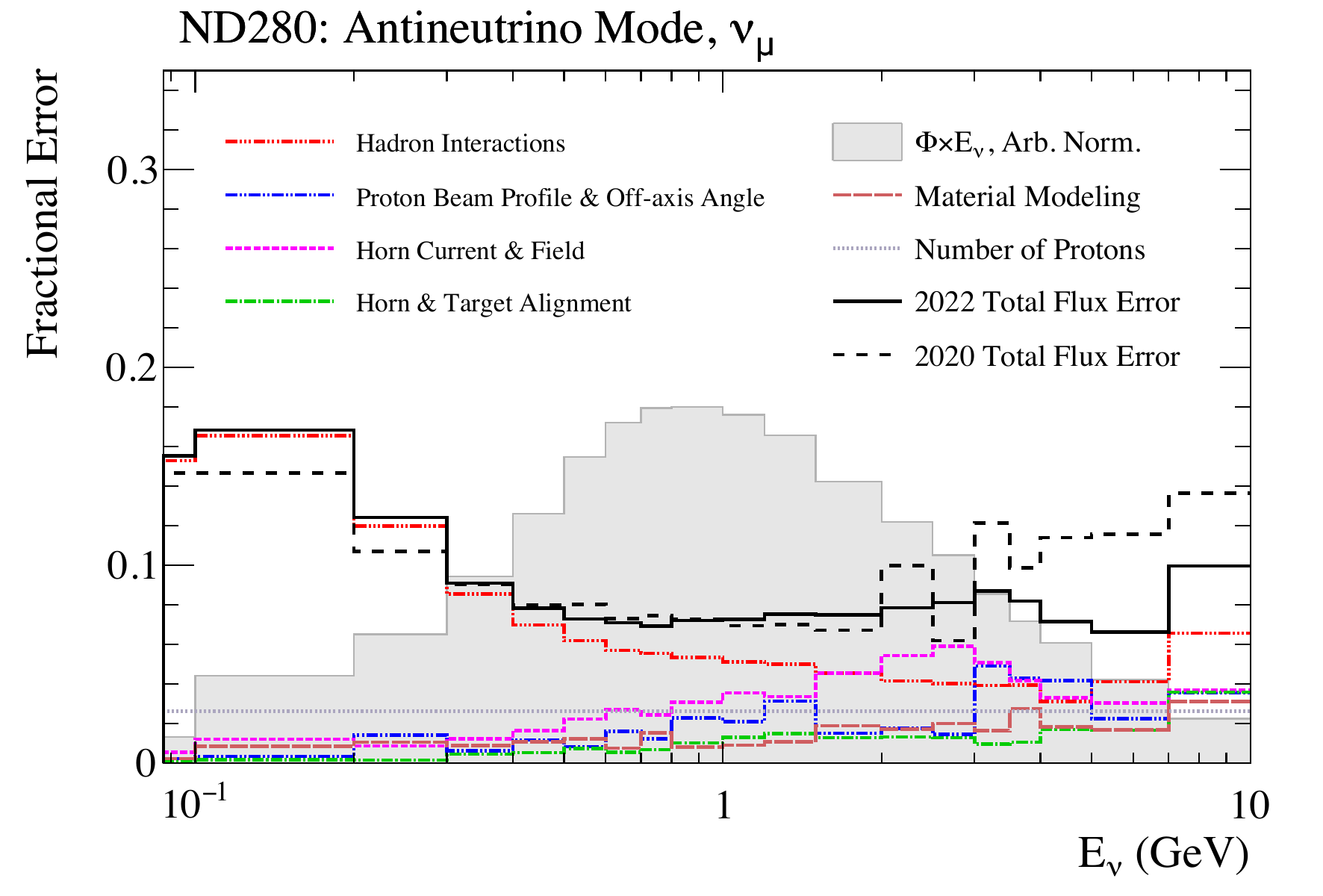}
    \caption{Fractional uncertainties on the \num flux in $\nu$-mode as well as on the \numb and \num fluxes in \nub-mode flux expected at ND280. The flux shapes of all three populations are also shown in grey. In the legend, ``2020'' refers to the previous tuning from~\cite{NA61SHINE:2016nlf}, while ``2022'' refers to the tuning used in this analysis based on Ref.~\cite{NA61SHINE:2018rhe}.}
    \label{fig::FluxUncer}
\end{figure}

This data set includes, for the first time, charged kaons and proton yields that allowed a further reduction of the flux uncertainty due to hadronic interactions, especially above 1$\sim$2 GeV where kaons exiting the target start to dominate the neutrino flux. This reduces the hadronic uncertainty to below 4\% for the energy range relevant to T2K, as shown in \autoref{fig::FluxUncer}. 

This analysis also implemented improvements to uncertainties related to the horn-cooling water and the strip lines supplying current to the horns. These uncertainties, previously treated as systematic uncertainties related to the material budget, are now explicitly included in the nominal flux simulation. For the horn-cooling water, the simulation was updated using measurements taken from a spare horn. This update to the nominal simulation enables the incorporation of hadronic uncertainties associated with interactions on water. These uncertainties have proven to be particularly significant at low energies, where they now represent the dominant source of uncertainty.

\section{Neutrino interaction model}
\label{sec:intModel}
A robust model of neutrino interactions, with uncertainties that parameterize its plausible variations, plays a crucial role in T2K neutrino oscillation analyses. A model is required to translate between measured final state particles and incoming neutrino energy, extrapolate constraints on neutrino cross sections between the near and far detectors, and predict unconstrained backgrounds to the neutrino oscillation analysis. The interaction model in this work maintains a very similar baseline model to T2K's previous analysis~\cite{T2K:2023smv} but has an improved treatment of its uncertainties.  
The most significant change is a move to a theory-inspired set of uncertainties to vary the nuclear model used for CCQE interactions, based on Ref.~\cite{Chakrani:2023htw}.

\subsection{Baseline neutrino interaction model}
\label{subsec:base_neut_int_model}

\begin{figure}[htbp]
    \centering
    \includegraphics[width=0.48\textwidth]{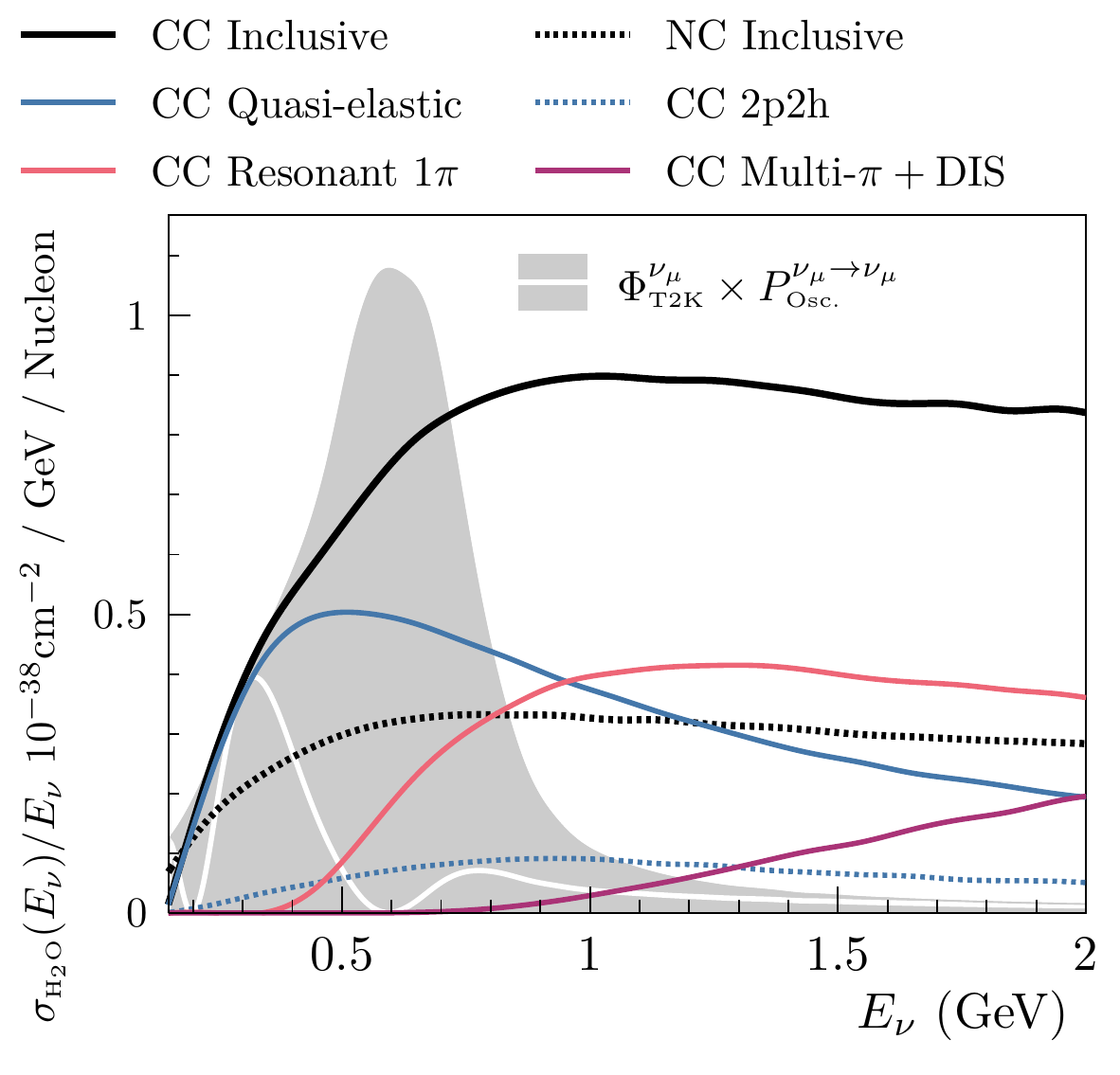}
    \caption{Neutrino cross sections for muon neutrinos interacting on a water target in NEUT, broken down by interaction mode as a function of neutrino energy. The predictions have been modified from their default to reflect the T2K nominal neutrino interaction model. The surviving muon neutrino flux as seen by the far detector is shown with a white line, and the unoscillated muon neutrino flux as seen by the near detector is shown as the gray-shaded region. The figure is adapted from Ref.~\cite{T2K:2023smv}.}
    \label{fig:xsecAndFlux}
\end{figure}

This analysis uses the NEUT~\cite{Hayato:2021heg} event generator v5.4.0 to simulate neutrino interactions. Like all neutrino interaction event generators, NEUT takes theory-inspired input cross-section calculations where they are available but makes a variety of approximations to produce the detailed predictions of all final state particle kinematics required in neutrino oscillation analyses (see, for example, Refs.~\cite{Nikolakopoulos:2022qkq, Nikolakopoulos:2023pdw}). The total cross sections predicted by NEUT as a function of neutrino energy are shown alongside the shape of the T2K flux before and after considerations of neutrino oscillations in \autoref{fig:xsecAndFlux}.

This section details the baseline input model used to generate events for T2K analyses. This is very similar to the model used in~\cite{T2K:2023smv}, but details are repeated here for completeness. 

\subsubsection{1p1h Interactions}
\label{sec:1p1h}

The one-particle one-hole (1p1h) channel, which is the dominant interaction for T2K's typical neutrino energies, describes charged current quasi-elastic (CCQE) and neutral current quasi-elastic (NCQE) neutrino interactions with a single nucleon bound inside a target nucleus. For the latest T2K oscillation analysis (as in the previous iteration~\cite{T2K:2023smv}) 1p1h interactions on carbon or oxygen targets, which are the vast majority of interactions selected in oscillation analysis event selections, are modeled according to the scheme presented in Ref.~\cite{Benhar:1994hw}, often called the Spectral Function model. 1p1h interactions on other targets default to an approach based on a relativistic Fermi-gas model (RFG). The Spectral Function model relies on the plane wave impulse approximation (PWIA) to `factorize' the 1p1h cross-section calculation into an expression containing a single nucleon cross-section alongside a `spectral function' (SF). The SF is a two-dimensional distribution describing the probability of finding a nucleon with some particular momentum ($|\vec{p}|$) and `removal energy' ($E_\mathrm{rmv}$), which corresponds to the energy required to remove the nucleon from the nuclear potential. Following the conventions in~\cite{Chakrani:2023htw}, these quantities can be written as:
\begin{equation}
    E_\mathrm{rmv} = E_\nu - E_\ell - T_{N'} - T_{A'} - \Delta m_N,
\end{equation}
\begin{equation}
    |\vec{p}| = \left|\vec p_\nu - \vec p_\ell - \vec p_{N'}\right|,
\end{equation}
where $E_\nu$ and $\vec p_\nu$ ($E_\ell$ and $\vec p_\ell$) are the energy and momentum of the incoming neutrino (outgoing charged lepton); $T_{N'}$ and $p_{N'}$ are the kinetic energy and momentum of the outgoing nucleon before FSI; $T_{A'}$ is the kinetic energy of the residual nucleus; and $\Delta m_N$ represents the mass difference between the initial- and final-state nucleon. $T_{A'}$ can be obtained from $p_{N'}$ under the assumption that the remnant nucleus has the mass of its ground state after the interaction.

\begin{figure}[htbp]
    \centering
    \includegraphics[width=0.48\textwidth]{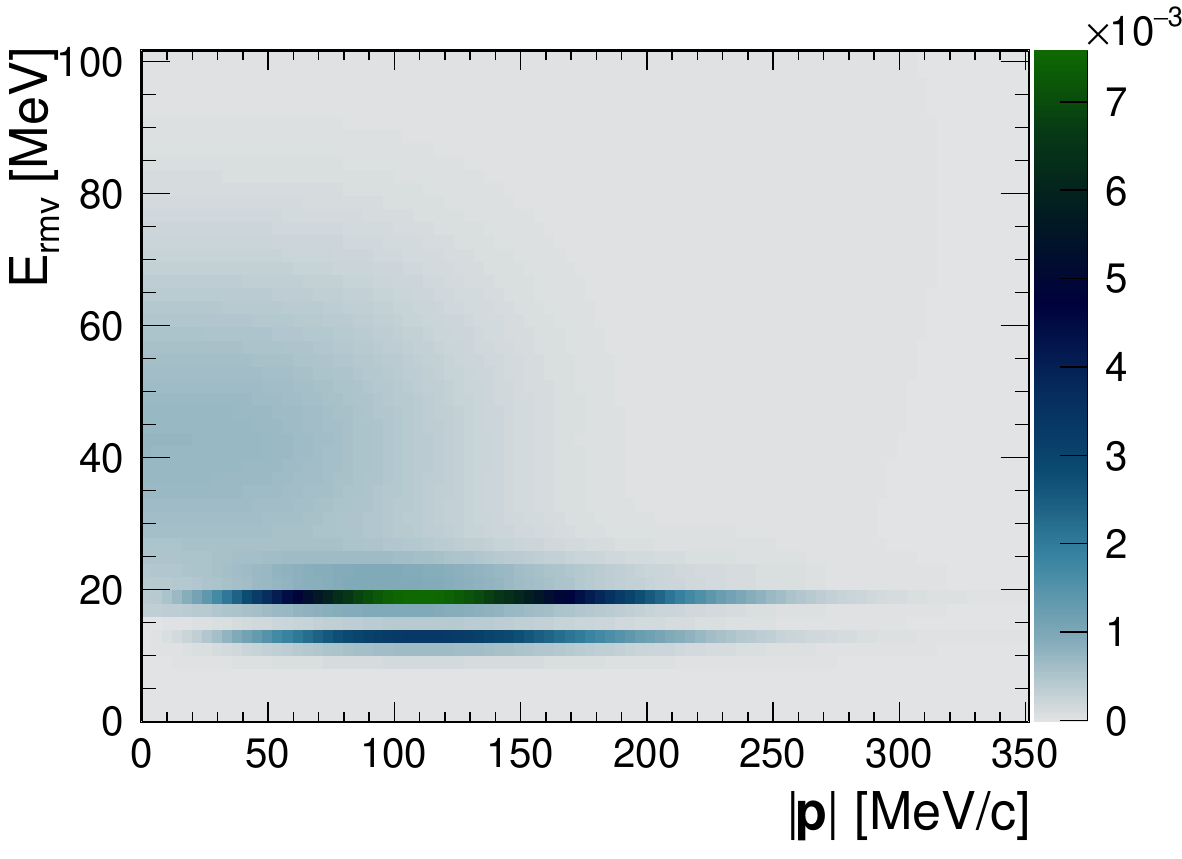}
    \caption{The two-dimensional probability density distribution for the spectral function for oxygen in NEUT~\cite{Benhar:1994hw}. The darker color represents a higher probability of finding an initial-state nucleon with a particular removal energy and momentum. The two sharp p-shells at $E_\mathrm{rmv}\sim12~\text{MeV}$ and $E_\mathrm{rmv}\sim18~\text{MeV}$, and the larger diffuse s-shell at $E_\mathrm{rmv}\sim20-65~\text{MeV}$ and $|\textbf{p}|\lesssim
150~\text{MeV/c}$, are visible. The figure is adapted from Ref.~\cite{T2K:2023smv}.}
    \label{fig:sf2DO}
\end{figure}

The SF provides a realistic shell-model-based description of the nuclear ground state and is largely built from exclusive measurements of 1p1h interactions in electron scattering, with additional theory-based contributions to account for initial-state correlations between neighboring nucleons. The two-dimensional SF for oxygen is shown in~\autoref{fig:sf2DO}. In NEUT, the single nucleon component uses the BBBA05~\cite{BRADFORD2006127} description for the vector component of the nucleon form factors, and a simple dipole model for the axial component using a nucleon axial mass value of $M_A^{\textrm{QE}}=1.03~\text{GeV}/c^2$.

The SF can be decomposed into Mean Field (MF) part which describes the nucleus in the shell model picture and is tightly constrained with electron scattering data and $P_\mathrm{corr}(\vec{p}, E_\mathrm{rmv})$ corresponding to the contribution from short-range correlated nucleons (SRC). This can be expressed as:
\begin{equation}
    P(\vec{p}, E_\mathrm{rmv})=P_\mathrm{MF}(\vec{p}, E_\mathrm{rmv})+P_{\text {corr}}(\vec{p}, E_\mathrm{rmv})
    \label{eq:SFMFSRC}
\end{equation}
SRC refers to pairs of correlated nucleons that can have a momentum higher than the Fermi surface momentum, thus contributing mainly to the tail of the nucleon momentum distribution. As such, the spectator nucleon of the correlated pair will likely leave the nucleus as well (as it becomes unbound once its pair is removed)\footnote{It is also important to note that, despite producing multiple nucleons, these interactions are not classified as 2p2h interactions (detailed in Sec.~\ref{sec:2p2h}). However, it is possible that by including SRCs in the SF framework, there is some double counting of multi-nucleon contributions, but that our large uncertainties on the normalizations of these components are expected to more than cover this.}. $P_\mathrm{corr}(\vec{p}, E_\mathrm{rmv})$ is computed theoretically in the local-density approximation~\cite{Benhar:1994hw}, and typically accounts for $\sim 20 \%$ of the total SF according to electron scattering experiments.

In NEUT, the full SF is provided by input tables from calculations by Benhar et al.~\cite{Benhar:1994hw}, in which the MF and SRC components are not formally separated. NEUT therefore defines its own convention to distinguish between the two contributions, using hard cuts on $E_\mathrm{rmv}$ and $|\vec{p}|$: nucleons with $E_\mathrm{rmv} <$~100~MeV and $|\vec{p}| < 300~\text{MeV}/c$ are classified as MF, with the remainder labeled as SRC. Consequently, if a neutrino interacts with a nucleon from the MF region, it will knock out one nucleon only, whereas if it interacts with a nucleon from the SRC region, an additional spectator nucleon is added with the knocked-out nucleon. In NEUT, this spectator is taken to have an opposite isospin and opposite momentum to the initial state nucleon in the rest frame of the nucleus. Using this implementation, NEUT predicts that around 5\% of nucleons are within the SRC region of the spectral function (for both carbon and oxygen). 

The SF model employed by NEUT incorporates a simplistic relativistic Fermi-gas (RFG) inspired approach to Pauli blocking where, as detailed in Ref.~\cite{Benhar:2006nr}, the cross section is set to zero 
for the portion of the phase space in which the pre-FSI outgoing primary nucleon has a lower momentum than some Fermi surface momentum ($p_F$, 209 MeV$/c$ is taken for carbon and oxygen in the latest T2K simulation). This procedure both reduces the cross section predicted by the model and induces significant shape changes at low momentum transfer (corresponding to low-momentum nucleons). 

By default, NEUT uses the same SF for neutrino and antineutrino interactions (i.e., assuming that the nuclear ground state for neutrons and protons is the same). Based on microscopic model predictions, this behavior is altered by shifting the SF used by neutrons (and thus less constrained by electron-scattering data) to have removal energies 2~MeV higher. This procedure is the same as in T2K's previous analysis, and further details on its treatment are available in~\cite{T2K:2023smv}.

\subsubsection{2p2h Interactions}
\label{sec:2p2h}

The two-particle two-hole (2p2h) interaction channel describes a variety of possible neutrino interactions, all of which involve two nucleons in the initial and final states. A final state topology with two nucleons can occur via different mechanisms -- for example, via an interaction of the neutrino with an SRC pair\footnote{The influence of SRCs is partially accounted for using the SF model.}, or with two independent nucleons via a ``meson exchange current'', which is usually mediated through a pion, although $\rho$ mesons are also considered. The implementation of 2p2h interactions in NEUT used for T2K analyses uses the Valencia group's model~\cite{Nieves:2011pp} to predict the cross section as a function of outgoing lepton kinematics. This model predicts a cross-section spread around two regions of energy-momentum transfer space, a higher lobe peaking at about 400 MeV energy transfer and a lower one at about 100 MeV (see Ref.~\cite{T2K:2021xwb} for more details). These broadly correspond to contributions to the cross section from diagrams containing or not containing a Delta baryon, respectively. Non-relativistic approximations in the model limit its applicability to energy transfers below 1.2 GeV. The cross section is approximated to zero above this energy transfer. 

In the NEUT implementation of the Valencia model used in this work, the nucleon pair types with which the neutrino interacts are fixed and constant for all lepton kinematics to match average predictions from the Valencia model ($\sim$70\% of proton-neutron initial state pairs)\footnote{A more advanced model implementation in NEUT allows the nucleon pair fraction to vary as a function of outgoing lepton kinematics, but this is not employed due to its limitation (at the time of the T2K simulation) of only applying to a limited set of nuclear targets, which made it unsuitable for simulation. This limitation has since been remedied, and so this new implementation will be used for future T2K analyses.}.

The Valencia model implemented in NEUT cannot directly predict outgoing nucleon kinematics. To calculate these for generated 2p2h interactions, NEUT uses a simple approximation that the momentum transfer from an interaction is shared equally between two nucleons independently sampled from a local Fermi-gas nuclear model (LFG)~\cite{Sobczyk:2012ms}, which are then individually propagated through NEUT's final state interaction intranuclear cascade (see~\autoref{sec:FSI}).

The Valencia 2p2h model, and by extension NEUT, accounts only for CC 2p2h interactions, whereas NC 2p2h interactions are unavailable. Therefore, in this paper, the terms ``2p2h'' and ``CC 2p2h'' are used interchangeably. Similarly to quasi-elastic interactions, NC 2p2h processes have a negligible impact on the T2K oscillation analysis, although they have been found to be relevant for neutron multiplicity measurements with the FD~\cite{T2K:2025ipr}, indicating the need for further model development.

\subsubsection{Single Pion Production}
\label{sec:SPP}

Single pion production (SPP) in T2K is dominated by resonant production, which in NEUT is described by the Rein–Sehgal model~\cite{Rein:1980wg} in the invariant hadronic mass region $W<2.0~\textrm{GeV}/c^2$. The NEUT model has additional improvements to the nucleon axial form factors~\cite{Graczyk:2014dpa,Graczyk:2007bc}, and includes effects from the final state lepton mass in the calculation~\cite{Berger:2007rq,Graczyk:2007xk,Kuzmin:2003ji}. Whilst $\Delta(1232)$ excitations are the dominant contributors to the cross section at T2K energies, a total of 17 baryon resonances are included in addition to a non-resonant process, following the description by Rein and Sehgal. Interferences between the resonances are included, but not between the resonant and non-resonant components. The form factor parameters are the resonant axial mass, $M_A^{RES}$, and the numerator of the axial form factor, $C_5^{A}(Q^2=0)$. Also, a parameter controls the amplitude of the non-resonant contribution. These parameters were tuned to the data, as detailed in \autoref{sec:SPPuncert}. The model describes neutrino interactions with free nucleons. To extend the calculation to nuclei, NEUT uses a simple relativistic Fermi-gas model for the nucleon's initial state, with no binding energy.

Coherent scattering off nuclei also contributes to the single pion production cross section, especially at low four-momentum transfer. NEUT describes coherent interactions with the Berger–Sehgal model~\cite{Berger:2008xs}.

\subsubsection{Shallow and Deep Inelastic Scattering}
\label{sec:DIS}

The NEUT implementation of inelastic processes beyond single pion production is split into a ``multi-$\pi$'' lower hadronic invariant mass ($1.3 < W < 2.0~\text{GeV}/c^2$) process, and a higher $W > 2.0~\text{GeV}/c^2$ Deep Inelastic Scattering (DIS) process. The cross sections for both sets of processes in NEUT are calculated using the GRV98~\cite{GRV98} Parton Distribution Functions (PDFs), which describe the probability density for finding a quark of a given type, as a function of Bjorken-$x$ -- the fraction of the nucleon momentum carried by the struck quark -- and the inelasticity variable $y$ -- the ratio of the energy transfer and neutrino energy. ``Bodek-Yang''~\cite{Bodek:2003wc, Bodek:2005de} modifications are made to extend the validity of this approach to the relatively low four-momentum transfers ($Q^2\lessapprox1.5~\text{GeV}^2/c^4$) typical of interactions at T2K (Ref.~\cite{Bodek:2005de}). The details of the generation of the hadronic state are where the processes differ. At lower $W$, the custom multi-$\pi$ model~\cite{Hayato:2021heg} is used to simulate interactions that produce more than one pion in the final state; thus avoiding double-counting the single pion channel in the $W$ range where the resonant (and non-resonant) single pion production processes are modeled. For $W>2~\text{GeV}/c^2$ interactions (DIS), PYTHIA 5.72 is used~\cite{SJOSTRAND199474}. It is worth noting that for $W>2~\text{GeV}/c^2$, PYTHIA may simulate single pion events, and these will be kept as there is no chance of double counting with single pion production at DIS kinematics. A simple relativistic Fermi-gas model without binding energy included is used to describe the initial state. 

\subsubsection{Final State Interactions}
\label{sec:FSI}

NEUT describes the re-interaction of outgoing hadrons with the nuclear medium, or FSI, using semi-classical intranuclear-cascade models in which hadrons are individually stepped through the nucleus, and a series of interactions is sampled based on hadron-nucleus cross-section measurements. Pion FSI are described based on the Salcedo and Oset model~\cite{Salcedo:1987md,Oset:1987re}, tuned to modern $\pi-A$ scattering data~\cite{PinzonGuerra:2018rju}. Nucleon FSI is described in an analogous cascade model~\cite{Hayato:2009zz} based on the model by Bertini~\cite{bertiniCasc}. More details of FSI modeling in NEUT and other generators can be found in Ref.~\cite{Dytman:2021ohr}.

\subsection{Uncertainty parameterization}
\label{subsec:xsecuncert}

The following section describes the parameterization of the systematic uncertainties applied to the baseline model. The purpose of this parameterization is to provide coverage of plausible variation of neutrino interaction cross sections and corresponding robustness for the near-detector analysis. Both nuclear model are nucleon-level uncertainties are also accounted for. A full list of all parameters is available in ~\autoref{tab:xsecparameters}. Throughout the text, the names given to each parameter (used in later sections) that are not contained within a single symbol are provided in \texttt{teletype font}. 

\subsubsection{CCQE Uncertainties}
\label{sec:1p1huncert}

The SF ground-state model used by T2K includes natural degrees of freedom that form the basis for the uncertainty parameterization employed. We consider uncertainties associated with:
\begin{itemize}
    \item The relative strength of the MF and SRC part of the SF (see \autoref{eq:SFMFSRC}),
    \item The strength of each nuclear shell (i.e. alterations to the normalization of components of the y-axis of \autoref{fig:sf2DO} corresponding to the different shells),
    \item The shape of the momentum distribution within each shell (i.e., altering the distribution of the x-axis for values of the y-axis corresponding to each shell in \autoref{fig:sf2DO}),
    \item Shifts to shell positions in $E_\mathrm{rmv}$ (i.e. shifts of \autoref{fig:sf2DO} along its y-axis).
\end{itemize}

The following details how each of these uncertainties is considered: \\
\paragraph{Short Range Correlations:} The SF model is also implemented in the NuWro Monte Carlo (MC) event generator~\cite{Golan2012nuwro}, which uses a different approach to identify nucleons belonging to SRC pairs. NuWro, unlike NEUT, applies non-rectangular cuts in $(E_\mathrm{rmv}, |\vec{p}|)$ covering a wider region of phase-space than NEUT, defines separate SRC cuts for each target nucleus, and requires the second nucleon to have sufficient energy (${>}14$~MeV) to escape the nuclear potential. As a result of these different kinematic selections, NuWro predicts that approximately 16\% of nucleons in the SF model belong to an SRC pair\footnote{Both NEUT and NuWro differ in the predictions for the fraction of nucleons in SRC pairs from the experimentally established value of ${\sim}20\%$~\cite{Hen:2016kwk}.}.

The discrepancy between NEUT and NuWro does not reflect genuine uncertainty in the SRC fraction itself, but rather differences in how the energy required to remove both nucleons of a correlated pair is modeled. This removal energy depends on collective nuclear effects and carries significant theoretical uncertainty --- the same correlated pair may be kinematically accessible in one generator but not in the other, depending on these assumptions. Since the SRC pair fraction provides a natural degree of freedom to encode this uncertainty, we introduce an effective normalization parameter on the SRC fraction in NEUT, with its uncertainty tuned to cover the difference between the NEUT and NuWro predictions. The uncertainty is set to cover a 15\% contribution, \emph{i.e.} 200\% of the NEUT nominal. Since nuclear effects are expected to differ between carbon and oxygen, separate parameters are introduced to allow independent freedom for each nucleus: \texttt{SRC norm $^{12}$C} and \texttt{SRC norm $^{16}$O} respectively.

It should be noted that coverage of the NuWro SRC normalization by this uncertainty does not imply that the full set of NEUT--NuWro differences is accounted for. Significant shape differences remain between the two generators in the nucleon momentum and energy distributions, which a simple SRC normalization scaling cannot address. Nevertheless, this uncertainty, acting in conjunction with the other parameters described below, provides additional freedom to accommodate differences between the two models. \\

\paragraph{Shell occupancy:}
Assuming isotropic Fermi motion of the nucleon within the nucleus, the missing energy distribution corresponds to 

\begin{equation}
P(E_\mathrm{rmv}) = \int d^3 \vec{p} P(\vec{p}, E_\mathrm{rmv}) =  4 \pi \int dp p^2 P(p, E_\mathrm{rmv}).
\end{equation}

As shown in Fig.~\ref{fig:sf2DO}, the missing energy distributions exhibit multiple peaks that correspond to shells to which nucleons can belong.

Following the prescription outlined in Ref.~\cite{Chakrani:2023htw}, we introduce parameters to control the occupancy of each shell. These occupancy dials only impact events from the MF region ($p_m <300~\text{MeV}/c$,  $E_\mathrm{rmv} <100~\text{MeV}$). 

To alter the occupancy of a given shell, the weights for the MF CCQE events are given by the following function:
\begin{align}
    f_{\text{shell}} (E_\mathrm{rmv}) &= 1 + N_{\text{shell}} \times \exp \left( - \frac{(E_\mathrm{rmv} - E_{\text{shell}})^2}{2 \sigma_{\text{shell}}^2} \right)\\
    &= 1 + N_{\text{shell}} \times g_{\text{shell}}(E_\mathrm{rmv})
\end{align}
where $N_{\text{shell}}$ is an occupancy parameter of a given shell and $E_{\text{shell}}$ and $\sigma_{\text{shell}}$ correspond to the center and the width of the Gaussian function, which are fixed for each shell. In total, this gives two ``shell normalization parameters'' for carbon and three for oxygen: \texttt{p-shell norm $^{12}$C}, \texttt{s-shell norm $^{12}$C}, \texttt{p$_{1/2}$-shell norm $^{16}$O}, \texttt{p$_{3/2}$-shell norm $^{16}$O}, \texttt{s-shell norm $^{16}$O}. The fixed values of $E_{\text{shell}}$ and $\sigma_{\text{shell}}$, derived from an analysis of NEUT $E_\mathrm{rmv}$ distributions, are shown for each shell in \autoref{tab:num}. The shell occupancy parameters alter the $\vec{p}$ distribution as well as the width of the shells since different shells contain different initial nucleon momentum distributions.

Whilst the shell normalizations are, in principle, constrained by external measurements (notably from electron-scattering data), in this analysis, the prior uncertainties for the shell normalization dials are left relatively free in the fit. This approach leaves the near-detector fit free to modify the shape of the input spectral function and uses these parameters and their uncertainties as effective degrees of freedom to account for some physics beyond PWIA. In order to keep consistent prior uncertainties across all shell normalization dials, they are chosen such that a one sigma variation in each of the dials causes a $\sim$10\% variation of the total CCQE cross section~\footnote{Within this work, the $p_{1/2}$-shell normalization for oxygen uses a prior uncertainty of 0.20 instead of 0.25. Though this deviates slightly from the original motivation for relative uncertainty sizing, the effect on the analysis remains minimal.}. The calculated prior uncertainties on the shell normalization dials are shown in \autoref{tab:num}. There are no prior correlations between the shell normalization dials. \\
\begin{table*}[ht]
    \centering
    \begin{tabular}{c|c|c|c|c}
        \hline\hline
        Target & Shell & Center (MeV) & Width (MeV) & Prior Norm. Uncert. \\
        \hline 
        \multirow{2}{*}{Carbon} & $p$-shell & 18 & 15 & 0.20\\ 
        \cline{2-5}
         & $s$-shell & 36 & 25 & 0.45\\
        \hline \hline
        \multirow{3}{*}{Oxygen} & $p_{1/2}$-shell & 12 & 8 & 0.20 \\ 
        \cline{2-5}
         & $p_{3/2}$-shell & 19 & 8 & 0.45\\
        \cline{2-5}
         & $s$-shell & 42 & 25 & 0.75\\
        \hline\hline
    \end{tabular}
    \caption{Energy levels of the different shells and the widths in NEUT in addition to the prior uncertainty set on the corresponding shell normalization dial (the central value of the dials is all 0).}
    \label{tab:num}
\end{table*}

\paragraph{Shell shape:}

Also following the prescription outlined in Ref.~\cite{Chakrani:2023htw}, we introduce parameters to control the shape of the $|\vec{p}|$ distribution of each shell. The SF in NEUT is compared with missing-momentum distributions from electron scattering in Ref.~\cite{JLabE91013:2003gdp}, available at different kinematics. By choosing two ``extreme'' distributions of the measured missing momentum in each shell (i.e. those that differ the most), parameters are defined to allow the shape of the missing momentum distribution to change: \texttt{p-shell shape $^{12}$C}, \texttt{s-shell shape $^{12}$C}, \texttt{p$_{1/2}$-shell shape $^{16}$O}, \texttt{p$_{3/2}$-shell shape $^{16}$O}, \texttt{s-shell shape $^{16}$O}. Each parameter acts as follows:

\begin{itemize}
    \item When it is equal to $1$, each $|\vec{p}|$ bin of NEUT is altered to match one of the measurements.
    \item When it is equal to $-1$, each $|\vec{p}|$ bin of NEUT is altered to match the other measurement.
    \item When it is equal to zero, the distribution is not altered, and it corresponds to the nominal NEUT.
    \item In each $|\vec{p}|$ bin, we interpolate the values of the weights using a second-order polynomial to obtain a smooth variation of the overall missing momentum distribution within the electron scattering boundaries. 
\end{itemize}

Whilst the parameter response is only well defined by data 
between -1 and 1, the parameter implementation extrapolates linearly beyond this. It should also be noted that the parameters are built for the two carbon shells (due to readily available data) and that the relative variations from these are then applied to the oxygen shells. \\

\paragraph{Mean removal energy shift:}

Adopting the same approach as in T2K's previous analyses~\cite{T2K:2023smv}, a global uncertainty of 6 MeV is applied on the mean removal energy. This constitutes an uncertainty envelope on the mean removal energy, whilst allowing the differences between carbon and oxygen removal energies to vary by 3 MeV, and that for neutrinos and antineutrinos by 4 MeV. These uncertainties are defined based on nuclear theory calculations and differences in measurements of shell positions from different electron-scattering experiments, as described in~\cite{Bodek:2018lmc}. They act via four parameters: \texttt{$\Delta E_\mathrm{rmv}$ $^{12}$C $\nu$}, \texttt{$\Delta E_\mathrm{rmv}$ $^{12}$C $\bar{\nu}$}, \texttt{$\Delta E_\mathrm{rmv}$ $^{16}$O $\nu$}, \texttt{$\Delta E_\mathrm{rmv}$ $^{16}$O $\bar{\nu}$}. It is worth noting that these effective parameters shift the SF distribution only along the $E_\mathrm{rmv}$ axis, therefore leaving the hadron kinematics unchanged. They do alter the predictions on the reconstructed neutrino energy.
Beyond the global oxygen-carbon difference and proton-neutron SF uncertainties motivated by the degrees of freedom within the SF model, we consider additional uncertainties associated with the PWIA, arising from the SF model's use of more \textit{ad hoc} approaches. \\

\paragraph{Pauli blocking:}
The SF model employed by NEUT uses a simplistic approach in which the cross section is set to zero for the portion of the phase space where the pre-FSI outgoing primary nucleon momentum has a lower value than the Fermi surface momentum ($p_F$, taken to be 209~MeV$/c$ for both carbon and oxygen in the latest T2K simulation). It both reduces the cross section predicted by the model and causes significant shape changes at low momentum transfer (which generally corresponds to low momentum nucleons).

Whilst the NEUT SF model uses a default $p_F$ of 209 MeV$/c$, different values are used elsewhere, varying depending on which model is fit to what data. For example, Ref.~\cite{Maieron:2001it} reports a $p_F$ for carbon 19 MeV$/c$ larger. Due to the fairly wide plausible range of $p_F$ values and the simplistic model of Pauli blocking used, a $\pm$30 MeV$/c$ uncertainty is adopted to achieve a primarily ND280 data-driven constraint. Separate parameters for varying $p_F$ are added for carbon and oxygen as well as for neutrino and antineutrino interactions, giving four parameters in total: \texttt{Pauli blocking $^{12}$C $\nu$}, \texttt{Pauli blocking $^{12}$C $\bar{\nu}$}, \texttt{Pauli blocking $^{16}$O $\nu$}, \texttt{Pauli blocking $^{16}$O $\bar{\nu}$}. The carbon and oxygen parameters are left uncorrelated, as there is no external data that would allow robust estimation of the correlations.

The neutrino and antineutrino parameters are given a $\pm9.5~\text{MeV}/c$ uncorrelated part of the uncertainty (in addition to the $\pm30~\text{MeV}/c$ considered fully correlated) corresponding to an 80\% correlation. This stems from the expected possible differences between the nuclear potentials felt by neutrons and protons and is calculated from a naive translation of the $\pm$2 MeV derived for neutron and proton removal energy differences. The nominal 209 MeV$/c$ Fermi surface momentum is reached for a nucleon with 23 MeV kinetic energy, making this a crude estimate for the binding energy in a Fermi-gas model with $p_F=209~\text{MeV}/c$. Changing this 23 MeV by $\pm 2~\text{MeV}$ translates back to a $\pm9.5~\text{MeV}/c$ uncertainty on $p_F$ (taking the larger of the two variations). \\

\paragraph{Optical potential correction:}

Since the SF model is built on top of the PWIA, FSIs (which cause the outgoing nucleon wave function to be a distorted wave and not a plane wave) are not included in NEUT's simulation of the total cross section\footnote{This deficiency is common to all models implemented in the version of NEUT used for this work (Ref.~\cite{McKean:2025khb} remedies this in NEUT's latest versions).}. The impact of FSI on outgoing hadron kinematics is modeled via NEUT's intranuclear cascade model (see \autoref{sec:FSI}). This approximates the distortion of the outgoing nucleon kinematics from FSI and further simulates extra hadron ejection, but it does not allow the total CCQE cross section to vary as a full treatment of the distortion of the outgoing nucleon wavefunction would. To account for this missing alteration to the cross section, an approach has been developed in Ref.~\cite{Ankowski:2014yfa} to modify the SF prediction accordingly. It should be noted that the impact of this alteration is calculated only for the outgoing lepton kinematics, and so the adjustment of nucleon kinematics via the FSI cascade is still required.

This approach introduces a ``folding function'' which is dependent on the energy transfer ($q_0$) into the SF model cross-section calculation, which allows the real part of the optical nuclear potential to be effectively included. 

The role of this real part can be interpreted as altering the inclusive cross section, while the imaginary part, which is not considered in this scheme, is expected to account for the role of inelastic FSI, which changes the hadronic final state (as is accounted for by an FSI cascade)~\cite{Nikolakopoulos:2022qkq,Nikolakopoulos:2023pdw}.

\begin{figure}[htbp]
    \centering
    \includegraphics[width=0.48\textwidth,page=3]{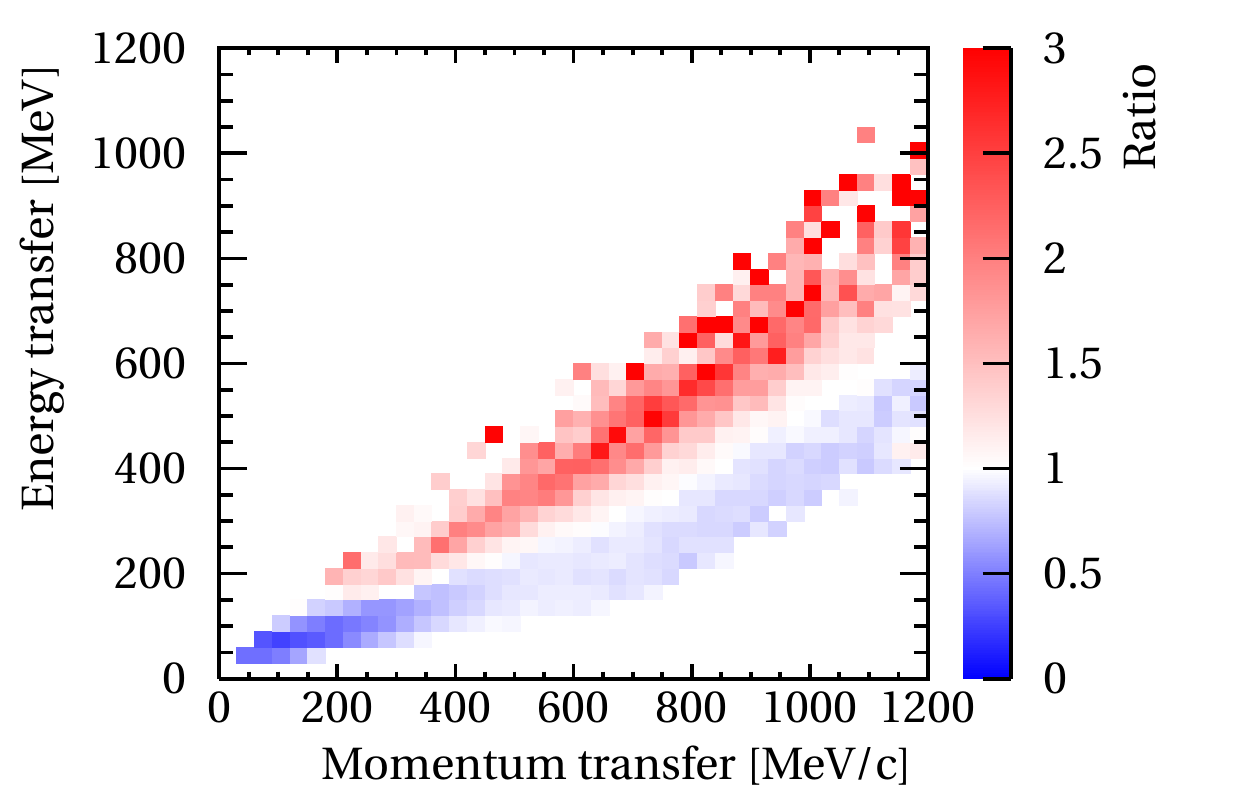}
    \caption{Ratio (with optical potential / without optical potential) of the double-differential T2K-flux-averaged cross section as a function of energy and momentum transfer for CCQE interactions on a carbon target with and without the optical potential correction, as predicted by the NuWro event generator.}
    \label{fig:q0q3_OP}
\end{figure}

The impact of this correction to the SF cross section is incorporated as an ``optical potential'' parameter. To calculate the impact of applying the correction, the NuWro event generator~\cite{Golan2012nuwro} (version 19.02.01) is used, as it has an option to include this effect. The impact of the optical potential correction on the T2K flux integrated cross section as a function of energy and momentum transfer is shown in \autoref{fig:q0q3_OP}, demonstrating the most substantial impact at low energy transfers. A simple linear interpolation between turning this correction on and off is used to build a continuous parameter. A calculation for this effect is only available for carbon. To include uncertainties for both targets, the carbon calculation is used to apply a correction for both, but separate carbon and oxygen parameters are introduced and left uncorrelated: \texttt{Optical Potential $^{12}$C}, \texttt{Optical Potential $^{16}$O}. We apply a flat (uniform) prior to these parameters. \\

\paragraph{Momentum transfer dependent $E_{rmv}$:}
The SF calculation of the differential cross section, built under PWIA, factorizes the nuclear ground state and the lepton-nucleon interaction. Therefore, by definition, the SF does not depend on the observable kinematics of the incoming or outgoing lepton. Models based on more sophisticated calculations, such as the relativistic mean field (RMF) model~\cite{Caballero:2006wi,Caballero:2005sj,Caballero:2007tz,Meucci:2009}, predict a strong momentum-transfer (\textit{q}) dependence of the CCQE peak position as a function of energy transfer. Within the SF model, this behavior can be incorporated by shifting the removal energy, which takes approximately a linear form and is supported by electron-scattering data~\cite{Bodek:2018lmc}. By including a \textit{q}-dependence of the removal energy in our model, we can take an explicit step beyond the limitations of the factorization approach. It should be noted that this aims to remedy similar limitations in the SF model as the application of the optical potential correction, and so the application of both may include some double-counting of updates to the SF model beyond PWIA.  

The exact correction is derived by comparing SF predictions to inclusive electron-scattering data and by studying the shift of the aforementioned CCQE peak from its expectation. This study involved implementing electron-scattering event generation in NEUT. This implementation and the derivation of the associated \textit{q}-dependent correction are detailed in Refs.~\cite{Dolan:2023iik, Abe:2024avs}. Similar to the optical potential case, a parameter is then introduced to linearly interpolate between the nominal and corrected model with no prior uncertainty: \texttt{$\alpha$ correction}. The same parameter affects neutrino and antineutrino interactions on carbon and oxygen nuclear targets.

\paragraph{Nucleon axial mass:}
As described in Ref.~\cite{T2K:2023smv}, the nucleon axial mass, \texttt{$M_A^{QE}$}, in the dipole parameterization of the axial form factor that is employed, is tuned to neutrino-deuterium scattering data~\cite{Miller:1982qi,Barish:1977qk,Baker:1981su,Allasia:1990uy,Kitagaki:1983px} using NUISANCE~\cite{Stowell:2016jfr}. Following the tune, the central value and its uncertainty are adjusted and inflated to ensure coverage of previous global fit results. The final central value and uncertainty used is $M_A^{QE}=1.03\pm0.06~\text{GeV}/c^2$. \\

\paragraph{High $Q^2$ modifications:}

Uncertainties on the higher $Q^2>0.25~\text{GeV}^2$ predictions of the neutrino-nucleon component of the CCQE model are driven by the axial component of the neutrino-nucleon interaction,
where the dipole model may be inadequate~\cite{Bhattacharya:2011ah, Meyer:2022mix}. To account for this, three \texttt{High Q$^2$ norm} parameters corresponding to $0.25 < Q^2 (\text{GeV}^2/c^4)< 0.5$, $0.5 < Q^2 (\text{GeV}^2/c^4)< 1$, and $Q^2(\text{GeV}^2/c^4) > 1$ are added to lessen the use of $M_A^{QE}$ as an effective parameter to correct for deviations from the dipole model. The $Q^2$ ranges and uncertainties of the new high $Q^2$ parameters are based on comparisons of the $Q^2$ shape of the dipole and z-expansion models~\cite{Bhattacharya:2011ah}.

\subsubsection{2p2h Uncertainties}
\label{sec:2p2huncert}

The uncertainties related to 2p2h interactions are similar to
those in T2K’s previous oscillation analysis~\cite{T2K:2023smv}. Parameters that alter the 2p2h normalization independently for neutrinos and antineutrinos, and for carbon and oxygen interactions, are used. The 2p2h normalizations, \texttt{2p2h norm $\nu$}, \texttt{2p2h norm $\bar{\nu}$}, are assigned flat (uniform) priors, and the carbon-oxygen scaling parameter, \texttt{2p2h norm $^{12}$C to $^{16}$O}, has a 20\% prior uncertainty. 

Uncertainties reflecting the shape of the energy dependence of 2p2h using three different plausible models of the process are used as well. The uncertainties span the maximal difference in 2p2h predictions from Martini et al.~\cite{Martini:2009uj}, Nieves et al.~\cite{Nieves:2011yp}, and SuSAv2~\cite{Megias:2016lke, RuizSimo:2016rtu} and are split into regions altering neutrino energies larger and smaller than 0.6 GeV. The parameters controlling the energy dependence are also split between neutrino and antineutrino, allowing freedom to cover differing energy-dependence behavior in different models~\cite{Dolan:2019bxf}. Overall the four parameters are: \texttt{2p2h Edep low E$_\nu$}, \texttt{2p2h Edep high E$_\nu$}, \texttt{2p2h Edep low E$_{\bar{\nu}}$}, \texttt{2p2h Edep high E$_{\bar{\nu}}$}.

The ratio of proton-neutron ($pn$) to neutron-neutron ($nn$) pairs for neutrinos, and correspondingly proton-neutron ($pn$) to proton-proton ($pp$) pairs for antineutrinos, in 2p2h interactions is not well known\footnote{Here $pn$, $nn$, and $pp$ refer to nucleon pairs in the nucleus prior to the 2p2h interaction.}. The fraction of pairs is model-dependent. In NEUT, $pn$ pairs account for $\sim70\%$ of the cross section, while within an alternative microscopic 2p2h model (the SuSAv2 2p2h implementation in GENIE~\cite{Dolan:2019bxf}), such a value is $\sim80\%$. SuSAv2 has a larger fraction of $pn$ pairs due to the inclusion of ``exchange interference terms'' in the calculation of 2p2h cross section, as is detailed in Ref.~\cite{RuizSimo:2016ikw}, which are neglected in the Valencia model. A new parameter, allowing a change of the ratio of $pn$ to $nn$ pairs, has been included in T2K's analyses: \texttt{PNNN shape}. The parameter smoothly varies the relative contribution of 2p2h pair contributions, and its uncertainty covers the difference between the Valencia and SuSA models. 

A previously used shape freedom shifting in the $\Delta$ and non-$\Delta$ contributions in the energy and momentum transfer to the nucleus, ($q_0$,$|q|$), of the Valencia model remains, but it has been divided to affect separately $pn$ to $nn$ for neutrinos (and correspondingly $pn$ to $pp$ pairs for antineutrinos) on top of the existing division on carbon and oxygen interactions. This gives in total 4 parameters controlling the shape of the 2p2h cross section in energy and momentum transfer for $np$ (neutron-proton) and $NN$ (proton-proton or neutron-neutron) pairs: \texttt{2p2h shape $^{12}C$ $NN$}, \texttt{2p2h shape $^{12}C$ $np$}, \texttt{2p2h shape $^{16}O$ $NN$}, \texttt{2p2h shape $^{16}$O $np$}.

\subsubsection{Single Pion Production Uncertainties}
\label{sec:SPPuncert}

The single pion production model remains the same as in the previous analysis~\cite{T2K:2023smv}, but with improved uncertainties to reflect the new $\nu_\mu$CC$1\pi^0$ sample at  ND280 and the $\nu_\mu$CC$1\pi^+$-like sample at FD.

First, bubble chamber data from ANL and BNL~\cite{ANL_CC1pi,ANL_NC1pi,BNL_CC1pi,BNL_CC1pi_isospin,ANL_BNL_corr} were used to extract $M_A^\textrm{RES}$, $C_5^A(Q^2=0)$, and the non-resonant $I_{1/2}$ scaling parameter, using NUISANCE~\cite{Stowell:2016jfr}. Minor updates to NEUT's single-pion production model yielded slightly different differential cross sections, necessitating this new tune.
As before, the $\sigma(E_\nu)$ and the distribution of the number of events in $Q^2$, or the shape of $d\sigma/dQ^2$ when available, were utilized. Investigations into the impact of adding other shape distributions, such as invariant masses and Adler angles, were conducted, but they had limited constraining power. ANL and BNL data with hadronic mass cuts of $W<1.4~\textrm{GeV}/c^{2}$ and $W<2.0~\textrm{GeV}/c^{2}$ were selected, tuning to the exclusive CC$1\pi^+1p$, CC$1\pi^+1n$ and CC$1\pi^0$ channels.
MiniBooNE data on the CC$1\pi^+$ cross section on CH~\cite{MiniBooNE:2010eis} was also studied, due to similarities in flux and interaction target to T2K. Data sets were fitted separately and then together, producing a model with central values and $1\sigma$ uncertainties that cover all datasets. The model updates led to a more consistent set of parameters across datasets, reducing the need for excessive inflation of the uncertainties. Furthermore, the new parameter central values were within the $1\sigma$ uncertainties of the previous analysis. In summary, the analysis of bubble chamber data and subsequent uncertainty inflation found $M_A^\textrm{RES}=0.91\pm0.10~\textrm{GeV}/c^2$, $C_5^{A}(Q^2=0)=1.06\pm0.10$ and $I_{1/2}=1.21\pm0.27$, with a correlation of $-0.11$ between $C_5^A(Q^2=0)$ and $M_A^\textrm{RES}$.

As in the previous analysis, a separate non-resonant $I_{1/2}$ background parameter is applied for antineutrinos, which produce pions below 200 MeV/$c$. This adds freedom to the single-pion backgrounds entering the single-ring antineutrino selections since negative pions below the Cherenkov threshold are challenging to detect at the far detector, and little external data exists. The parameter is not constrained by the ND280 analysis and has a 100\% uncertainty. 

An uncertainty related to the energy and momentum sharing of the pion-nucleon system in resonance decays was developed~\cite{Rein:1980wg,Schreiner:1973mj,Ravndal:1973xx,Rein:1987cb}. The default model in NEUT considers the non-uniform dynamics of the $\Delta(1232)$-decay according to the Rein--Sehgal model, and uses a uniform distribution (i.e., a flat distribution in the Adler angles) for all other cases. Three additional options were implemented: one flat in the Adler angles for all resonances, one treating all resonance decays like a $\Delta(1232)$ decay, and one including multiple interfering resonances. For the multiple interfering resonances, the contributions are different for the mixed isospin ($\textrm{CC}1\pi^{+}1n$, $\textrm{CC}1\pi^01p$) and $I_{3/2}$ ($\textrm{CC}1\pi^+1p$) channels. For the mixed isospin channel, the $P_{33}(1232)$, $P_{11}(1450)$, $D_{13}(1525)$, and $S_{11}(1540)$ resonances were considered; for the $I_{3/2}$ channel, the $P_{33}(1232)$ and $P_{33}(1640)$ resonances were considered\footnote{This nomenclature follows the convention in Rein--Sehgal, where $P_{33}(1232)$ is the $\Delta(1232)$.}. 

\begin{figure*}[htpb]
 \centering 
\subfloat{
  \includegraphics[page=6, trim={0mm 0mm 15mm 8mm}, clip, width=0.3\textwidth]{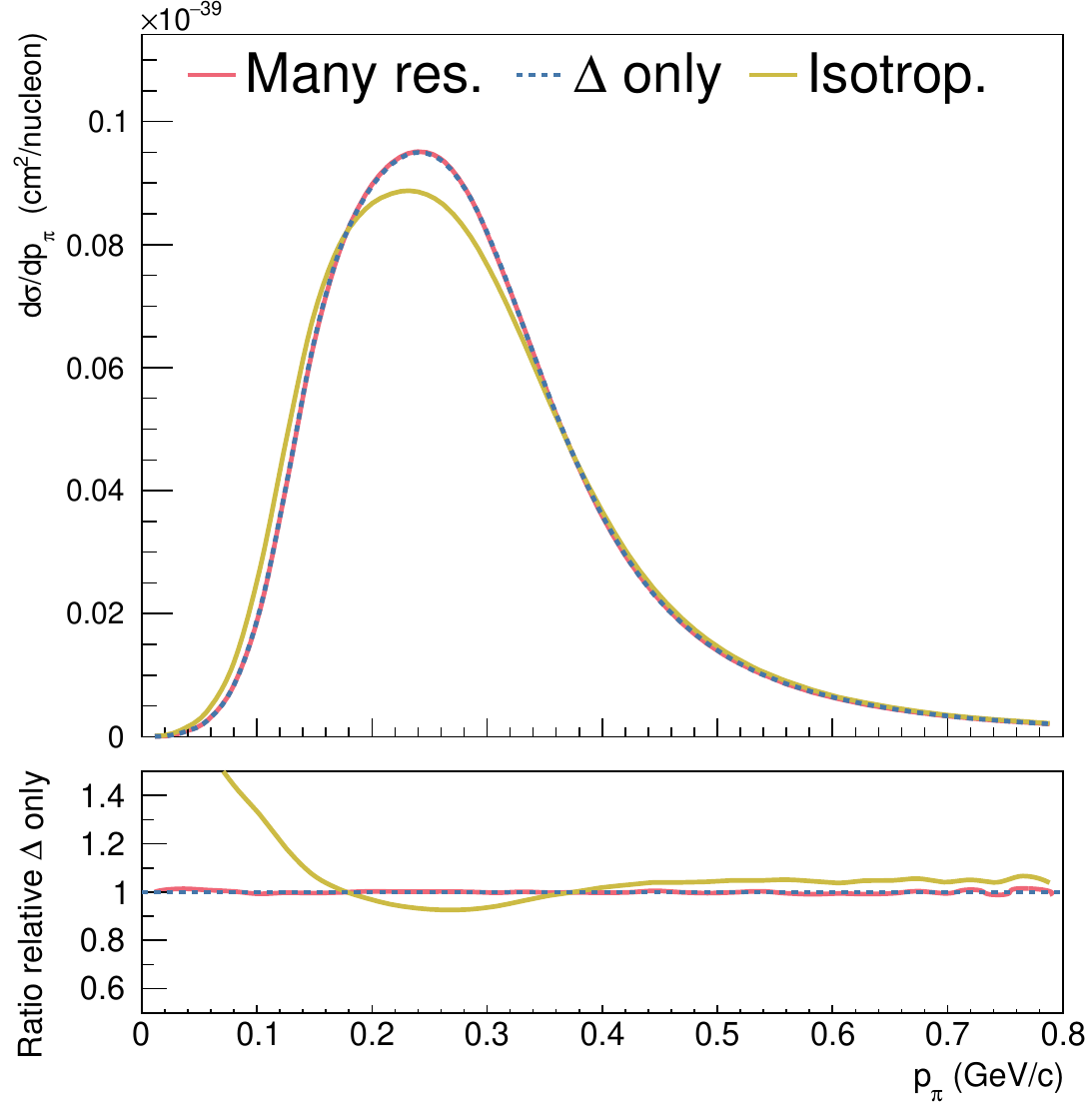}
}
\subfloat{
  \includegraphics[page=16, trim={0mm 0mm 15mm 8mm}, clip, width=0.3\textwidth]{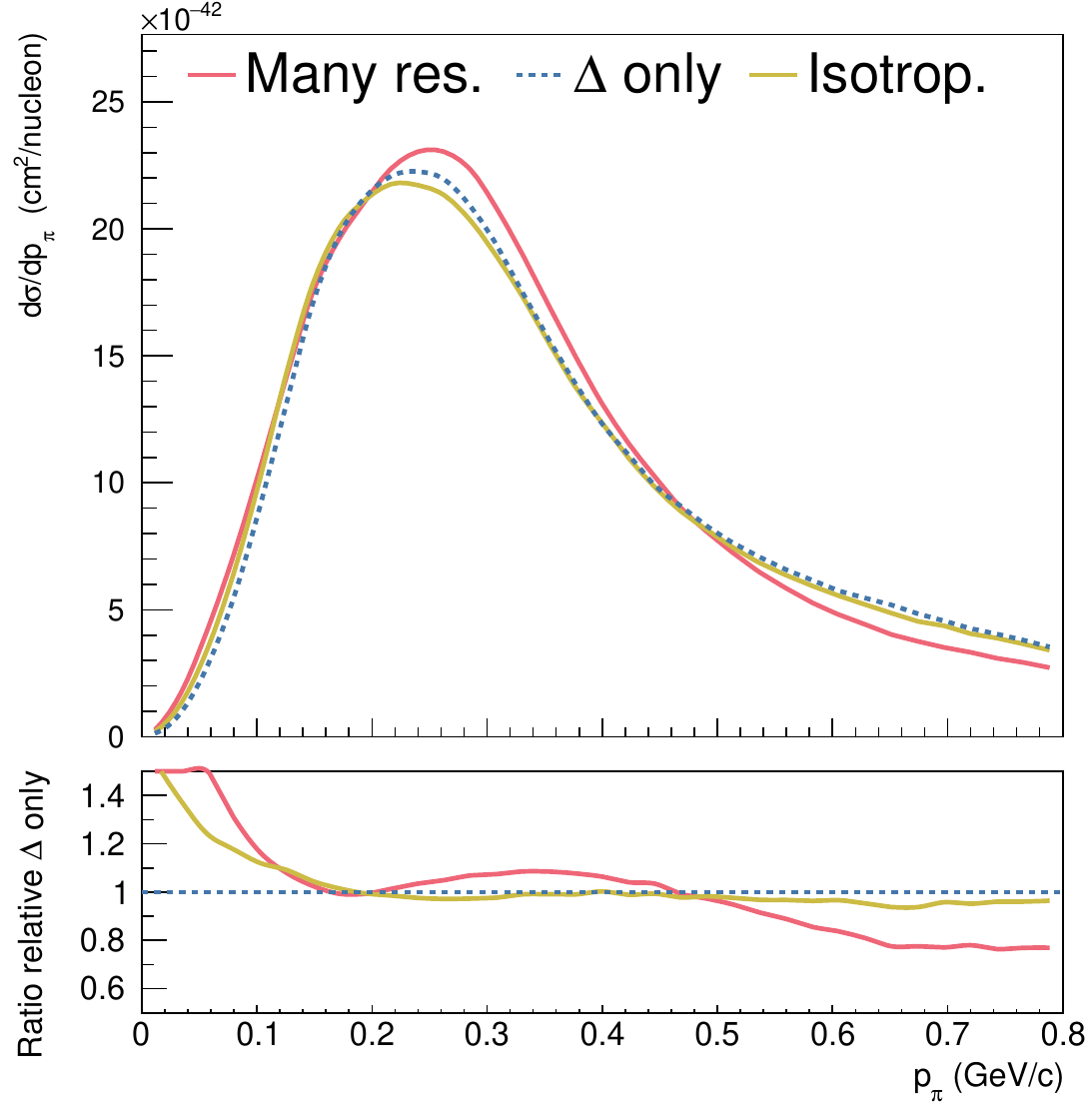}
}
\subfloat{
  \includegraphics[page=19, trim={0mm 0mm 15mm 8mm}, clip, width=0.3\textwidth]{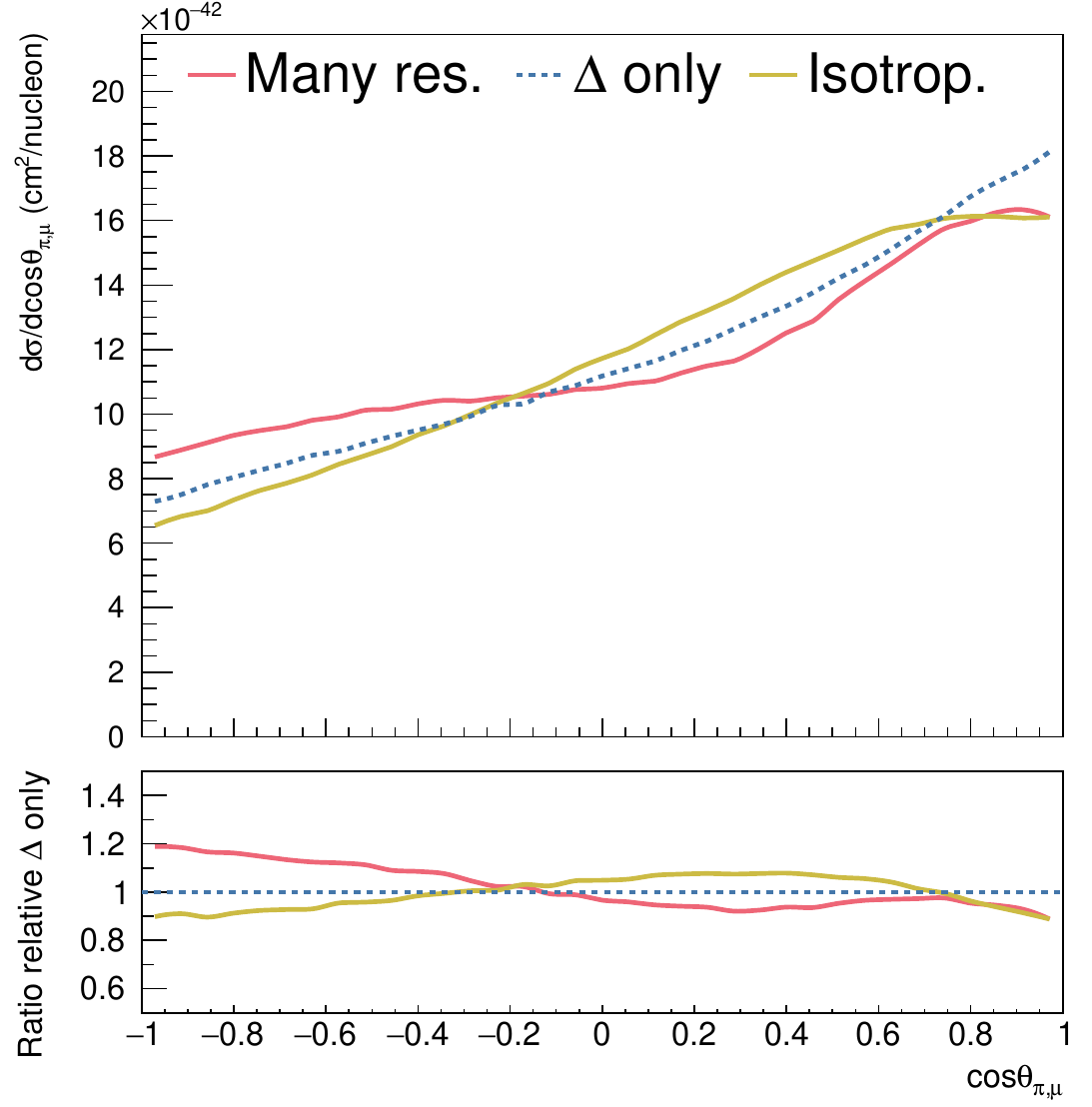}
}
\caption{Distributions of pion momentum for $\nu_\mu\textrm{CC}1\pi^{+}1p$ (left), pion momentum for $\nu_\mu\textrm{CC}1\pi^+1n$ (center), and cosine of the angle between the outgoing pion and muon for $\nu_\mu\textrm{CC}1\pi^+1n$ (right), using $N^{*}\rightarrow N\pi$ decay kinematics ($N^*$ being a resonance) based on calculations using many resonances, $\Delta$ only, and isotropic, for interactions on a nucleon target for the T2K $\nu_\mu$ in $\nu$-mode flux.}
\label{Fig::piej}
\end{figure*}

The main effect of the resonance decay treatments is to redistribute the cross section at high $p_\pi$ to low $p_\pi$ by up to 40\%, below the tracking threshold around $0.2~\textrm{GeV/c}$, as shown in \autoref{Fig::piej}. The uncertainty impacts the kinematics of both the pion and nucleon. It affects all the resonant interactions producing a single pion and for all flavors, and is not the same across channels, neutrino type, or flavor, and is also different between charged- and neutral-current interactions.  

As expected, the effects from the $\Delta$-only and many-resonance calculations are very similar for the pure $I_{3/2}$ channel due to the $\Delta$-dominance, which is less true in the mixed isospin case, where larger differences are observed. However, due to T2K's neutrino energy, the $\Delta(1232)$ is still dominant in the mixed isospin channels.

These modifications to the decay kinematics of the resonances are invariant in lepton, initial state, and interaction variables, like $E_\nu$, $E_\text{lep}$, $Q^2$, $W$, $p_N$, and only affect the \textit{shape} of the outgoing nucleon-pion kinematics. In other words, they do not change the differential cross section in these variables; they only change how the resonance decays into the pion and nucleon, and so directly affects the \textit{shape} of the differential cross section in pion and nucleon kinematics.
Data from ANL and BNL on hydrogen and deuterium in the $\nu_\mu\textrm{CC}1\pi^+1p$ channel indicate a non-flat, $\Delta$-dominated distribution in the Adler angles, but no data exists for the mixed isospin channels. Scant nuclear target data exists for the Adler angles~\cite{T2K:2016cbz}. Comparisons to neutrino-induced single-pion production data from T2K~\cite{T2K:2016cbz}, MiniBooNE~\cite{MiniBooNE:2010eis}, and MINER$\nu$A~\cite{MIN_CC1pip,MIN_pion_2016,MIN_CC1pim_nubar} showed only minor effects since the detectors were not sensitive to the lowest momentum pions, where the effect is largest. As a result, the uncertainty is left unconstrained, since it consistently was the most significant difference, due to the dominance of the $\nu_\mu\textrm{CC}1\pi^+1p$ channel and the beam energy of the T2K experiment. It is defined as a linear interpolation between the $\Delta$-only and isotropic distributions, and the associated parameter is referred to as \texttt{RS $\Delta$ decay}.

An alternative treatment, scaling the strengths of the different matrix elements contributing to the $\Delta$-only amplitude~\cite{Ravndal:1973xx,Feynman:1971wr}, was also developed. 
The calculation was used as a ``simulated data study'' to evaluate potential biases arising from model selection, as discussed in \autoref{sec:FDS}.

The impact of various nuclear effects on single-pion production was investigated, including Pauli blocking, Fermi momentum, initial-state models, and removal energy. The removal energy was found to be the most significant effect, and the others were small compared to existing uncertainties. To account for this, the removal energy was varied in the NuWro generator~\cite{Golan2012nuwro}, and the variations were applied in four dimensions: the neutrino energy $E_\nu$, the four-momentum transfer squared, $Q^2$, the absolute value of the three-momentum transfer, $|q_3|$, and the invariant hadronic mass, $W$. They are split into carbon and oxygen parameters for neutrino and antineutrino single-pion production interactions and are referred to as \texttt{RES $E_\mathrm{rmv}$}. The effect causes a relatively flat suppression of the cross section throughout $E_\nu$. In $Q^2$, it suppresses the cross section uniformly above $Q^2=0.2~\text{GeV}^2/c^4$, with a decreasing impact at lower $Q^2$. This is also reflected in the lepton kinematics, where the effect is the least pronounced in the most forward region, and becomes sizeable below $\cos\theta_\mu<0.6$.

Finally, the new $\nu_\mu$CC$1\pi^0$ sample at the near detector prompted
a survey of MINERvA~\cite{MIN_CC1pi0, MIN_CC1pi0_2, MIN_CC1pip, MIN_pion_2016, MIN_CC1pim_nubar} and MiniBooNE~\cite{MiniBooNE:2010eis, MiniBooNE:2010cxl} $\numuany$CC$1\pi^0$ and $\numuany$CC$1\pi^\pm$ data on carbon-based targets. NEUT and other generators generally do not consistently describe data on both topologies simultaneously: CC$1\pi^+$ data is generally overestimated, and CC$1\pi^0$ data is generally underestimated. Since the majority of constraints on the resonant model come from CC$1\pi^+$ interactions in the near detector, due to the larger interaction cross section, two new uncorrelated normalization parameters for CC$1\pi^0$ $\nu_\mu$ and \numb interactions were introduced; \texttt{RES $\pi^0$ norm $\nu_\mu$} and \texttt{RES $\pi^0$ norm \numb}. The uncertainty was conservatively set to 30\%, which covers all available external data in all kinematic variables. This parameter thus allows for decoupling of charged- and neutral-pion interactions via resonances.

\subsubsection{Shallow and Deep Inelastic Scattering Uncertainties}
\label{sec:DISuncert}

The uncertainties related to shallow and deep inelastic scattering are extended from
those in T2K’s previous oscillation analysis~\cite{T2K:2023smv}.

The uncertainty in the Bodek-Yang (BY) correction, based on Ref.~\cite{Bodek:2013}, is parameterized as a linear interpolation of the difference between using the GRV98 PDFs with and without the BY corrections. In previous T2K analyses, the parameters doing this were split only between $W < 2~\textrm{GeV}/c^2$ (multi-$\pi$) and $W > 2~\textrm{GeV}/c^2$ (DIS) interactions. For this analysis, Bodek-Yang corrections for multi-$\pi$ interactions (but not DIS interactions) have been divided into axial and vector components, giving three parameters: \texttt{M$\pi$ BY Vector}, \texttt{M$\pi$ BY Axial}, \texttt{DIS BY}. 

In addition, a previously used parameter accounting for the differences in the total cross section between a custom model~\cite{Hayato:2021heg} and AGKY~\cite{Yang:2009zx} hadron multiplicity and kinematic models has been supplemented by another that alters the shape of the cross section in the two-dimensional space of $W$ and N$\pi$. These parameters are: \texttt{M$\pi$ Multi TotXSec} and \texttt{M$\pi$ Multi Shape}.

Two normalization uncertainties are also included, \texttt{CC DIS M$\pi$ Norm $\nu$} and \texttt{CC DIS M$\pi$ Norm \nub}, motivated by comparing the NEUT CC-inclusive cross section to the world average of measurements at higher neutrino energies~\cite{ParticleDataGroup:2018ovx}. The uncertainties are 3.5\% for neutrino interactions and 6.5\% for antineutrino interactions, and the two are uncorrelated.

\subsubsection{Final State Interaction Uncertainties}
\label{sec:FSIuncert}

The uncertainties related to pion FSI are very similar to
those in T2K’s previous oscillation analysis~\cite{T2K:2023smv}, where parameters are included to alter the probabilities of different types of pion interactions in the FSI cascade by changing the mean free paths, based on Ref.~\cite{PinzonGuerra:2018rju}. The only exception is that a parameter that alters the charge-exchange probability for high pion momenta ($>500~\text{MeV}/c$) is now included (\texttt{$\pi$ FSI CEX high E}), in addition to the previous five parameters detailed in~\cite{T2K:2023smv}. The full set of parameters are: \texttt{$\pi$ FSI QE low E}, \texttt{$\pi$ FSI QE high E}, \texttt{$\pi$ FSI Prod.}, \texttt{$\pi$ FSI Abs.}, \texttt{$\pi$ FSI CEX low E}, \texttt{$\pi$ FSI CEX high E}.

Previous analyses did not include an uncertainty on nucleon FSI; in this work, we introduce one. Whilst ideally a complete set of parameters to alter interaction probabilities within NEUT's cascade would be used (as is the case for pions), the necessary tools to implement this were not available at the time of the analysis (although it should be noted that they are used for subsequent T2K analyses). Instead, an alternative approach is taken, which introduces a single ``nucleon fate FSI'' parameter: \texttt{Nucleon FSI}. Each event is classified as either impacted or not impacted by FSI by comparing particles before and after FSI.

The dial then alters the number of ``FSI'' events up and down, and correspondingly moves the ``no FSI'' events as required to maintain the inclusive cross section.

The new nucleon fate FSI dial effectively varies the ``nuclear transparency'', defined as the probability that a struck nucleon escapes the nucleus without significant reinteractions. Based on a study presented in~\cite{Niewczas:2019fro}, it was decided to assign 30\% uncertainty on nuclear transparency. 

\subsubsection{Other Uncertainties}
\label{sec:miscuncert}

Additional uncertainties are applied to processes with small
contributions to the analysis. These remain essentially the same as in the previous T2K analysis~\cite{T2K:2023smv}.

There is one parameter controlling the normalization of the electron neutrino cross section relative to muon neutrinos, and another controlling the same for electron antineutrinos: \texttt{$\nu_{e}$/$\nu_{\mu}$}, \texttt{\nueb/\numb}. Following T2K's previous work, the uncertainties are composed of two parts: a 2\% uncorrelated part and a 2\% anti-correlated part, mainly based on Ref.~\cite{Day:2012gb}. Following Ref.~\cite{Tomalak:2021hec, Dieminger:2023oin}, such uncertainties may overestimate the impact on the double ratio of the muon to electron, neutrino to antineutrino cross section.

The total cross sections of CC resonant single-photon production, CC resonant kaon production, CC resonant eta production, and CC diffractive pion production are controlled by a single new parameter referred to as \texttt{CC Misc}, which has a 100\% normalization uncertainty, and such interactions are not affected by other model parameters.

Additional uncertainties are applied to other processes that make only small contributions to the analysis. As in previous analyses, the NC1$\gamma$ production cross section has a parameter controlling the process' normalization with a 100\%  uncertainty \texttt{NC 1$\gamma$}. The NCQE, NC resonant kaon, eta production, and NC DIS interactions are grouped together and referred to as ``NC other'' interactions, which have a parameter controlling their normalization with a 30\% uncertainty. In neutrino oscillation analyses, this parameter is split between the near and far detectors to give two parameters: \texttt{NC Other near}, \texttt{NC Other far}. A similar parameter controlling the normalization of processes is applied separately for charged coherent pion production on carbon and oxygen targets with a 30\% uncertainty: \texttt{CC Coh $^{12}$C}, \texttt{CC Coh $^{16}$O}. A single parameter controlling neutral-current coherent pion production with a 30\% uncertainty is also defined: \texttt{NC Coh}. 

Two parameters accounting for normalization alterations of the total charged current cross section are slightly modified and affect $0.3<E_{\nu}<0.6$ GeV instead of $0.4<E_{\nu}<0.6$ used in the previous analysis. Uncertainties are motivated by Coulomb corrections; the selected range is the consequence that, with an increase in energy, the lepton energy becomes much larger than the Coulomb energy, and the Coulomb correction should approach unity. We defined a 2\%(1\%) uncertainty for (anti)neutrinos, 100\% anti-correlated: \texttt{CC norm $\nu$}, \texttt{CC norm \nub}.

\section{Near-detector analysis}
\label{sec:NDfit}

\subsection{Selections and detector systematics}
\label{subsec:ndsel}
In this section, we summarize improvements made to the ND280 selection, first describing the samples used and later introducing new detector systematic uncertainties and other improvements. Outside of the differences highlighted in this section, the evaluation of detector systematic uncertainties is the same as described in~\cite{PhysRevD.91.072010}.

Following improvements to the cross-section systematic uncertainty model, the ND280 selections have been expanded by adding proton and photon tagging to the $\nu$-mode samples. In the previous analysis, samples were divided only by pion tagging. Events are first categorized by whether a reconstructed photon is present; events without photon candidates are then further classified based on pion multiplicity. Lastly, events with no reconstructed pions are divided into two samples: those with zero and at least one proton.

\subsubsection{Proton-tagged samples}

\begin{figure}[htbp]
 \centering
   \includegraphics[page=1, width=0.98\columnwidth]{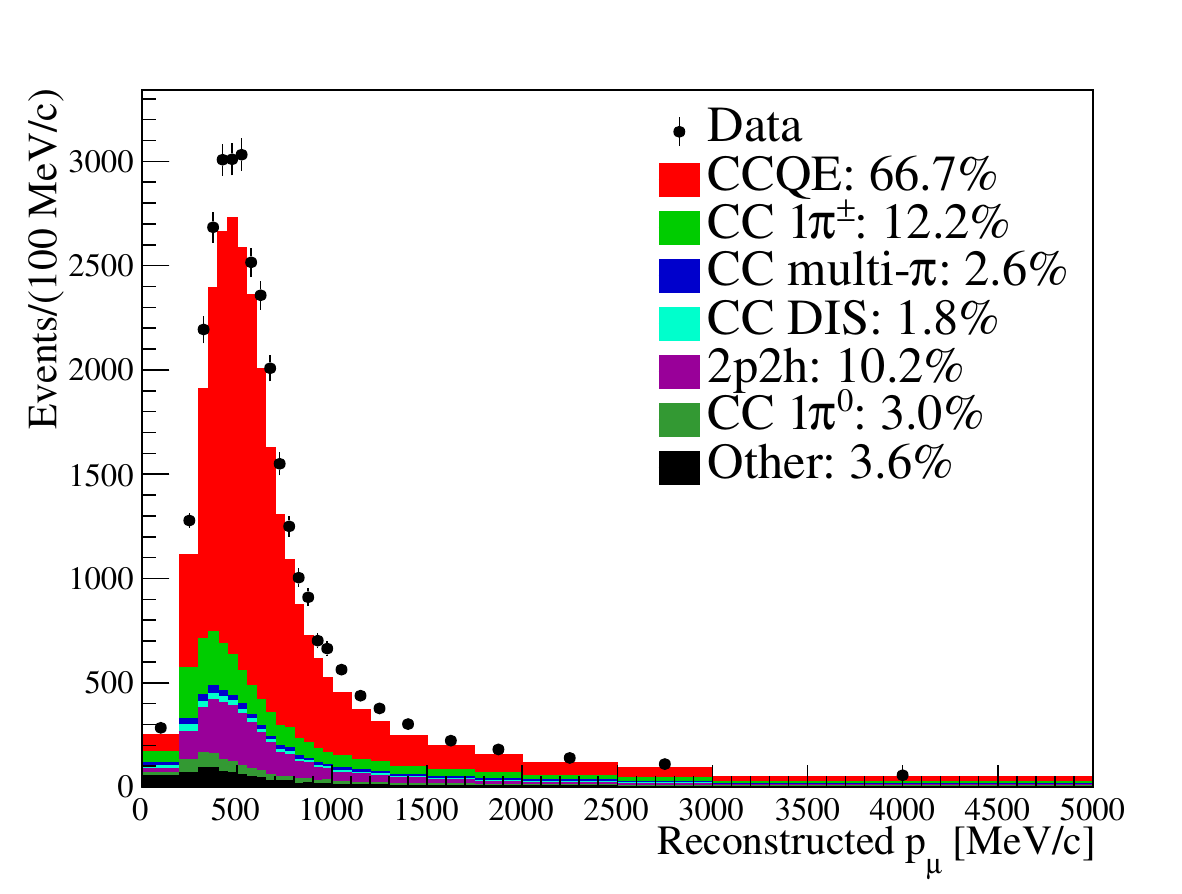}
   \includegraphics[page=2, width=0.98\columnwidth]{Figures/ndselection/ND280_Samples_noGrid_Tittle.pdf}
 \caption{Data and prefit prediction for ND280 FGD1 $\nu$-mode CC0$\pi$ samples without (top) and with (bottom) tagged protons. The prefit prediction is broken down by the contributions of different interaction channels.}
 \label{Fig::NDSamplesProton}
\end{figure}

At T2K energies, CCQE interactions produce more low-momentum protons than 2p2h, and 2p2h has a higher probability of producing multi-proton states. In ND280, which has a proton tracking threshold of approximately 450 MeV/c, most protons produced from CCQE interactions fall below this threshold~\cite{T2K:2018rnz}. As a result, the pionless zero-proton sample has a larger fraction of CCQE events than the sample with protons, as shown in~\autoref{Fig::NDSamplesProton}. The zero-proton sample also preferentially selects CCQE interactions with a lower energy transfer.

\begin{figure}[htbp]
 \centering
   \includegraphics[page=1, width=0.98\columnwidth]{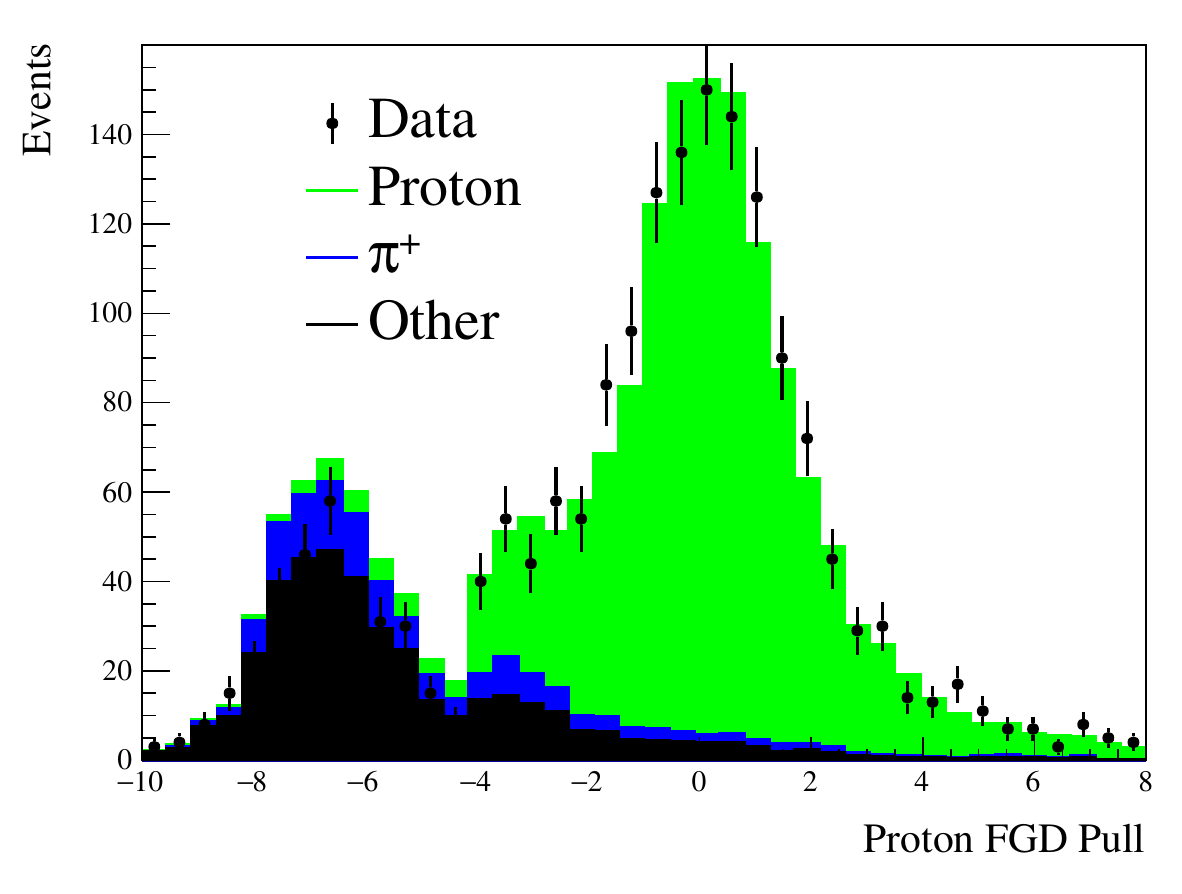}
   \includegraphics[page=3, width=0.98\columnwidth]{Figures/ndselection/Proton_Pull_Likelihood_newer.pdf}
 \caption{The proton pull for FGD-contained tracks (top), and the proton score for tracks with a TPC segment (bottom) for the FGD1 \cczeropizerop sample.}
 \label{Fig::ProtonPID}
\end{figure}

A proton candidate starting in the FGD can either be tagged by the TPC if the track is long enough, or tagged in the FGD if the track is contained. For TPC-tagged protons, a ``pull'' ($\delta$) is calculated using the measured ($m$) and expected ($e$)  energy loss per unit length ($dE/dx$) for particle type $i$,
\begin{equation}
    \delta_i^\text{TPC} = \frac{\frac{dE}{dx}^\text{m} - \frac{dE}{dx}_i^\text{e}}{\sigma\left(\frac{dE}{dx}_i^\text{m}\right),}
\end{equation}
whereas for FGD-contained particles, the total energy deposited $E$ along the track length ($L_\text{m}$) is instead used to calculate the pull,
\begin{equation}
    \delta^\text{FGD}_i = \frac{E^\text{m} - E_i^\text{e}(L_\text{m})}{\sigma \left(E_i^\text{e}\right)}.
\end{equation}
For proton candidates in the TPC, the normalized score for each particle type is calculated,
\begin{equation}
    \mathcal{L}_i = \frac{e^{-\delta_i^2}}{\sum_j{e^{-\delta_j^2}}}.
\end{equation}
Proton candidates are required to originate in the same FGD as the muon candidate with a shared vertex, have $\mathcal{L}_{p} > 0.5$ for TPC-tagged protons, and $\delta_{p}^\text{FGD} > -4$ for FGD-contained protons. The distributions of $\mathcal{L}_{p}$ and $\delta_{p}^\text{FGD}$ are shown in~\autoref{Fig::ProtonPID}.

The samples of TPC-tagged proton candidates have a 98\% purity for interactions in both FGDs, with positively charged pions constituting 1\% of each sample. For the FGD-contained proton candidate samples, the purities are 94\% and 74\% in FGD1 and FGD2, respectively. Positively charged pions constitute 4\% and 16\% of the proton candidates in FGD1 and FGD2, with FGD2 being markedly worse due to the passive water layers. The passive water layers can cause low-momentum FGD-contained pions---which would be tagged by the Michel electron from the $\pi^+$ decay chain---to be tagged as protons instead, since the Michel electron has a higher probability to be missed. The passive water layers also cause fewer protons to be tagged in FGD2 than in FGD1. However, there are generally more reconstructed protons that enter the TPCs than are contained in the FGDs. Tagged protons have relatively high momenta, mostly distributed between $p=0.45-1.4~\text{GeV}/c$. The sample with at least one proton has a higher fraction of resonant events than the zero-proton sample, since protons produced in such events typically have higher momenta.

The antineutrino CCQE interaction, dominant at T2K's energies, does not produce any outgoing protons in the absence of nuclear effects. The protons can only be produced by FSI or non-CCQE processes, such as 2p2h or SPP. Thus, these samples are sensitive to interesting nuclear effects and non-CCQE backgrounds. However, due to the rarity of the processes and lower efficiencies due to the larger wrong-sign background, the proton-tagged subset of the $\bar{\nu}_\mu$CC$0\pi$ sample would have limited statistics ($\sim6\%$ of the integrated $\bar{\nu}_\mu$CC$0\pi$ sample). Similarly, the statistics of the antineutrino samples are smaller at the far detector and are less sensitive to the mismodelling of neutrino interactions than those of the neutrino samples. Therefore, proton tagging for antineutrino selections is not included in this analysis, but will provide an interesting probe of nuclear effects as statistics increase in the coming years.

\subsubsection{Photon-tagged samples}

\begin{figure}[htbp]
 \centering
   \includegraphics[page=3, width=0.45\textwidth]{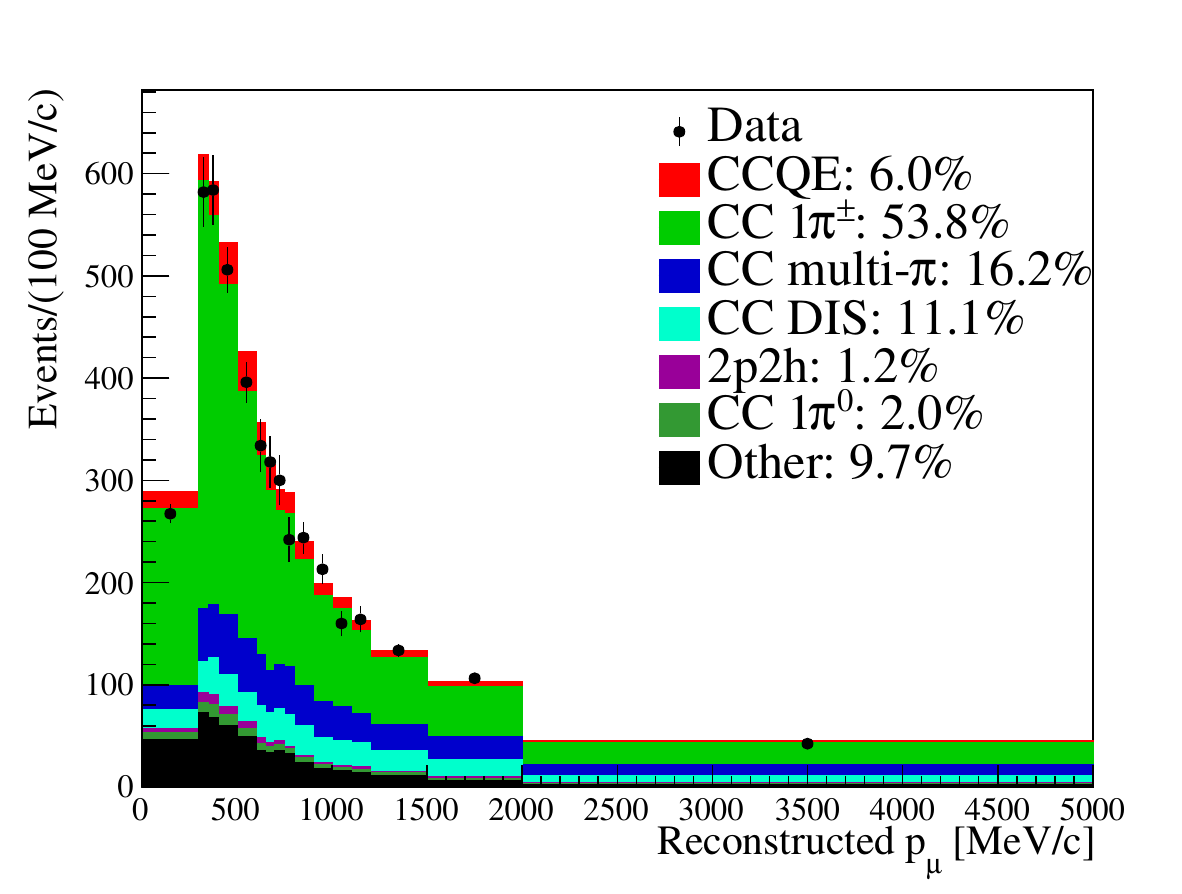}
   
   \includegraphics[page=4, width=0.45\textwidth]{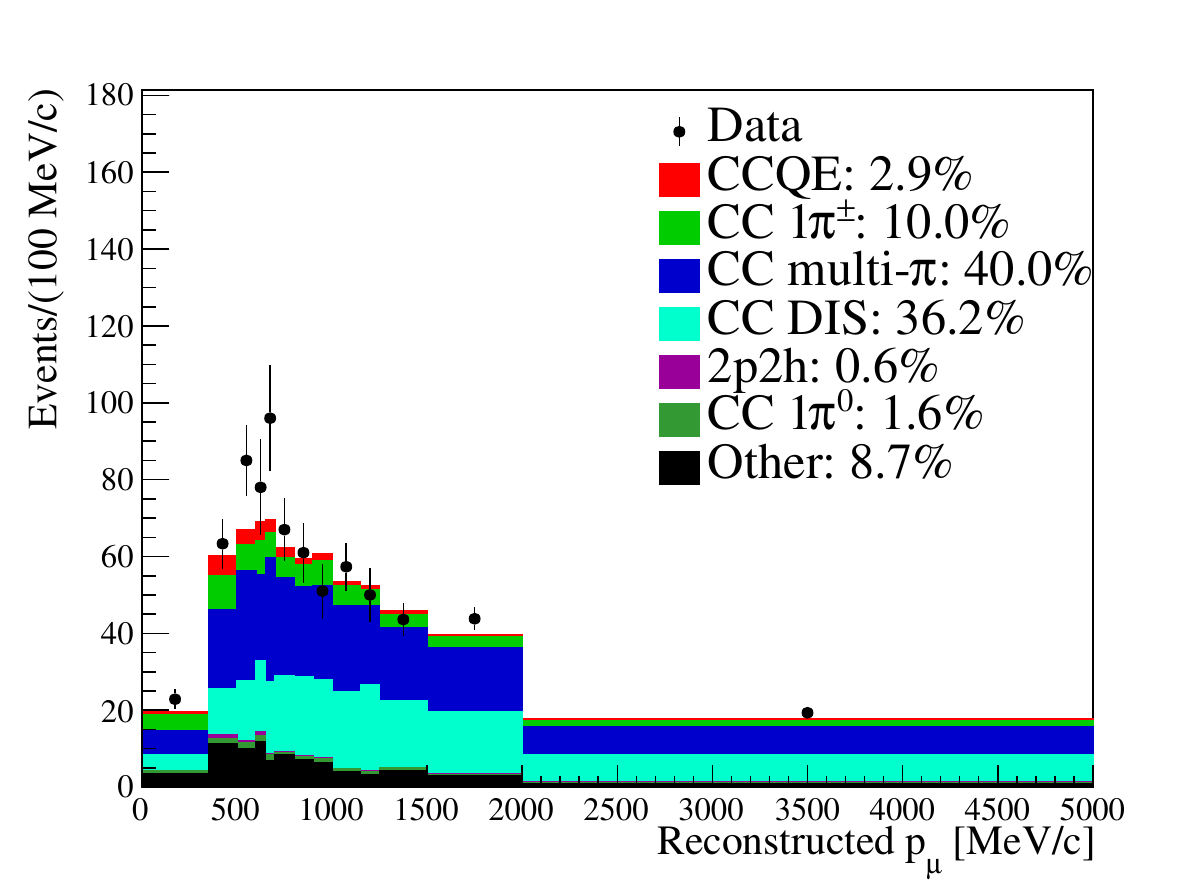}
   
   \includegraphics[page=5, width=0.45\textwidth]{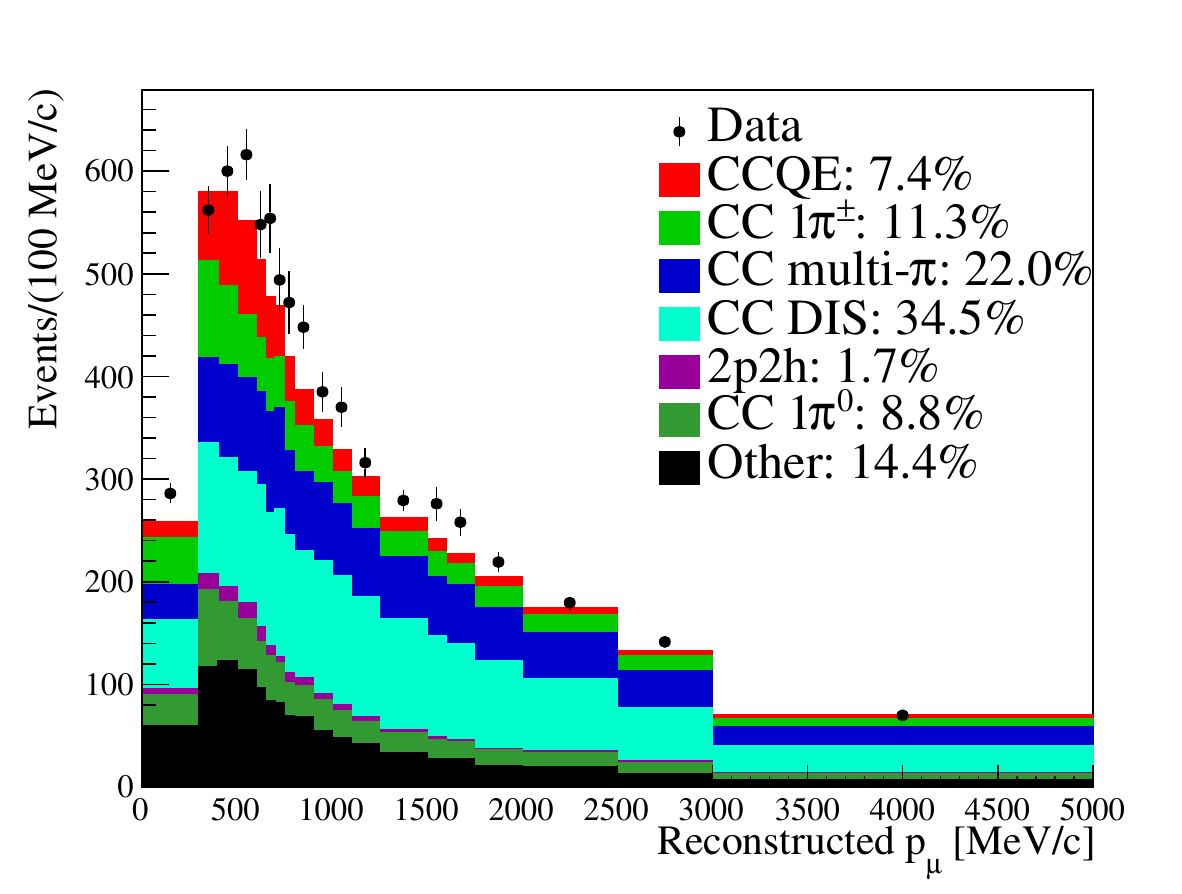} 
 \caption{Data and prefit predictions for ND280 FGD1 $\nu$-mode samples \cconepi (top), \ccother (center), \ccphoton(bottom). The prefit prediction is broken down by the contributions of different interaction channels.}
 \label{Fig::NDSamplesPhoton}
\end{figure}

The photon-tagged samples were primarily added to separate CC$\pi^0$ interactions from CC multi-$\pi$ events, which previously would both enter the \ccotherold selection. The separation was needed because the corresponding FD analysis introduced a $\nu_\mu$CC1$\pi$ sample where the pion is above the Cherenkov threshold, which has important contributions from CC multi-$\pi$, but not CC$1\pi^0$, events, and those backgrounds would be better constrained by moving events with photons into a separate sample in the near-detector analysis.
The \ccphoton sample is dominated by resonant $\pi^0$ production and DIS interactions, and the new \ccother sample is composed primarily of multi-$\pi$ and DIS interactions without $\pi^0$'s in the final state, shown in~\autoref{Fig::NDSamplesPhoton}. Additionally, requiring that no reconstructed photons are present in the main selections increases their purities. The CC0$\pi$ and CC1$\pi^+$ samples increase in purity by approximately 5\% and 7\%, respectively, compared to the previous analysis.

Photons are tagged using two methods that aim to select photons that convert in different regions of ND280. Photons that convert in the tracker region are tagged using the TPC PID to identify at least one electron or positron candidate. The intention is to collect signatures of photons from a $\pi^0$ decay. In previous analyses, these events were placed in the \ccotherold category.

The second method identifies photons that enter the ECal surrounding the tracker region. Two conditions are placed on isolated objects in the ECal: they must start in one of the innermost six layers and must be identified as being more electromagnetic-like than ionizing, while muons are the opposite. Since the lead layers of the ECal are approximately one radiation length thick, photons coming from the tracker are likely to shower in the first layers. In contrast, objects in the outer layers are more likely to be a result of accidental activity that is in time with the neutrino interaction of interest. Interestingly, CCQE events with a photon tag in the ECal often do not contain a true neutral pion, and instead include a neutron, which provides a means to separate neutron interactions to explore in the future.

The \nub-mode photon-tagged samples are not included in this work, but have been explored in the context of future T2K analyses when more \nub-mode data will be available. 

\subsubsection{Summary of ND280 samples}

The events in the near detector are categorized into 22 samples, divided into eleven equivalent FGD1 and FGD2 samples to separate neutrino interactions in plastic scintillator (FGD1), and plastic scintillator and water (FGD2). The \nub-mode samples remain unchanged since the last analysis~\cite{T2K:2023smv}. All samples and data event rates are summarized in~\autoref{Tab:NDsamples}. 

Neutrino oscillations are dependent on the neutrino energy, which can be inferred using lepton momentum and angle~\cite{T2K:2023smv}, as is done at the far detector. Similarly, all ND280 samples are binned in muon momentum, $p_\mu$, and the cosine of the muon's angle with respect to the detector $z$-axis, \cosmu. Accordingly, the cross-section model primarily focuses on predicting lepton variables as a function of incident neutrino energy and the relative amounts of different neutrino interaction modes (e.g., CCQE, 2p2h, SPP), which bias the neutrino energy reconstruction differently.

\begin{table}[htbp]
\centering
\begin{tabular}{ l | c | c ||c }
\hline \hline
Selection                   & Topology              & Target & Data Events \\ \hline
\multirow{12}{*}{\num in $\nu$-mode} 
                            & \multirow{2}{*}{\cczeropizerop} & FGD1 & 21329 \\
                            &                                        & FGD2 & 22935 \\ \cline{2-4}
                            & \multirow{2}{*}{\cczeropiNp} & FGD1 & 9257 \\
                            &                                        & FGD2 & 7373 \\ \cline{2-4}
                            & \multirow{2}{*}{\cconepi}  & FGD1 & 6224 \\
                            &                                        & FGD2 & 5099 \\ \cline{2-4}
                            & \multirow{2}{*}{\ccphoton}             & FGD1 & 11156 \\
                            &                                        & FGD2 & 10460 \\ \cline{2-4}
                            & \multirow{2}{*}{\ccother}       & FGD1 & 1737 \\
                            &                                        & FGD2 & 1620 \\ \cline{2-4}
                            & \multirow{2}{*}{Total}                 & FGD1 & 49703 \\
                            &                                        & FGD2 & 47487 \\ \hline
\multirow{8}{*}{\numb in \nub-mode} 
                            & \multirow{2}{*}{CC0$\pi$}     & FGD1 & 8676 \\
                            &                             & FGD2 & 8608 \\ \cline{2-4}
                            & \multirow{2}{*}{CC1$\pi^{-}$} & FGD1 & 719 \\
                            &                             & FGD2 & 660 \\  \cline{2-4}       
                            & \multirow{2}{*}{\ccotherold}      & FGD1 & 1533 \\
                            &                             & FGD2 & 1396 \\    \cline{2-4} 
                            & \multirow{2}{*}{Total}      & FGD1 & 10928 \\
                            &                             & FGD2 & 10664 \\ \hline
\multirow{8}{*}{\num in \nub-mode} 
                            & \multirow{2}{*}{CC0$\pi$}       & FGD1 & 3714 \\
                            &                               & FGD2 & 3537 \\ \cline{2-4}
                            & \multirow{2}{*}{CC1$\pi^{+}$}   & FGD1 & 1147 \\
                            &                               & FGD2 & 955 \\  \cline{2-4}    
                            & \multirow{2}{*}{\ccotherold}        & FGD1 & 1425 \\
                            &                               & FGD2 & 1334 \\ \cline{2-4} 
                            & \multirow{2}{*}{Total}        & FGD1 & 6286 \\
                            &                               & FGD2 & 5826 \\ \hline \hline
\end{tabular}
\caption{Total number of events for each ND280 sample as well as the total number of events in each subsample. A total of 130,894 events are used in this analysis.}
\label{Tab:NDsamples}
\end{table}

\subsubsection{Detector uncertainties}

The introduction of new samples requires the consideration of new detector systematic uncertainties or improvements to existing ones. Methods for evaluating systematic uncertainties remain unchanged from the previous analysis~\cite{T2K:2023smv}, and dedicated control samples have been developed to assess the ND280 response and quantify these uncertainties.

A proton Secondary Interaction (SI) uncertainty was employed in the previous analysis~\cite{T2K:2023smv}, but it only affected events where the true proton was reconstructed as a pion, and thus the overall impact was negligible. To properly account for the inclusion of proton tagging, this systematic uncertainty was updated to apply to all events with at most two proton tracks. The SI uncertainty is estimated from GEANT4 comparisons to available data on proton inelastic interactions on carbon. In general, fairly good agreement is found, and so a conservative 10\% uncertainty is assigned~\cite{GEANT4:2002zbu,Allison:2016lfl}. The inelastic cross section is taken from the Axen-Wellisch parameterization used by the GEANT4 Bertini cascade physics model~\cite{Geant4HadronicWorkingGroup:2011kka}.

With the inclusion of the \ccphoton sample, four additional detector systematic uncertainties were added related to reconstruction in the ECal and the variables used in the ECal-tagged photon selection.

First, an uncertainty to account for ECal reconstruction efficiencies was included for events where the success or failure of tagging a photon in the ECal would change the event categorization. An uncertainty related to ECal-tracker matching was also included to account for the possibility of isolated objects in the ECal being the result of broken tracks in the global reconstruction or cases where objects in the ECal are incorrectly associated with tracker objects. Although the selection of objects in the first layers of the ECal removes backgrounds from accidental energy from other neutrino interactions in the outer layers, this does not remove accidental activity in the layers accepted in the selection. Therefore, an uncertainty is included to account for this possibility. Finally, an uncertainty associated with the PID variable used to tag electromagnetic-like objects in the ECal is included to cover data-MC differences in the PID discriminator around the cut value.

The errors on the event rates from detector uncertainties for each sample are presented in~\autoref{Tab:NDSystError}. The secondary interaction uncertainties are large, therefore samples with more protons and pions have much higher uncertainties. The four ECal systematic uncertainties that were added contribute 0.1-4.3\% of the total uncertainty for each sample. Inclusion of these new systematic uncertainties increases the total uncertainty compared to the previous analysis~\cite{T2K:2023smv} from $\sim$1.2\% to $\sim$1.8\% for the 0$\pi$-0$\gamma$ samples, which is mainly driven by proton Secondary Interaction uncertainties. The largest change is visible in the \ccother samples, from around $\sim$2\% to $\sim$6\%. This increase is a consequence of enhancing the CC DIS purity within this sample and thus increasing the impact of SI-related systematics, which are by far the most impactful.

\begin{table}[htbp]
\centering
\begin{tabular}{ l | c | c ||c }
\hline \hline
Selection                   & Topology              & Target & Uncertainty (\%) \\ \hline
\multirow{10}{*}{$\nu_{\mu}$ in $\nu$-mode} 
                            & \multirow{2}{*}{\cczeropizerop} & FGD1 & 1.8 \\
                            &                                        & FGD2 & 2.1 \\ \cline{2-4}
                            & \multirow{2}{*}{\cczeropiNp} & FGD1 & 3.5 \\
                            &                                        & FGD2 & 3.9 \\ \cline{2-4}
                            & \multirow{2}{*}{\cconepi}  & FGD1 & 3.0 \\
                            &                                        & FGD2 & 3.6 \\ \cline{2-4}
                            & \multirow{2}{*}{\ccphoton}             & FGD1 & 2.8 \\
                            &                                        & FGD2 & 2.5 \\ \cline{2-4}
                            & \multirow{2}{*}{\ccother}       & FGD1 & 5.2 \\
                            &                                        & FGD2 & 6.3 \\ \hline
\multirow{6}{*}{\numb in \nub-mode} 
                            & \multirow{2}{*}{CC0$\pi$} & FGD1 & 1.9 \\
                            &                         & FGD2 & 2.1 \\ \cline{2-4}
                            & \multirow{2}{*}{CC1$\pi^{-}$} & FGD1 & 4.2 \\
                            &                         & FGD2 & 3.9 \\  \cline{2-4}               
                            & \multirow{2}{*}{\ccotherold}  & FGD1 & 3.5 \\
                            &                         & FGD2 & 2.9  \\    \hline      
\multirow{6}{*}{$\nu_{\mu}$ in \nub-mode} 
                            & \multirow{2}{*}{CC0$\pi$} & FGD1 & 2.2 \\
                            &                         & FGD2 & 2.2 \\ \cline{2-4}
                            & \multirow{2}{*}{CC1$\pi^{+}$} & FGD1 & 3.3 \\
                            &                         & FGD2 & 3.2 \\  \cline{2-4}                 
                            & \multirow{2}{*}{\ccotherold}  & FGD1 & 2.6  \\
                            &                         & FGD2 & 2.1 \\ \hline \hline                            
\end{tabular}
\caption{Errors on the total number of events in the ND280 analysis from detector uncertainties only, broken down by selection.}
\label{Tab:NDSystError}
\end{table}

\subsection{Fitter details}
\label{subsec:fitter}
The near-detector data are analyzed using two independent frameworks: one based on a hybrid-frequentist approach and the other on Bayesian methods. Both approaches aim to measure the neutrino flux and cross section at the near detector and constrain their uncertainties. To accomplish this, they evaluate the following extended binned negative log-likelihood:

\begin{widetext}
\begin{align}
    -\ln\mathcal{L}_\textrm{Total} &= -\ln\left(\mathcal{L}_\textrm{Stat} \mathcal{L}_\textrm{MC stat} \mathcal{L}_\textrm{Syst}\right)\\ 
    &= \sum^{\textrm{Samples}}_s \sum^\textrm{Bins}_{j,s} \Bigg[ \bigg( N^\textrm{MC}_{j,s} - N^\textrm{Data}_{j,s} + N^\textrm{Data}_{j,s} \ln{\frac{N^\textrm{Data}_{j,s}}{N^\textrm{MC}_{j,s}}} \bigg) + \frac{\left(\beta_{j,s}-1\right)^{2}}{2\sigma^{2}_{\beta_{j,s}}} \Bigg] + \frac{1}{2}\left(\vec{x}-\vec{\mu}\right)^T \mathbf{V}^{-1} \left(\vec{x}-\vec{\mu}\right)
\label{eq:barlowbeeston_total}  
\end{align}
\end{widetext}

where the total likelihood is composed of three terms: $\mathcal{L}_\textrm{Stat}$ is the statistical likelihood, $\mathcal{L}_\textrm{MC stat}$ accounts for the finiteness of the generated MC using the Barlow-Beeston formalism and Conway's approximation~\cite{Barlow-Beeston, Conway:2011in}, and $\mathcal{L}_\textrm{Syst}$ is the likelihood associated with systematic uncertainties.

The first term is derived under the assumption that the observed event counts follow a Poisson distribution. $\beta_j$ is a scaling parameter that relates the generated MC to the expected event rates, for a given bin $j$, while $\sigma_{\beta_j}$ is the relative statistical uncertainty on the prediction in bin $j$. The penalty term ($-ln\mathcal{L}_\textrm{Syst}$) assumes Gaussian prior uncertainties and penalizes the parameters $\vec{x}$ from moving away from their prior central values $\vec{\mu}$. The covariance matrix of the systematic parameters is encoded in the matrix $\mathbf{V}$.

It is worth noting that the sum of the squares weights ($\sum w^2$, where the weight, $w$ is defined in \autoref{eq:nevtpara}) used for the calculation of $\sigma_{\beta_j}$ had to be fixed to their prior values to aid fit stability. We suspect that, for example, the FSI parameters have large variances or non-Gaussian distributions, which violate the assumptions of the Conway calculations. To ensure this does not bias the results, alternative fits have been performed using the likelihood outlined in Ref.~\cite{Arguelles:2019izp}. This alternative likelihood did not suffer from the same instabilities, and the results are identical, as presented later in this paper. Therefore, we are confident this minor modification does not affect the analysis.\footnote{We did not use the alternative likelihood, as it is more challenging to calculate the likelihood for the null hypothesis, which prevents the total log-likelihood from reaching zero in the ideal case. This does not affect the fitting procedure, but it makes the interpretation of multiple validation steps more challenging.}

The bin predictions $N^\text{MC}_j$ are built using a set of generated events passing the same reconstruction methods and sample selection criteria as the data. The predicted number of events in bin $j$ is parameterized as:

\begin{equation}
    N^\textrm{MC}_j = w_{POT} d_j \sum^{E_{\nu} \text{ bins}}_k f_k \sum^\text{int. modes}_i N_{j,k,i}^{\textrm{Nom}}\;\prod_{p} \omega_i(x_p).
\label{eq:nevtpara}
\end{equation}

Here, $k$ runs over neutrino energy bins, and $i$ over interaction modes. Response functions that encode how the weight of each event changes as a function of cross-section parameters, such as the quasi-elastic axial mass (M$_A^{QE}$), are denoted with $\omega_i(x_p)$. They are pre-calculated on an event-by-event basis for a discrete set of variations of the cross-section parameters, and the resulting weights are then interpolated either linearly or cubically to obtain the value of $\omega_i(x_p)$ for arbitrary parameter values. The term $d_j$ is a detector-related weight, and $f_k$ accounts for the flux normalization in energy bin $k$. $N_{j,k,i}^\mathrm{Nom}$ represents the nominal number of events in bin $j$, energy bin $k$, and interaction mode $i$.

Both the Bayesian and hybrid-frequentist analyses employ the same likelihood function. The term ``best fit'' is used to denote the parameter values providing the most representative description of the data, although the statistical procedures used to obtain these values differ between the two approaches. Based on the results from each analysis, the covariance matrix between the parameters is obtained. Both the central values and covariances are propagated to the far detector analyses, including cross-correlations between different classes of systematic uncertainties.

The likelihood minimization with the hybrid-frequentist approach is performed using the MIGRAD algorithm implemented in Minuit2~\cite{James:1994vla}. MIGRAD numerically computes the second derivative of each pair of systematic uncertainties to find a better point in the parameter phase space. This feature requires the likelihood surface to be smooth over the explored phase space. Once the best fit has been reported, the HESSE algorithm from Minuit2 is employed to compute the Hessian matrix. The inverse of the Hessian gives an estimate of the postfit error covariance matrix, assuming the likelihood has a Gaussian profile. Non-Gaussianities are expected due to the parametrization of some cross-section systematic uncertainties. These include the CCQE removal energy systematic parameters since they affect the reconstructed energy of the outgoing lepton. Propagating this uncertainty to the observed samples requires shifting MC events between bins of reconstructed momentum. However, since the hybrid-frequentist fitting approach relies on gradient descent -- which requires a continuous likelihood -- direct event migration is not feasible. Instead, the migration was modeled by applying response functions that reweight and interpolate the impact of variations of individual CCQE events within their respective kinematic bins. To ensure the robustness of this approach, both linear and cubic interpolation methods were tested. The differences observed in the final results were found to be negligible.

Some additional measures were taken to aid the convergence of the hybrid-frequentist fit due to known issues of the fitting algorithms near phase space boundaries. Specifically, the response functions for cross-section parameters whose prior central values were at physical boundaries were ``mirrored'' across the boundary~\footnote{Mirroring involves copying a parameter response across a physical boundary. For instance, if the boundary is at value 0, the response at 1 is copied to -1.}.

The Bayesian approach uses a Markov Chain Monte Carlo (MCMC) approach implemented in Ref.~\cite{mach3ref}. The Metropolis–Rosenbluth–Rosenbluth–Teller–Teller (MR$^{2}$T$^{2}$) algorithm~\cite{Metropolis:1953am, Hastings:1970aa, Gubernatis:2005zz} is used to explore the posterior phase-space and extract the N-dimensional posterior distributions, where each dimension corresponds to a model parameter. The N-dimensional distribution is marginalized to obtain one-dimensional distributions for each parameter. The advantage of MCMC is that it does not rely on assumptions about the underlying distribution, allowing it to explore non-Gaussian parameter spaces and handle discontinuous likelihoods. This capability enables a more accurate understanding of the cross-section parameters and their correlations. In addition, the MCMC approach does not require a smooth likelihood surface since no gradient has to be computed. For example, the MCMC framework directly implements the removal energy shift parameters for CCQE interactions~\footnote{Removal energy for CC resonant pion production is a weight systematic and is implemented in the same way in both the hybrid-frequentist and Bayesian approaches.} that migrate MC events between bins during the fit. The $\alpha$ correction is only implemented within the MCMC fitter; however, as discussed in the following sections, the impact of $\alpha$ on the results is negligible (See \autoref{Fig::SavageDickey} in \autoref{subsec:dataresults}).

Despite the advantages of MCMC methods, MINUIT-based analyses remain indispensable, as they are far less computationally intensive than running MCMC and provide frequentist goodness-of-fit measures. This makes them a key cross-check for the robustness of T2K analyses.

\subsection{Goodness of fit}
\label{subsec:pvalue}
A goodness-of-fit test is performed with the hybrid-frequentist framework to provide a measure of consistency between the fitted model and experimental observations. The metric used is the \textit{p}-value, which is obtained by generating pseudo-experiments and fitting them. Each pseudo-experiment corresponds to a plausible range of variations expected from the model when sampling all systematic categories, including the flux, cross-section, and detector systematics. Additionally, data and MC statistical fluctuations are included via Poisson variations. The parameter values are randomly sampled from their prior covariance matrices, accounting for uncertainties and correlations. If the new parameter value is outside its validity range, the draw is repeated. A \textit{p}-value is calculated as the fraction of pseudo-experiments whose best-fit log-likelihood is worse than the data best-fit log-likelihood. A high \textit{p}-value indicates that the model adequately describes the data, whereas a low \textit{p}-value may signal a potential discrepancy that requires further investigation, for example, a lack of freedom in the uncertainty model. Cross-section parameters that are not varied in the fit, and those with flat priors, are not varied in the \textit{p}-value calculation. This fixed parameter list is reported in \autoref{tab:xsecparameters} in \autoref{app::more_plots}. The parameters with flat priors are: \texttt{2p2h norm $\nu$}, \texttt{2p2h norm $\bar{\nu}$}, and Rein--Sehgal $\Delta$ decay.

\begin{figure}[htp!]
    \centering
    \includegraphics[width=\linewidth]{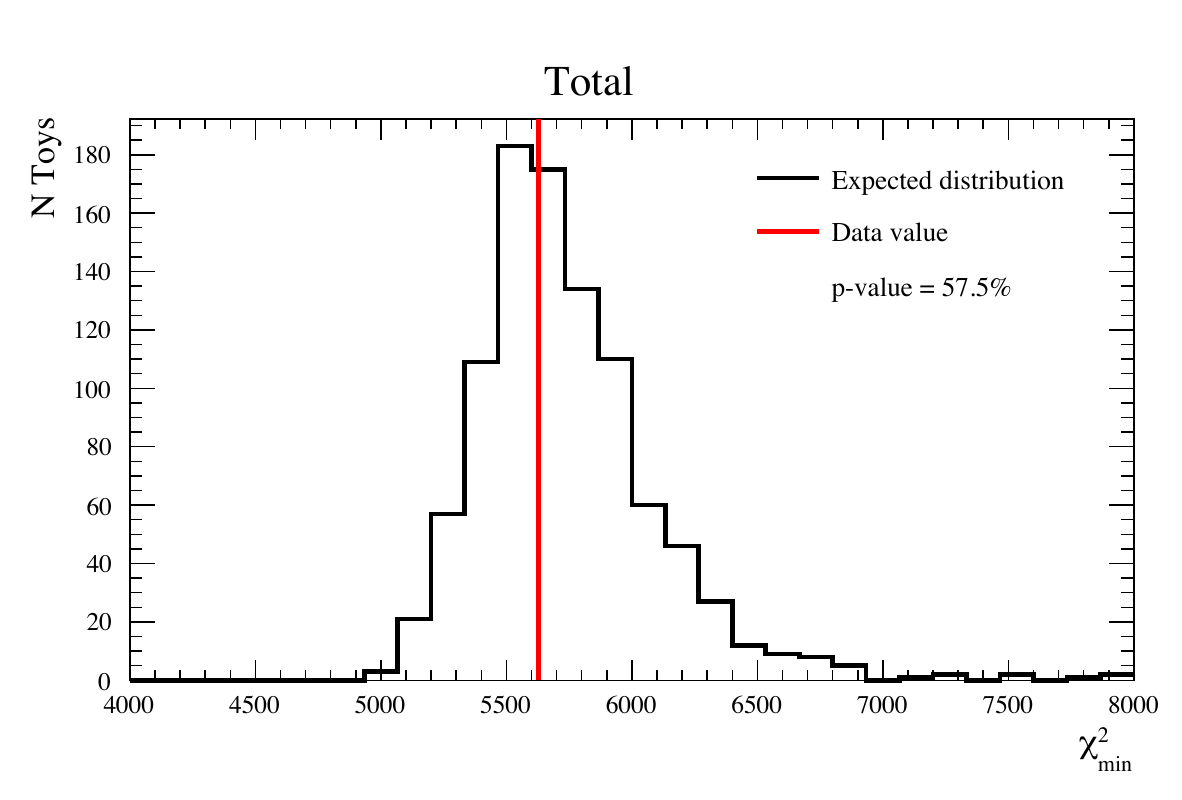}
    \caption{$\chi^2$ distribution for the pseudo-experiments. The $p$-value is calculated as the fraction of fits with a $\chi^2$ equal to or greater than that observed in the data, relative to the total number of fits. The minimum $\chi^2$ obtained from the fit to the data is $\chi^2_{\min} = 5628.82$.}
    \label{fig:pvalue}
\end{figure}

To calculate the prior \textit{p}-value, a total of 2000 pseudo-experiments were randomly generated and fitted. We compute the fraction of pseudo-experiments whose minimum $\chi^2$ relative to the nominal MC is greater than that observed in data. The total \textit{p}-value for the data fit is 57.5\%, given a $\chi^2$ of 5628.82 (see \autoref{fig:pvalue}). It was observed that the \textit{p}-value stabilized after fitting approximately 800 pseudo-experiments. This value is well above the conventional threshold of 5\%, indicating that the data are adequately consistent with the input model. Notably, this \textit{p}-value is lower than that obtained in the previous analysis, which was 74\%. This reduction may be due to the modifications introduced in the selection and systematic treatment to probe the input physics models, for example, by adding proton and photon tagging in the event selections and replacing ad-hoc systematic uncertainties with more theoretically motivated alternatives.

The results of the \textit{p}-value studies are summarized in \autoref{tab:likelihoods}. We also break down the likelihood distribution into its components: the statistical term for each sample and its total, and the systematic penalty term. Overall, the \textit{p}-value obtained for each component is reasonable and indicates that our Monte Carlo model describes the data with sufficient accuracy. The penalty contribution from the constrained nuisance parameters was found to be low (3.3\%), indicating tensions between the fitted nuisance parameters and their external constraints.

\begin{table}[h]
    \centering
    \begin{tabular}{l | c | c || c |c}
        \toprule  \hline \hline 
         \multicolumn{3}{c||}{} & Data postfit $\chi^2$ & \textit{p}-value(\%) \\
         \hline
        \midrule
        \multirow{10}{*}{$\nu_{\mu}$ in $\nu$-mode}
            & \multirow{2}{*}{\cczeropizerop} & FGD1 & 706.8 & 67.6 \\
            &                                & FGD2 & 697.9 & 77.7 \\ \cline{2-5} 
            
            & \multirow{2}{*}{\cczeropiNp} & FGD1 & 378.4 & 78.0 \\
            &                             & FGD2 & 341.3 & 94.5 \\ \cline{2-5} 
            
            & \multirow{2}{*}{\cconepi} & FGD1 & 315.9 & 61.1 \\
            &                          & FGD2 & 284.3 & 78.2 \\ \cline{2-5} 
            
            & \multirow{2}{*}{\ccother} & FGD1 & 145.3 & 97.4 \\
            &                          & FGD2 & 151.8 & 96.6 \\ \cline{2-5}  
            
            & \multirow{2}{*}{\ccphoton} & FGD1 & 489.4 & 46.8 \\
            &                           & FGD2 & 425.9 & 80.4 \\ \hline 
        
        \multirow{6}{*}{\numb in \nub-mode} 
            & \multirow{2}{*}{CC$0\pi$} & FGD1 & 376.1 & 21.8 \\
            &                           & FGD2 & 373.3 & 9.6 \\ \cline{2-5} 
            
            & \multirow{2}{*}{CC$1\pi^{-}$} & FGD1 & 66.0 & 44.5 \\
            &                           & FGD2 & 56.1 & 54.8 \\ \cline{2-5} 
            
            & \multirow{2}{*}{CC-Other} & FGD1 & 99.8 & 46.3 \\
            &                          & FGD2 & 98.6 & 62.0 \\ \hline 
            
        \multirow{6}{*}{$\nu_{\mu}$ in \nub-mode} 
            & \multirow{2}{*}{CC$0\pi$} & FGD1  & 144.5 & 38.3 \\
            &                           & FGD2 & 142.3 & 32.4 \\ \cline{2-5} 
             
            & \multirow{2}{*}{CC$1\pi^{+}$} & FGD1 & 59.9 & 9.6 \\
            &                           & FGD2 & 61.5 & 11.0 \\ \cline{2-5} 
  
            & \multirow{2}{*}{CC-Other} & FGD1 & 68.1 & 38.7 \\
            &                          & FGD2 & 60.8 & 58.0 \\ \hline
        \multicolumn{3}{c||}{Statistical term} & 5543.81 & 72.5 \\ \hline     
        \multicolumn{3}{c||}{Penalty term} & 85.01 & 3.3 \\ \hline \hline 
        \bottomrule
    \end{tabular}
    \caption{Contribution of each sample to the total likelihood, including the statistical term and any associated penalty terms.}
    \label{tab:likelihoods}
\end{table}

\subsection{Data results}
\label{subsec:dataresults}
The results of the fit for both the cross-section and flux are discussed in this section, as these parameters are being propagated to FD. We also briefly address the correlation between flux and cross-section, as it is crucial for the final results.

\begin{figure}[htb]
 \centering 
  \includegraphics[page=2, width=\columnwidth]{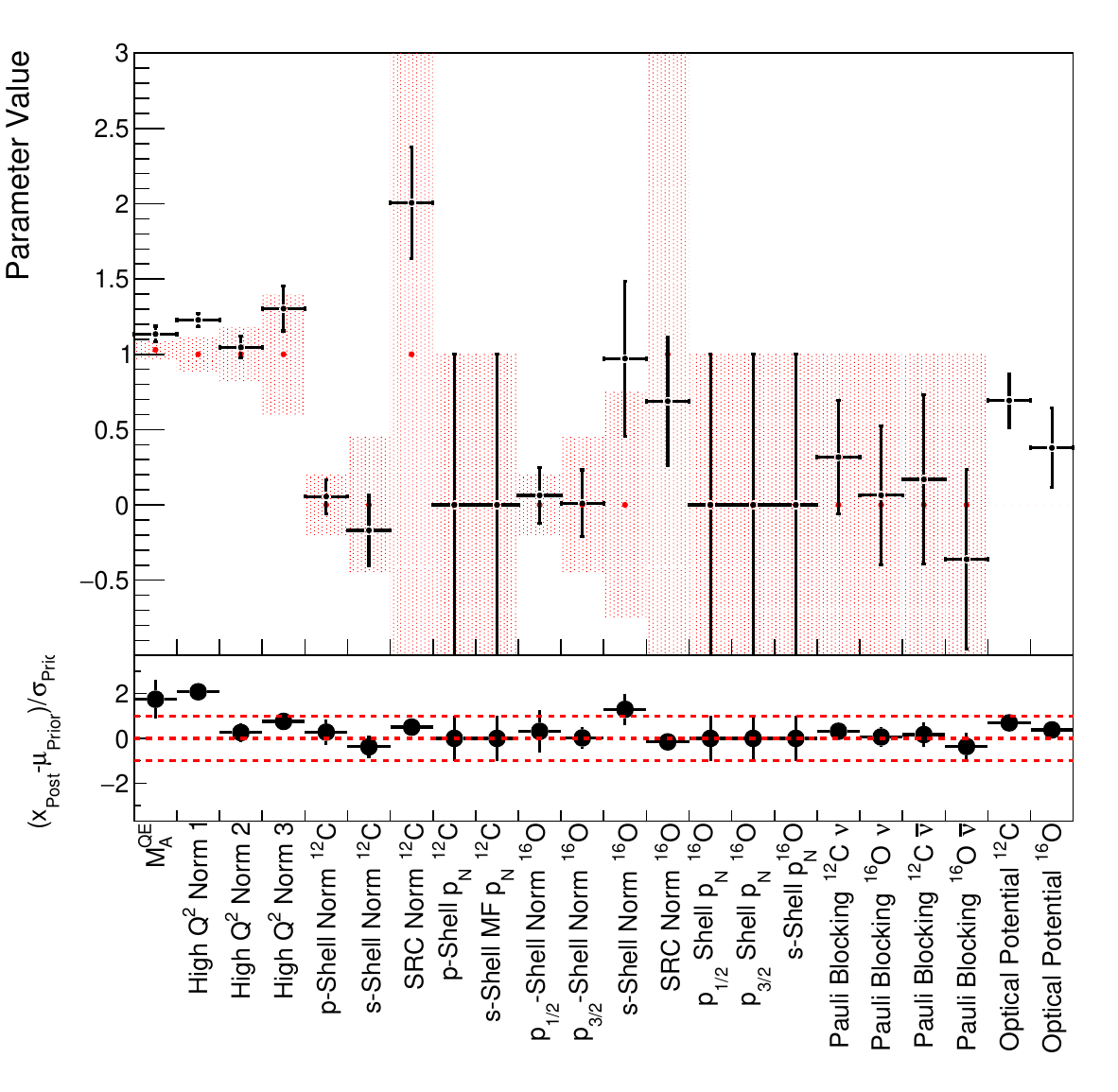}
  \includegraphics[page=3, width=\columnwidth]{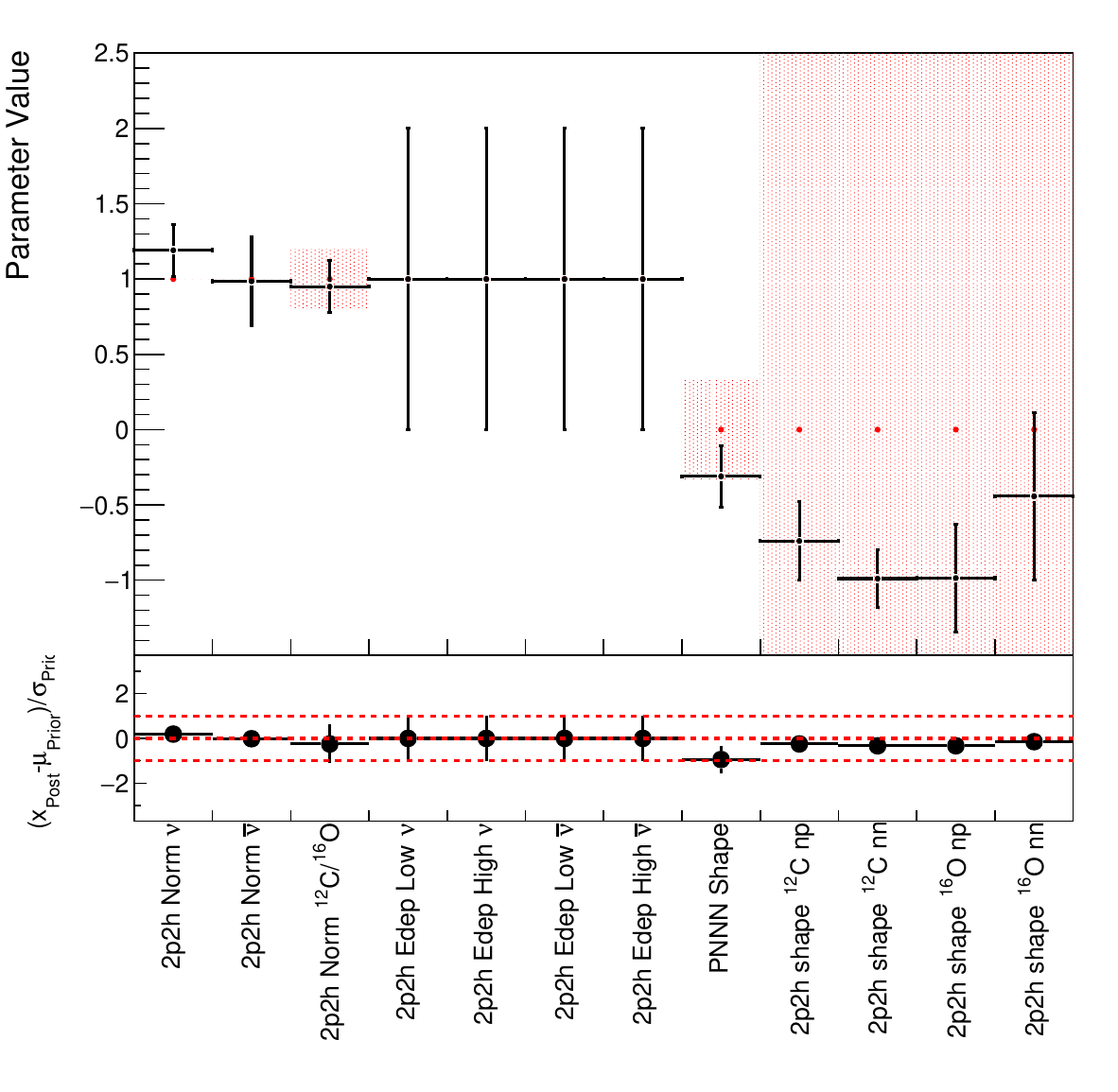}
 \caption{Postfit values and uncertainties (black) for CCQE (top) and 2p2h (bottom) cross-section parameters from the data fit, compared with the corresponding prefit values and uncertainties (red).}
 \label{Fig::DataFit_1D_Xsec_CCQE}
\end{figure}

The results of the fit to data for interaction parameters related to CCQE and 2p2h are shown in \autoref{Fig::DataFit_1D_Xsec_CCQE}. $M^{\textrm{QE}}_{A}$ is pulled significantly above its prior value, which is consistent with the results from the previous ND280 fit~\cite{T2K:2023smv}. The four-momentum transfer ($Q^2$) parameters are also pulled away from their priors, with {\tt high~$Q^2$~norm~1} (relevant for the region of $0.25<Q^2<0.5~\text{GeV}^2/c^4$) being pulled most significantly. Overall, this corresponds to a reshaping of the CCQE cross section as a function of $Q^2$ (enhancing the high $Q^2$ tail with respect to the nominal model).

The {\tt s-shell MF Norm $^{12}$C} parameter is pulled in the opposite direction to the {\tt p-shell MF Norm $^{12}$C} parameter, resulting in an increase of s-shell contribution. In contrast, the p-shell contribution is weakened, thereby reshaping the nucleon momentum distribution. Note that both parameters are within their prior uncertainty bands. Given that both highly affect CCQE normalization, their opposite movement leads to a reshaping without a large impact on the CCQE cross-section normalization. The oxygen-related shell dials, {\tt $p_{1/2}$} and {\tt $p_{3/2}$ Shell MF Norm $^{16}$O}, are very close to their priors; however, {\tt s-Shell MF Norm $^{16}$O} is pulled $1\sigma$ away from the prior, increasing the contribution from this shell. {\tt SRC Norm $^{12}$C} is strongly pulled away from the prior, increasing the relative strength of the SRC region in carbon, while {\tt SRC Norm $^{16}$O} is pulled slightly in the opposite direction.

Slightly different postfit values for $\nu$ and \nub 2p2h normalizations are observed, but these are within each others’ postfit uncertainties and consistent with the nominal model. All 2p2h shape parameters are shifted toward values of $-1$, which corresponds to an increase in the strength of the QE-like peak in the energy and momentum transfer phase space of the Valencia model. The {\tt PNNN Shape} parameter favors an increase in the number of $NN$ pairs. 

\begin{figure}[htb]
 \centering 
  \includegraphics[page=3, width=\columnwidth]{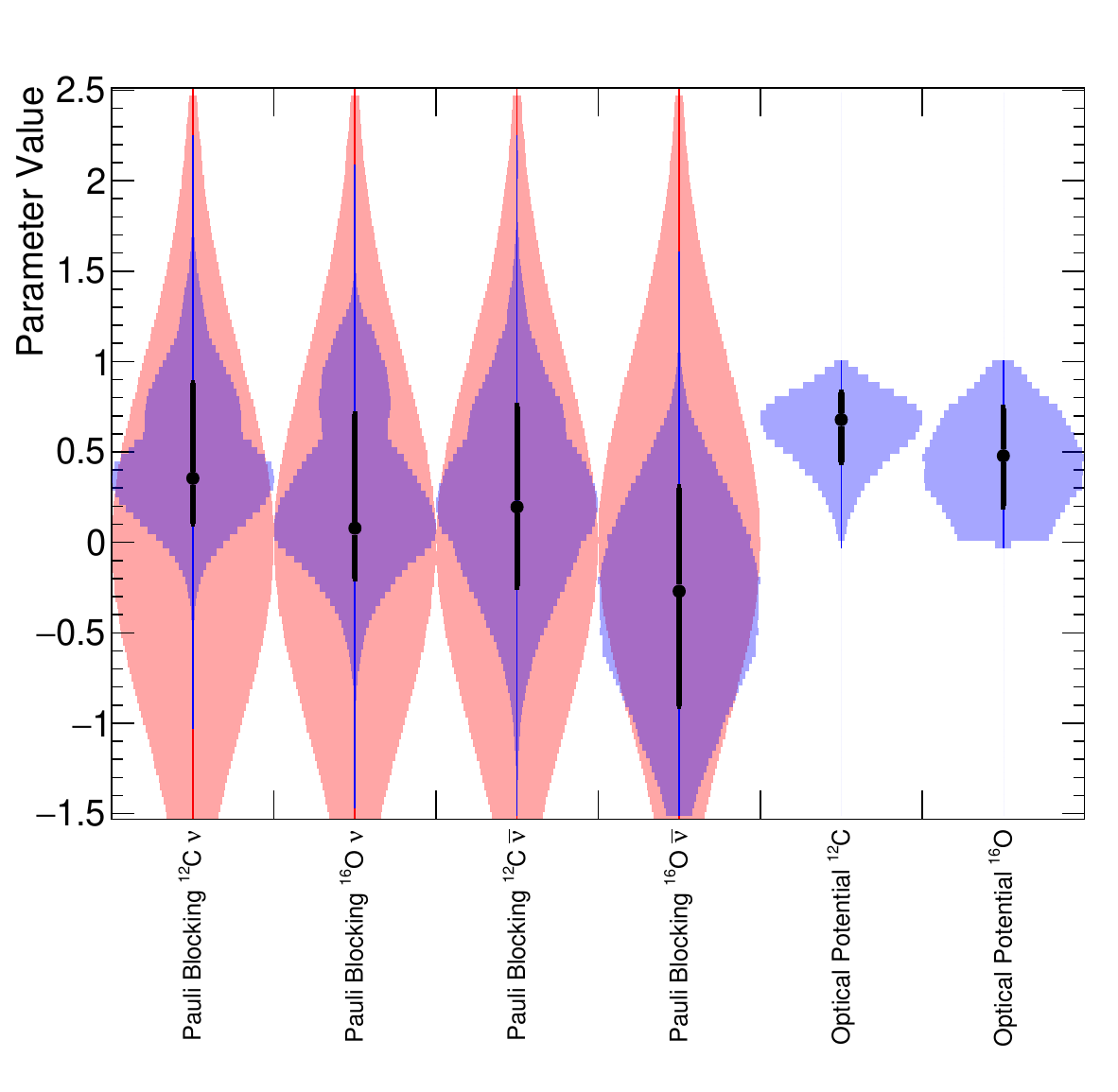}
  \includegraphics[page=8, width=\columnwidth]{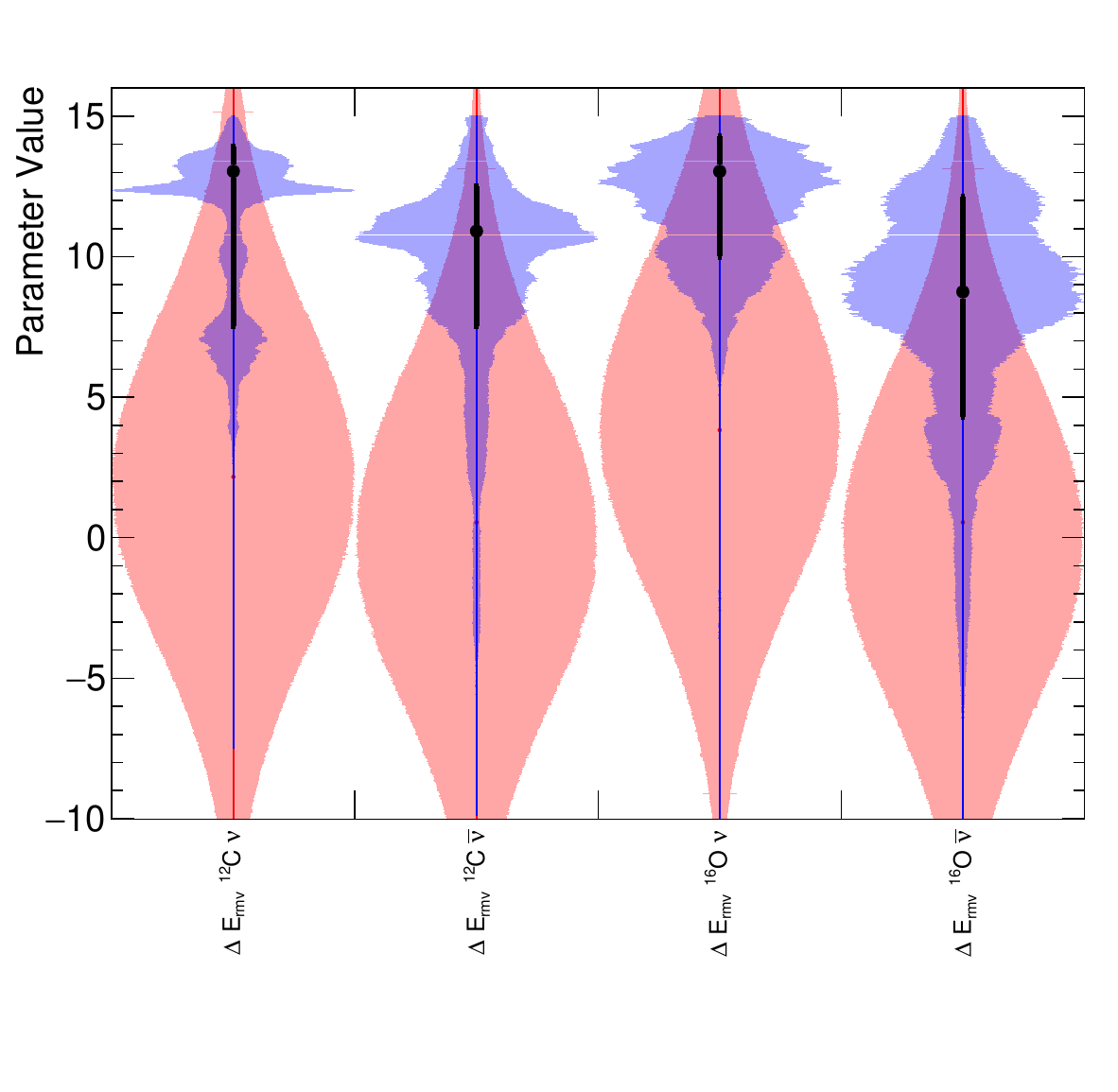}
 \caption{Posterior distributions (violin format) for selected parameters: red (blue) bands represent prior (posterior) probabilities, and black points indicate the Highest Posterior Density value and uncertainties.}
 \label{Fig::Violin_CCQE}
\end{figure}

Almost all of the Pauli blocking dials are pulled to higher values with respect to their prior central values. This translates to an increased probability of events being Pauli blocked, leading to a suppression in the low energy transfer CCQE region. Furthermore, we observe a preference for an application of some of the optical potential strength, further suppressing the low energy transfer region. The Pauli blocking parameters have highly non-Gaussian posterior distributions, which can be better visualized in \autoref{Fig::Violin_CCQE}, which shows their individual posteriors.

Non-Gaussianity is also present for the CCQE removal energy parameters (also shown in \autoref{Fig::Violin_CCQE}). These show a relatively strong preference for larger removal energies than in the nominal model, further reducing and reshaping the low-energy transfer region of the CCQE cross section. All the removal energy parameters are fit to larger values than the nominal model (although broadly compatible with their prior uncertainties), indicating a need to shift lepton momentum in the CCQE-sensitive ND280 samples to lower values.

\begin{figure}[htb]
 \centering 
 \subfloat{
  \includegraphics[page=1, width=0.6\columnwidth]{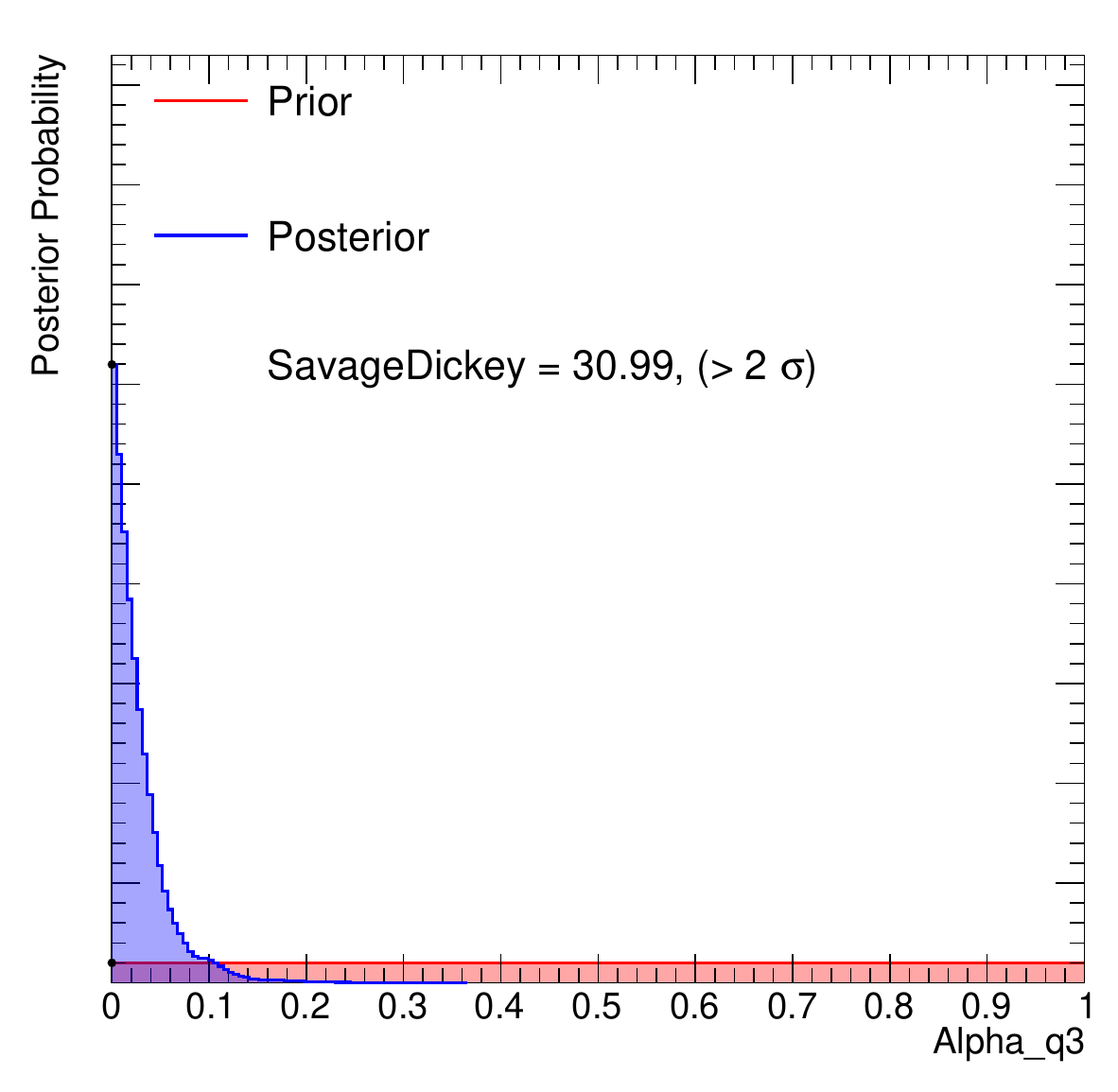}
 }
 \caption{Posterior and prior distributions for $\alpha$ correction. Using the Savage-Dickey method, the Bayes factor was calculated and is equal to 31.}
 \label{Fig::SavageDickey}
\end{figure}

Concerning the momentum-transfer dependence of the removal energy, the data fit indicates a strong preference for $\alpha = 0$. Within the Bayesian framework, the strength of this preference can be quantified using the Bayes factor. For point-like hypotheses, such as in the case of $\alpha$, the Savage-Dickey formalism provides a convenient method to compute the Bayes factor as the ratio of the posterior to the prior distribution evaluated at the given value—in this case, $\alpha = 0$. Following this approach, we obtain a Bayes factor of 31, indicating very strong evidence in favor of $\alpha = 0$. This result is illustrated in \autoref{Fig::SavageDickey}. This may seem like a contradiction with the work presented in Ref. ~\cite{Dolan:2023iik, Abe:2024avs}, which suggests $\alpha$ is close to 1. One potential factor that could contribute to this apparent disagreement is the lack of consideration of optical potential systematic uncertainty in Ref. ~\cite{Dolan:2023iik, Abe:2024avs}. Both the optical potential and the $\alpha$-correction are related to similar underlying physics, which could lead to comparable effects.

\begin{figure}[htb]
 \centering 
 \subfloat{
  \includegraphics[page=2, width=\columnwidth]{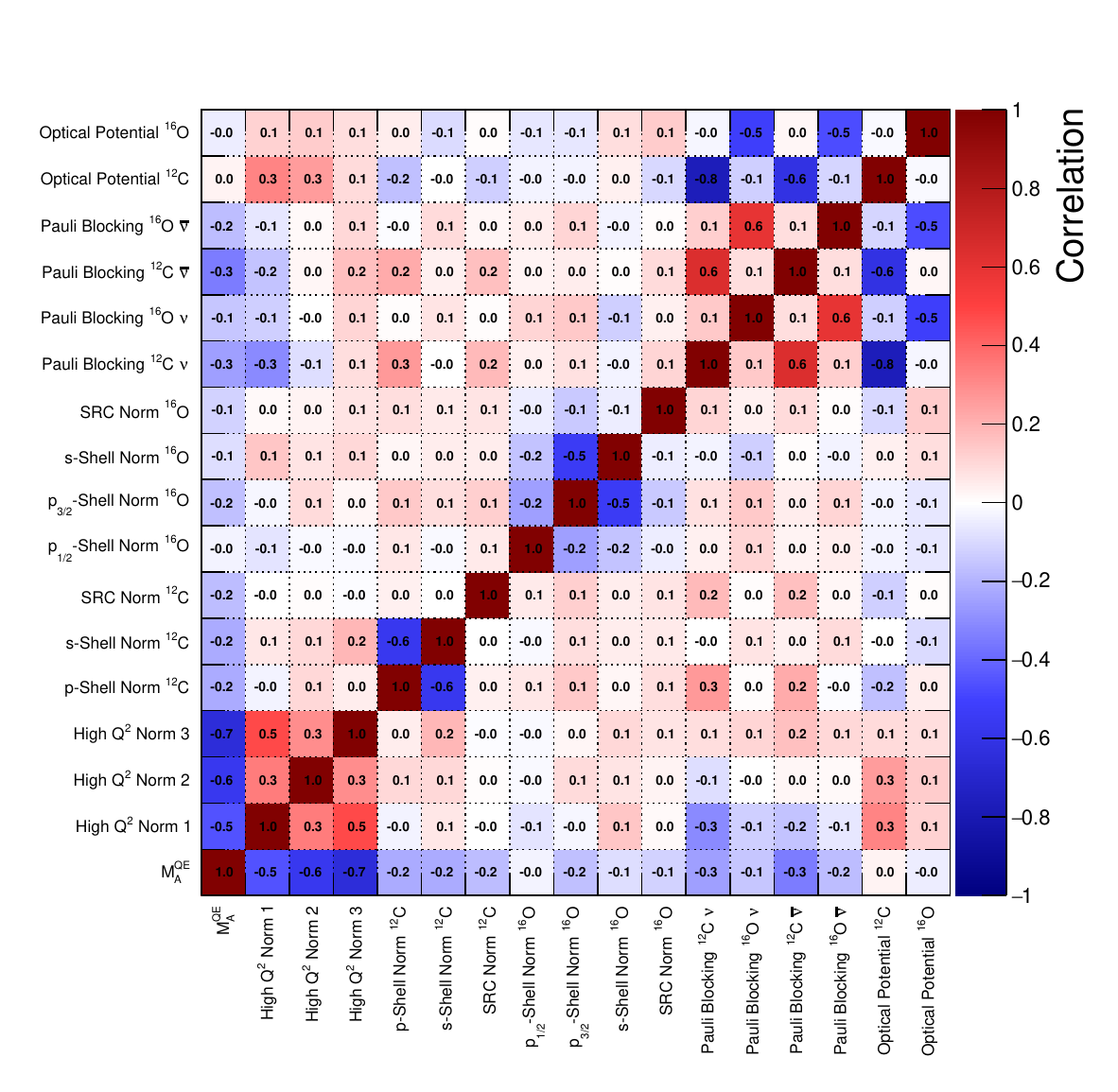}
 }
 \caption{Posterior correlation from matrix from MCMC for CCQE parameters.}
 \label{Fig::CorrMatrixCCQE}
\end{figure}

\autoref{Fig::CorrMatrixCCQE} presents the posterior correlation sub-matrix for CCQE parameters only. First of all, there are strong anti-correlations between the $M^{QE}_{A}$ and the $Q^2$ normalization parameters, which is expected as they both predominantly affect the same region of kinematic phase space (high $Q^2$). Given that $Q^2$ parameters serve as extensions to the dipole model, this might suggest that normalization of the single nucleon cross section is well constrained, but the shape as a function of $Q^2$ retains significant uncertainty even after the fit. 

Secondly, there are very strong anti-correlations among shell normalization dials, as expected, since both strongly affect the normalization of the total cross section. The anti-correlations, therefore, indicate a relatively constrained cross-section normalization but significant uncertainty in the exact shape of the spectral function. Lastly, the strong anti-correlation between Pauli blocking and optical potential parameters is noteworthy. Both sets of parameters affect low energy transfer regions in a similar way, resulting in such high anti-correlation. A complete degeneracy is avoided since these parameter sets modify the lepton kinematics within the zero-proton samples in slightly different ways (Pauli blocking is a pure suppression of the event rate at forward outgoing lepton angles, whilst the optical potential parameters tend to shift strength from forward angles to larger angles, as indicated by~\autoref{fig:q0q3_OP}). These observations are consistent with fits to cross-section measurements presented in Ref.~\cite{Chakrani:2023htw}.

\begin{figure}[htb]
 \centering 
 \subfloat{
  \includegraphics[page=1, width=\columnwidth]{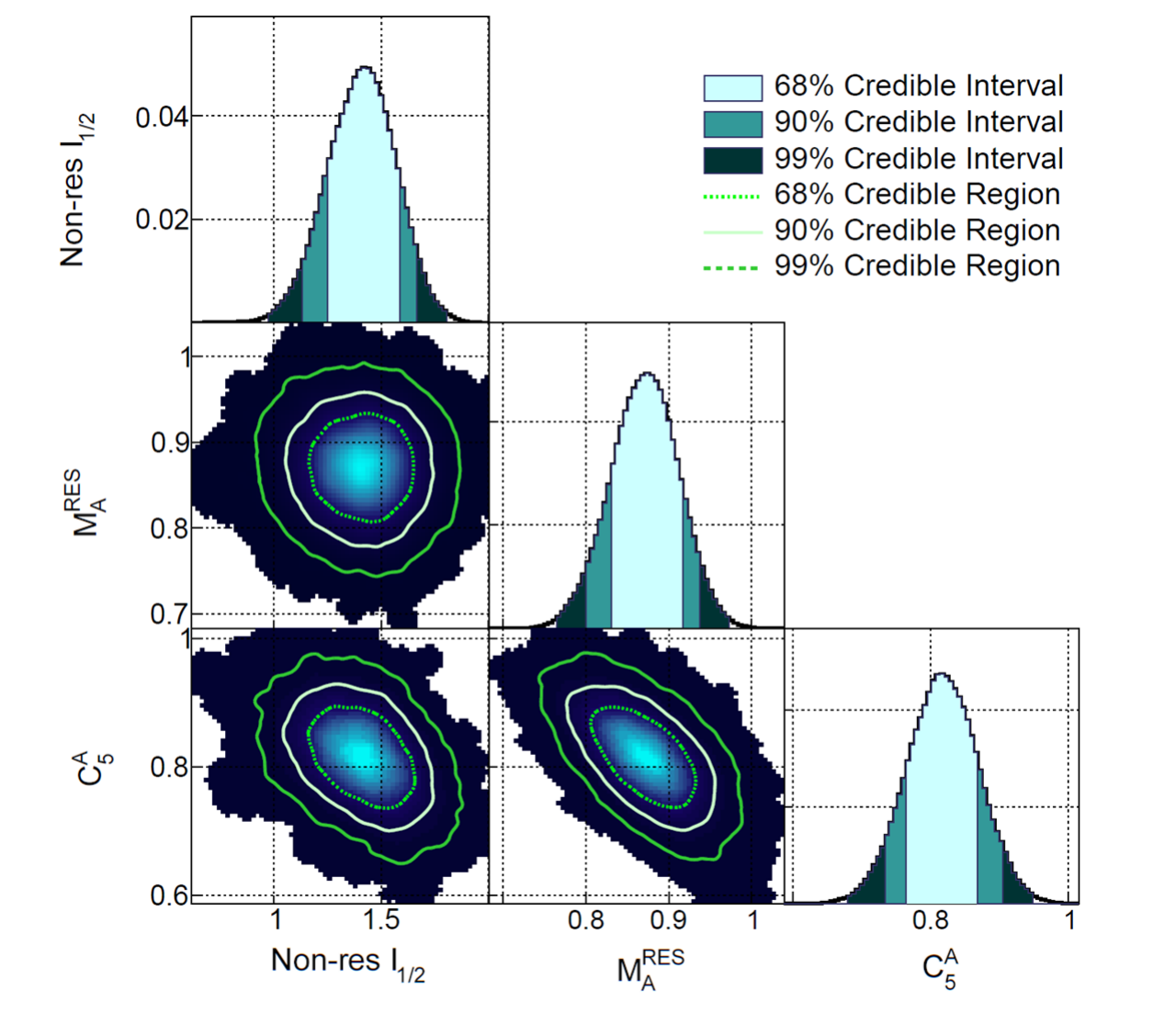}
 }
 \caption{Individual posterior distribution and joint distributions for Single Pion Production parameters with marked credible regions and intervals.}
 \label{Fig::TrianglePlot}
\end{figure}

\begin{figure}[htb]
 \centering 
  \includegraphics[page=4, width=\columnwidth]{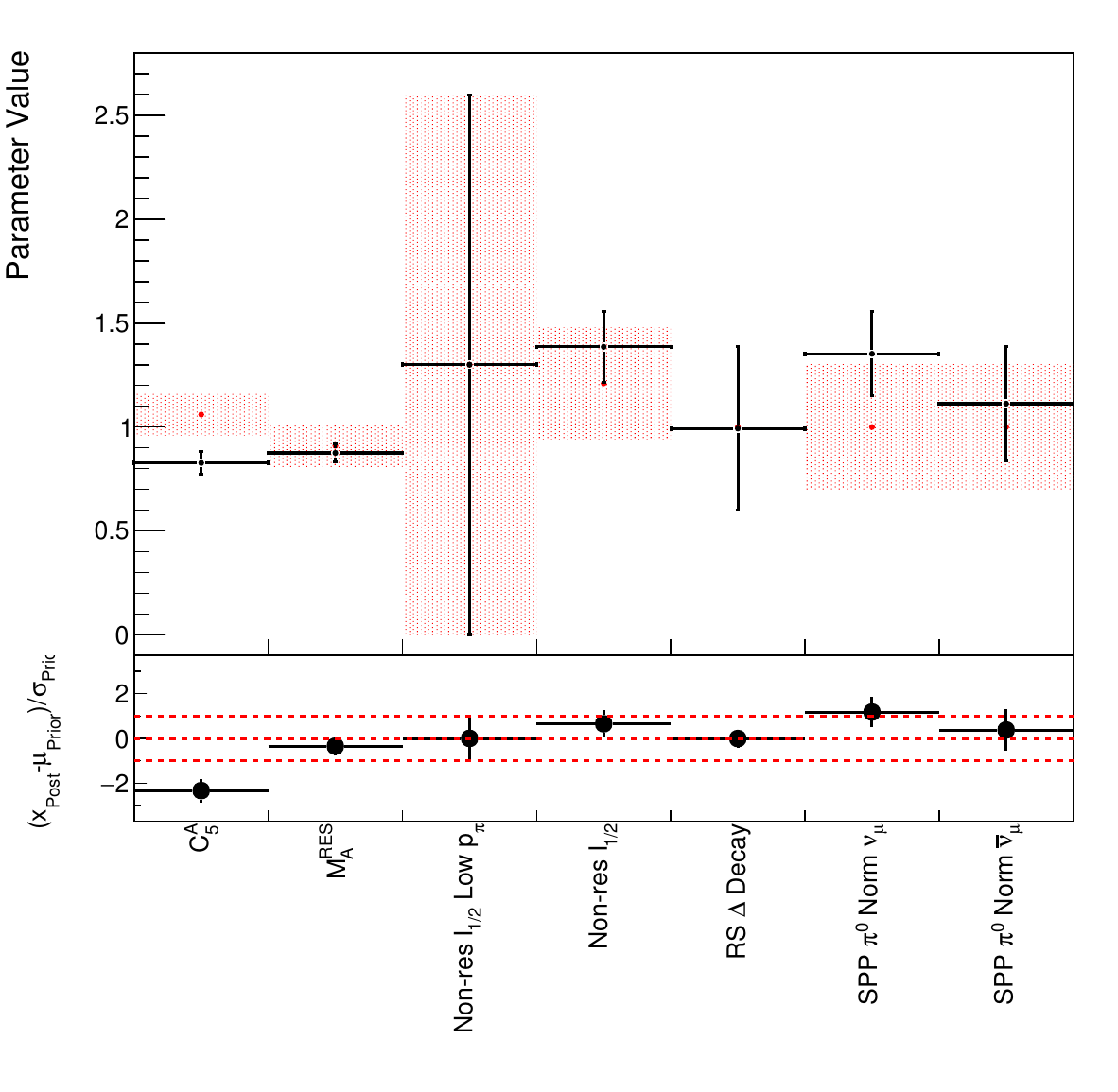}
 \caption{Postfit values and uncertainties (black) for SPP cross-section parameters from the data fit, compared with the corresponding prefit values and uncertainties (red).}
 \label{Fig::DataFit_1D_Xsec_SPP}
\end{figure}

\autoref{Fig::TrianglePlot} depicts individual 1D and 2D posteriors between selected single $\pi$ production parameters. We note an anti-correlation between the form factor parameters $C^{A}_{5}$ and $M_A^\textrm{RES}$, as well as between {\tt non-res $I_{1/2}$} and $C^{A}_{5}$, but a very weak correlation between {\tt non-res $I_{1/2}$} and $M_A^\textrm{RES}$. This is interesting because in the new SPP tune, the prior correlation between $C^{A}_{5}$ and $M^{RES}_{A}$ was reduced from $\sim$-80$\%$ to $\sim$-10$\%$ with respect to the previous analysis~\cite{T2K:2023smv}. In the previous analysis, the anti-correlations were mostly driven by prior, but in this analysis, it is clear that the strong anti-correlations are mostly driven by T2K data. This correlation arises because of the parametrization of $C^{A}_5(Q^2)$, which depends on both $C^{A}_5(Q^2=0)$ and $M_A^{RES}$, with $C^{A}_5(Q^2=0)$ largely controlling the normalization of the cross section, and $M_A^{RES}$ controlling the shape in $Q^2$. Within the new tune, anti-correlations between {\tt non-res $I_{1/2}$} and $C^{A}_{5}$ have been reduced from $\sim$-30\% to $\sim$-3\% in this new analysis. The posterior correlation is equal to $\sim$-30\%, which is a very similar value to the previous analysis, despite the great change in prior correlation, once again indicating correlations are driven by T2K data and not by prior correlations. Concerning {\tt non-res $I_{1/2}$} and $M_A^\textrm{RES}$ correlations, both prior and posterior are consistent with previous results, and the latter one is \text{$\sim$-10\%}. For the full picture, we include SPP parameters with prior values in \autoref{Fig::DataFit_1D_Xsec_SPP}.

\begin{figure}[htb]
 \centering 
  \includegraphics[page=6, width=\columnwidth]{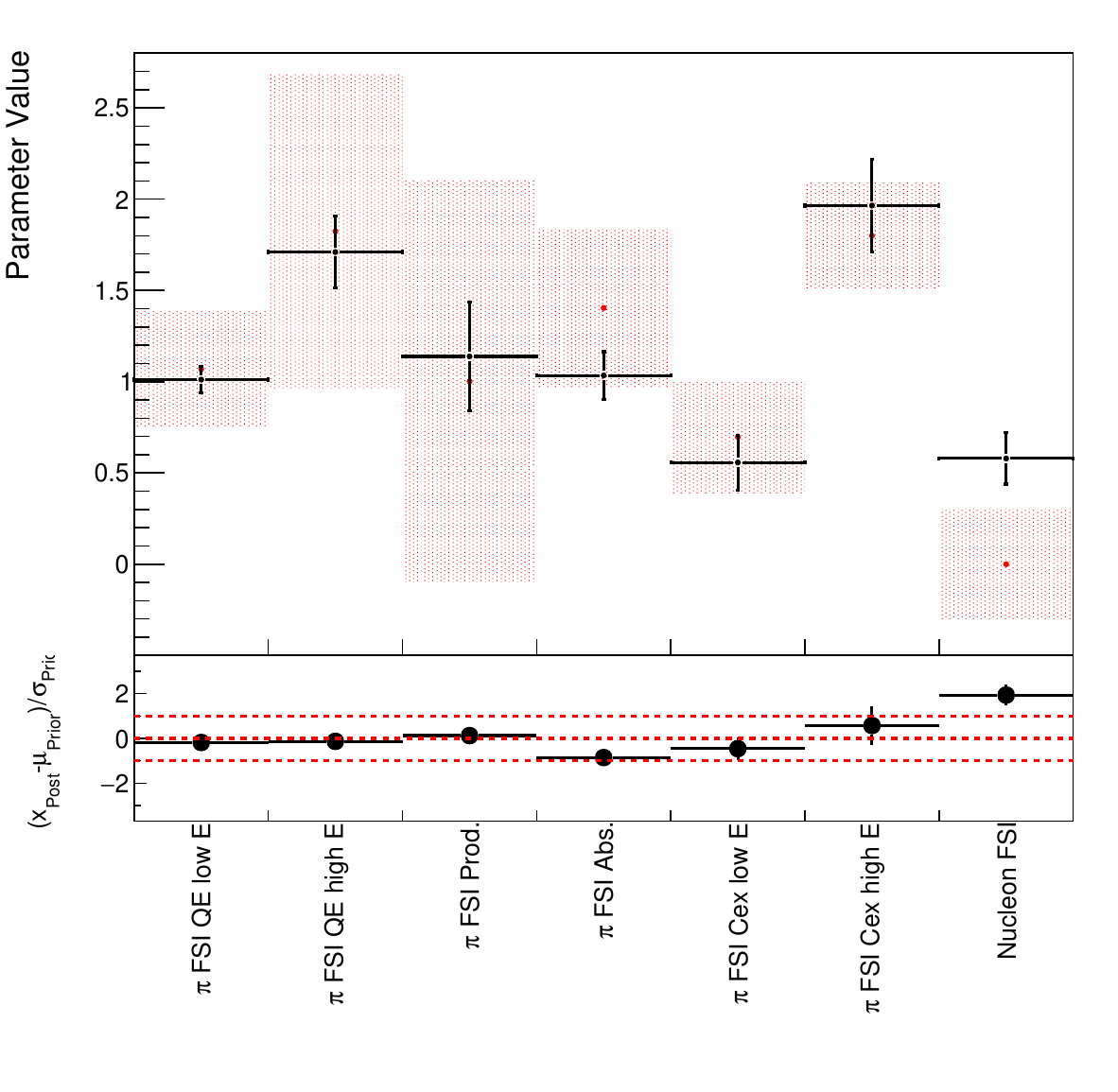}
  \includegraphics[page=7, width=\columnwidth]{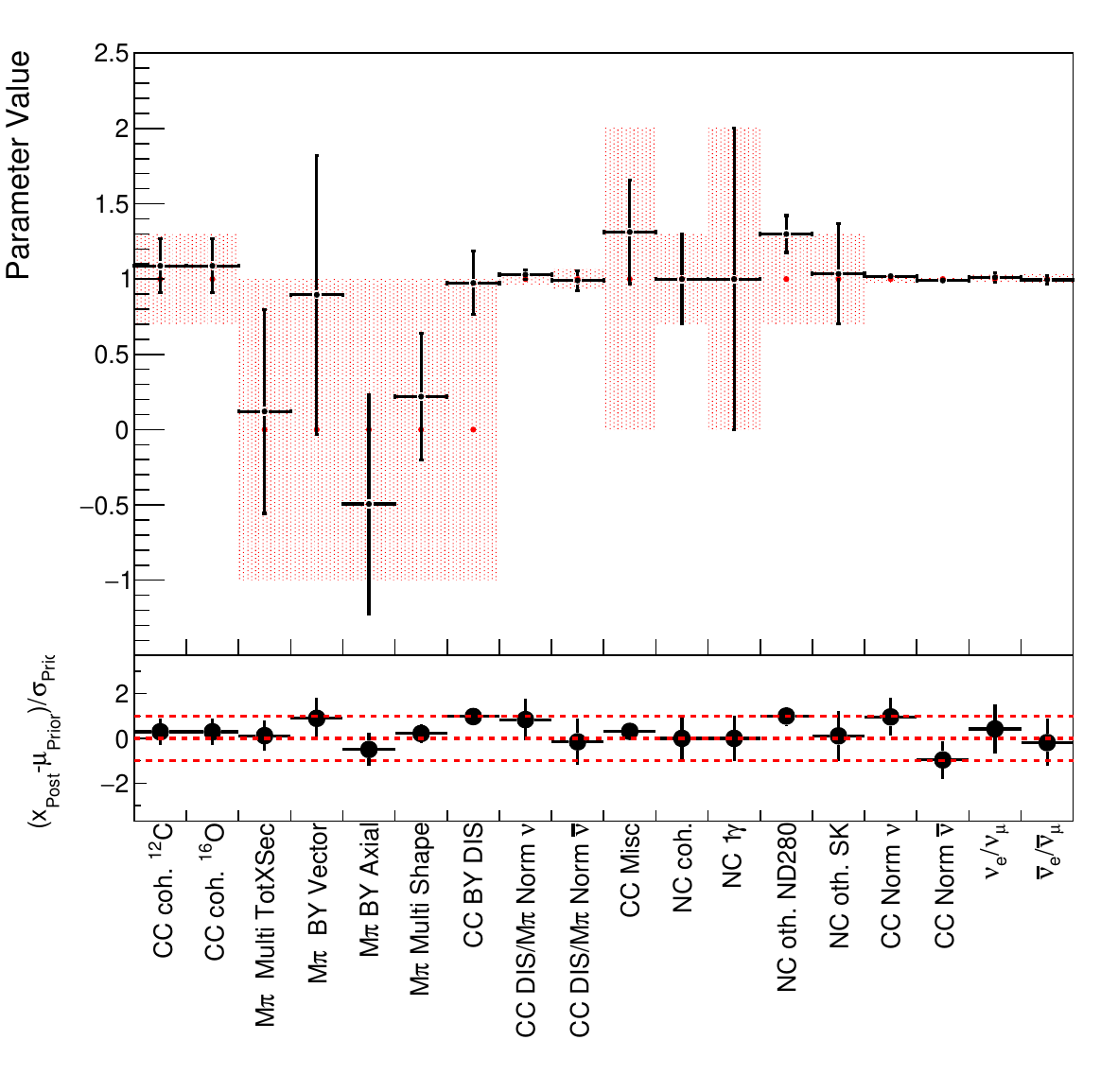}
 \caption{Postfit values and uncertainties (black) for FSI (top) and other cross-section parameters (bottom) from the data fit, compared with the corresponding prefit values and uncertainties (red).}
 \label{Fig::PostFit_Other}
\end{figure}

The postfit values of the final state interaction parameters are presented in \autoref{Fig::PostFit_Other}, most of $\pi$ FSI are consistent with external fits to FSI~\cite{PinzonGuerra:2018rju} except for the absorption parameter, which is consistent with the prior uncertainty. The newly added nucleon FSI is strongly pulled up, increasing the probability of nucleon FSI happening, thereby lowering proton momenta and making it less likely for protons to be reconstructed in ND280. This is likely an important handle for the fit to correct the initial poor agreement seen in the CC0$\pi$-0p-0$\gamma$ sample (seen in \autoref{Fig::NDSamplesProton}).

\autoref{Fig::PostFit_Other} presents postfit values of DIS and other cross-section parameters. Most parameters are poorly constrained and comparable with their prior values at the one-sigma level. The two exceptions are the {\tt CC BY DIS} being strongly pulled toward a value of one, indicating that the Bodek-Yang correction to the cross section is largely disabled, and the {\tt NC Other} normalization being pulled to increase NC events in the ND280 selections by $\sim30$\%. The latter is consistent with a previous analysis~\cite{T2K:2023smv}, where a similar value was preferred even in the presence of other charged-current models. Note that in the ND280 analysis, there are no dedicated NC samples, and this constitutes a change to a small background.

\begin{figure}[htb]
 \centering 
 \subfloat{
  \includegraphics[page=9, width=\columnwidth]{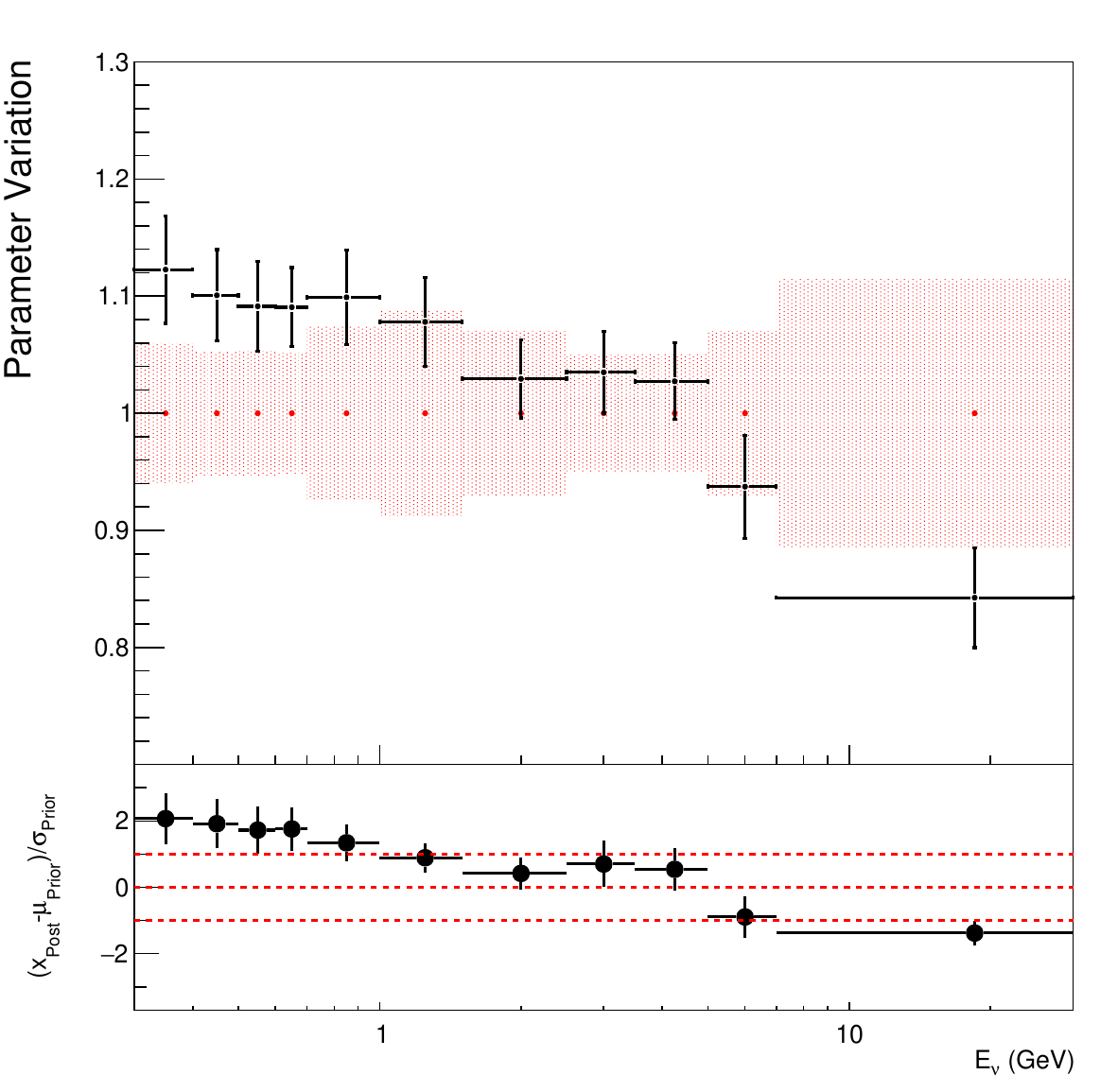}
 }
 \caption{Postfit values and uncertainties (black) describing the flux of $\nu_{\mu}$ in $\nu$ beam mode, obtained from the data fit, compared with the corresponding prefit values and uncertainties (red).}
 \label{Fig::PostFit_Flux}
\end{figure}

\autoref{Fig::PostFit_Flux} presents prefit and postfit values for $\nu_{\mu}$ flux parameters in $\nu$ beam mode. One can observe there is a pull of about 10\% for neutrino energies below 1 GeV. This effect decreases as the energy increases. For energies above $E_{\nu} =$ 4 GeV, the postfit values fall below the priors. Therefore, the normalization of events is increased for low $E_{\nu}$ and is decreased for high $E_{\nu}$. Overall, this represents a significant reshaping of the flux. 
It is unclear whether this shape arises from the internal structure of the flux correlation matrix or from missing degrees of freedom in the cross-section part of our model. Nevertheless, a similar shape has been observed in previous analysis~\cite{T2K:2023smv}, despite differences in both the cross-section model and the flux matrix (see \autoref{sec:fluxpred}).

\begin{figure}[htb]
 \centering 
 \subfloat{
  \includegraphics[page=1, width=1.0\columnwidth]{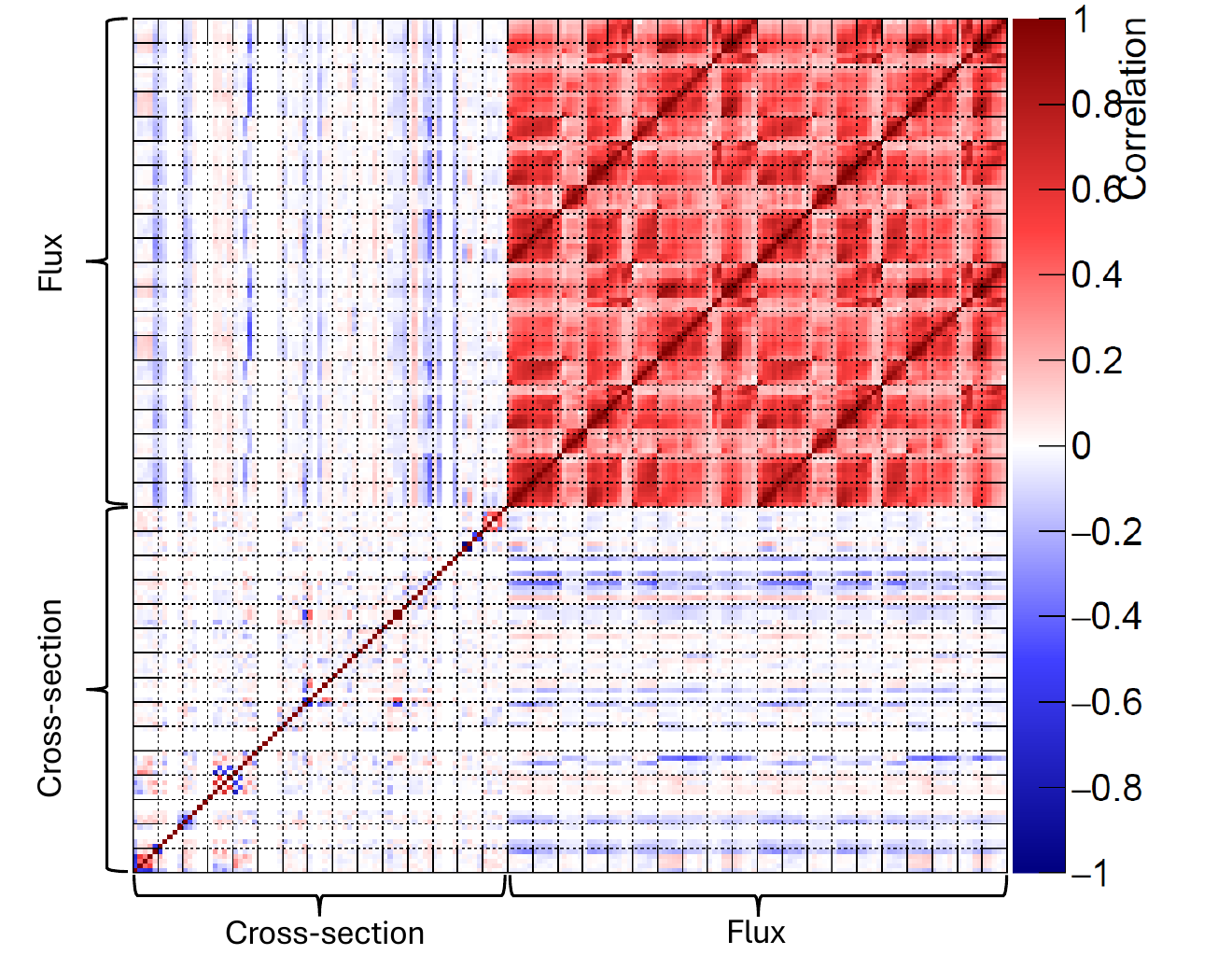}
 }\
 \caption{Posterior correlation matrix of cross-section and flux parameters.}
 \label{Fig::PosteriorMatrixCropped}
\end{figure}

\autoref{Fig::PosteriorMatrixCropped} depicts a matrix representing the correlation between cross-section and flux parameters. This matrix was derived from an MCMC chain by calculating the correlation factors from the two-dimensional marginalized posterior distributions. The flux matrix has a very strong correlation, which is mostly driven by NA61/SHINE replica target tuning encoded in the prior constraint. It is important to notice high anti-correlations between the flux and the cross-section parameters. The flux parameters act as a normalization as a function of $E_{\nu}$, thus anti-correlating with many cross-section uncertainties, especially those strongly affecting overall cross-section normalization.

\autoref{Tab::PriorPostErrorMode} presents the fractional uncertainties in the total number of selected events for various reaction modes, arising from different sources of uncertainty (cross section, flux, and ND280 detector). It can be seen that cross-section systematic uncertainties dominate the total uncertainty. Furthermore, one can observe that the prior uncertainty for 2p2h processes is significantly larger than that for CCQE interactions. This significant difference is due to the prior uncertainties being dominated by the normalization uncertainty in 2p2h processes. The uncertainty for CC 1$\pi^{0}$ interactions is larger than CC 1$\pi^{\pm}$, where most of the parameters are 100\% correlated. This difference stems from the $\pi^{0}$ normalization uncertainty. At higher energies, the prior uncertainty for DIS processes is much larger before the fit than for multi-$\pi$ interactions. After the fit, this trend reverses. 

The prior flux uncertainties are highest for DIS interactions, as flux uncertainties are highest at high neutrino energies, where DIS Processes are more important. However, after the fit, the uncertainty becomes much more similar across interactions (around 3\%), due to strong correlations within the flux model. For ND280 detector systematics, one can see that the prior uncertainty is higher for CC 1$\pi^{\pm}$ followed by CC DIS. This is mostly driven by uncertainties in pion secondary interactions.

\begin{table*}[htb!]								
\centering								
\begin{tabular}{ l | cc | cc | cc  | cc }  
\hline\hline
\multirow{3}{*}{\textbf{Mode}}  	& \multicolumn{8}{c}{$\delta N / N (\%)$} \\ 
\hline
    & \multicolumn{2}{c|}{\textbf{Cross section}} & \multicolumn{2}{c|}{\textbf{Flux}} &	\multicolumn{2}{c |}{\textbf{ND280}} &	\multicolumn{2}{ c}{\textbf{Total}} \\ 
\hline
 & Prefit	& Postfit & Prefit	& Postfit & Prefit	& Postfit & Prefit	& Postfit	\\ 
 \cline{2-9}	
CCQE &	 12.2 &	 3.4 &	 4.7 &	 2.6 &	1.9 &	 1.1 &	13.2 &	 2.0 \\
2p2h &	 52.3 &	 12.3 &	 4.7 &	 2.7 &	2.2 &	 1.2 &	50.5 &	 12.2 \\
CC 1$\pi^{\pm}$  &	 14.6 &	 4.0 &	 5.0 &	 2.7 &	 2.6 &	 1.2 &	15.7 &	3.7 \\
CC 1$\pi^{0}$ &	 28.7 &	 12.3 &	 5.0 &	 2.8 &	 2.3 &	 1.3 &	31.4 &	11.1 \\
CC multi-$\pi$ &	 8.1 &	 4.8 &	 4.6 &	 2.6 &	 2.4 &	 1.4 &	9.7 & 4.1 \\
CC DIS &	 17.3 &	 3.6 &	 5.6 &	 2.8 &	 2.4 &	 1.5 &	19.3 &	 3.0 \\
\hline\hline
\end{tabular}
\caption{Prefit and postfit uncertainties on the number of events for selected interaction modes arising from various uncertainty sources. Uncertainties were calculated using all events that passed ND280 selections.}
\label{Tab::PriorPostErrorMode}
\end{table*}

In addition, in \autoref{tab:cross_section} we report how much the T2K flux integrated cross section for each process has changed after the fit. For example, we can observe a factor of 1.22 increase for the CCQE cross section on CH, which should not be a surprise after seeing the pull in $M^{QE}_{A}$. Another interesting example is the difference between CC 1$\pi^{\pm}$ and  CC 1$\pi^{0}$. Both channels are being pulled down, for example, by $C^{A}_{5}$; however, there are dedicated normalizations for $\pi^{0}$ production, which increase the cross section.

The change in the DIS cross section for antineutrino interactions is particularly striking. This comes from the modification to the Bodek-Yang correction strength, where the ND280 tune turns the correction off almost entirely. Due to the different structure and kinematics of neutrino and antineutrino interactions, this is much more impactful in the latter than in the former. This parameter is fully correlated between neutrinos and antineutrinos, and it is likely that the statistically dominant neutrino samples at ND280 are driving this change. In future analysis, we will study the potential to decorrelate the corrections associated with neutrino and antineutrino interactions.

\begin{table}[ht]
    \centering
    \begin{tabular}{l | c | c | c | c}
        \toprule
        \hline\hline
        Process & CH $\nu$ & CH \nub & H$_2$O $\nu$ & H$_2$O \nub\\
        \hline        
        CCQE            & 1.22 & 1.06 & 1.30 & 1.15\\
        2p2h            & 1.30 & 1.07 & 1.29 & 1.06\\
        CC 1$\pi^{\pm}$ & 0.87 & 1.07 & 0.86 & 1.07\\
        CC 1$\pi^{0}$   & 1.27 & 1.25 & 1.26 & 1.21\\
        CC multi-$\pi$  & 1.08 & 1.01 & 1.08 & 1.01\\
        CC DIS          & 1.22 & 2.13 & 1.22 & 2.11\\
        \hline\hline
        \bottomrule
    \end{tabular}
    \caption{T2K flux integrated cross-section scaling for different processes after applying ND280 tuning.}
    \label{tab:cross_section}
\end{table}

\begin{figure*}[htb]
 \centering 
  \includegraphics[page=1, width=0.32\textwidth,trim=3mm 2mm 9mm 15mm, clip]{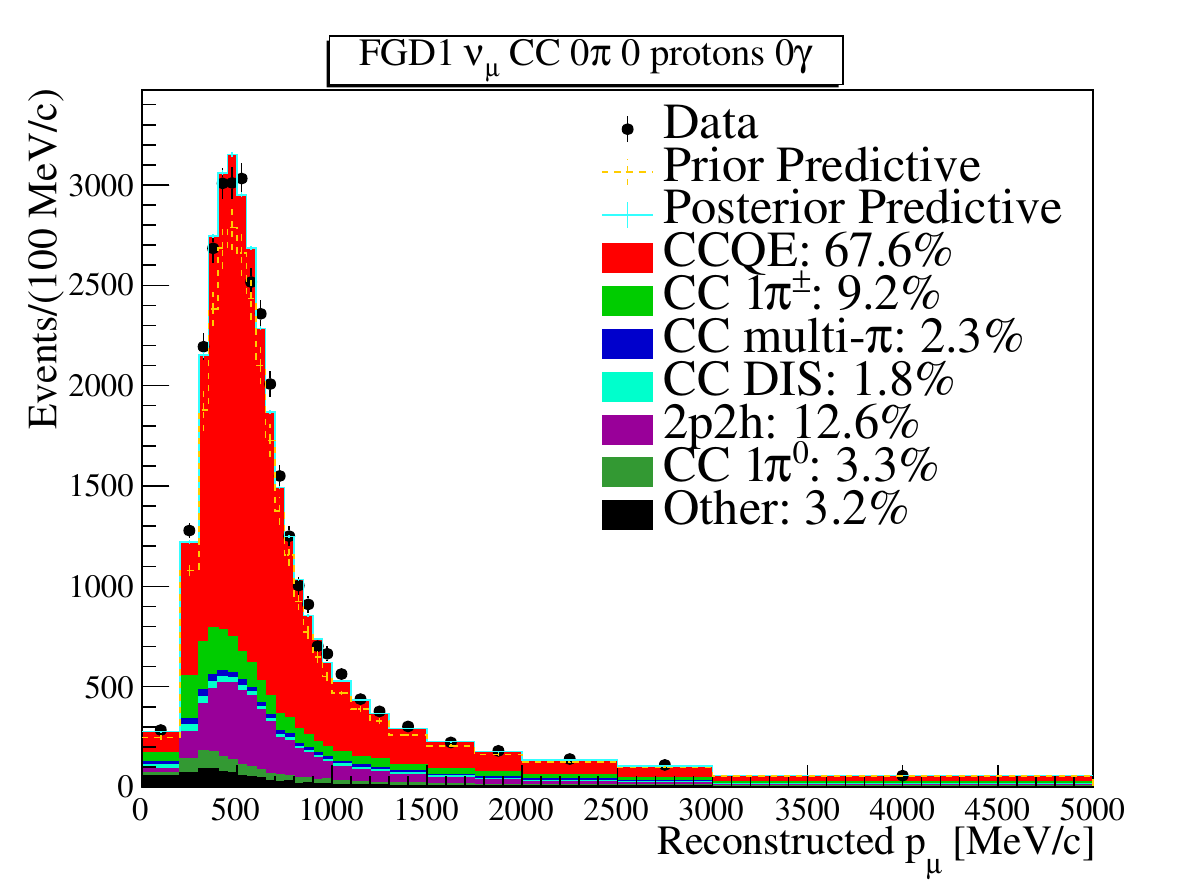}
  \includegraphics[page=2, width=0.32\textwidth,trim=3mm 2mm 9mm 15mm, clip]{Figures/DataFitResults/PosteriorPredictive_Distribution_NoGrid.pdf}
  \includegraphics[page=14, width=0.32\textwidth,trim=3mm 2mm 9mm 15mm, clip]{Figures/DataFitResults/PosteriorPredictive_Distribution_NoGrid.pdf}
  \caption{The data and predictions for the FGD1 $\nu_\mu$ \cczeropizerop (left), FGD1 $\nu_\mu$ \cczeropiNp (center) and FGD2 \numb CC$0\pi$ (right) selections, shown as a function of the reconstructed momentum of the muon candidate. The prediction after the fit (``posterior predictive'') is broken down by interaction channel, and the overall prediction before the fit (``prior predictive'') is overlaid.}
 \label{Fig::PosteriorPredictive_Distribution}
\end{figure*}

The predicted and observed events in the FGD1 $\nu_\mu$CC0$\pi0p$, FGD1 $\nu_\mu$CC0$\pi Np$, and FGD2 $\numb$CC0$\pi$ samples are projected onto $p_{\mu}$ in \autoref{Fig::PosteriorPredictive_Distribution} before and after the fit to data. 
Before the fit, there is a notable under-prediction in the peak region. In contrast, an under-prediction is not observed for the equivalent antineutrino selections, suggesting modeling issues specific to the neutrino selection. After the analysis, the discrepancy is resolved, showing how the flux and interaction model is able to accommodate different modifications to neutrino and antineutrino event rates. Overall, the prediction after the fit is in good agreement with the data for both selections. Furthermore, the uncertainties in the event counts become significantly smaller, as shown in \autoref{Tab::PriorPostErrorMode}.
The feature of increasing the rate of $\nu_\mu $CC0$\pi 0p$ events whilst keeping the rate of $\nu_\mu$ CC0$\pi Np$ and \nub CC$0\pi$ events roughly constant showcases the ability of the model to modify and constrain neutrino and antineutrino interactions separately. Overall, the \textit{p}-value is between $\sim$67\% and $\sim$95\% for the neutrino samples, while is $\sim$22\% ($\sim$10\% in FGD2) for the antineutrino sample (see \autoref{tab:likelihoods}). The slightly lower \textit{p}-value in the antineutrino case is compatible with statistical fluctuations, but could also suggest missing degrees of freedom in the fit. Further investigation would be needed to confirm this argument.

As discussed, the main analysis bins events in muon kinematics, separated by proton, photon, and pion multiplicities, and the charge of the outgoing muon. Nonetheless, these results can be used to predict proton and pion kinematic variables and to compare with observed data. This benchmark aspects of the neutrino interaction modeling, notably the correlations between the leptonic and hadronic systems. It also highlights regions that may need improvement for future analyses that utilize hadron kinematics, for instance, in neutrino energy reconstruction or improving the separation of neutrino interaction modes.

\begin{figure}[htb]
 \centering 
  \includegraphics[page=9, width=\columnwidth, trim=0mm 0mm 9mm 21mm, clip]{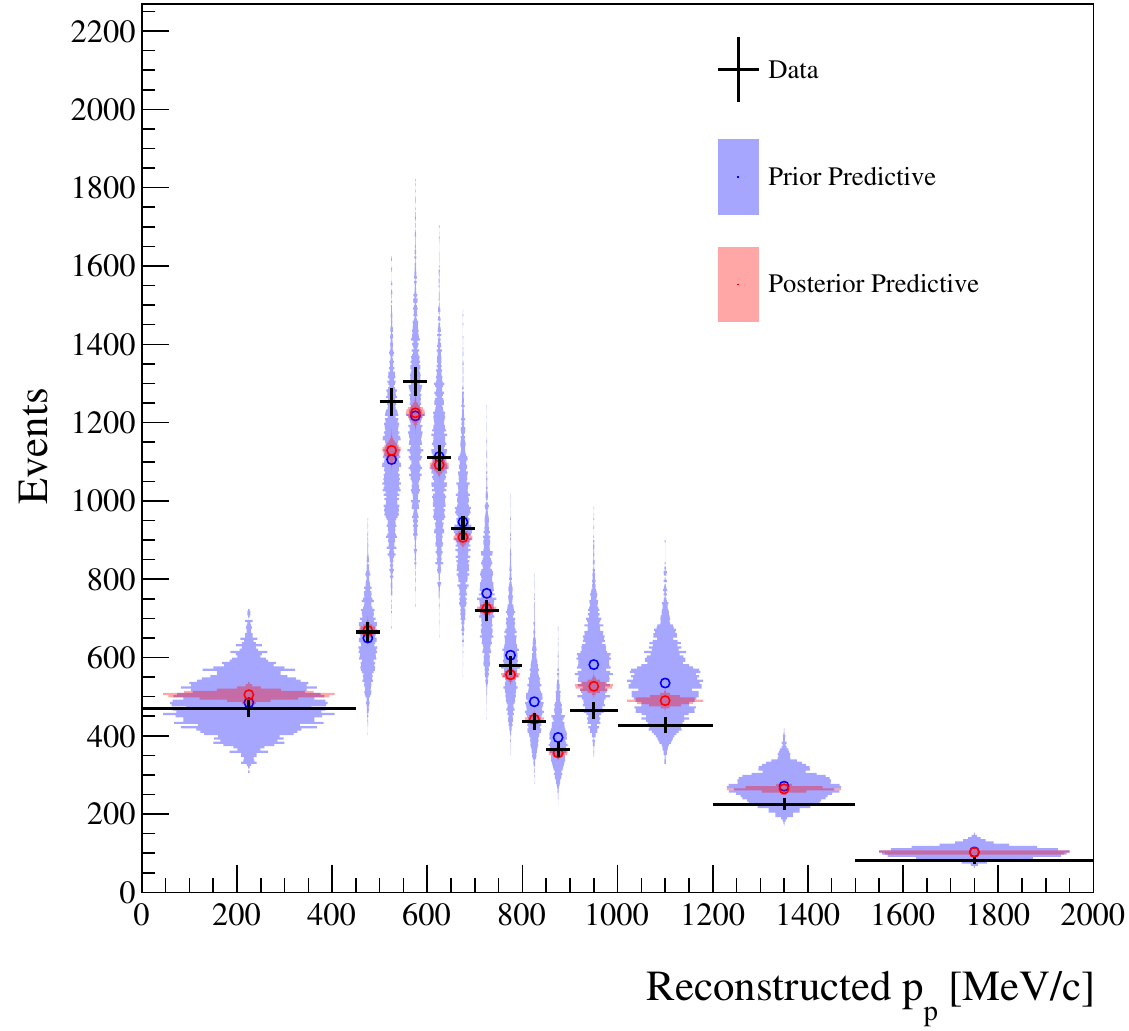}
 \caption{The reconstructed momentum of the highest momentum proton candidates in the FGD1 \cczeropiNp sample, showing the predicted distributions, including all uncertainties, before and after the fit in muon kinematics. Similar results were found for FGD2.}
 \label{Fig::ProtonKinemPosteriorPredictive}
\end{figure}

\autoref{Fig::ProtonKinemPosteriorPredictive} shows the distributions of the reconstructed highest momentum proton candidates' momentum in FGD1 for both FGD-TPC and FGD-contained protons, obtained by sampling the prior and posterior distributions from the ND280-only data fit in muon kinematics, as detailed in the previous section. The detector resolution for proton momenta above 400 MeV/c is around 15\%, which is sufficient to observe the improved agreement between the posterior predictive distributions and the data relative to the prior predictive distributions. Even though the proton kinematics were not used in the fit, the posterior predictive distributions are much closer to the data than the prior predictive distribution. This shows that the separation in proton multiplicities allows some probing of the proton kinematics. The same observation also supports that the uncertainty model has predictive power to capture at least some of the correlation between lepton and proton kinematics, although not the finer details, as demonstrated in \autoref{subsec:benchmark}. 

\begin{figure}[htb]
 \centering 
  \includegraphics[page=11, width=\columnwidth, trim=0mm 0mm 9mm 21mm, clip]{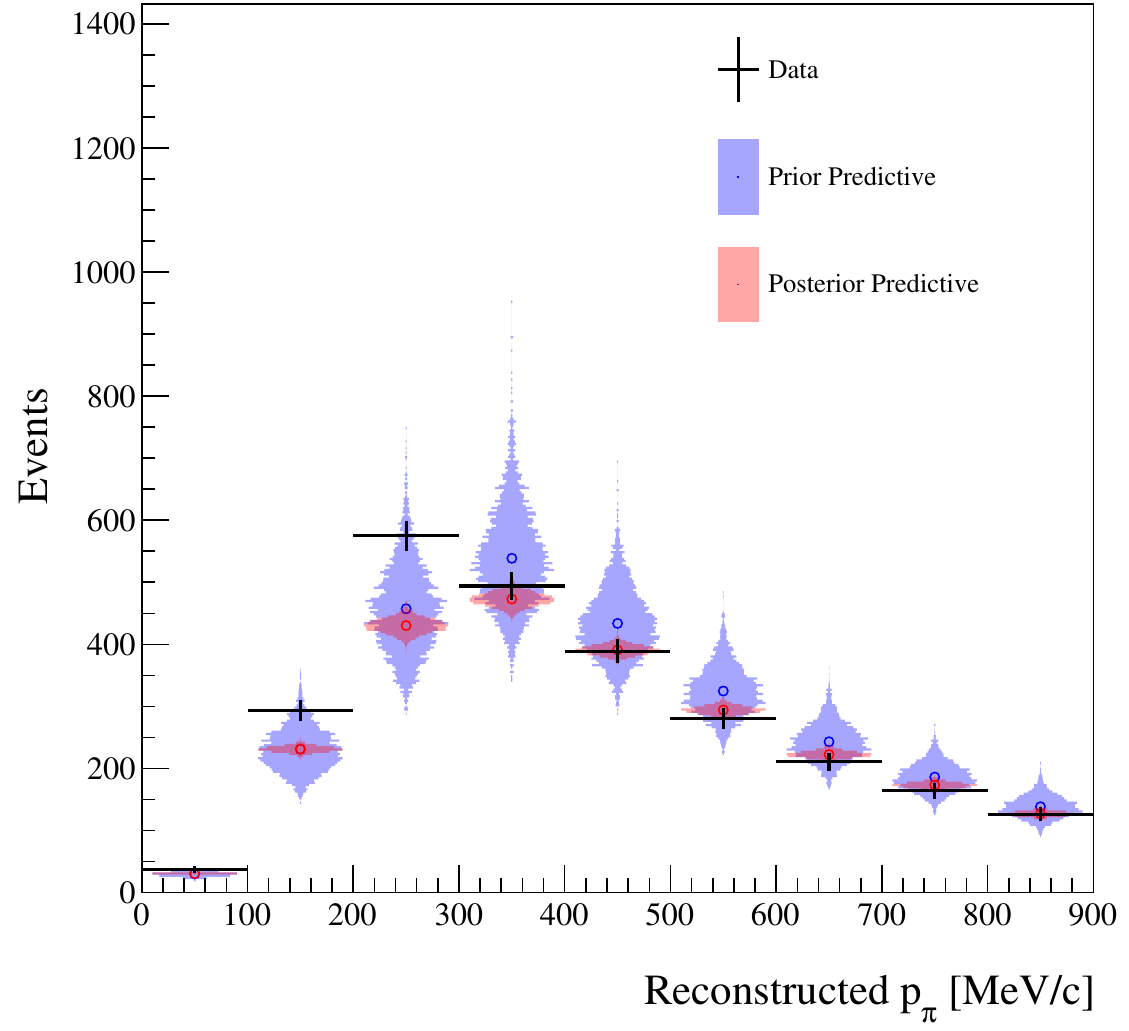}
 \caption{Reconstructed momentum of the leading pion candidate in the FGD1 $\nu_\mu$ \cconepi sample for FGD-TPC and FGD-contained pions, showing the predicted distributions, including all uncertainties, before and after the fit in muon kinematics. Similar results were found for FGD2.}
 \label{Fig::Pion}
\end{figure}

\autoref{Fig::Pion} presents the highest momentum pion candidate's reconstructed momentum for the FGD1 CC$1\pi$ sample, showing the FGD and TPC-tagged pions, due to the Michel-tagged pions not having a momentum estimator for this analysis. Above 300~MeV/c, the prediction after the near-detector analysis describes the data significantly better, and the uncertainty band narrows considerably. Below 300~MeV/c, there is a significant underestimation of the data, which does not improve after the near-detector analysis. A similar low-momentum excess has been observed in T2K's far-detector sample with one electron-like ring and one decay electron~\cite{T2K:2023smv} and FD's atmospheric sub-GeV sample that also has one electron-like ring and one decay electron~\cite{T2K:2024wfn}. 

Importantly, \autoref{Fig::Pion} only shows FGD-TPC and FGD contained pions, whereas for Michel-tagged pions---whose momentum is typically $<200~\text{MeV}/c$---there is instead an over-estimate of the data. Complicating this further, the accuracy of the particle identification of tracked charged pions at low momentum worsens significantly. Nonetheless, accurately modeling muon-pion correlations and developing adequate model freedom in the low pion momentum region will need further study before such features are exploited in the analysis.

\subsection{Benchmarking the near-detector analysis with cross-section measurements}
\label{subsec:benchmark}
As a benchmark for the near-detector analysis, we compare the cross-section prediction from the neutrino interaction model before and after the near-detector fit to a variety of T2K cross-section measurements. This section presents a subset of comparisons with more available in \aref{app::more_plots}. The list of cross-section measurements analyzed in this section, and their corresponding $\chi^2 / \mathrm{dof}$, can be found in \autoref{Tab::benchmark_chi2}.

Neutrino interaction events are generated with NEUT 5.6.2~\cite{Hayato:2009zz, Hayato:2021heg}\footnote{This is different from \autoref{subsec:base_neut_int_model} due to software compatibilities, but the physics in these NEUT versions were validated as identical.}. NUISANCE~\cite{Stowell:2016jfr} is used for the comparisons to data, and interfaces with T2K software to reweight NEUT predictions following parameter variations. The near-detector analysis includes 75 cross-section parameters, of which 63 are used in this study. The unused parameters are either those that are expected to have an almost negligible impact on the cross sections considered or are parameters that are expected to only affect a small portion of the variance within the uncertainty model. The parameters are: the normalization for CC resonant single photon, kaon and eta production, combined with the normalization of CC diffractive; the normalization of NC coherent events; the normalization of NC single photon production; the normalization of NC diffractive and resonant single eta production; the electron (anti)neutrino scaling parameters; and the CCQE removal energy parameters\footnote{CC Misc, NC coh, NC 1$\gamma$, NC other near, NC other far, $\nue / \num$, $\nueb / \numb$, $\Delta E_{rmv}$$^{12}\mathrm{C} \, \nunu$, $\Delta E_{rmv}$$^{12}\mathrm{C} \, \nub$, $\Delta E_{rmv}$$^{16}\mathrm{O} \, \nunu$, $\Delta E_{rmv}$$^{16}\mathrm{O} \, \nub$.}.

A comparison of the NEUT predictions to cross-section measurements is made with parameters fixed at their values before (``prefit'') and after (``postfit'') the near-detector analysis. In each case, a $\chi^2$ test-statistic is calculated between the predicted and measured cross sections using the full covariance matrices provided with the measurements. However, the uncertainty on the prefit and postfit parameters is not considered. Therefore, whilst good agreement between a model and measurement demonstrates compatibility, a poor agreement does not necessarily imply the opposite. Fits to some of the cross-section measurements shown here are available in Ref.~\cite{Chakrani:2023htw} using a small subset of the cross-section parameters, alleviating this issue.

\begin{figure}[hptb]
    \centering
        \includegraphics[ width=\columnwidth]{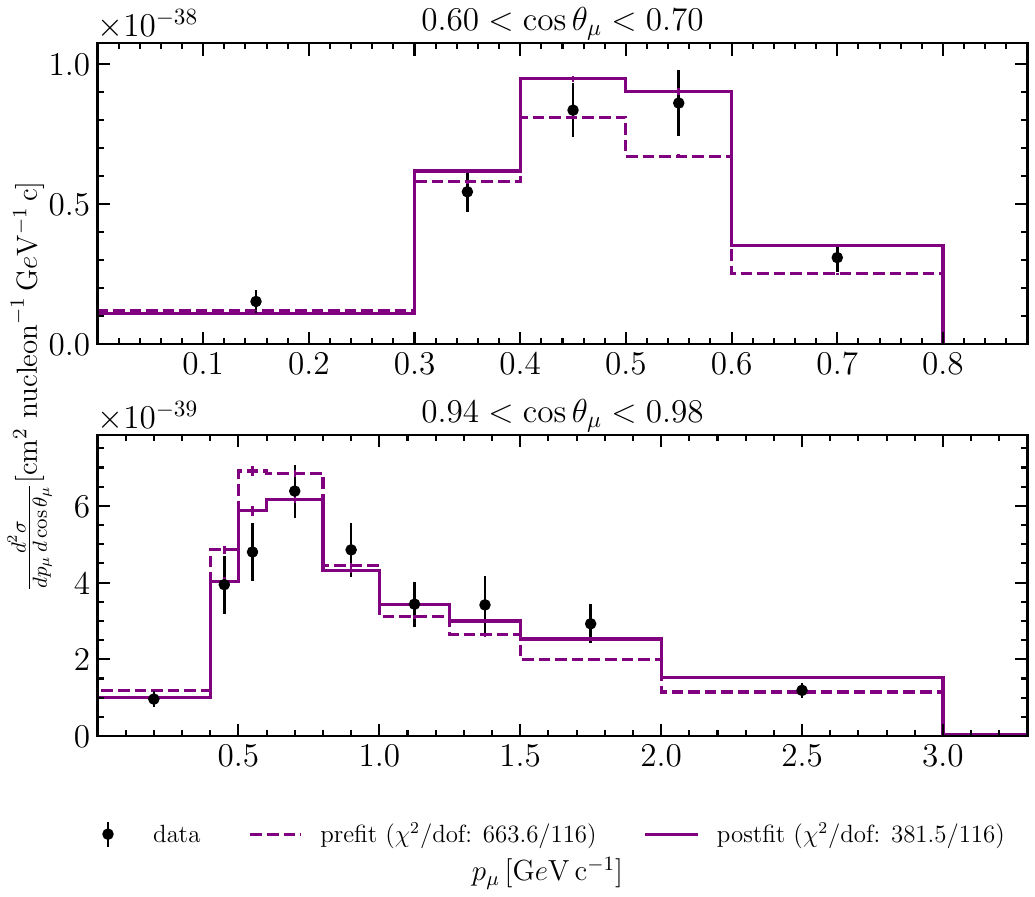}
        \includegraphics[ width=\columnwidth]{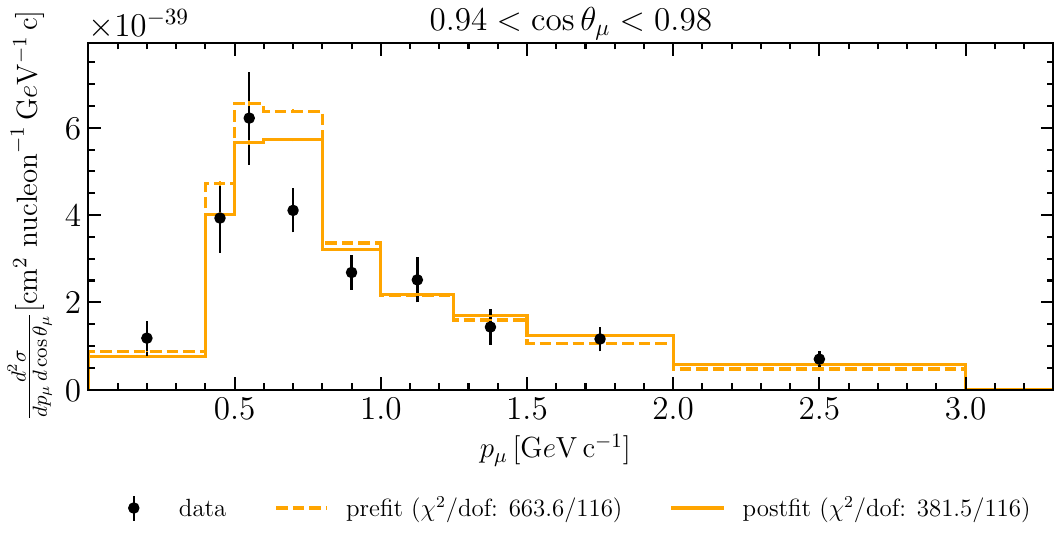}
    \caption{Predictions from the interaction model for selected slices of the differential cross sections for T2K's joint \num-\numb CC0$\pi$ measurement on a CH target~\cite{T2K:2020sbd}, before and after the near-detector fit presented in this paper. The upper panels in purple are for muon neutrinos, and the lower panel in orange is for muon antineutrinos. The reported $\chi^2/\mathrm{dof}$ is calculated for the full joint measurement.
}
    \label{Fig::BANFF_NuAnu_2Dpcos_sel}
\end{figure}

The T2K measurement of the \num CC0$\pi$ cross section, which simultaneously extracts the neutrino and antineutrino cross section on a CH target, provides a benchmark of the process that comprises the largest population of the far-detector events. Comparisons between the prefit and postfit interaction models are shown in \autoref{Fig::BANFF_NuAnu_2Dpcos_sel}, and the full comparison, including all angular panels, is shown in \autoref{Fig::BANFF_NuAnu_2Dpcos} of \aref{app::more_plots}. In the prefit case, the postfit simulation shows suppressed peaks and enhanced tails in the $0.94 < \cosmu < 0.98$ bins, bringing the model closer to the measurement. There is an enhancement of the cross section in all bins of muon momentum for $0.2 < \cosmu < 0.6$, partially driven by the changes to $M^{QE}_{A}$ and the \texttt{High $Q^2$ norm} parameters, but also enabled by the alteration to the CCQE and SRC normalizations. This also brings the postfit model closer to the measured cross section. Additionally, the bins at the most forward angles, which correspond to lower energy-transfer CCQE interactions, show an improvement in almost all momentum bins. This is due to suppression of the cross section in the postfit model, likely driven by the alterations the near-detector fit makes to the optical potential and Pauli blocking parameters.

The comparison shows a significantly improved $\chi^2$ for the postfit model, going from $\chi^2/\mathrm{dof}=663.6/116$ to $381.5/116$. However, the final $\chi^2$ remains high, demonstrating a poor agreement with the measured cross section for both the prefit and postfit models.
Breaking the $\chi^2$ down into neutrino and antineutrino contributions, the contribution from the neutrino measurement is $\chi^2/\mathrm{dof}=286.3/58$ for the prefit model, while it is $130.4/58$ for the postfit model. The contribution from the antineutrino measurement is $\chi^2/\mathrm{dof}=207.7/58$ for the prefit model, and $168.8/58$ for the postfit model. Note that the total $\chi^2$ is not the sum of the aforementioned two $\chi^2$ as the experimental measurement provides correlations between the neutrino and antineutrino cross sections. The poor agreement between the postfit model and this measurement motivates future developments, but it is worth reiterating that this shows only incompatibility with the prefit and postfit model central values, not with the full uncertainty model, as the $\chi^2$ does not account for uncertainties in the model parameters.

\begin{figure}[hptb]
    \centering
        \includegraphics[ width=\columnwidth]{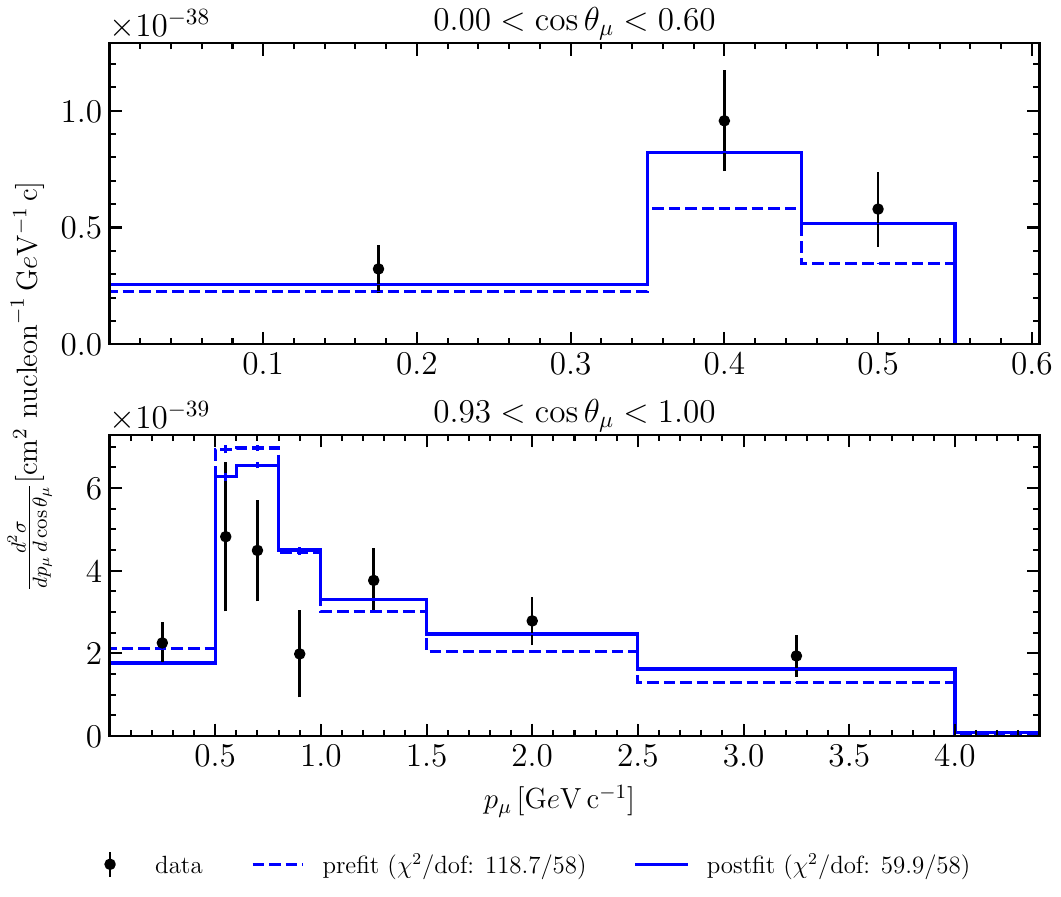}
        \includegraphics[ width=\columnwidth]{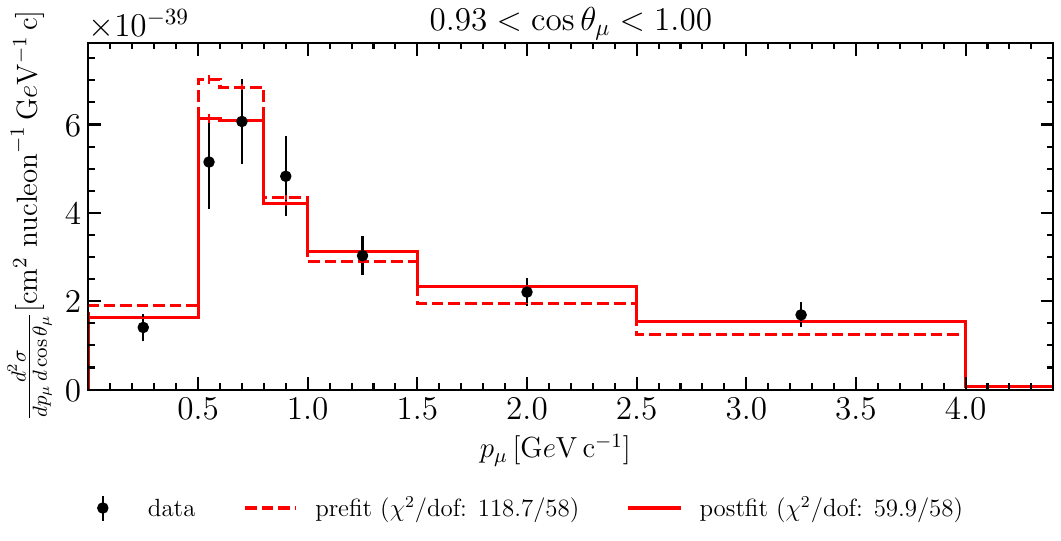}
    \caption{Predictions from the interaction model for selected slices of the differential cross sections for T2K's \num CC$0\pi$ joint measurement on C and O targets~\cite{T2K:2020jav}, before and after the near-detector fit presented in this paper. The upper panel in blue is for oxygen, and the lower panel in red is for carbon.}
    \label{Fig::BANFF_OC_2Dpcos_sel}
\end{figure}

To explore how the T2K uncertainty model covers differences in cross sections on carbon and oxygen (crucial for the extrapolation between T2K's near and far detectors), we compare the prefit and postfit models to the T2K \num CC0$\pi$ joint cross-section measurement on the two targets~\cite{T2K:2020jav}.
A significant improvement of the postfit model with respect to the prefit model is evident, as shown for a subset of the measurement's angular slices in \autoref{Fig::BANFF_OC_2Dpcos_sel}. The $\chi^2/\mathrm{dof}$ decreases from $118.7/58$ for the prefit model to $59.9/58$ for the postfit model. 
The full comparison, including all angular panels, is shown in \autoref{Fig::BANFF_OC_2Dpcos} of \aref{app::more_plots}. The same broad alterations to the postfit cross section discussed for the simultaneous neutrino and antineutrino analysis are observed for carbon and oxygen, again bringing the postfit model closer to the measurement. When splitting the $\chi^2/\mathrm{dof}$ values into components, the oxygen measurement for the prefit model is $56.2/29$, decreasing to $29.8/29$ for the postfit model. For carbon, the $\chi^2/\mathrm{dof}$ for the prefit model is $40.1/29$, which reduces to $26.3/29$ for the postfit model. As with the joint neutrino and antineutrino cross section, the $\chi^2$s for the individual target measurements do not sum to the total $\chi^2$ as T2K's measurement provides correlations between the two cross sections.

Although there is significant overlap between the data used for the \num CC$0\pi$ O-C measurement and the \num portion of the joint $\num + \numb$ CC0$\pi$ measurement, the $\chi^2$s for the two similar measurements are very different. The $\chi^2/\mathrm{dof}$ for the carbon component of the O-C measurement is $29.8/29$---signifying good agreement---whereas the \num component of the $\num + \numb$ measurement is $130.4/58$---signifying poor agreement. This is because the latter measurement has significantly finer binning in the high-statistics forward-going region, where the worst agreement is found, which the O-C measurement smooths over, effectively missing the details in the mismodeled region. The coarser binning was chosen so that direct comparisons between the differential cross sections on carbon and oxygen could be performed, as the binning of the cross section on carbon is limited by the statistics of interactions on oxygen. Both measurements are shown over the entire phase space in \aref{app::more_plots}.

\begin{figure}[hptb]
    \centering 
    \includegraphics[ width=0.96\columnwidth]{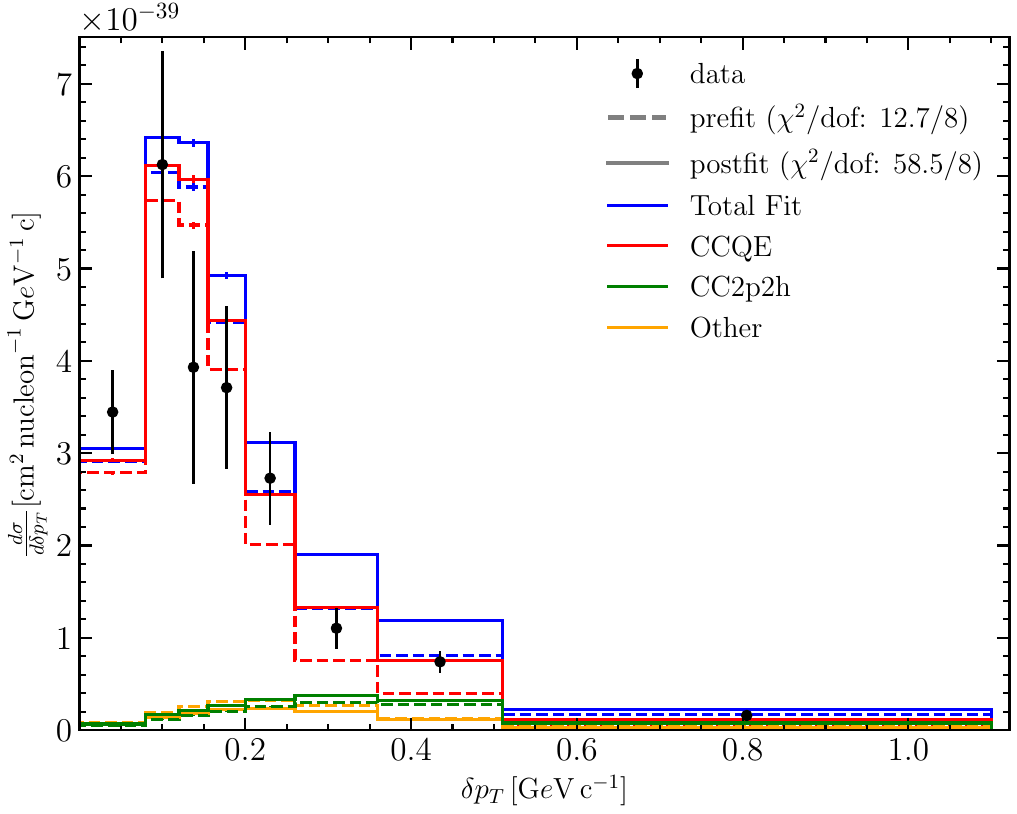}
    \includegraphics[ width=0.96\columnwidth]{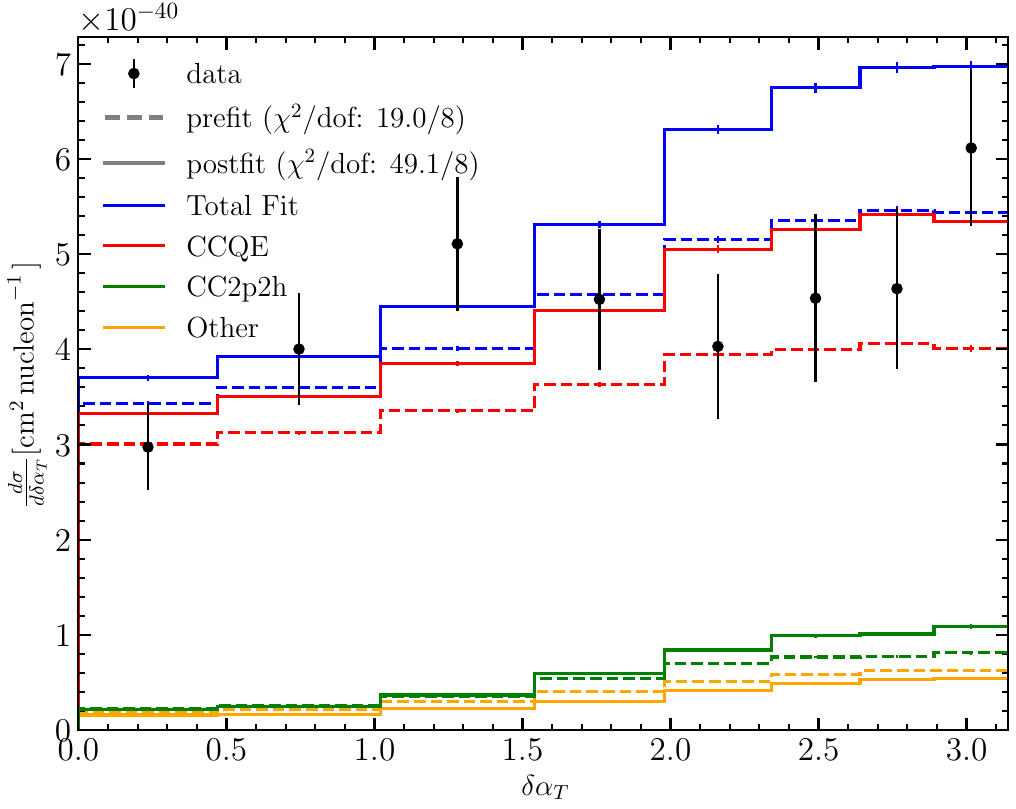}
    \includegraphics[ width=0.96\columnwidth]{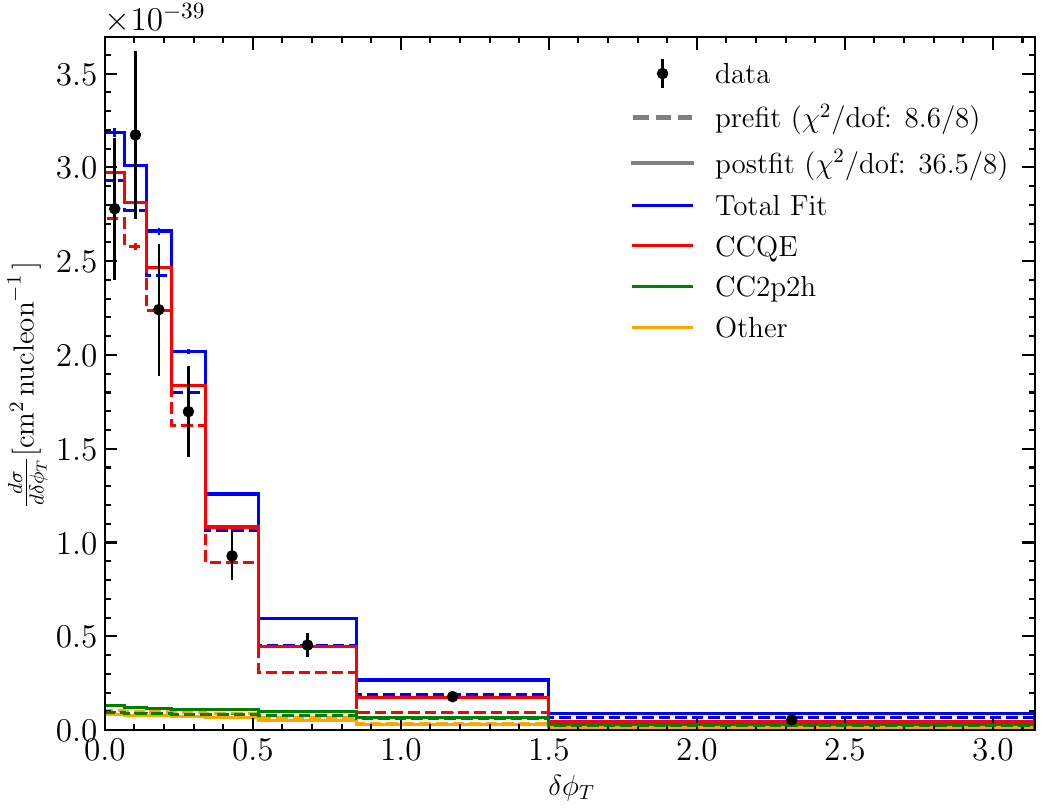}
    \caption{Predictions from the interaction model of the differential cross sections for T2K's \num CC0$\pi$Np measurement as a function of variables characterizing kinematic imbalance on the plane transverse to the incoming neutrino~\cite{T2K:2018rnz}, before and after the near-detector fit presented in this paper.
    The two largest interaction modes (CCQE and CC2p2h) are shown along the sum of all other modes. 
    }
    \label{Fig::BANFF_STV_1Ddpt_other}
\end{figure}

While the previous comparisons focus on analysis of outgoing muon kinematics, as is the case for the near detector fit and oscillation analysis, the prefit and postfit models can also be compared to measurements of other kinematic variables. Projecting the results of the muon-focused near-detector analysis to predict T2K CC0$\pi$ results, which include hadron kinematics, provides a benchmark of the model in kinematic variables sensitive to nuclear effects that the model has not directly fit, giving an indication of the model's predictive power. \autoref{Fig::BANFF_STV_1Ddpt_other} shows the result of the prefit and postfit models when compared to the CC$0\pi$ measurements of three variables characterizing kinematic imbalance on the plane transverse to the incoming neutrino~\cite{T2K:2018rnz}. 
The transverse momentum imbalance (\dpt) measures the missing momentum in the plane transverse to the neutrino direction between the highest-momentum outgoing proton and the muon, and is sensitive to a variety of nuclear effects relevant for T2K's analysis of neutrino oscillations~\cite{Lu:2015tcr}. 
Of particular interest is that \dpt separates different interaction mechanisms contributing to the CC0$\pi$Np final state, where $\dpt< 250~\text{MeV}/c$ is CCQE dominated, whilst higher \dpt has significant contributions from 2p2h and CC1$\pi$1p interactions where the pion is absorbed in the nuclear medium.

The T2K measurements of \dpt are described significantly better by the prefit model than the postfit model, especially in the tails of the distributions. The impact of the near-detector fit is to increase the CCQE and 2p2h contributions to the cross section to the extent that the CCQE contribution alone after the fit is roughly equivalent in normalization to the total prediction from all interaction channels before the fit. This leads to an overall overestimate of the cross-section measurement, particularly in the tail. This was, however, not the case for the measurements in muon kinematics, presented earlier. 

A similar conclusion can be reached considering the other transverse variables: \dphit and \dat, also shown in \autoref{Fig::BANFF_STV_1Ddpt_other}. In each case, the postfit model describes the measurement worse than the prefit model. In particular, the significant increase in the postfit prediction at large $\delta \alpha_T$, which is especially sensitive to 2p2h and FSI, moves the simulation further from the measurement. For \dphit, like for \dpt, the postift increase in 2p2h and CCQE contributions, which together cause a fairly uniform enhancement of the cross section, degrades the ability of the simulation to describe the data. 

Although the data included in these measurements are similar to that fitted in the near-detector analysis, these measurements have a wider muon phase space by including muons emitted at high and backward angles and those contained inside FGD1. The measurements are also constrained to a restricted region of proton kinematic phase space, representing less than half of all CC0$\pi$ interactions. The requirement for a minimum proton momentum of 450~MeV/c effectively removes all events with low energy transfer, leading the optical potential and Pauli blocking parameters, which suppress the cross section, to have very little impact. The data in the measurements have the most overlap with the CC0$\pi$Np selection, which was overestimated in \autoref{Fig::PosteriorPredictive_Distribution}. 

Even after removing the normalization change with a shape-only comparison, worse agreement is still observed for the postfit model, as shown in \autoref{Tab::benchmark_chi2}. We observe suppression in bins below 0.2~GeV/c and an enhancement in bins above that. This result is not necessarily unexpected, since the pre- and postfit models are built and fit in lepton kinematics---not hadron kinematics---as previously discussed. Thus, the current model likely needs improvements to its predictive power for proton kinematics before their direct inclusion in future oscillation analyses. Other comparisons of NEUT to these measurements have shown potential to improve agreement via alterations to the nuclear ground state model (with improvements seen using a local Fermi gas model with respect to spectral function) or to FSI~\cite{Filali:2024vpy,Dolan:2018zye,Chakrani:2023htw}.

\begin{table*}[hptb]
    \centering
    \begin{tabular}{l|l|l||l|l}
        \hline\hline
        \multirow{2}{*}{Sample Name} & \multirow{2}{*}{Target} & \multirow{2}{*}{Fit Type} & \multicolumn{2}{ c }{$\chi^2/\mathrm{dof}$} \\
         & & & Prefit & Postfit \\ \hline 
         Joint \num-\numb CC0$\pi$ ~\cite{T2K:2020sbd} & CH & Norm + Shape & 663.6/116 & 381.5/116\\ \hline
         \num CC0$\pi$ Joint O-C~\cite{T2K:2020jav} & O and C & Norm + Shape & 118.7/58 & 59.9/58 \\ \hline
         \multirow{2}{*}{CC0$\pi$Np \dpt ~\cite{T2K:2018rnz}} & \multirow{2}{*}{CH} & Norm + Shape & 12.7/8 & 58.5/8 \\
         & & Shape only & 10.6/7 & 35.3/7 \\ \hline 
         \multirow{2}{*}{CC0$\pi$Np \dat~\cite{T2K:2018rnz}} & \multirow{2}{*}{CH} & Norm + Shape & 19.0/8 & 49.1/8 \\
         & & Shape only & 18.0/7 & 29.8/7 \\ \hline 
         \multirow{2}{*}{CC0$\pi$Np \dphit~\cite{T2K:2018rnz}} & \multirow{2}{*}{CH} & Norm + Shape & 8.6/8 & 36.5/8 \\
         & & Shape only & 8.2/7 & 25.7/7 \\ \hline
         CC1$\pi^+$ $p_\mu \cosmu$~\cite{T2K:2019yqu} & CH & Norm + Shape & 12.8/16 & 11.4/16 \\ \hline
         CC1$\pi^+$ $p_\mu$~\cite{T2K:2016cbz} & H$_2$O & Norm + Shape & 14.8/15 & 12.6/15 \\ \hline
         CC1$\pi^+$ $p_\pi$~\cite{T2K:2019yqu} & CH & Norm + Shape & 30.9/17 & 22.0/17 \\ \hline
         CC1$\pi^+$Np $p_N$~\cite{CC1pipNp_1DSTV} & CH & Norm + Shape & 14.2/4 & 3.5/4 \\ \hline
         CC1$\pi^+$Np \dptt~\cite{CC1pipNp_1DSTV} & CH & Norm + Shape & 12.7/5 & 4.9/5 \\ \hline
         CC1$\pi^+$Np \dat~\cite{CC1pipNp_1DSTV} & CH & Norm + Shape & 1.9/3 & 1.9/3 \\
         \hline\hline
    \end{tabular}
    \caption{Predictions from the interaction model of the differential cross sections for the various measurements discussed in \autoref{subsec:benchmark}, before and after the near-detector fit presented in this paper.}
    \label{Tab::benchmark_chi2}
\end{table*}

\begin{figure}[hptb]
	\centering
	\includegraphics[ width=\columnwidth]{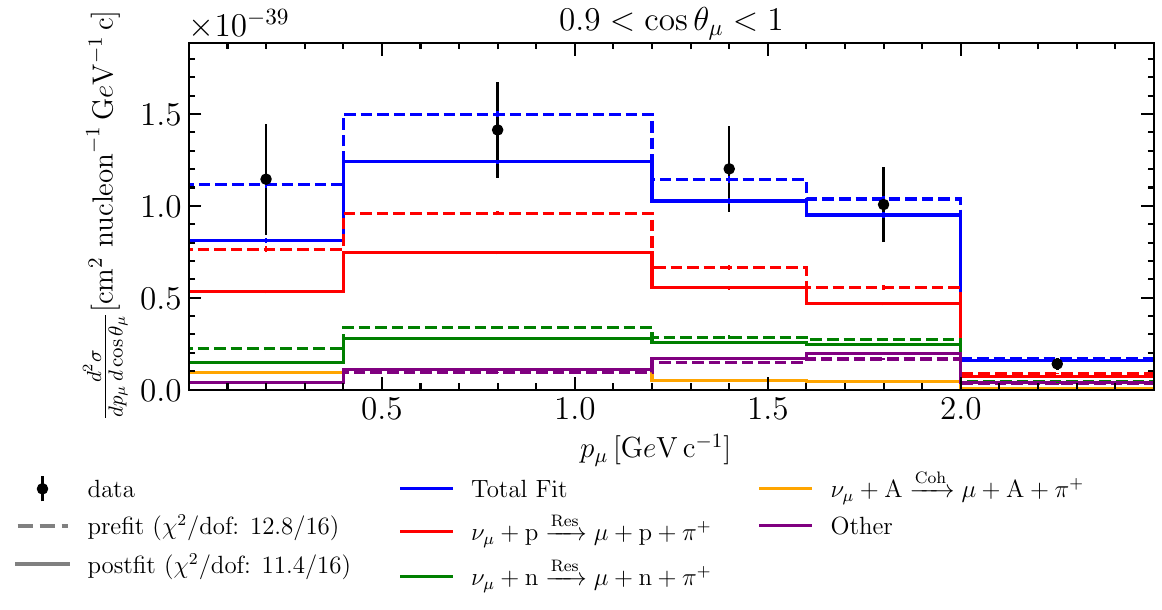}
	\caption{Predictions from the interaction model of the differential cross sections for T2K's CC1$\pi^+$ measurement on a CH target in $p_\mu \cosmu$~\cite{T2K:2019yqu}, before and after the near-detector fit presented in this paper. The right-most bin extends out to a muon momentum of 15 GeV/c. The calculated $\chi^2$ and number of bins include all bins of the measurement.
    The largest interaction modes, resonant and coherent pion production channels, are shown along with the sum of all other modes.}
	\label{Fig::BANFF_2Dpmucosmu_sel}
\end{figure}

Although the dominant event topology at T2K energies is CC0$\pi$, the near-detector and far-detector inputs to neutrino oscillation analysis also include CC$1\pi$ backgrounds in CC0$\pi$ selections in addition to dedicated selections which are rich in CC$1\pi^+$ events, necessitating a robust modeling of them. T2K's cross-section measurement of the CC$1\pi^+$ topology on a plastic scintillator (CH) target~\cite{T2K:2019yqu} is shown in \autoref{Fig::BANFF_2Dpmucosmu_sel} in bins of muon momentum for the most forward-going muons. In addition to tracked pions, this projection includes events in which the single charged pion is below the tracking threshold and is instead tagged via a delayed Michel electron, thereby covering a wide phase space of pion kinematics. A suppression of the postfit model cross section with respect to that of the prefit is evident, as is a shape change in the muon momentum distribution. The near-detector fit improves the $\chi^2$ by 1.4 units, from 12.8 to 11.4 for 16 degrees of freedom, although both the prefit and postfit models adequately describe the measurement. Comparisons for $\cos{\theta_\mu} > 0$ are shown in \autoref{Fig::BANFF_2Dpmucosmu} in \aref{app::more_plots}.

\begin{figure}[hptb]
	\centering
	\includegraphics[ width=\columnwidth]{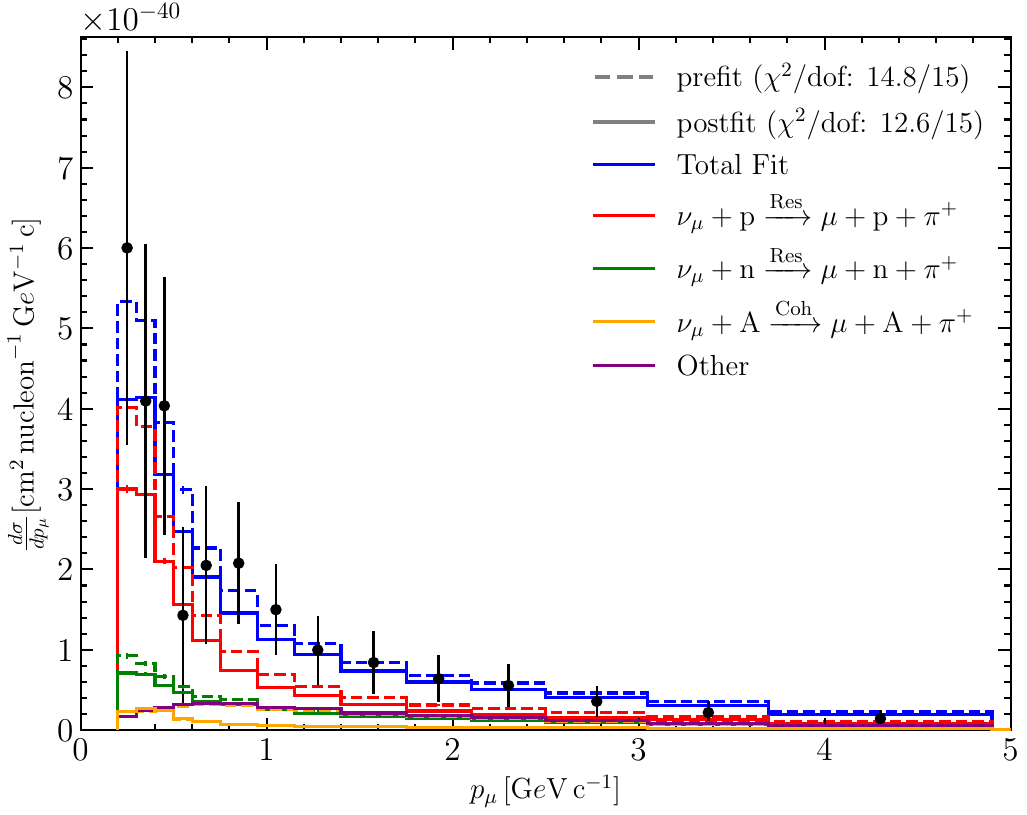}
	\caption{Predictions from the interaction model of the differential cross sections for T2K's CC1$\pi^+$ measurement on a water target~\cite{T2K:2016cbz} projected onto the muon momentum, before and after the near-detector fit presented in this paper.
	The largest interaction modes, resonant and coherent pion production channels, are shown along with the sum of all other modes.
	}
	\label{Fig::BANFF_H2O_1D_other}
\end{figure}

Similar results are observed when comparing the models to the CC$1\pi^+$ measurement on ND280's H$_2$O target in muon momentum~\cite{T2K:2016cbz}, shown in \autoref{Fig::BANFF_H2O_1D_other}. A suppression of the cross section is again evident in the postfit model; this translates to a small reduction in an already good $\chi^2$ (shown in the figure) with respect to the prefit model. The first bin in muon momentum is under-predicted, which was also observed for the measurement on CH, although this bin has the largest uncertainty.

The prediction is further divided into the true interaction modes contributing to the CC$1\pi^+$ topology. Most of the improvement comes from a suppression of the resonant pion production on a proton, which is the dominant interaction mode. This is largely due to the decrease in $C_5^A$, which also impacts the sub-dominant neutron channel. The contributions from other interaction modes are much less affected by the near-detector fit.

\begin{figure}[hptb]
	\centering
	\includegraphics[ width=\columnwidth]{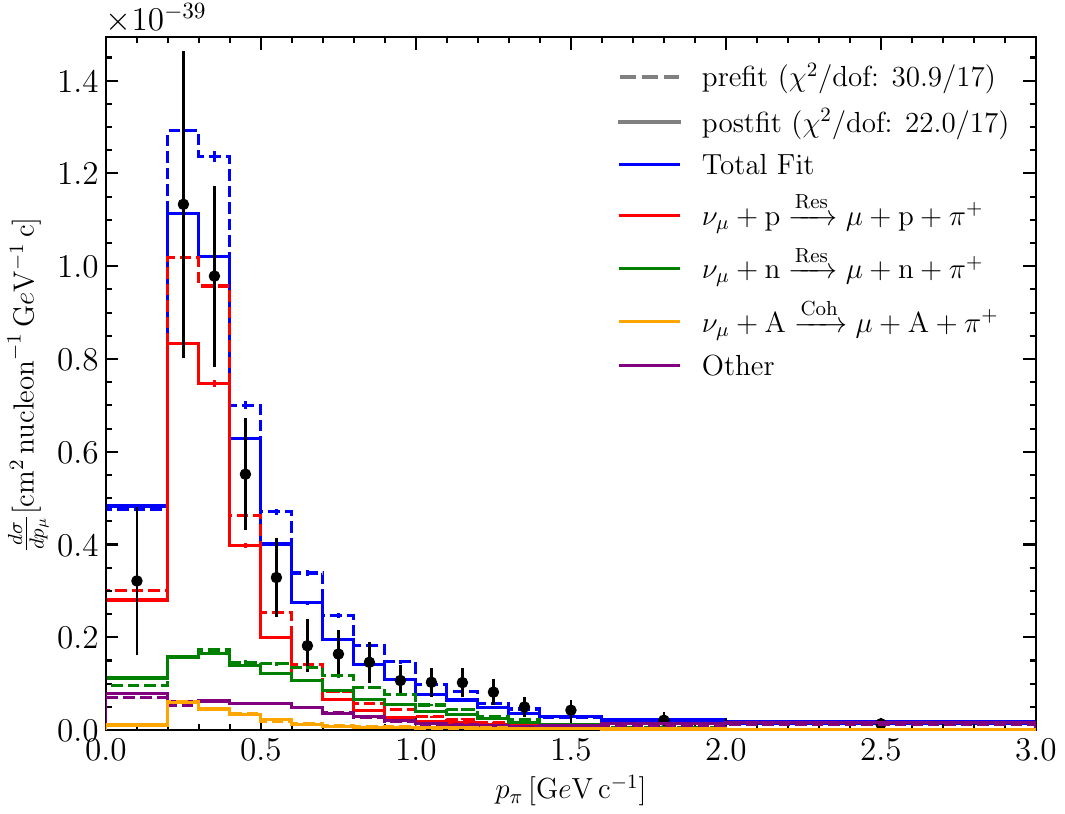}
	\caption{Predictions from the interaction model of the differential cross sections for T2K's CC1$\pi^+$ measurement on a CH target in $p_\pi$~\cite{T2K:2019yqu}, before and after the near-detector fit presented in this paper. The largest interaction modes, resonant and coherent pion production channels, are shown along with the sum of all other modes.}
	\label{Fig::BANFF_CC1pip_ppi}
\end{figure}

Similar to the CC0$\pi$ analysis presented earlier, the portion of the near-detector fit concerning single pion events uses samples binned in muon kinematics, so an improvement in the agreement between the postfit model and the CC1$\pi^+$ cross-section measurements in muon kinematics is generally expected. The near-detector fit does not explicitly include pion kinematics, and comparing the postfit model to measurements in pion kinematics provides a benchmark of how well lepton-pion correlations are modeled. \autoref{Fig::BANFF_CC1pip_ppi} shows the differential cross section in pion momentum on a plastic scintillator (CH) target, which was previously shown in muon momentum\footnote{Although the pion momentum measurement does not include the Michel tagged pions, whereas the measurements of muon kinematics do.}, and the near-detector analysis clearly improves the prediction. The lowest momentum bin is overestimated, as is the fall-off region between 0.5-0.8 GeV/c, but the remaining distribution is well described. The near-detector fit changes the shape of the pion momentum distribution, and the prediction in the first bin sees little impact from the near-detector fit due to pion FSI uncertainties and the new resonance decay uncertainty. The $\chi^2$ for the prefit model is relatively poor, although the postfit model improves the prediction to an acceptable level.

\begin{figure*}[hptb]
    \centering
    \includegraphics[ width=0.425\linewidth]{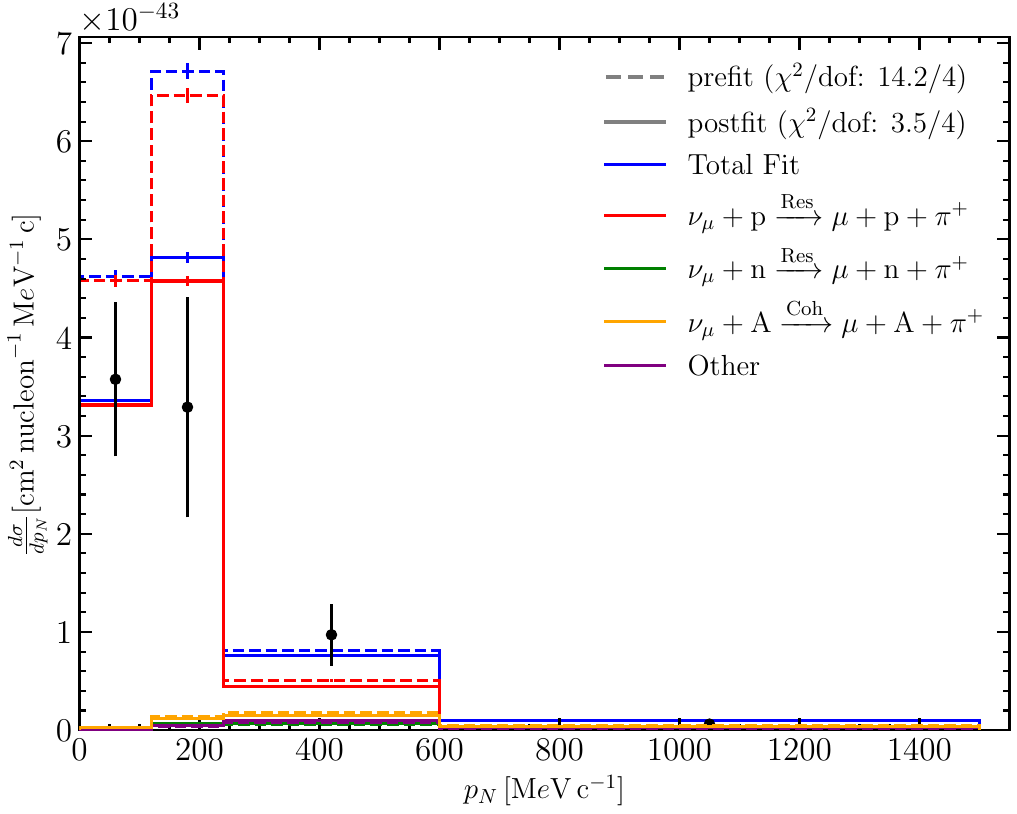}
    \includegraphics[ width=0.425\linewidth]{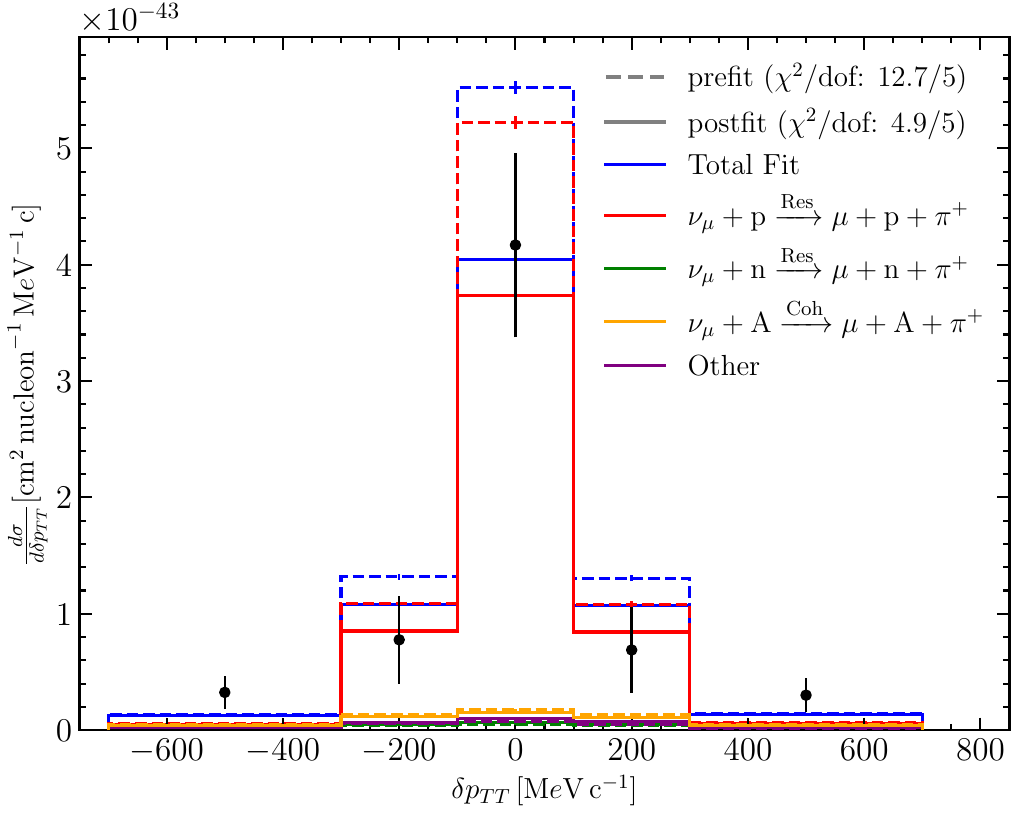}
    \hfill \\
    \includegraphics[ width=0.457\linewidth]{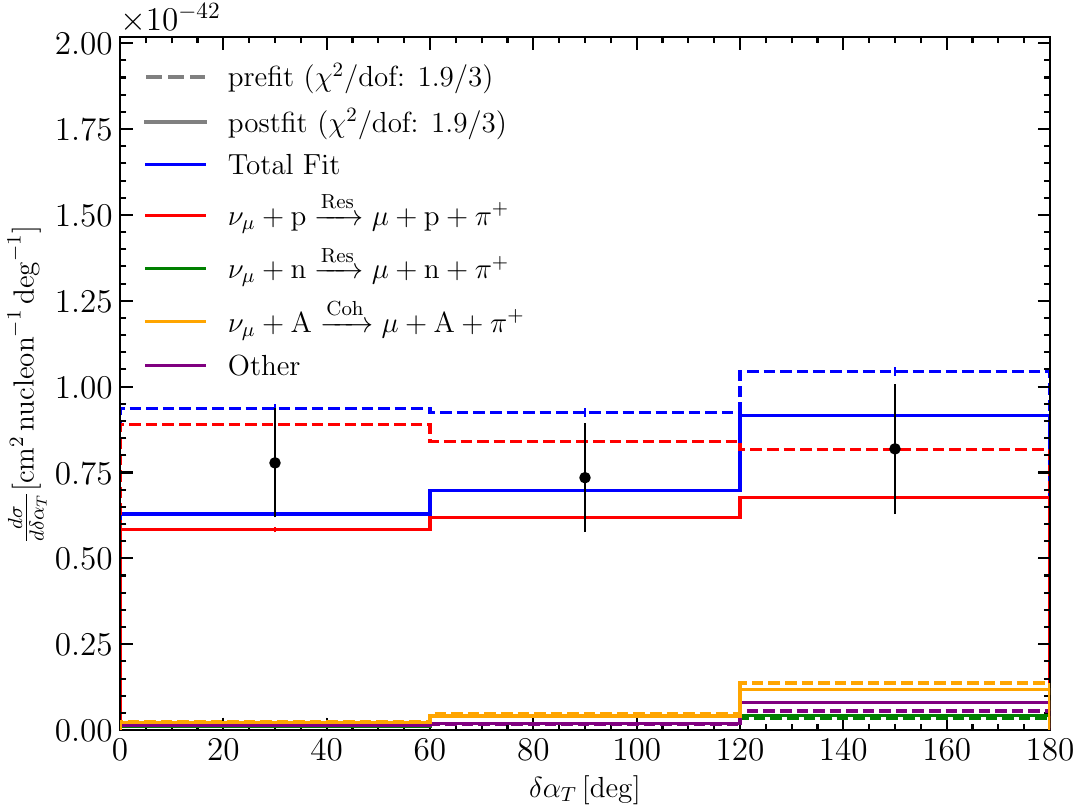}
    \caption{Predictions from the interaction model of the differential cross sections for T2K CC1$\pi^+$Np measurements on a CH target of the transverse variables $p_N$, \dptt, and \dat ~\cite{CC1pipNp_1DSTV}, before and after the near-detector fit presented in this paper. The largest interaction modes, resonant and coherent pion production channels, are shown along with the sum of all other modes.
    }
    \label{Fig::BANFF_CC1pipNp_oth}
\end{figure*}

T2K has also made measurements of pion production events~\cite{CC1pipNp_1DSTV} as a function of variables that characterize kinematic imbalance. In this measurement, charged-current events with a final state containing a single charged pion, as well as at least one proton track, are selected, and the highest momentum proton track is used to construct the transverse variables. As such, these variables probe muon, pion, and proton kinematic correlations, and therefore the nuclear modeling and the relative contributions of resonant single pion production and other pion production modes to the CC1$\pi^+$Np final state. Generally, these measurements have relatively low statistical power due to a lower number of events in the data, but broad conclusions on model-measurement compatibility are still possible. \autoref{Fig::BANFF_CC1pipNp_oth} shows T2K's measurement of \num $\mathrm{CC}1 \pi^+$Np on CH as a function of three different variables. 

The reconstructed initial state momentum, $p_N$~\cite{CC1pipNp_1DSTV}, provides insight into the Fermi motion of nucleons within the nucleus~\cite{Furmanski:2016wqo}. We observe a drastic improvement in the postfit model compared to the prefit case, particularly at the peak of the distribution, largely due to changes in $C^A_5$ and $M_A^{RES}$. A reasonable postfit $\chi^2$ is found, even though the nuclear model for single-pion production is a simple global relativistic Fermi gas, which is known to be inadequate to describe higher statistics CC0$\pi$ cross-section measurements~\cite{T2K:2018rnz}.

Similar improvements are seen in the \dptt distribution, which measures the momentum imbalance between the hadronic and leptonic system~\cite{Lu:2015hea}. The momentum imbalance can be non-zero due to the Fermi motion of the bound nucleon or FSI of outgoing protons and pions. The postfit reduction of the central bins brings the simulation into much better agreement with the measurement. The high momentum imbalance region appears poorly modeled and is barely affected by the near detector fit. However, the small cross section and large uncertainties in this region mean it has a negligible impact on the overall $\chi^2$. 

For the differential cross section in \dat, the $\chi^2$ before the near-detector fit is already satisfactory, and the postfit shows no significant improvement. However, the shape of the resonant pion production channel with an outgoing proton significantly changes, from monotonically decreasing with \dat in the prefit to the opposite in the postfit. This is primarily due to alterations of the removal energy parameters.

The near-detector fit in the oscillation analysis includes three general sources of systematic uncertainties---neutrino flux, interaction, and ND280 detector and reconstruction---and separating the effects from each other is nontrivial. This section has shown that the neutrino interaction model constrained by the near-detector fit has good general agreement with cross-section measurements at ND280 as a function of outgoing muon kinematics, most often improving the agreement with data. This supports that the constrained interaction model from the near-detector fit is reasonable for T2K energies and selections, and that it is not modified in unphysical ways by the fitting procedure. It is not surprising that agreement with cross sections differential in muon kinematics generally improves, and it is encouraging that comparisons to data in pion kinematics and CC$1\pi^+$Np transverse variables show considerable improvement, even though pion and proton kinematics are not included in the fit (although it should be noted that these measurements have large statistical uncertainties and so have limited sensitivity to potential mismodeling). However, when comparing to CC0$\pi$ measurements which probe muon-proton correlations, like those made as a function of \dpt, it is clear that the constrained model does not improve details related to nuclear physics. The poor postfit model agreement with the T2K joint neutrino and antineutrino CC0$\pi$ measurement is also of concern, implying that the postfit model may struggle to describe finely-binned details of the evolution of the CC0$\pi$ cross section even in muon kinematics. Whilst tests of the plausible bias caused by such mismodeling show a likely small impact on current statistics-limited oscillation analyses, further discussed in \autoref{sec:FDS}, improvements to nuclear modeling and measurements that probe them are clearly needed for future analyses, such as the final T2K analyses using the near-detector upgrade and analyses with Hyper-Kamiokande.

\section{Assessing out-of-model effects through simulated data studies}
\label{sec:FDS}
The systematic uncertainties described in \autoref{sec:intModel} are constructed in order to account for known uncertainties in the chosen neutrino interaction model. However, they do not represent an exhaustive set of all known plausible variations away from the nominal model, as suggested by the postfit model's imperfect description of the cross-section measurements presented in~\autoref{subsec:benchmark}. To assess the robustness of the analysis against alternative models or effects that cannot be easily tested during the fit, several simulated data sets (SDSs) are used to verify that the uncertainty model can cover alternative model predictions.

SDS studies investigate the potential impact of alternative interaction models and data-driven tunes at both the ND280 and FD. They are conducted by generating toy data sets replicating the physical effect that is being tested. For each SDS, the entire neutrino oscillation analysis chain is repeated, including the near- and far-detector fits, treating the SDS as if it were data. 

The result of the near-detector fit to the SDS yields a modified prediction at the far detector and a modified correlation matrix. This extrapolation of the result of the near-detector fit to the SDS may (and often does) produce a different prediction of the far-detector simulation than the one obtained by modifying the far-detector MC according to the predictions of the SDS directly. Obtaining such a result implies an incorrect ND280 to FD extrapolation and a potential subsequent bias on oscillation parameters if the SDS under consideration were closer to reality. In addition to the extrapolation check, the far detector fit is performed to the SDS. The oscillation parameter contours obtained as a result of this fit are then compared to the ones obtained using the Asimov maximum-sensitivity fit, where the Asimov data set is constructed by replacing the observed data with the predicted event rates calculated using the pre-fit values of all parameters~\cite{Cowan:2010js}. By quantifying the difference between the oscillation parameter contours and central values, the results of the SDS are used to assess whether the systematic error model has enough freedom to cover the variations in the alternative data set and whether the uncertainty model's incompleteness has an impact on the extraction of neutrino oscillation parameters.

Some of the simulated data sets considered in this analysis are similar to those presented in T2K’s previous analyses~\cite{T2K:2023smv}; others are tested here for the first time. The studies are updated relative to the previous analysis to reflect the significant changes in the uncertainty model alongside recent developments in neutrino interaction theory. They are selected to cover potential mis-modeling of several interaction types and effects, varying features of the nominal interaction model from nuclear ground states to changes in pion kinematics from resonance baryon decays. \autoref{Tab::FDSlist} summarizes all SDS performed for this analysis. In this section, emphasis is placed on discussing new and updated SDS from the previous analysis~\cite{T2K:2023smv}, which are detailed below.

\begin{table*}[htb]								
\centering								
\begin{tabular}{ l | c | c | c }  \hline\hline
Study &	 Model Component & Status & References\\ \hline
Alternative CCQE form factors & CC0$\pi$ &	Repeated &  \\
Local Fermi Gas & CC0$\pi$ & New & \cite{Nieves:2011pp} \\
Hartree-Fock CRPA &  CC0$\pi$ & New &  \cite{Jachowicz:2002rr,Pandey:2014tza} \\
Removal energy &  CC0$\pi$ & Repeated &  \\
Removal Energy with different interpolation methods  &  CC0$\pi$ & New &  see \autoref{subsec:fitter}\\
Non-CC-Quasi-Elastic (non-CCQE) contributions &  CC0$\pi$ & Updated &  \\ \hline

$\Delta(1232)$ matrix element uncertainties & CC1$\pi$ & New & \cite{Gottfried:1964nx,Rein:1980wg} \\
Martini~\em{et al.} 1$\pi$ & CC1$\pi$ & New & \cite{Martini:2009uj,Martini:2014dqa}\\
MINER$\nu$A pion suppression & CC1$\pi$ & Repeated & \cite{MINERvA:2019kfr} \\
ND280 data-driven pion momentum modification & CC1$\pi$ & Updated & \\
Pion secondary interactions & CC1$\pi$ & Repeated & \\ \hline

Radiative Corrections & Other & New & \cite{Tomalak:2021hec, Day:2012gb} \\
\hline\hline
\end{tabular}
\caption{Summary of SDS performed for this analysis. ``Repeated'' means nothing changed with respect to~\cite{T2K:2023smv}, while  ``updated'' means there are changes with respect to the previous analysis. ``New'' means SDS are used here for the first time; see the text for details.}
\label{Tab::FDSlist}
\end{table*}

\subsection{New and updated simulated data studies}
\label{sec:newfdslist}

\textbf{CC0$\pi$ simulated data sets:}
\begin{itemize}
\item \textit{Local Fermi Gas (LFG)}---This SDS transforms the nominal SF model for 1p1h interactions to an LFG+RPA model based on the calculations of the Valencia group~\cite{Nieves:2011pp}. The LFG model~\cite{Nieves:2011pp} is a Fermi gas nuclear model that assumes that nucleons are bound within a box-like potential inside the nucleus. The value of the potential well has a radial dependence on the nuclear density. At low values of energy and momentum transfer, nuclear screening effects become non-negligible and are included via a Random Phase Approximation (RPA) treatment. Whereas this model does not provide a full prediction for the outgoing hadron kinematics, inclusive measurements performed by the T2K collaboration~\cite{T2K:2020jav} and semi-exclusive measurements performed by T2K and MicroBooNE (\cite{MicroBooNE:2023cmw, Filali:2024vpy}) show good agreement with this model. The LFG+RPA model is implemented in the NEUT event generator~\cite{Hayato:2021heg}. To produce the LFG SDS, the nominal T2K simulation, generated according to the Benhar SF model, is reweighted to the LFG+RPA model by applying the ratio of the cross section predicted by the two models as a function of neutrino flavor, nuclear target, neutrino energy, outgoing lepton momentum, and outgoing lepton angle. 

\item \textit{Hartree-Fock Continuous Random Phase Approximation}---For this SDS the 1p1h neutrino-interaction cross section is altered to that of the Hartree-Fock Continuum Random Phase Approximation (HF CRPA) model~\cite{Jachowicz:2002rr, Pandey:2014tza}. However, the non-relativistic assumptions in the HF+CRPA model are not valid at high energy transfer. Therefore, for energy transfers above 1 GeV, we alter the cross section to be the SuSAv2~\cite{Gonzalez-Jimenez:2014eqa,Megias:2016fjk} model prediction. Between 500 MeV and 1 GeV, a transition between HF-CRPA and SuSAv2 is employed.

The HF-CRPA treatment of nucleon-nucleon long-range correlations and of final state interactions (FSI) via a distortion of the outgoing nucleon wave function leads to significantly different predictions for muon and electron neutrino cross sections at low energy transfers compared to the widely used plane wave impulse approximation (PWIA) models (which do not include FSI), such as the Spectral Function. With respect to the baseline model, HF-CRPA predicts a different cross-section evolution as a function of neutrino energy, neutrino/antineutrino cross-section ratios, carbon/oxygen cross-section ratios, and muon/electron neutrino cross-section ratios. This SDS is produced by reweighting the nominal SF model using the same five variables as in the LFG case.

\item \textit{Non-CC-Quasi-Elastic (non-CCQE) contributions}---The nominal systematic error model contains significant freedom in the prediction for quasi-elastic (QE) interactions, but this study allows a more loose parametrization of the uncertainties that affect the background processes in the CC0$\pi$ sample (predominantly 2p2h and pion absorption FSI). While lacking strong theoretical guidance on how this could be implemented, we test an alternative data-driven model in which the postfit discrepancy between the ND280 data and the simulation is assigned to non-QE processes. Practically, this corresponds to altering the postfit model so that the QE parameters are set to their nominal values, and varying the non-QE component of the simulation as a function of reconstructed momentum transfer ($Q^2_{QE,\mathrm{rec}}$) until an agreement is reached with the ND280 data. This procedure is applied separately to neutrino and antineutrino samples but does not discriminate between nuclear targets (i.e., the same reweighting is applied for both carbon and oxygen). This SDS follows the same philosophy as the one described in~\cite{T2K:2023smv}, but the exact values of the scaling factors as a function of $Q^2_{QE,\mathrm{rec}}$ are different in this updated analysis.
\end{itemize}

\textbf{CC1$\pi$ simulated data sets:}

\begin{itemize}
\item \textit{$\Delta(1232)$ matrix element uncertainties}---The differential cross section for resonant CC1$\pi$ interactions is modified by altering the generalized density matrix elements, $\rho$, in the Rein--Sehgal model~\cite{Rein:1980wg}. 

These elements are contracted with the relevant spherical harmonics, $Y_l^m$, to form the angular distribution, $W(\theta,\phi)$, where $\theta$ and $\phi$ are the Adler angles~\cite{Adler:1968tw}. These are defined in the resonance rest frame, and $W(\theta,\phi)$ controls how the pion and nucleon are distributed when the resonances decay. Modifications to $W(\theta,\phi)$ only affect the distribution of the nucleon and pion and leave leptonic variables, such as $Q^2$ or $E_\textrm{lep}$, unchanged. This feature is particularly important, since the ND280 analysis is performed in muon kinematics but has indirect sensitivity to hadron kinematics through the use of samples with or without a reconstructed pion. This also applies to FD, where, for instance, the change in the pion kinematics can determine if a pion is above or below the Cherenkov threshold, and thus can cause an event to migrate between selection samples or exit all selected samples. At T2K's energies and current selections, the CC$1\pi^+$ final state is dominated by the single $P_{33}(1232)$ resonance, which simplifies $W(\theta,\phi)$ to~\cite{Gottfried:1964nx,Rein:1980wg}

\begin{widetext}
\begin{equation}
    W^{\Delta}(\theta,\phi) = \frac{1}{\sqrt{4\pi}} \frac{1}{\Tilde{\rho}} \left\{
    Y^0_0 \Tilde{\rho} - \frac{2}{\sqrt{5}} Y^2_0\left( \Tilde{\rho}_{+3,+3}-\frac{1}{2}\Tilde{\rho} \right) + \frac{4}{\sqrt{10}} \Big( \text{Re} \left( Y^{2}_1 \right) \text{Re} \left( \Tilde\rho_{+3,+1} \right) 
    - \text{Re} \left( Y^2_2 \right) \text{Re} \left( \Tilde{\rho}_{+3,-1} \right) \Big) \right\} \\
\label{eq:spp_w}
\end{equation}
\end{widetext}

The $\rho$ parameters in \autoref{eq:spp_w} control the relative strengths of the different spherical harmonics,
\begin{align}
\label{eq:spp_rho}
    \Tilde{\rho} &= \sum_{m=-3/2}^{m=+3/2} \rho_{m, m} \\
    \Tilde{\rho}_{+3,+3}                            &= \rho_{+3/2,+3/2}+\rho_{-3/2,-3/2} \\
    \Tilde{\rho}_{+3,+1}                            &= \rho_{+3/2,+1/2}-\rho_{-1/2,-3/2} \\
    \Tilde{\rho}_{+3,-1}                            &= \rho_{+3/2,-1/2}+\rho_{+1/2,-3/2}
\end{align}

where $\rho_{m,m'}$ are the generalized density elements for a $m\rightarrow m'$ transition, $l$ is the azimuthal quantum number and $m$ is the magnetic quantum number, which runs from $m=-l,-l+1,\dots,l-1,l$. These are calculable through the Feynman--Kislinger--Ravndal (FKR) model~\cite{Feynman:1971wr}. Reasonable variations of $\rho$ are estimated to be 30\%, from comparing predictions of the FKR model against data, which here is defined as a \lq\lq $1\sigma$'' variation. However, the FKR formalism is quoted as \lq\lq a shadow of the truth... a partial reflection of reality''~\cite{Feynman:1971wr} in the original paper, and thus is treated cautiously here. Hence, the definition of \lq\lq$1\sigma$'' is based on comparisons to data, and acts to define a known reference value. The focus of this SDS is to test the methodology of abruptly changing the pion-nucleon kinematics and to learn what biases may arise from such mis-modeling. The three most important density matrix elements are $\rho_{-3,-3}, \rho_{-1,-1}, \rho_{-1,-3}$, which were set to $\mp3\sigma, \pm3\sigma, \pm3\sigma$, respectively, to maximize the impact, defining two SDSs.

\begin{figure*}[htpb]
 \centering 
\subfloat{
  \includegraphics[page=2, trim={5mm 0mm 5mm 5mm}, clip, width=0.3\textwidth]{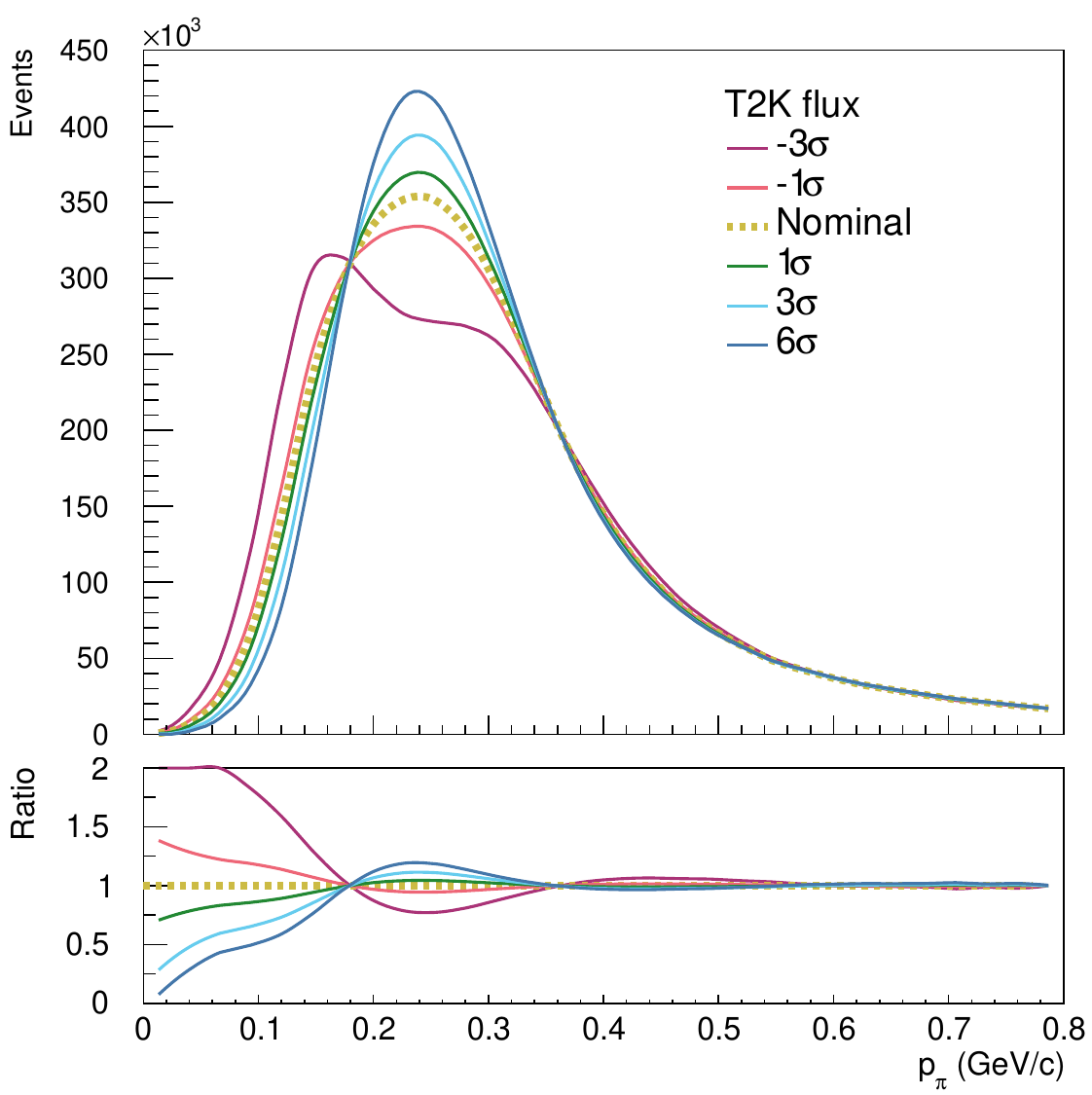}
}
\subfloat{
  \includegraphics[page=3, trim={5mm 0mm 5mm 5mm}, clip, width=0.3\textwidth]{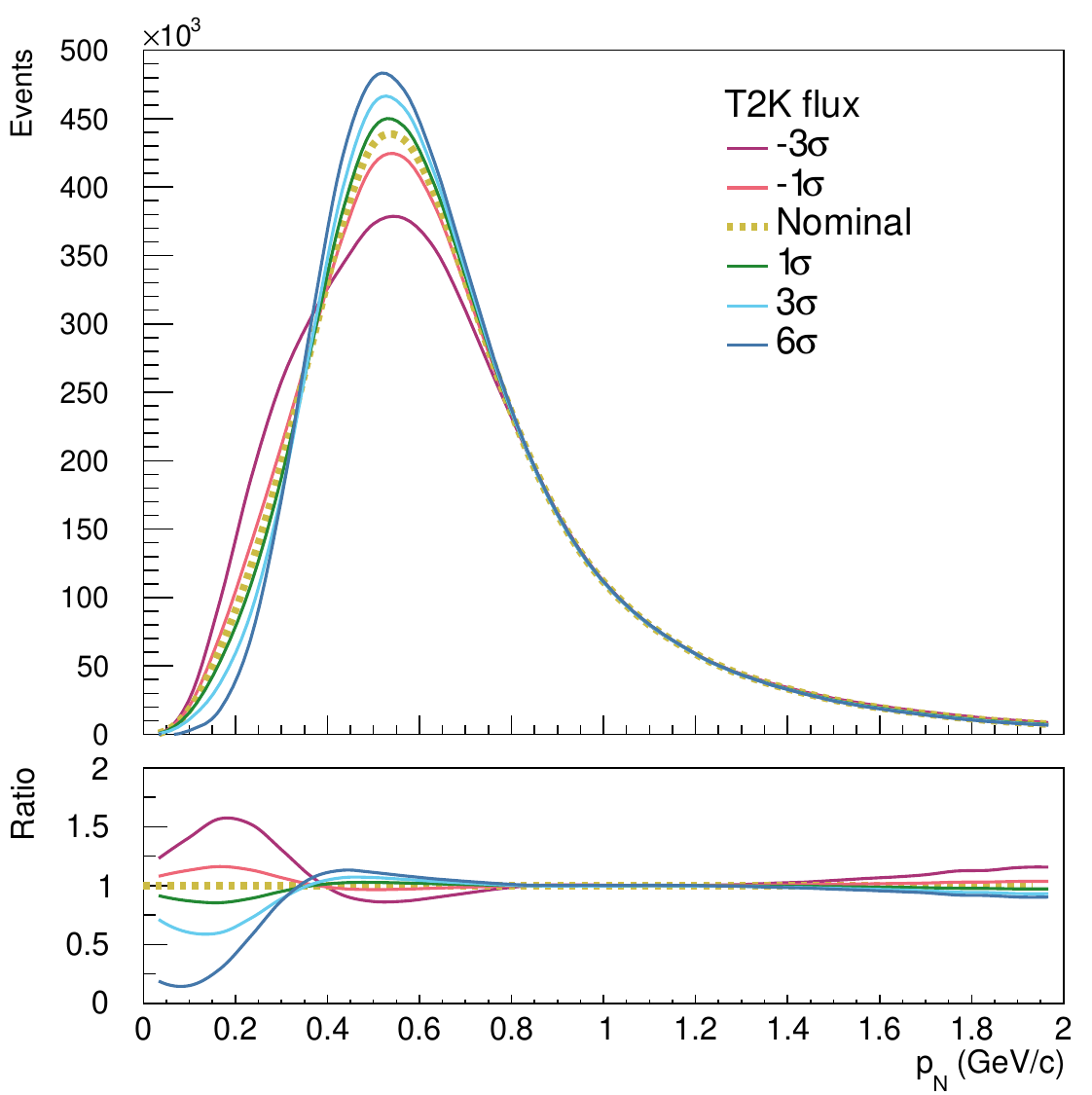}
}
\subfloat{
  \includegraphics[page=4, trim={5mm 0mm 5mm 5mm}, clip, width=0.3\textwidth]{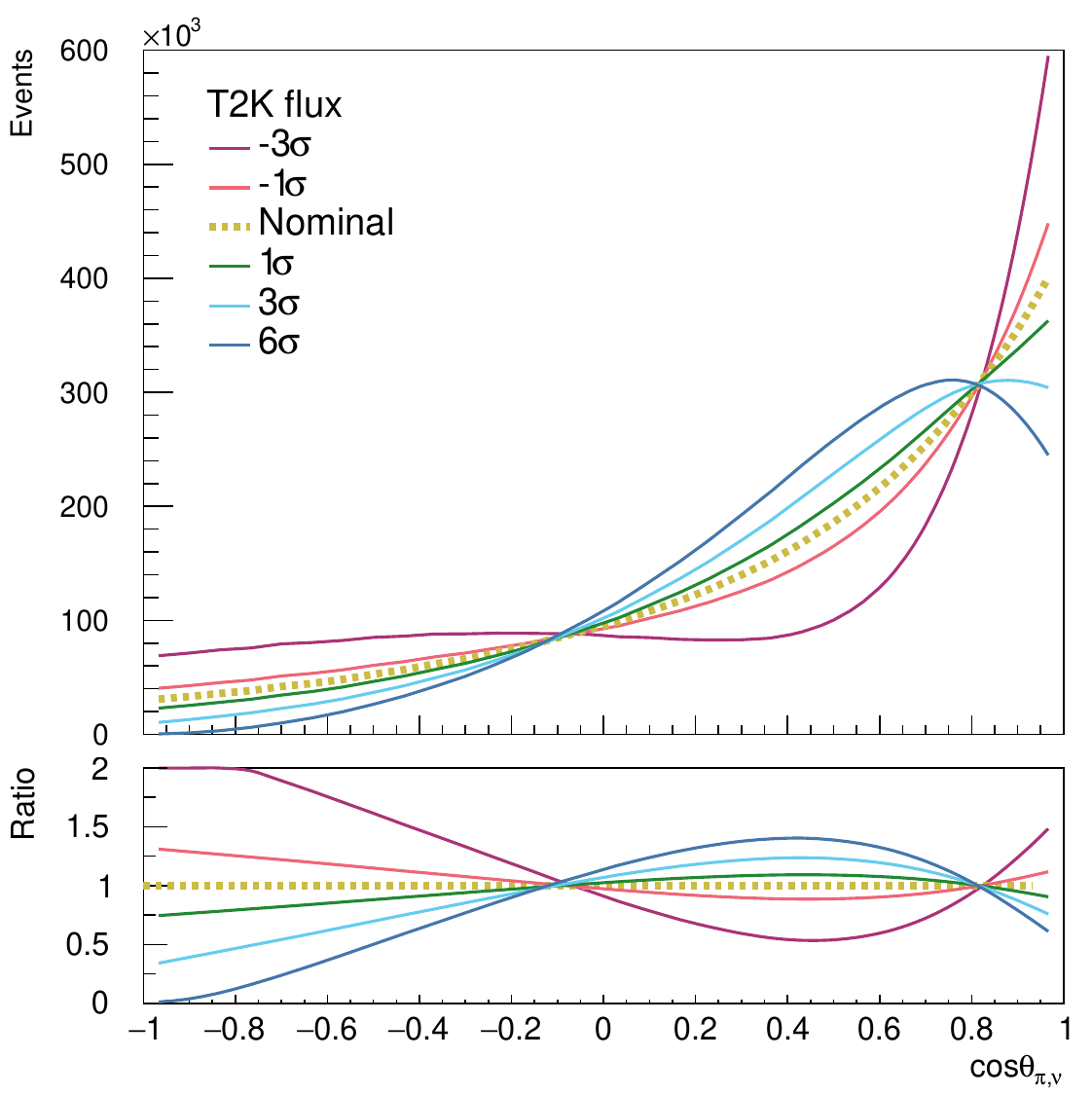}
}
 \caption{Distributions of outgoing pion momentum (left), nucleon momentum (center), and the cosine of the angle between the pion and incoming neutrino (right), under simultaneous variations of the density matrix parameters $\rho_{-3,-3},\rho_{-1,-1},\rho_{-1,-3}$. The $-3\sigma$ and $3\sigma$ variations define the two SDSs, which set $\rho_{-3,-3},\rho_{-1,-1},\rho_{-1,-3}=\{-3\sigma, +3\sigma, +3\sigma\}, \{+3\sigma, -3\sigma, -3\sigma\}$, respectively. The summed cross section for $\nu_\mu\textrm{CC}1\pi^+1p$ and $\nu_\mu\textrm{CC}1\pi^+1n$ resonant interactions are shown. Interactions on free protons and nucleons are only considered in these figures, without any nuclear effects applied.}
 \label{Fig::Rho}
\end{figure*} 

The variations for neutrino-nucleon interactions are shown in \autoref{Fig::Rho}, where a significant shift in the cross section from high to low pion and nucleon momentum is seen before consideration of nuclear effects. The largest effect is seen at the pivot points of $p_\pi\sim0.18~\text{GeV/c}$ and $p_N\sim0.4~\text{GeV/c}$, close to the tracking thresholds in the near detector of the respective particles. The study, which concerns the resonance decay kinematics, barely affects the ND280 samples, since the $p_{\mu}$ and $\cos{\theta_{\mu}}$ distributions are invariant under these parameter variations. However, it has an important effect on the far-detector samples, as it increases the number of low momentum pions below the Cherenkov threshold, which is an important cut quantity for the $\nu_e$CC1$\pi^+$ selection.

\item \textit{Martini {\em et al.} 1$\pi$}---This SDS replaces the Rein--Sehgal single-pion production model with the Martini~\em{et al.} 1$\pi$ model~\cite{Martini:2009uj, Martini:2014dqa}. The Martini~\em{et al.} model describes neutrino-nucleus interactions by employing the concept of nuclear response functions handled using RPA and including Delta-resonance excitations. It affects $\num/\numb$ as well as $\nu_{e}/\nueb$. It is included to test the robustness of the modeling for the new $\nu_{\mu}$CC1$\pi^+$ sample at FD.

\item \textit{ND280 data-driven pion momentum modification}---The near-detector analysis is performed in $p_\mu, \cos\theta_\mu$, with the only sensitivity to pion kinematics coming through the different pion tagging methods. This SDS uses the output of the near detector analysis to make predictions for the pion kinematics in the FGD-contained and TPC-tagged CC$1\pi^+$ samples and compares them to the data, similar to the previous analysis~\cite{T2K:2023smv}.

However, in this analysis, the ratio between prediction and data in $p_\pi$ and $\cos\theta_\pi$ serves as a template, rather than a single normalization parameter, and the SDS is applied to both ND280 and FD. The SDS implicitly assumes perfect reconstruction, i.e., that a ratio in reconstructed pion kinematics can be applied as a weight as a function of true pion kinematics, both at ND280 and FD. Therefore, applying the SDS to the model prediction does not perfectly describe the ND280 data in pion kinematics, but significantly improves it.
\end{itemize}

\textbf{Other simulated data sets:}

\begin{itemize}
     \item \textit{Radiative Corrections}---When calculating CCQE-like (1$\mu$1p) cross sections, contributions beyond the tree-level amplitude must be considered. A notable example is the emission of a real photon, but other contributions are also possible. These radiative corrections are non-negligible for the CCQE-like cross section~\cite{Tomalak:2021hec, Day:2012gb}, and differ between electron and muon neutrinos due to the lepton mass dependence of the corrections. If the emitted photon is sufficiently energetic, it can produce an additional ring in FD that may be reconstructed, causing events to migrate between or out of FD selections. In particular, the photon can cause interactions of a muon neutrino to be reconstructed as an electron neutrino. In this analysis, the size of this effect is conservatively estimated through a dedicated SDS, which is only applied to FD samples\footnote{Future iterations of this work will endeavor to complete an equivalent SDS using an ND280 constraint. However, the impact on the ND280 is expected to be small as the muon remains easily identifiable even in the presence of a collinear photon.}. To build this SDS, a simple simulation of lepton and lepton+photon pairs at FD is generated, with the lepton momentum following the NEUT CCQE prediction at FD and the photon energy following a one-over-photon-energy distribution to broadly match expectations from radiative correction calculations. The photon direction is sampled uniformly in solid angle relative to the lepton direction, up to a maximum opening angle of 180° (i.e., isotropic). The events are weighted such that radiative (lepton+photon) events contribute approximately 3\% of the total event rate. The samples are then passed through the FD simulation and reconstruction suites, and a linear weighting that accounts for the expected sample migration from additional real photon production is derived as a function of true lepton energy. These weights are then applied to the nominal MC prediction at FD to form the SDS. Although the weights are extracted from CCQE events, they are applied to all CC events. The net effect of the SDS is to decrease the number of candidates in the single ring samples by $\sim$2-3\%. In addition to the decrease in the single-ring samples, the $\nu_{\mu}$CC1$\pi^+$ sample at FD is weighted up to account for events where the radiated photon can be mis-reconstructed as a decay electron. Note that this SDS is evaluated only at FD. 
\end{itemize}

\subsection{Results of simulated data studies}
\label{subsec:fds-results}
This section describes the procedure used to apply contour smearing based on SDS results, and summarizes the final smearing used. We identify the most impactful effect, based on the size of the implied bias in the extraction of the neutrino oscillation parameters.

\subsubsection{Procedure, Criterion, and Results}

The SDS procedure allows us to quantify two types of effects. First, the difference between the FD prediction following the extrapolated ND280 fit to the SDS and the directly modified FD prediction: this highlights limitations in the extrapolation procedure between the near and far detectors. Second, the difference between the oscillation parameter values resulting from the SDS fit and the nominal Asimov fit allows us to quantify the bias and uncertainty modifications introduced by an alternative model. 

\begin{table}[htb]								
\centering								
\begin{tabular}{ l | c  }  \hline\hline
Study &	 $\chi^2$ \\ \hline
Alternative CCQE form factors & 137.6 \\
Local Fermi Gas & 39.8 \\
HF CRPA &  104.2 \\
Removal Energy with different interpolation methods  &  55.4 \\
Non-CC-Quasi-Elastic (non-CCQE) contributions &  56.8 \\

$\Delta(1232)$ matrix element uncertainties & 16.4 \\
Martini~\em{et al.} 1$\pi$ & 13.5 \\
MINER$\nu$A pion suppression & 26.5 \\
ND280 data-driven pion momentum modification & 3.8 \\
\hline\hline
\end{tabular}
\caption{Values of post-fit $\chi^2$ for selected SDS for ND280 only fits.}
\label{Tab::SDSChi2}
\end{table}

\autoref{Tab::SDSChi2} summarizes the $\chi^2$ (taken to be $-2\log\mathcal{L}_{\mathrm{Total}}$ defined in \autoref{eq:barlowbeeston_total}) values for a selection of ND280-only SDSs. While $\chi^2$ can be a useful indicator, it comes with significant caveats when applied to SDSs. This predominantly comes from the fact that the simulation used to create an SDS is generated from the same simulation as the one that is used during the fit, so they are statistically fully correlated. Indeed, most of the obtained $\chi^2$ values are below 100, despite the number of degrees of freedom exceeding 4000. Nevertheless, we consider it useful to show the $\chi^2$, particularly for comparisons: SDSs with higher $\chi^2$ are generally more likely to exhibit a bias in the oscillation parameters, as they indicate a larger source of tension in the near-detector fit.

Assessing the impact of bias induced by an SDS through the ND280 fit on the size of the uncertainties extracted on oscillation parameters requires care. Each SDS likely has a different intrinsic statistical sensitivity to oscillations than the nominal Asimov data --- for instance, an SDS that raises the overall cross section produces more FD events, which artificially tightens the oscillation contours. To prevent this statistical effect from contaminating our estimate of the systematic bias, we compare two fits that share the same SDS-modified FD event rate:
\begin{enumerate}
    \item \textbf{SDS fit (Fit 1):} the SDS is treated as fake data at ND280 and fit with the nominal MC; the obtained post-fit ND constraint is propagated to FD, where the SDS is also taken as the FD fake data, and oscillation parameters are extracted.
    \item \textbf{Scaled Asimov fit (Fit 2):} the SDS modification is applied directly to the FD MC, and the SDS is fit to itself at FD with no ND step. By construction, the model reproduces the data at FD, so this fit is bias-free with respect to the SDS, but it inherits the same modified event rate, and therefore the same statistical sensitivity, as Fit 1.
\end{enumerate}
Because the two fits share the modified FD event rate, the statistical-sensitivity change is common to both and cancels in their difference. Any residual difference is attributable to misfitting at ND, i.e., the ND fit's inability to fully absorb the SDS into its nuisance parameters. The same approach was employed in the previous analysis detailed in Ref.~\cite{T2K:2023smv}. 

For all SDSs, tests were first conducted to assess whether the extrapolated ND-to-FD model is in good agreement with the FD prediction obtained by applying the SDS weights directly to the oscillated spectrum. Then, following the explained procedure, SDS fits were performed and oscillation parameters were extracted, based on which biases were calculated. In total 19 SDS fits were performed\footnote{Some SDSs, such as the $\Delta(1232)$ matrix element uncertainties, were performed for multiple configurations (e.g., shifts of $+3\sigma$ and $-3\sigma$). This results in a total count higher than the number of distinct SDS listed in \autoref{Tab::FDSlist}.}, the sizes of the bias introduced by the most impactful effects are listed in \autoref{Tab:FakeData}\footnote{All SDS are listed in \autoref{tab:bias_table_fakedata_full}.}. The metric reported in \autoref{Tab:FakeData} is defined as the fractional shift in the center of the $2\sigma$ intervals between the Scaled Asimov and SDS fits, $\Delta_{2\sigma} = \hat{x}^{2\sigma}_\mathrm{Scaled\,Asimov} - \hat{x}^{2\sigma}_\mathrm{SDS}$, relative to the total and systematic intervals extracted from the Scaled Asimov fit. The systematic error is extracted from the difference in quadrature of the total and statistical errors, $\sigma^{2}_\mathrm{syst.} = \sigma^{2}_\mathrm{tot.} - \sigma^{2}_\mathrm{stat.}$, and is recomputed for each SDS since the statistical component changes with the SDS event rate.

\begin{table*}[htbp]
\centering
\begin{tabular}{ l | l c c c }
\hline\hline
Simulated data set & Relative to & $\sin{\theta_{23}}$ & $\Delta m^2_{32}$ & $\delta_{\text{CP}}$ \\
 \hline
 \multirow{2}{*}{CCQE z-exp upper var.$^\ast$} & Total & $-0.5$\% & $-9.5$\%  & $-0.5$\%\\
                                  & Syst. & $-1.0$\% & $-24.1$\% & $-2.2$\%\\
\hline
\multirow{2}{*}{HF-CRPA}       & Total & $-11.7$\% & $33.8$\% & $-2.8$\% \\
                               & Syst. & $-25.1$\% & $84.9$\% & $-11.2$\% \\
\hline
\multirow{2}{*}{Martini~\em{et al.} 1$\pi$} & Total & $-1.5$\% & $-7.3$\%  & $-0.4$\% \\
                                & Syst. & $-3.2$\% & $-18.5$\% & $-1.7$\% \\
\hline
\multirow{2}{*}{Non-CCQE}       & Total & $4.9$\%  & $-30.0$\% & $-0.1$\% \\
                                & Syst. & $10.4$\% & $-76.3$\% & $-0.5$\% \\
 \hline
 \multirow{2}{*}{Pion SI}        & Total & $-4.8$\% & $20.3$\% & $0.5$\% \\   
                                 & Syst. & $-10.1$\% & $51.6$\% & $2.1$\% \\
\hline\hline
\end{tabular}
\caption{Differences in the oscillation parameter constraints observed in a selection of the new and most impactful SDSs. Both changes to the center of the 2$\sigma$ confidence interval relative to the systematic and total uncertainties are given. $^\ast$The Z-exp upper variation corresponds to one extreme configuration based on fits to bubble chamber data.}
\label{Tab:FakeData}
\end{table*}

\subsubsection{Discussion of the Results}

Biases in the fitted value of $\Delta m^2_{32}$ are observed in several SDSs, indicating potential limitations in the current model. However, these effects do not change the overall conclusions of the oscillation analysis; it is important to account for them to ensure a conservative interpretation of the results. Given the approximately quadratic shape of the log-likelihood profile of $\Delta m^2_{32}$, we decided to apply the effect of all SDS as an additional smearing on this parameter by adding their effect in quadrature. 

The total smearing obtained and applied on $\Delta m^2_{32}$ is $0.03 \times 10^{-3}$~eV$^2/c^4$. For reference, the absolute smearing motivated here is three times larger than the one in the previous analysis~\cite{T2K:2023smv}. There is no additional smearing for $\sin{\theta_{23}}$ and $\delta_{\text{CP}}$. The performed SDSs indicate that none of the tested out-of-model effects have an impact that would alter the conclusions on the oscillation parameter measurements. However, they provide valuable indications about areas in which the model should be improved in the future. For instance, the impact of low-energy transfer physics effects highlighted by the HF-CRPA SDS can be addressed by using alternative models to bracket the size of these effects and implement new parameterized uncertainties in the analysis, which will be included in the next iteration of the analysis. 

\begin{figure}[htb]
 \centering 
  \includegraphics[page=1, width=\columnwidth]{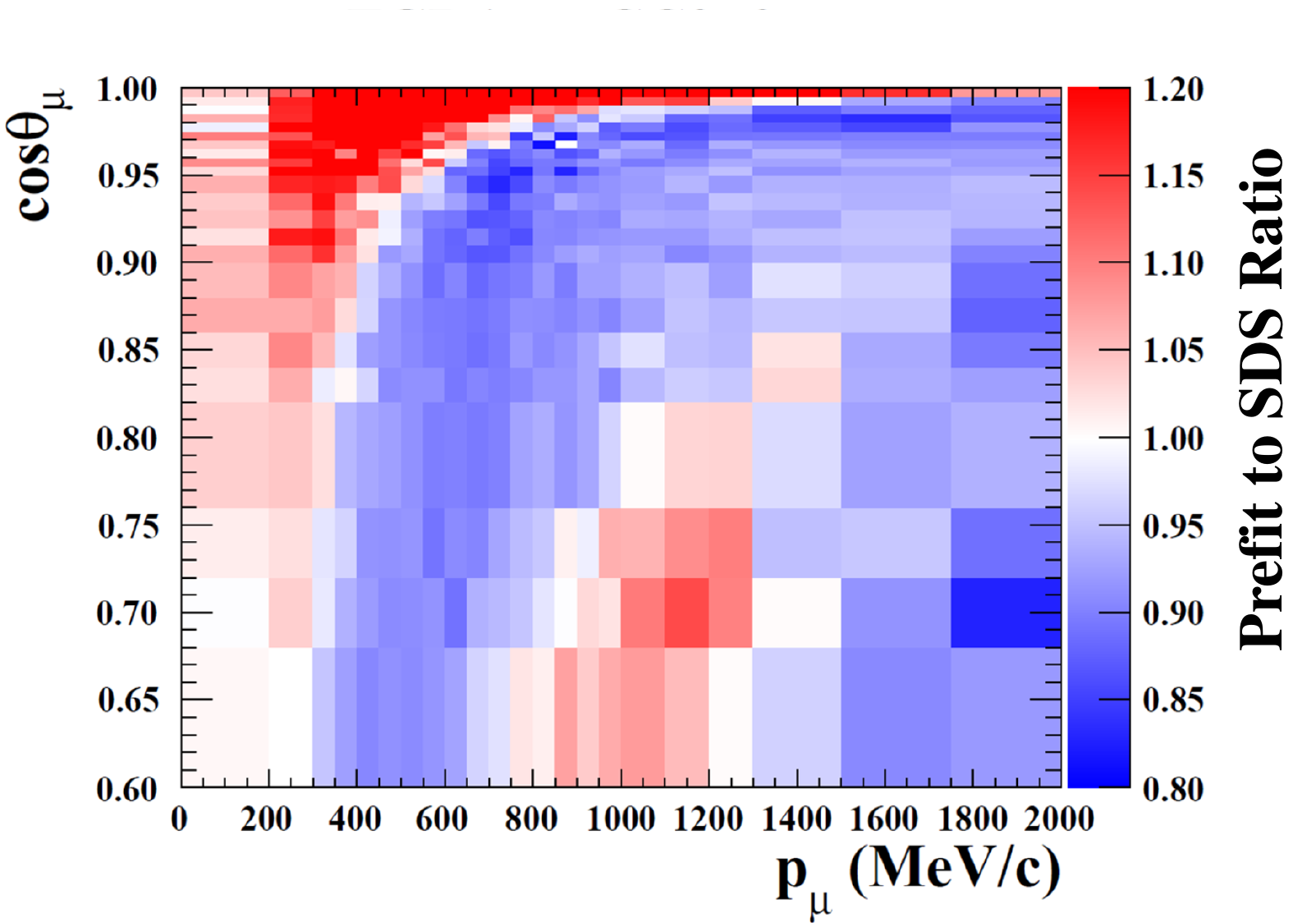}
  \includegraphics[page=1, width=\columnwidth]{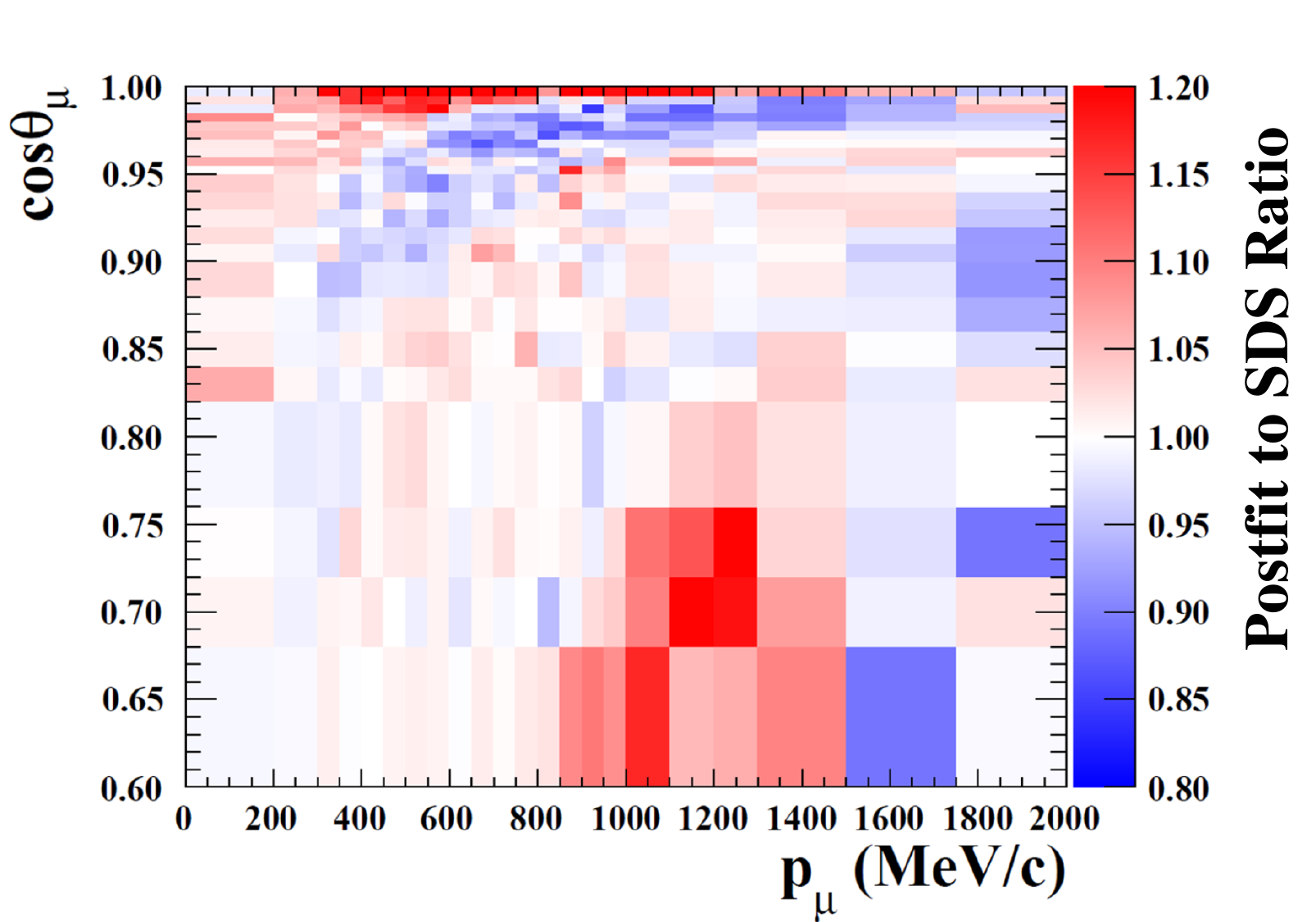}
 \caption{Ratios of the prefit (top) and postfit (bottom) event rates to the HF-CRPA SDS for the FGD1 \cczeropizerop sample as a function of the reconstructed outgoing muon momentum and angle with respect to the incoming neutrino beam.}
 \label{fig:CRPA-pmudis}
\end{figure}

Indeed, among the biases presented here, the HF-CRPA SDS shows the largest effect. Therefore, we investigate it in more detail. First, we conduct the ND280 analysis in accordance with the method employed to obtain the SDS results. Then, we construct the corresponding FD SDS and extrapolate the obtained ND constraint. The impact of HF-CRPA on the ND280 sample is presented in \autoref{fig:CRPA-pmudis}. As can be seen from the 2D distribution of the muon $\cos\theta$ versus the momentum before the fit for the \cczeropizerop sample (top plot), the effect of this SDS, with respect to the nominal MC, is to decrease the cross section at forward going and low momentum region (corresponding to low-Q$^2$), and increase for higher momenta and more backward going muons (corresponding to higher-Q$^2$). 

The corresponding ratio between the postfit 2D distribution and the SDS prediction is reported in the bottom plot in \autoref{fig:CRPA-pmudis}. This shows that, whilst the fit generally improves the agreement with the simulated data, there remain regions at very forward angles and low momenta where sizable discrepancies are observed, as well as at higher angles around 1 GeV$/c$ momentum, potentially indicating that the uncertainty parameterization is not able to describe the SDS in some parts of the kinematic phase space.

\begin{figure}[htb]
 \centering 
 \subfloat{
  \includegraphics[page=1, width=0.5\textwidth]{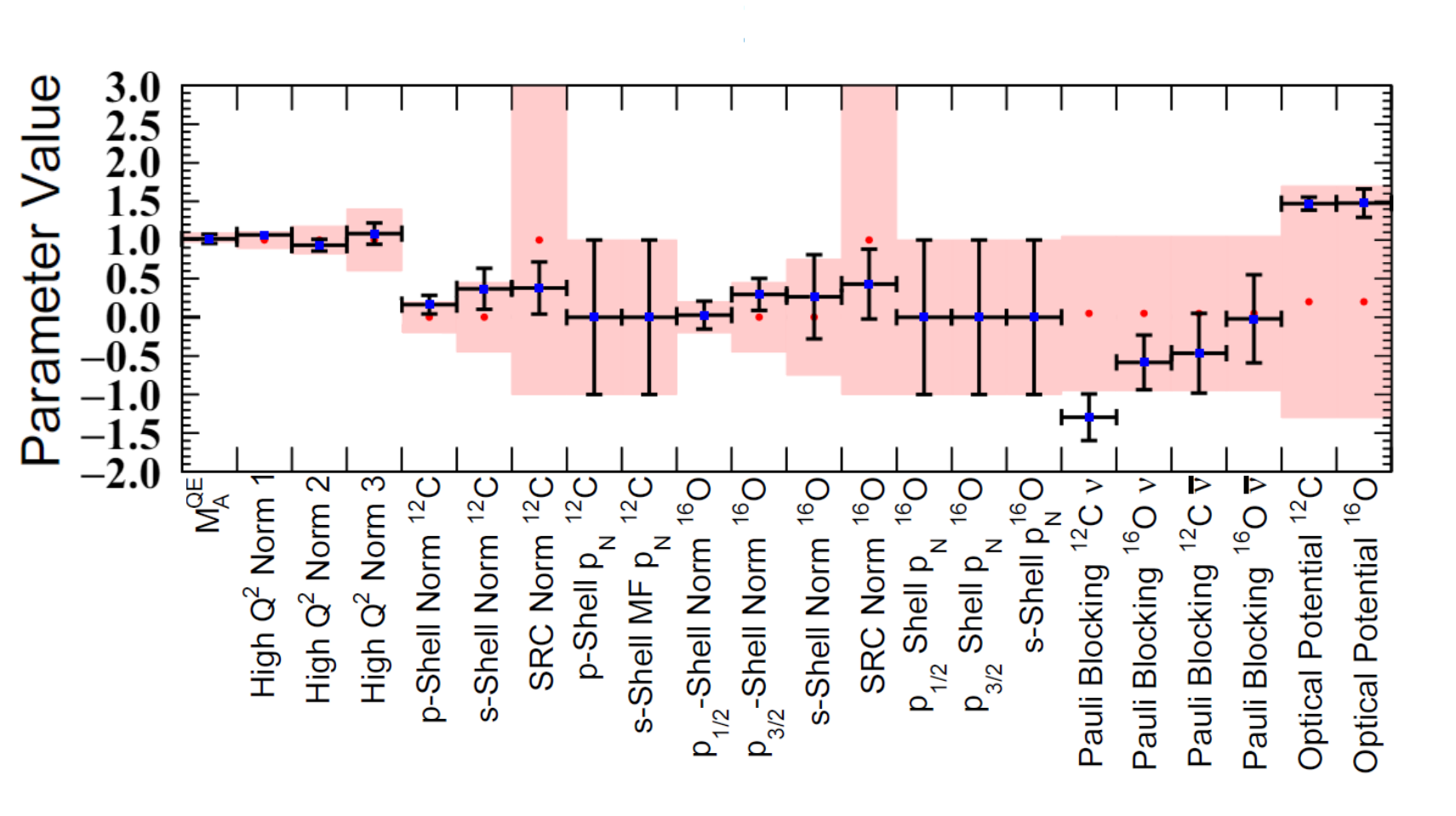}
  }
 \caption{Prefit (red band) and postfit (blue dots and black error bars) cross-section CCQE parameters from the fit to the HF-CRPA SDS.}
 \label{fig:CRPA-xsec-param}
\end{figure}

The central values and uncertainty for the CCQE parameters, as obtained from the fit to the HF-CRPA SDS, are reported in \autoref{fig:CRPA-xsec-param}. They are also compared with the prior central values and uncertainties (red band). For brevity, the fit results for other parameters are not presented, unlike for the main fit. 

To achieve the best agreement between the simulated data and the nominal MC, the fit adjusts the parameters that affect the CCQE cross section at low energy transfer, in particular, Pauli blocking and the optical potential.

When propagated to FD, a discrepancy between the extrapolated ND-to-FD spectrum and the FD-only spectrum at FD is observed, notably close to the oscillation dip in the muon (anti)neutrino disappearance samples. This indicates a potential missing freedom within the T2K uncertainty model to reliably extrapolate if the HF-CRPA is the model that most closely resembles reality

The shapes of the muon neutrino disappearance samples have a leading-order effect on the determination of $\Delta m^2_{32}$, and the discrepancy introduced by the HF-CRPA SDS has been found to introduce the largest bias on this oscillation parameter in this analysis, corresponding to 84.9\% of the size of its $1\sigma$ systematic uncertainty, and 33.8\% of the total uncertainty. The same features were observed on the electron (anti)neutrino samples, but they contribute substantially less to the relative observed bias than the disappearance samples due to the much higher statistical uncertainty. 

\begin{figure}[htb]
 \centering 
  \includegraphics[page=1, width=\columnwidth]{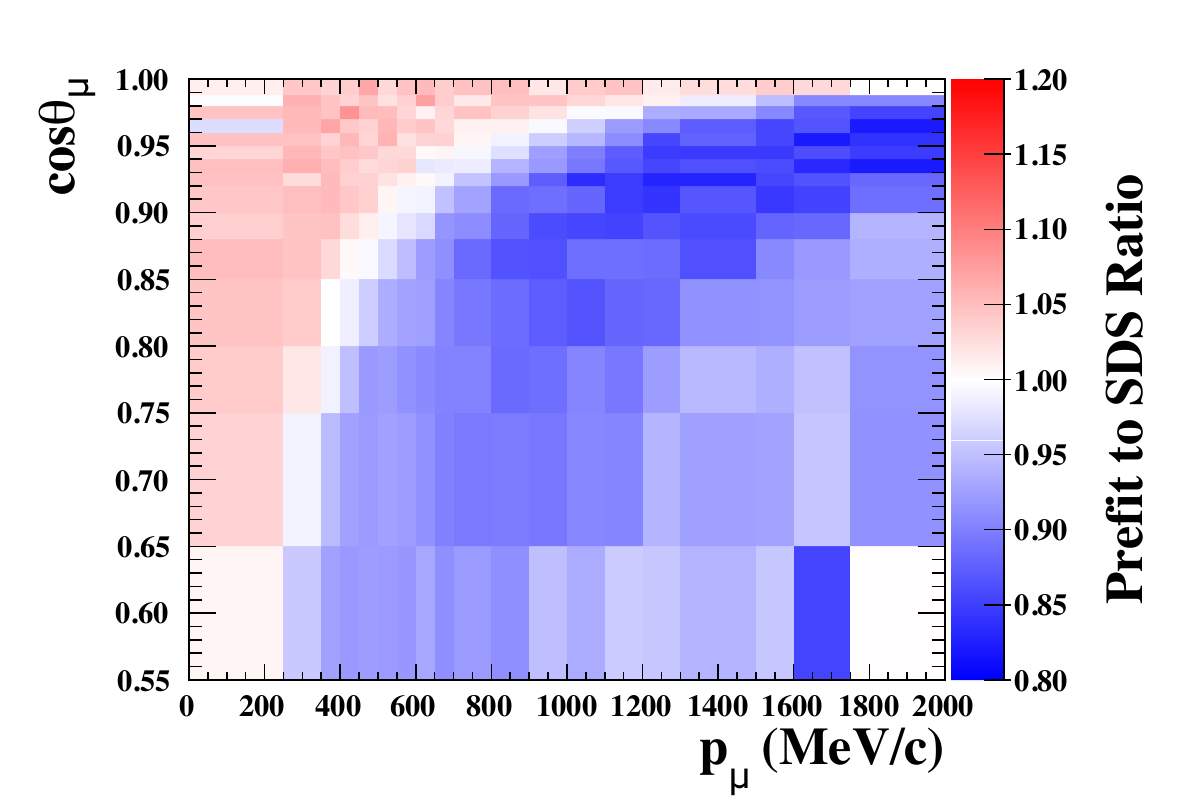}
  \includegraphics[page=1, width=\columnwidth]{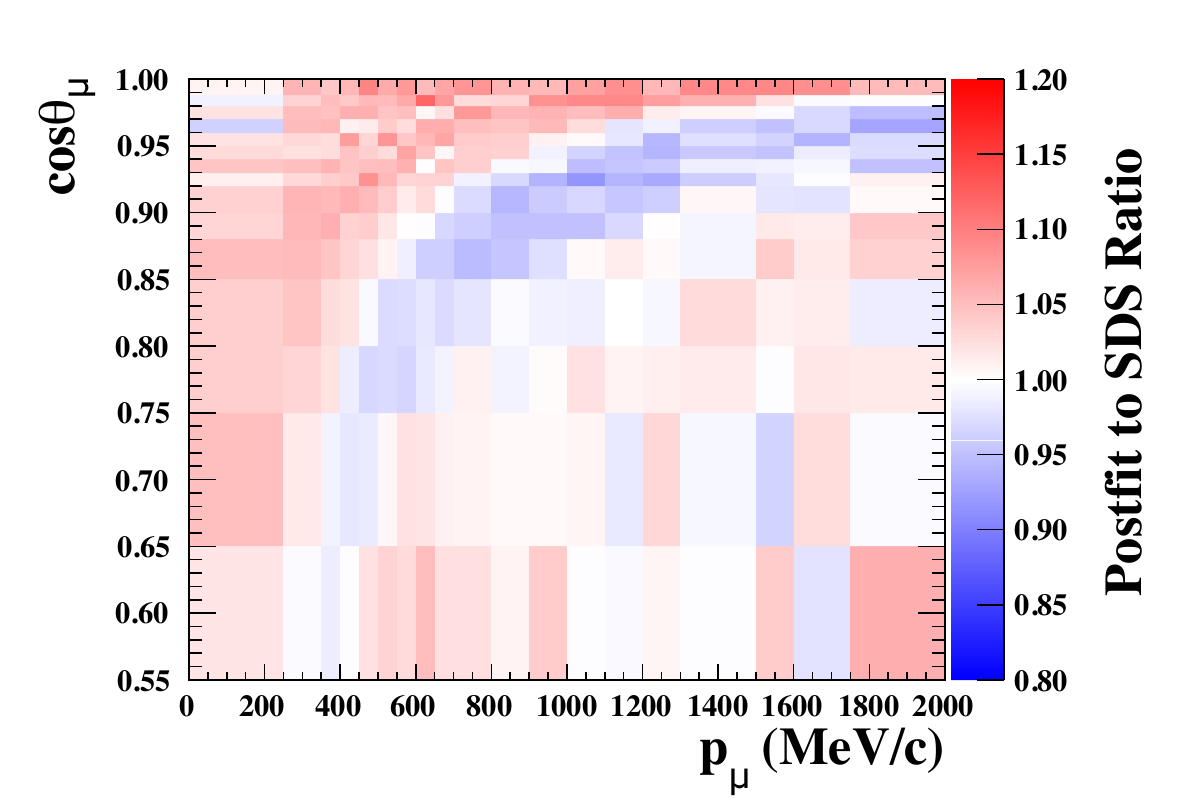}
 \caption{Ratios of the prefit (top) and postfit (bottom) event rates to the non-QE SDS for the FGD1 \cczeropiNp sample as a function of the reconstructed outgoing muon momentum and angle with respect to the incoming neutrino beam.}
 \label{fig:nonQE-pmudis}
\end{figure}

\begin{figure*}[htb]
 \centering 
  \includegraphics[width=0.4\linewidth]{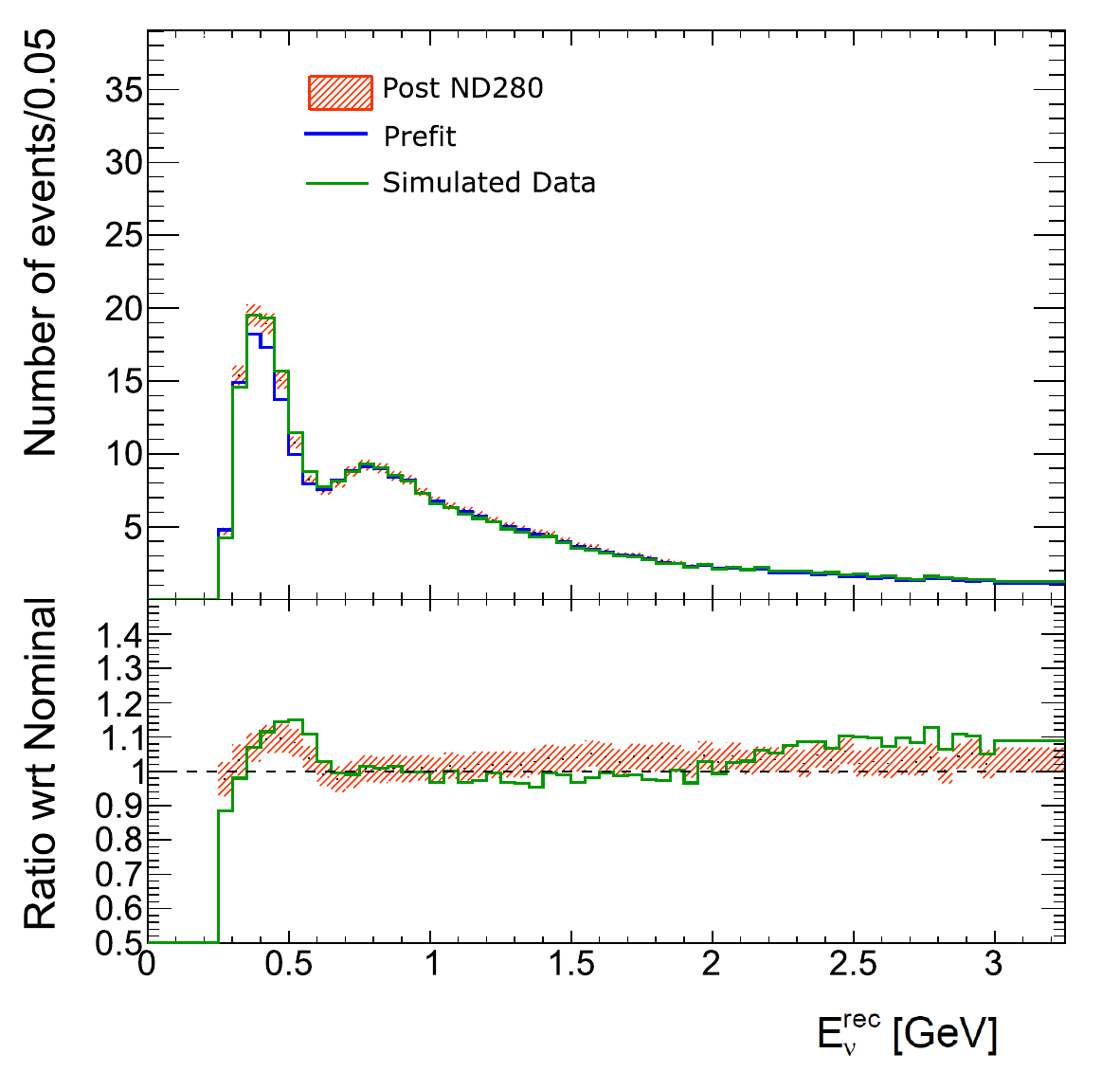}
  \includegraphics[width=0.4\linewidth]{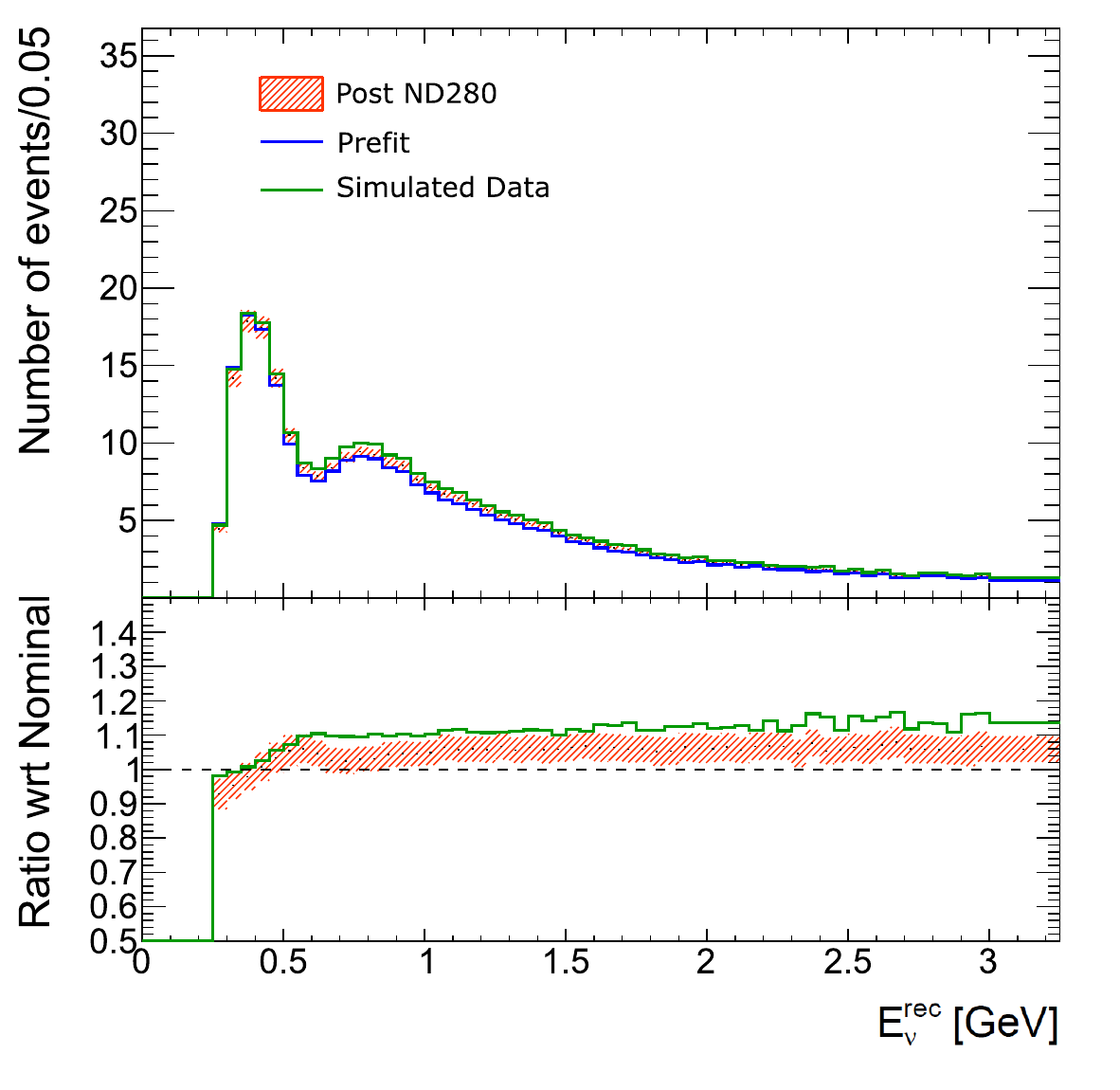}
 \caption{
 Number of predicted events at FD as a function of the reconstructed neutrino energy for the single-ring muon neutrino samples. 
 The red band corresponds to the spectrum as predicted as a result of the ND280 fit to the SDS, along with its 1$\sigma$ error. 
 The green line is the prediction obtained only by applying the weights to the oscillated FD spectra. The nominal prediction from the ND280 Asimov fit is shown for reference. The left figure presents results for HF-CRPA SDS, while the right one shows results for the non-QE SDS.}
 \label{Fig::CRPA_SK_predictions}
\end{figure*}

The SDS with the second largest impact on oscillation parameters is the non-CCQE SDS. As detailed in~\autoref{sec:newfdslist}, the effect of this SDS at FD is to significantly enhance the proportion of non-QE events in the single-ring or mesonless samples, which significantly changes the biases on the neutrino energy reconstruction. The dominant non-QE contribution in these samples comes from 2p2h interactions and pion-production processes. Although the near-detector fit has freedom to accommodate these differences by varying systematics parameters related to 2p2h, FSI, and pion-production processes (as can be seen in \autoref{fig:nonQE-pmudis}), the extrapolation to FD energies fails to reproduce the spectrum when the same set of weights is applied at the far detector. This is illustrated in the right panel of \autoref{Fig::CRPA_SK_predictions}, which shows a discrepancy between the predicted ND280 event rate and error band and the corresponding SDS at FD, notably in the high-energy tail of the 1R$\mu$ distribution. The effect is only present in the neutrino disappearance sample, which drives the statistics in the analysis. Since the SDS inputs are constructed separately for neutrino and antineutrino samples, this indicates a potential missing freedom related to the species dependence of the uncertainty model. 

\section{Discussion}
\label{sec:discussion}
The current T2K baseline model and associated uncertainty parameterization have been shown to be able to well-describe the 22 ND280 samples ($\sim$130,000 events across the $\sim$5000 kinematic bins) used for the near detector constraint within T2K's neutrino oscillation analyses. The \textit{p}-value for all samples is above 5\%. The relatively poor \textit{p}-values for the separate flux, cross-section, and detector response parameterizations may suggest that the total uncertainty parameterization is incomplete (such that the existing parameters are effectively accounting for missing degrees of freedom), which warrants attention for future higher-statistics analyses. A wide range of plausible simulated data studies gives us confidence that the current model and uncertainty parameterization are sufficient for T2K's current oscillation analysis. 

In achieving agreement with the data, the ND280 fit alters the model to reshape the CCQE cross-section as a function of $Q^2$, raising the tail of the distribution, as well as altering the distribution of $p_\mathrm{miss}$ for neutrino-carbon interactions. Alterations of parameters controlling Pauli blocking and the size of an optical potential correction allow the fit to further make a low energy-transfer suppression of the cross section, as seems to be required in the analysis of T2K cross-section measurements (\autoref{subsec:benchmark}). This suppression appears more prevalent for interactions on carbon (preferring a stronger optical potential correction) and is least required for antineutrino interactions on oxygen (which have a smaller optical potential correction and prefer less Pauli blocking than is in the nominal model, albeit with a considerable uncertainty). An increase in the nucleon FSI strength shifts the proton momentum distribution to lower values, implying more CCQE interactions in which the outgoing proton is not detected (swapping strength from the CC0$\pi$ ND280 sample with detected protons to the sample without). Two-nucleon final states from SRCs prefer a significant increase in strength. The net impact of all changes is a considerable reshaping and an increase in the normalization of the CCQE cross section with respect to the nominal model (see \autoref{tab:cross_section}).

The 2p2h cross section is also reshaped, for both oxygen and carbon, for all 2p2h pair types, to move strength from the $\Delta$ to non-$\Delta$ contributions, effectively moving the energy transfer distribution for 2p2h interactions more under the QE peak than the resonance peak. Overall, the 2p2h cross section is also increased in strength by $\sim$30\% for neutrinos and $\sim$10\% antineutrinos. Especially when considered in conjunction with the aforementioned increase in SRC interactions, this leads to a significant increase in multi-nucleon final states but concentrated at lower energy transfers than in the nominal model. 

The pion production parameters also reshape the cross section, reducing the contributions at low $Q^2$ (through the reduction of the $C_{5}^{A}$ parameter), enhancing the $\pi^0$ production over the charged-pion production, and increasing the non-resonant background. Pion FSI parameters are not pulled far from their nominal values. Whilst the resultant model can describe the pion production ND280 data well as a function of lepton kinematics, the model under-predicts the fraction of low momentum pions (as seen in \autoref{Fig::Pion}), indicating the need for model improvement.  

All variations of the cross-section parameters should be considered in the context of the fit's preference to also reshape the incoming neutrino flux, such that it gains a larger contribution at lower energies, which may partially explain the need to lower the cross section at low four-momentum transfer. Anti-correlations between the flux and cross-section parameters suggest that a smaller modification to the cross section may be plausible if the flux was fixed close to its prior shape.

Tests for potential bias from out-of-model variations show that, in general, the uncertainty model has the flexibility to describe alternative baseline neutrino interaction models and to propagate the constraints to the far detector with acceptably low impact on oscillation parameter sensitivities. These include gross variations to the CCQE, 2p2h, and pion production models. Some bias is identified in the neutrino oscillation analysis in the propagation of a constraint between the near and far detectors for the simulated data studies using the HF-CRPA model and for the ``Non-CCQE'' study. The bias redistributes model alterations from CCQE to 2p2h and pion production channels, causing a bias in subsequent measurements of $\Delta m^2_{32}$ corresponding to more than 75\% of the systematic error budget. Given the significant statistical uncertainties, the bias on the total uncertainty is less than 35\%, mitigating the impact on the present analysis. Still, the demonstrated inability of the uncertainty model to correctly propagate constraints from the near to the far detector highlights areas in which the uncertainty model could be improved in preparation for future oscillation analyses. These include, but are not limited to, increasing the flexibility of the parameterization of effects impacting the low-energy transfer region (such as Pauli blocking, optical potential corrections, long-range correlations) and potential missing freedoms in the energy dependence of these effects. Finally, the observed bias is accounted for via an uncertainty inflation.

Comparisons of the prefit and postfit models to neutrino cross-section measurements highlight further areas of improvement. Whilst the model can describe some CC0$\pi$ and CC1$\pi$ T2K measurements as a function of only outgoing lepton kinematics, with a clear improvement in agreement seen in the postfit model, the description of T2K's joint neutrino and antineutrino CC0$\pi$ measurement is poor. Moreover, the postfit model describes T2K's \dpt measurement significantly worse than the prefit model. This indicates further model development is required, especially before samples with more finely grained hadron kinematic sensitivity are used. One recent development in this direction is a microscopic QE model with distorted waves, known as ED-RMF~\cite{McKean:2025khb}.

Beyond drastically reducing uncertainties on measurements of the neutrino oscillation parameters, the impact of the ND280 fit on T2K's current and future neutrino oscillation analysis can be assessed through other means. One key element of the neutrino oscillation analysis is neutrino energy reconstruction. A reliable reconstruction is required to interpret event rate spectra at the far detector in terms of neutrino oscillation probability. Neutrino energy is reconstructed for the statistically dominant 1R$\mu$ samples at the far detector from only outgoing lepton kinematics using

\begin{align}
        E_\nu^{QE} &= \frac{1}{2} \frac{m^2_\ell+(m_{N}^\mathrm{eff})^ 2-m^2_{N'} -2E_\mu m_N^\mathrm{eff}}{E_\ell - |\vec{p_{\ell}}|\cos\theta_\ell - m_N^\mathrm{eff}} \text{ and} 
    \label{eq:enurec} \\
        m_N^\mathrm{eff} &=m_N-E_b, \nonumber
\end{align}

where $m_\ell$, $E_\ell$, $p_\ell$, and $\theta_\ell$ are the outgoing lepton mass, total energy, momentum, and angle to the incoming neutrino, respectively. $m_N$ and $m_{N'}$ are the incoming and outgoing nucleon masses, assumed to be a neutron and proton, respectively, for neutrino and vice versa for antineutrino interactions. $E_b$ is an assumed single ``binding energy'', taken to be 27 MeV, which is approximately the average removal energy in the baseline nuclear ground state model (shown in \autoref{fig:sf2DO}). This equation assumes a CCQE interaction with a stationary target nucleon, subject to a fixed nuclear binding energy. Deviations from this due to the Fermi motion of nucleons, a continuous spectrum of binding energies, and the presence of non-CCQE interactions introduce considerable smearing between true and reconstructed neutrino energy.

The distribution of the neutrino energy reconstruction bias ($\frac{E_\nu^{QE}}{E_\nu^\mathrm{true}}-1$) gives a metric for how precisely the true neutrino energy can be inferred from the reconstructed neutrino energy. However, the uncertainty on the bias from potential mismodeling of neutrino-nucleus interactions is much more important than the bias itself. A significant uncertainty indicates potential degeneracy in assigning distortions to the expected spectra of events, to alterations of oscillation probabilities, or neutrino interaction properties. 

\begin{figure}[htbp]
 \centering
   \includegraphics[page=1, width=0.92\columnwidth]{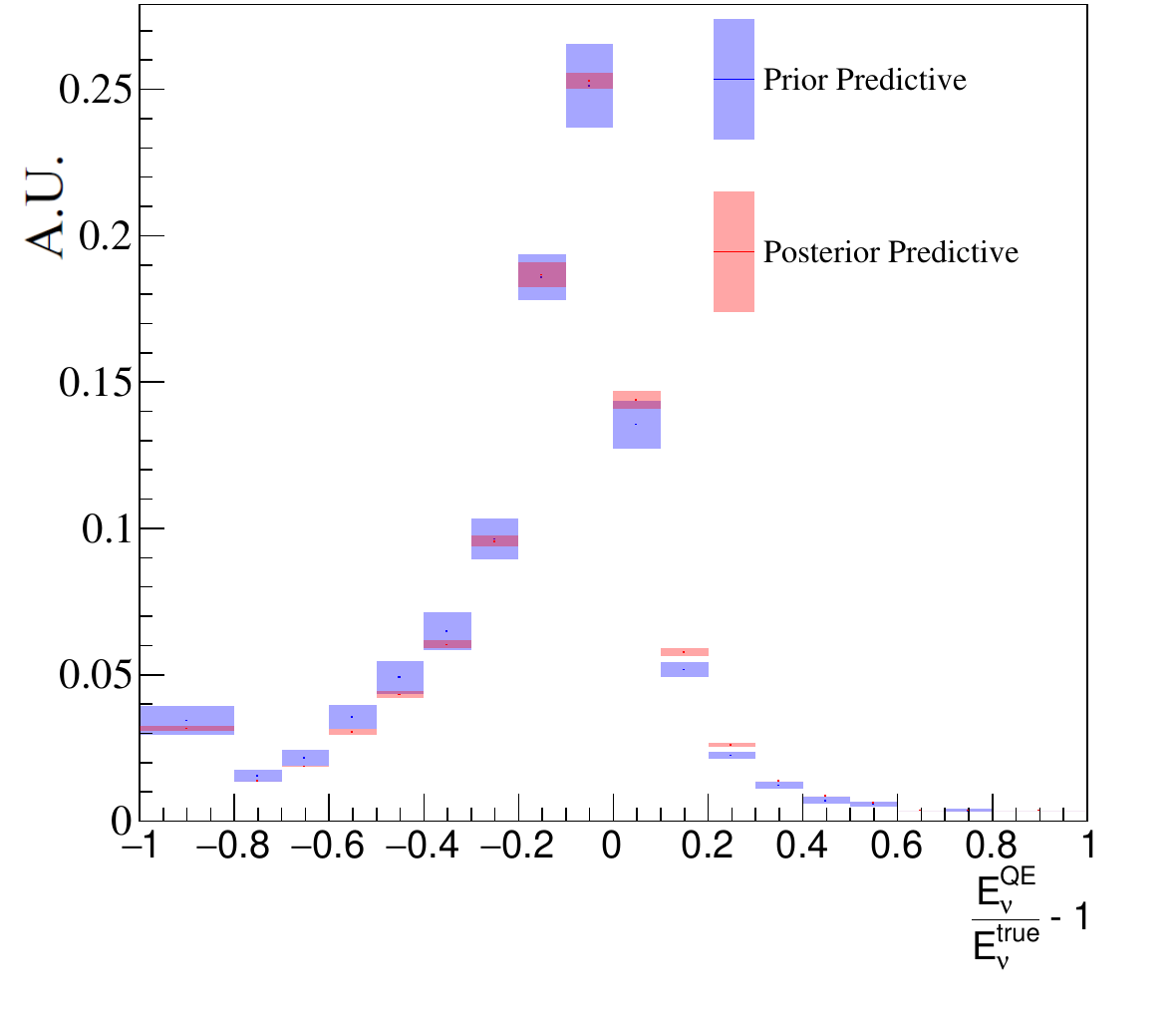} 
   \includegraphics[page=3, width=0.92\columnwidth]{Figures/Bias/Bias_BySample_New2_NoGrid.pdf} 
 \caption{Bias on reconstructed neutrino energy for \cczeropizerop FGD1 sample for the whole sample (top) and for only on 2p2h component (bottom).}
 \label{Fig::Bias}
\end{figure}

\begin{figure}[htbp]
 \centering
   \includegraphics[page=1, width=0.92\columnwidth]{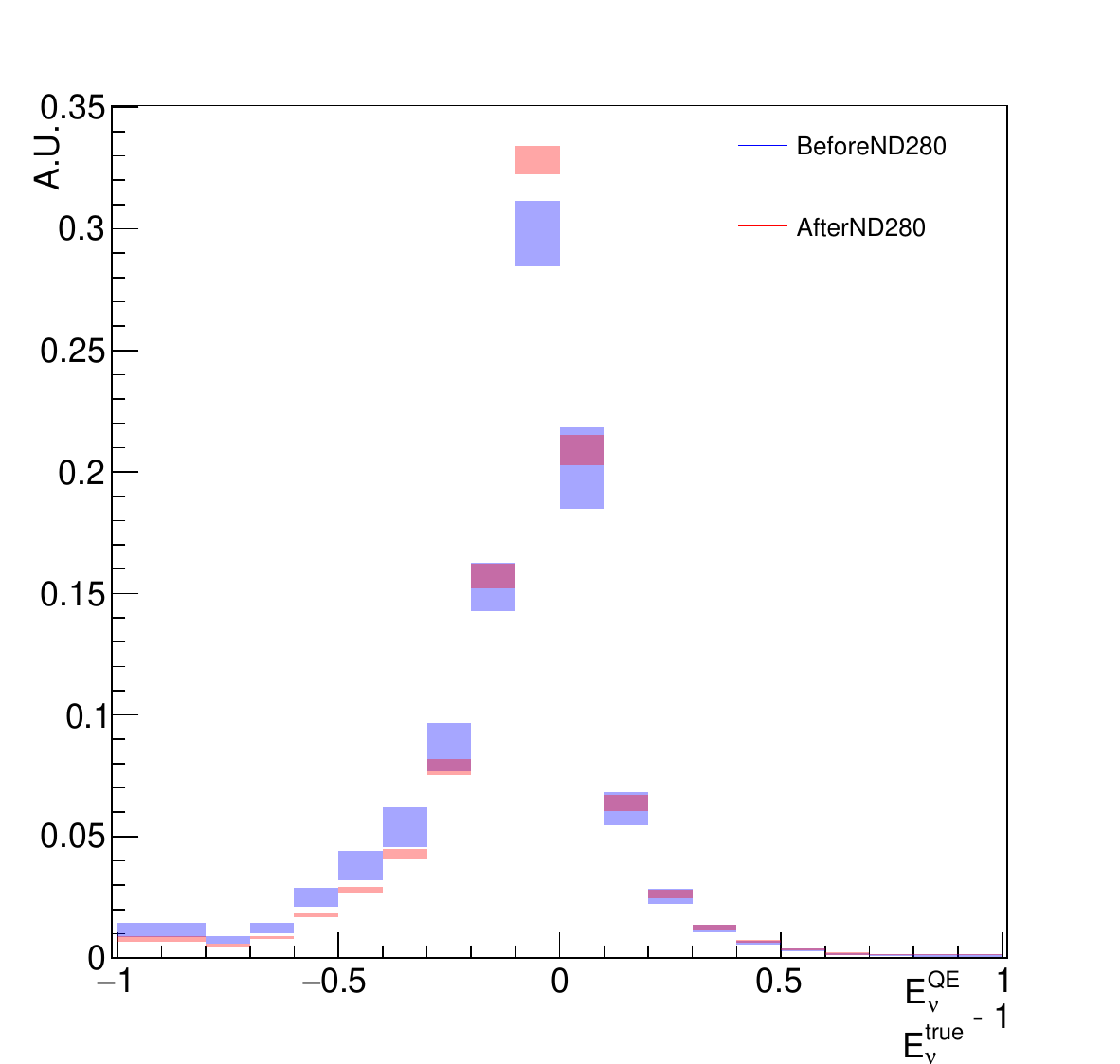}  
   \includegraphics[page=1, width=0.92\columnwidth]{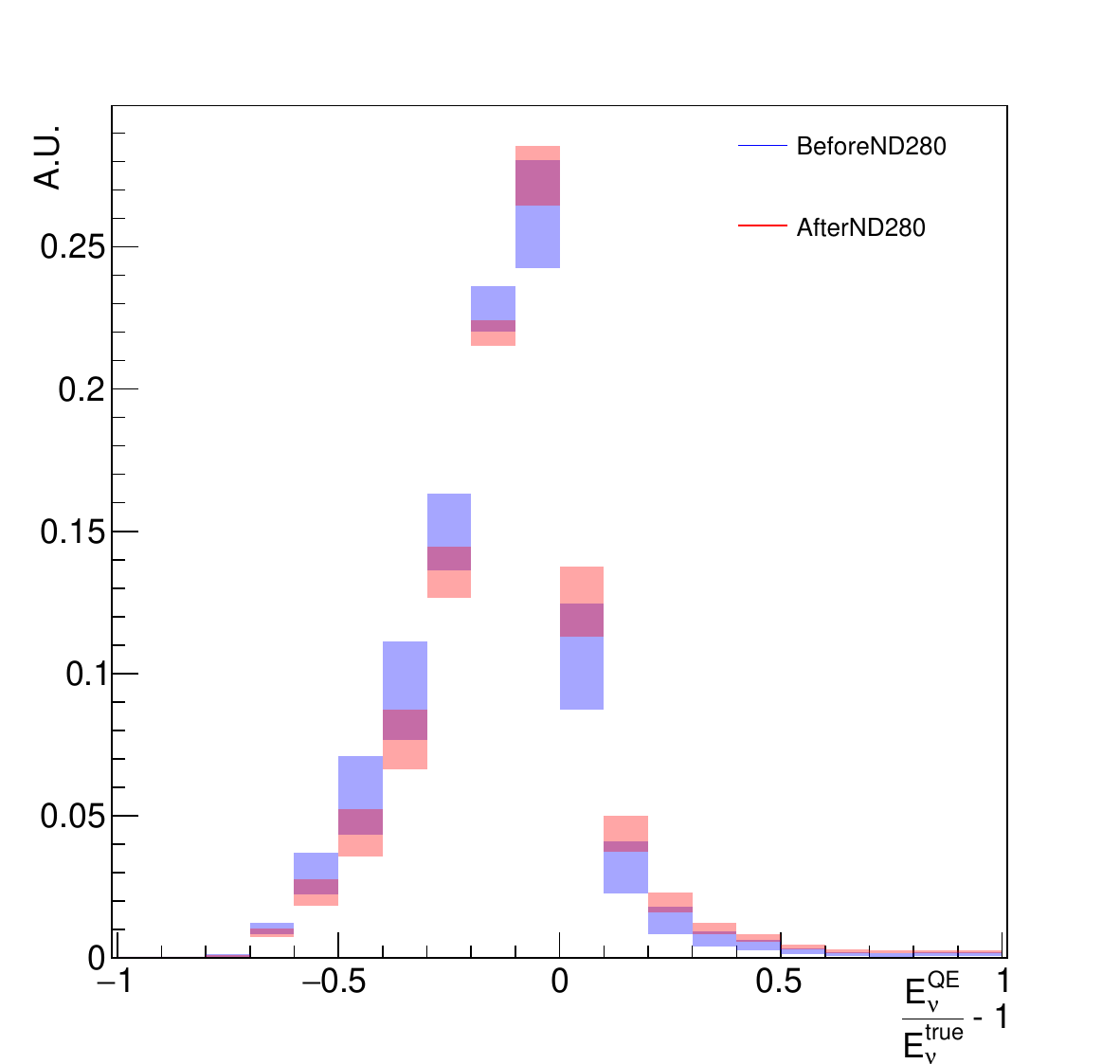} 
\caption{Bias in reconstructed neutrino energy for the FD 1R$\mu$ sample. The top panel shows the bias for the entire sample, while the bottom panel displays the bias for the 2p2h component only.}
  \label{Fig::BiasSK}
\end{figure}

\autoref{Fig::Bias} and \autoref{Fig::BiasSK} present the bias on energy reconstruction for one of the ND280 input samples and one of the input samples at the far detector, respectively, before and after the constraints from the ND280 fit for all interactions passing the sample selection and for only 2p2h interactions. The mean and width of the neutrino energy bias, as well as the uncertainty on them, before and after the fit are presented in \autoref{Tab::EnuBias} and \autoref{Tab::EnuBiasRMS}. Crucially, it is clear that the uncertainty on the shape of the bias is significantly reduced following the fit, indicating a better known inference of the neutrino energy in the neutrino oscillation analysis. A much more significant shift is observed for 2p2h compared with CCQE. As discussed in \autoref{sec:intModel}, 2p2h interactions are simulated, including contributions from non-$\Delta$-like and $\Delta$-like kinematics, the latter producing a larger bias due to production of an intermediate $\Delta$. As presented in \autoref{Fig::DataFit_1D_Xsec_CCQE}, all 2p2h shape parameters are pulled towards the non-$\Delta$-like case, shifting the bias distribution considerably to lower absolute values. 

The alteration of the neutrino energy reconstruction bias distributions and the associated uncertainty reduction on the bias are similar between the near and far detectors. Unsurprisingly, the postfit uncertainties on the distributions at the far detector are slightly larger as the ND280 samples do not have perfect kinematic overlap with them. This is due to differing detector acceptance, differing nuclear target contributions, and because neutrino oscillations change the distributions of interaction modes and particle kinematics. Still, the correlations within the baseline model allow an extrapolation of constraints from ND280, albeit with larger uncertainties than when applying them directly at ND280.  

\begin{table*}[htb!]								
\centering								
\begin{tabular}{ l | c | c | c | c }  \hline\hline
 & \multicolumn{ 2}{ c| }{ND280} & \multicolumn{ 2}{ c}{FD} \\ \hline
Mean$\pm$Error & Prior  	& Posterior  & Prior  	& Posterior \\	  \hline
All  & -0.147 $\pm$ 0.006 & -0.131 $\pm$ 0.002 & -0.096 $\pm$ 0.006 & -0.074 $\pm$ 0.002 \\ 
CCQE & -0.051 $\pm$ 0.004 & -0.050 $\pm$ 0.002 & -0.001 $\pm$ 0.003 & -0.007 $\pm$ 0.002 \\
2p2h & -0.166 $\pm$ 0.012 & -0.113 $\pm$ 0.003 & -0.145 $\pm$ 0.008 & -0.114 $\pm$ 0.005 \\ \hline\hline
\end{tabular}
\caption{Mean and error on E$^{\nu}_\mathrm{bias}$ distribution for selected ND280 (\cczeropizerop) and FD (1R$\mu$) sample. Values are for the whole sample as well as CCQE and 2p2h contributions.}
\label{Tab::EnuBias}
\end{table*}

\begin{table*}[htb!]								
\centering								
\begin{tabular}{ l | c | c | c | c }  \hline\hline
 & \multicolumn{ 2}{ c| }{ND280} & \multicolumn{ 2}{ c}{FD} \\ \hline
RMS$\pm$Error & Prior  	& Posterior  & Prior  	& Posterior \\	  \hline
All  & 0.279 $\pm$ 0.004 & 0.275 $\pm$ 0.001 & 0.275 $\pm$ 0.005 & 0.208 $\pm$ 0.002 \\
CCQE & 0.200 $\pm$ 0.003 & 0.206 $\pm$ 0.001 & 0.153 $\pm$ 0.002 & 0.150 $\pm$ 0.002 \\ 
2p2h & 0.248 $\pm$ 0.009 & 0.235 $\pm$ 0.002 & 0.195 $\pm$ 0.006 & 0.206 $\pm$ 0.004 \\ \hline	
\end{tabular}
\caption{RMS and error on E$^{\nu}_\mathrm{bias}$ distribution for ND280 (\cczeropizerop) and FD (1R$\mu$) sample. Values are for the whole sample as well as CCQE and 2p2h contributions.}
\label{Tab::EnuBiasRMS}
\end{table*}

\begin{figure*}[htpb]
 \centering 
   \includegraphics[page=1, width=0.88\columnwidth]{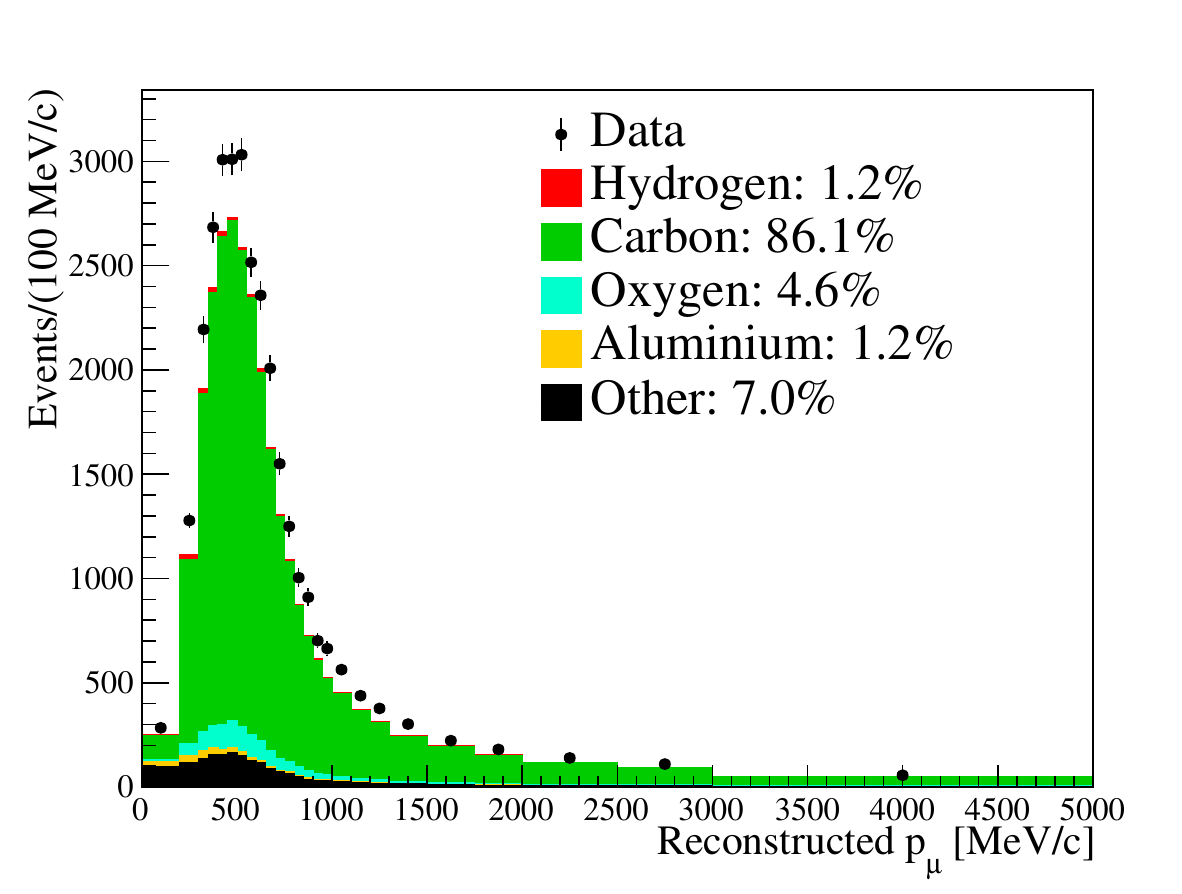}
    \includegraphics[page=1, width=0.88\columnwidth]{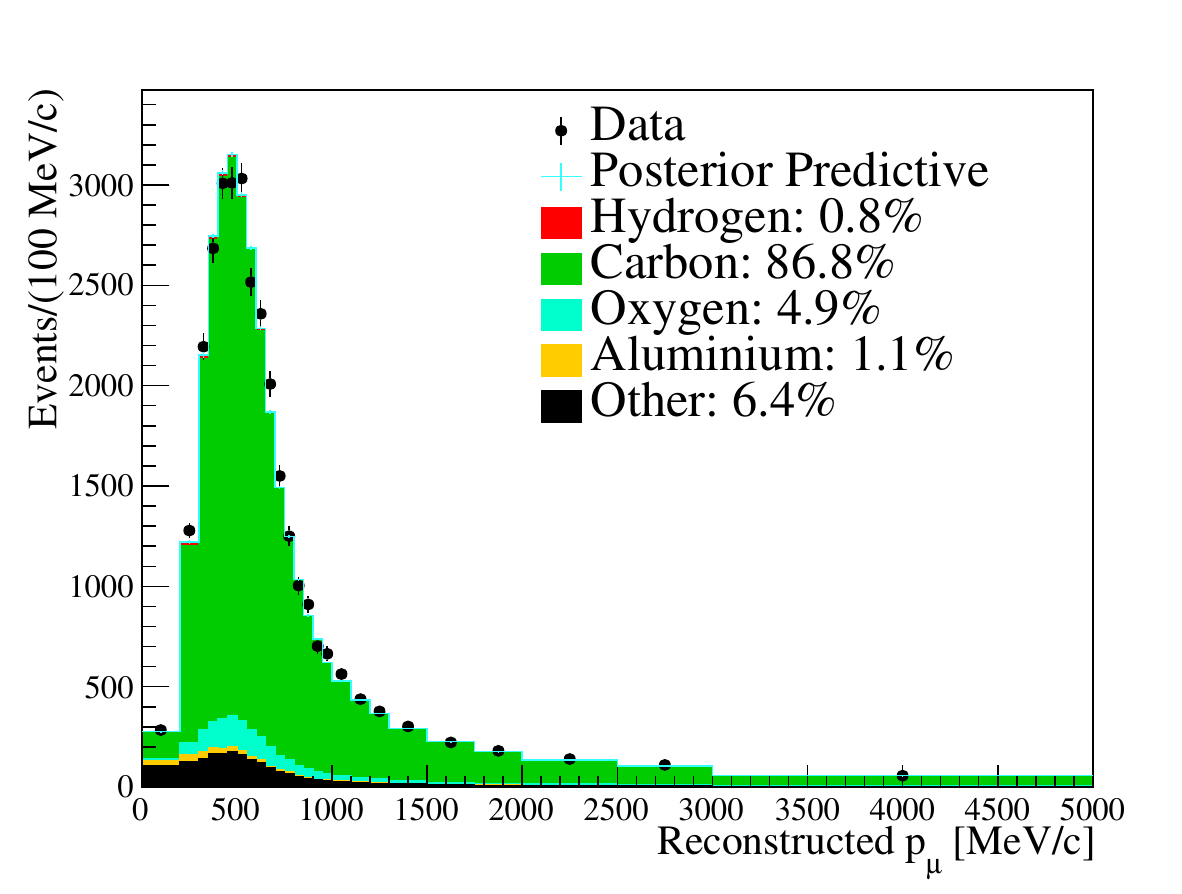} 
 
   \includegraphics[page=6, width=0.88\columnwidth]{Figures/Breakdown/TargetNominal_NOgRID_NoTittle.pdf} 
    \includegraphics[page=6, width=0.88\columnwidth]{Figures/Breakdown/TargetPosterior_NoGrid_NoTitle.pdf} 
 \caption{Target breakdown for before (left) and after (right) the fit for FGD1 (top) and FGD2 (bottom) \cczeropizerop sample.}
 \label{Fig::TargetBreakdwons}
\end{figure*}

In addition to analyzing how the ND280 constraint helps to characterize the neutrino energy reconstruction bias, it is also interesting to consider how it constrains the cross section on oxygen compared to carbon. The breakdown of nuclear targets in one of the FGD1 and FGD2 samples is shown in \autoref{Fig::TargetBreakdwons}, illustrating the dominance of carbon interactions in the former and a fairly even split between oxygen and carbon in the latter. This is shown before and after the fit, demonstrating that the breakdown of interactions as a function of nuclear target remains essentially unchanged. \autoref{Tab::TargetError} presents the uncertainty on the selected interaction rate at ND280 for different nuclear targets before and after the fit. It can be seen that the prior error on oxygen is similar to that of carbon before the fit. The difference is due to the different treatments of nuclear ground-state uncertainties for the two targets. However, the postfit uncertainty is about twice as large for oxygen compared to carbon. This demonstrates that the current model does not entirely propagate constraints on carbon to those on oxygen (for example, the nuclear ground-state uncertainties for the two targets are entirely uncorrelated). This is partially driven by the fact that many nuclear effect uncertainties between carbon and oxygen, such as Pauli blocking, optical potential, or shell strength, are treated as uncorrelated in the model, with only a few effects assumed to be partially correlated, like $\Delta E_\mathrm{rmv}$, or fully correlated, like FSI. This highlights the importance of the oxygen in the FGD2 samples to directly constrain oxygen cross section, but also demonstrates how improved modeling to robustly infer alterations of the oxygen cross sections from $^{12}$C data would allow a significant additional reduction of uncertainty.

\begin{table}[htb]
\centering
\begin{tabular}{ l | c | c }
\hline\hline
Mode & \multicolumn{2}{ c }{$\delta N / N (\%)$}   \\ \hline
 & Prefit & Postfit  \\ \hline
Hydrogen & 7.46 & 2.05 \\
Carbon & 10.11 & 0.50 \\
Oxygen & 11.35 & 1.19 \\ \hline\hline
\end{tabular}
\caption{Fractional uncertainties on event rates for neutrino interactions on different targets used in ND280 before and after the fit.}
\label{Tab::TargetError}
\end{table}

\section{Future sensitivity}
\label{sec:sensitivity}
To overcome the current ND280 limitations and further reduce flux and neutrino interaction uncertainties, an upgrade of the near detector has just been completed~\cite{T2K:2019bbb}. 

\begin{figure}[htbp]
 \centering
    \includegraphics[page=1, width=1\columnwidth]{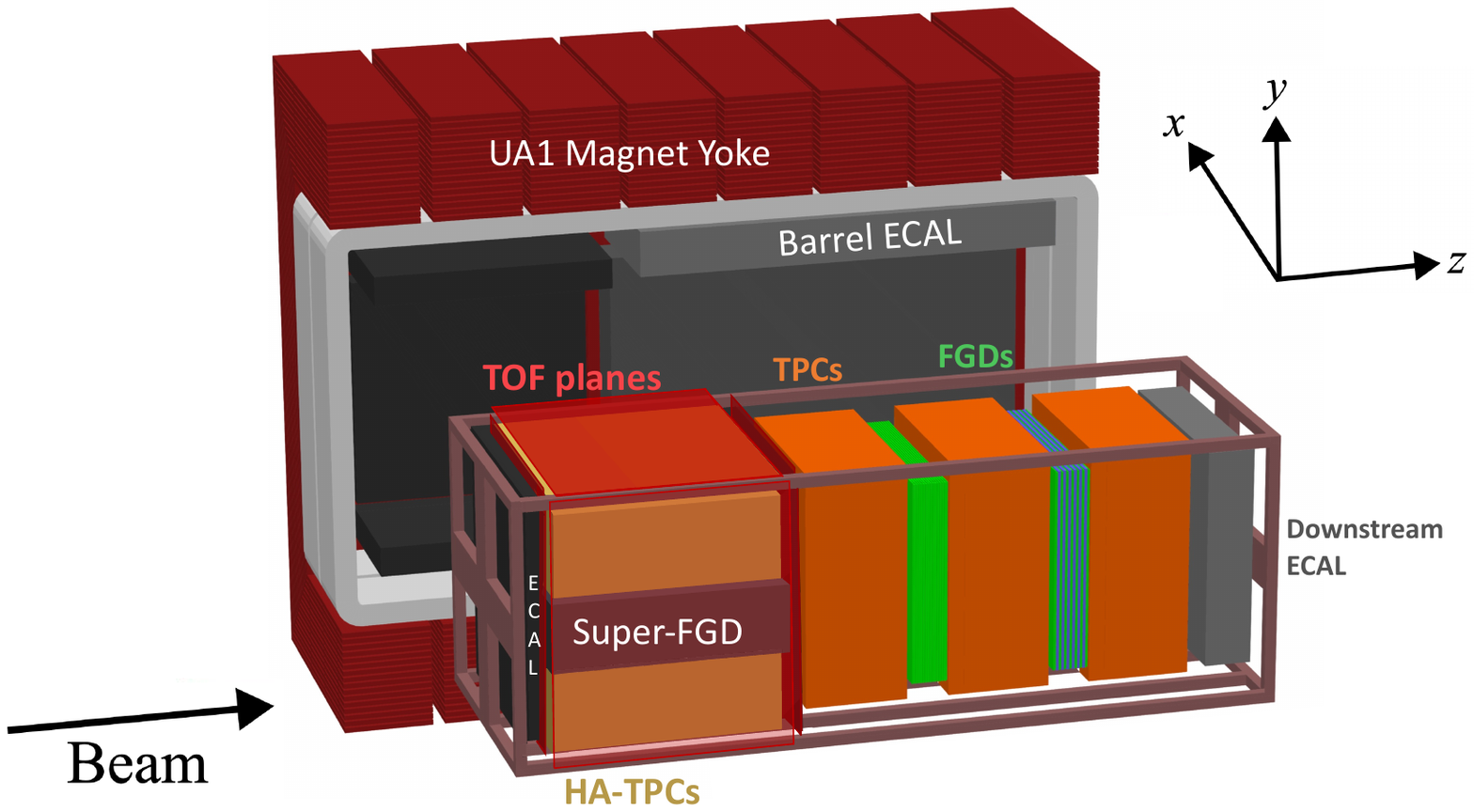} 
 \caption{Scheme of the ND280-Upgrade design: the P$\emptyset$D (shown in \autoref{fig:nd280}) has been replaced by the Super-FGD and two horizontal TPCs, all surrounded by a Time Of Flight detector. }
 \label{Fig::NDUp}
\end{figure}

The ND280-Upgrade allows for a full polar angle acceptance for muons produced in charged current interactions \cite{T2K:2019bbb}, as well as the ability to reconstruct low momentum outgoing protons (i.e., those with $p_p > 300~\text{MeV}/c$) and to detect outgoing neutrons \cite{PhysRevD.101.092003}. To reach this goal, as shown in \autoref{Fig::NDUp}, the P$\emptyset$D detector has been dismantled and replaced by a new highly granular, 3D
scintillator detector \cite{Blondel_2018}, called the Super-FGD \cite{Blondel_2020,AGARWAL2023137843,Alekseev_2023}, contained between two new horizontal TPC \cite{Attie:2022smn,Ambrosi:2023smx}, named High-Angle (HA) TPCs, able to reconstruct tracks at a high angle relative to the neutrino direction. An additional new Time Of Fight detector surrounds HATPCs and Super-FGD. The Super-FGD is composed of 2 million 1\,cm plastic scintillator cubes, read out by three planes of wavelength-shifting fibers in three orthogonal directions. This results in roughly doubling the current target mass of the ND280 tracker whilst dramatically improving its track resolution and acceptance.

In this section, we report the expected improvement in our ability to constrain neutrino interaction parameters with the newly upgraded ND280. This work completes preliminary studies previously published, where a simplified near-detector analysis was performed of the ND280-Upgrade \cite{Dolan:2021hbw}. This time, a complete near-detector analysis replicating what is described in the present paper is applied. This implies the simultaneous use of the  \num and \numb ND280-Upgrade CC0$\pi$ samples, in addition to all the event samples described in \autoref{sec:NDfit}, as well as the implementation of an identical parameterization of the flux and neutrino interaction uncertainties also for the \ndup samples.

The upgrade of the near detector is accompanied by improvements to the beam power of the J-PARC accelerator. The maximum power that was reached in currently used T2K data runs is $\sim 500$~kW, while most recent runs were taken with a beam power around 800~kW. 

This is expected to steadily increase further over the upcoming years to reach the goal of 1.3~MW in 2027, before the start of Hyper-Kamiokande. It will result in an increased delivered POT for T2K. In particular, within three years of data taking\footnote{Each year consists of around four months of continuous data taking.}, T2K will reach over double the delivered POT collected so far. The POT assumed for the next years in this study are summarized in \autoref{tab:t2k2POT}; they are separately reported for FD and ND280, according to the current difference between the accumulated statistics at the two detectors.
 
The Monte Carlo samples used for the \ndup are the same as those described in Ref.~\cite{Dolan:2021hbw}. These samples are obtained by using the same NEUT model as the one in \autoref{sec:intModel}; detector effects on reconstructed quantities are then applied on a particle-by-particle basis as Gaussian smearing on the angle and momentum for protons, muons, and charged pions. For neutrons, the same pseudo-reconstruction as detailed in Ref.~\cite{PhysRevD.101.092003} is used. 

In this study, only CC0$\pi$ samples are used for the upgraded part of  ND280. 
 
Thanks to the expected ability to reconstruct neutrons in the \sfgd, it is not only possible to further split the CC0$\pi$ samples depending on the number of reconstructed protons or neutrons in the final state, but also to exploit their kinematics variables. Thus, in addition to all the samples listed in \autoref{Tab:NDsamples}, four additional ones are considered for the \sfgd target: CC0$\pi$-0$p$, CC0$\pi$-N$p$ for the $\nu$-mode and CC0$\pi$-0$n$ and CC0$\pi$-N$n$ for the $\overline\nu$-mode. As reported in \autoref{Tab:NDupsamples}, where 1.08 $\times$ 10$^{21}$ POT are assumed, the \ndup samples with and without reconstructed final state hadrons are expected to be roughly equally populated. 

The ability of the \ndup to precisely reconstruct the kinematics of low momentum protons ($p>300~\text{MeV}/c$) and neutrons opens the door to the use of more sophisticated variables exploiting both the lepton and the hadron kinematics information. For samples with reconstructed protons or neutrons, among the combinations proposed in Ref.~\cite{Dolan:2021hbw}, the potential of a 2D binning in \dpt and \evis has been explored. While \dpt is the transverse momentum imbalance as defined in \cite{Lu:2015tcr}, \evis is defined here as the sum of the muon energy and of the final nucleon kinetic energy. This variable pair is very powerful in constraining 1p1h, 2p2h, and FSI systematics. For samples without a reconstructed final nucleon, the standard \pmu and \cosmu binning is used, replicating the selection used for the ND280 FGD1 CC0$\pi$-0$p$-0$\gamma$ sample given in~\autoref{subsec:ndsel}.

\begin{table}[]
\centering
\begin{tabular}{ c | c | c | c}
\hline\hline  
{Power [kW]}  & {POT/year}  &  {Accumulated }&  {Accumulated }\\
 &  &   {ND280 POT} &{FD POT}\\
\hline    -- & 0 & $2.02 \times 10^{21}$ & $3.82 \times 10^{21}$ \\
 757 & $1.08 \times 10^{21}$ & $3.10 \times 10^{21}$ & $4.90 \times 10^{21}$\\
 830 & $1.18 \times 10^{21}$ & $4.28 \times 10^{21}$ & $6.08 \times 10^{21}$\\
 928 & $1.32 \times 10^{21}$ & $5.60 \times 10^{21}$ & $7.40 \times 10^{21}$\\
 1160 & $1.65 \times 10^{21}$ & $7.25 \times 10^{21}$ & $9.06 \times 10^{21}$\\
 1226 & $1.75 \times 10^{21}$ & $9.00 \times 10^{21}$ & $10.8 \times 10^{21}$ \\
\hline\hline
\end{tabular}
\caption{Summary of the accumulated POT as in the analysis presented in this paper (first line) and of the assumed delivered POT in the next years, per each step of increased beam power, assuming 4 months of data taking per step. The POT are separately given for the near detector (ND280), used for the presented studies, and the far detector (FD).}
\label{tab:t2k2POT}
\end{table}

\begin{table}[]
\centering
\begin{tabular}{ l | c | c | c }
\hline\hline
Selection                   & Topology              & Target & Expected Events \\ \hline
\multirow{2}{*}{$\nu_{\mu}$ in $\nu$-mode} 
                            &    {0$\pi$-0$p$}       & \sfgd & 49789 \\ \cline{2-4}
                            &    {0$\pi$-N$p$} & \sfgd & 46821  \\\cline{2-4}
\hline                          
\multirow{2}{*}{\numb in \nub-mode} 
                            & {0$\pi$-0$n$} & \sfgd & 12808 \\\cline{2-4}
                            &  {0$\pi$-N$n$}& \sfgd & 10975 \\ \cline{2-4}
\hline\hline                                                        
\end{tabular}
\caption{Expected number of events for each \ndup sample, for $1.08 \times 10^{21}$~POT in the $\nu$-mode or in $\overline\nu$-mode.}
\label{Tab:NDupsamples}
\end{table}

To probe the potential of the additional variables the \ndup gives access to, three configurations have been considered and compared:
\begin{itemize}
	\item FGD1+2 only: this corresponds to the 22 samples already treated in the paper, binned in (\pmu, \cosmu), without any additional samples.
	\item Super-FGD+FGD1+2 $\mu$ only: to these 22 samples the four \sfgd CC$0\pi$ samples are added, all of them binned in (\pmu, \cosmu).
	\item Super-FGD+FGD1+2 $\mu+N$: the four additional \sfgd samples are binned in (\dpt, \evis) when a nucleon is reconstructed, and in  $(\pmu, \cosmu)$ when no nucleon is detected.
\end{itemize} 

\begin{figure*}[]
    \centering
    {\includegraphics[width=0.45\linewidth,page=1,trim={0 0 1cm 1cm},clip]{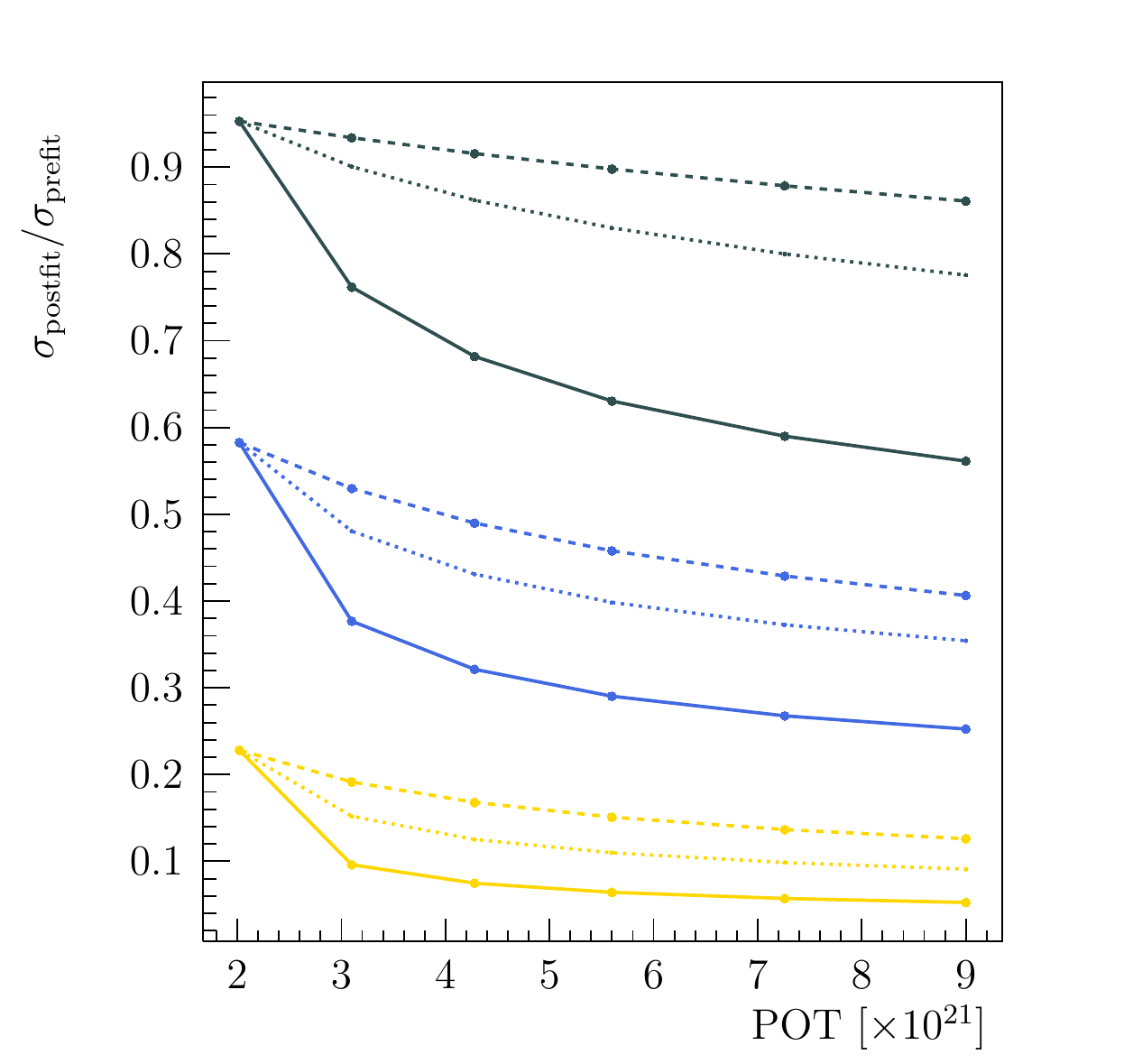}} 
    \hfill
    {\includegraphics[width=0.45\linewidth,page=2,trim={0 0 1cm 1cm},clip]{Figures/sensitivity/xsec_params_forNDPaper_NDPOT.pdf}} 
    {\includegraphics[width=0.45\linewidth,page=3,trim={0 0 1cm 1cm},clip]{Figures/sensitivity/xsec_params_forNDPaper_NDPOT.pdf}} 
    \hfill{\includegraphics[width=0.45\linewidth,page=4,trim={0 0 1cm 1cm},clip]{Figures/sensitivity/xsec_params_forNDPaper_NDPOT.pdf}} 
    {\includegraphics[width=0.45\linewidth,page=5,trim={0 0 1cm 1cm},clip]{Figures/sensitivity/xsec_params_forNDPaper_NDPOT.pdf}} 
    \hfill
    {\includegraphics[width=0.45\linewidth,page=6]{Figures/sensitivity/xsec_params_forNDPaper_NDPOT.pdf}}
    \caption{Expected constraints on some 1p1h- and 2p2h-related parameters for the FGD1+2 only (dashed), Super-FGD+FGD1+2 $\mu$ only (dotted) and  Super-FGD+FGD1+2 $\mu+N$ (full) configurations described in the text. Plots are a function of the accumulated POT at the near detector.} 
    \label{fig:sensi_CCQE1_t}
\end{figure*}

Asimov near-detector fits are then performed assuming the expected POT for each increasing beam power step as reported in \autoref{tab:t2k2POT}. At each step, 50\% of $\nu$-mode and 50\% of $\overline\nu$-mode beam are considered. The sensitivity to constrain systematic uncertainties with the near-detector analysis in each of the three configurations is then quantitatively estimated by quoting the ratio of the postfit parameter errors to the prefit values. Results for some 1p1h and 2p2h parameters are reported in \autoref{fig:sensi_CCQE1_t}.

As shown in this figure, the expected reduction in postfit uncertainties is driven not only by the increased number of collected events but also by the choice of variables used for the analysis. A notable improvement in the near-detector constraints can be seen once using the (\evis, \dpt) combination for \sfgd samples with reconstructed proton/neutrons, for all the SF shell-related carbon parameters (top left panel of \autoref{fig:sensi_CCQE1_t}). Also, a strong sensitivity to the carbon $p_N$ shape parameters, which describe the uncertainties on the initial-state nucleon momentum $\vec{p}$, is observed, as shown in \autoref{tab:xsecparameters}. Indeed, data coming from the \ndup will allow constraints on such parameters, especially when using variables like \dpt and \evis specifically adopted to gain sensitivity on the initial nucleon kinematics. 

Projected uncertainties on Pauli blocking and optical potential parameters for carbon are displayed in the top right panel of \autoref{fig:sensi_CCQE1_t}. The reduction of the uncertainties is clear and is due to the increased number of carbon interactions expected with the addition of the \ndup. These parameters are not better constrained when binned in (\evis, \dpt). This is expected since these parameters specifically affect the low momentum transfer region, roughly corresponding to the high \cosmu regions and to low momentum hadron events, that are typically selected in the 0$\pi$-0$p/n$ samples. The standard (\pmu, \cosmu) binning is thus particularly effective in constraining these parameters that instead don't strongly affect the  0$\pi$-N$p/n$ samples.

Oxygen SF shell-related parameters can only be constrained by neutrino interactions on oxygen, thus happening in the FGD2. In the current configuration, where no prior correlations are assumed between carbon and oxygen SF shell-related parameters, the addition of the \ndup samples into T2K's oscillation analysis cannot help in reducing the oxygen parameter uncertainties (middle left panel of \autoref{fig:sensi_CCQE1_t}). Indeed, as can be seen in \autoref{Fig::CorrMatrixCCQE} and in Fig. 14 of Ref.~\cite{Chakrani:2023htw}, also in data fits, postfit correlations between SF oxygen and carbon parameters remain weak.

Similar considerations apply for 2p2h parameters (bottom right panel of \autoref{fig:sensi_CCQE1_t}). For them, an improved constraint is observed for carbon, more so when CC0$\pi$-N$p$ samples can be fully exploited, while the same is not true for oxygen.

The situation is different for $E_\mathrm{rmv}$ parameters (middle right panel of \autoref{fig:sensi_CCQE1_t}), where the existing prior oxygen/carbon correlations allow the additional samples from \ndup to constrain both carbon and oxygen-related parameters better.

\begin{figure}[]
 \centering 
 \includegraphics[width=0.9\columnwidth]{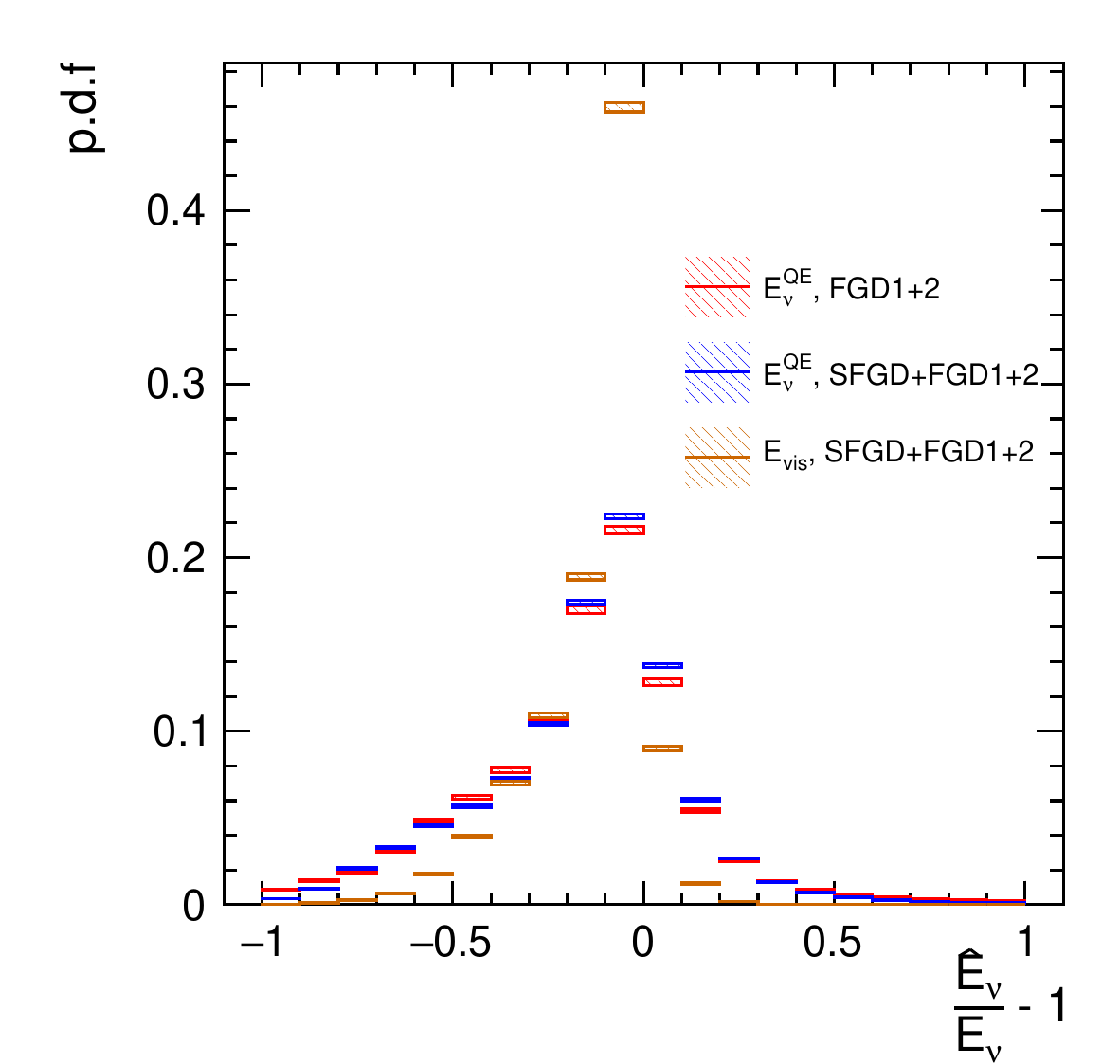} 
 \caption{Postfit bias on the reconstructed neutrino energy when using different energy estimators (collectively denoted as $\hat{E_\nu}$) using 9 $\times$ 10$^{21}$ at ND280 and considering only neutrino-carbon interactions and  CC0$\pi$ samples. The different energy estimators consider: using FGD1+2 samples (red) and E$_\nu^{QE}$, adding \sfgd samples and using E$_\nu^{QE}$, and adding \sfgd samples and using \evis for samples with reconstructed protons/neutrons (brown). The corresponding biases are respectively $-0.16 \pm 0.28$, $-0.15 \pm 0.26$, and $-0.13 \pm 0.15$.}
 \label{fig:ereso_ndup}
\end{figure}

The effect of the improved constraints on neutrino interaction parameters can be visualized in \autoref{fig:ereso_ndup}, where the bias on the neutrino reconstructed energy is reported for CC0$\pi$ carbon events, showing a dramatic reduction when the full potential of the \ndup is used.

In \autoref{tab:upgradeSK}, the effect of the near-detector constraints on flux and neutrino interaction uncertainties is evaluated on the two main appearance channels at FD. While the inclusion of \ndup samples in the ND280 fit causes a relatively modest additional constraint, this grows dramatically with the use of an assumed correlation between oxygen and carbon SF shell parameters.

Future analyses could therefore benefit from further development of the systematics model to provide theory-driven prior correlations between oxygen and carbon, as well as from a broader use of neutrino–oxygen interaction data.

Although Asimov fits are a useful way to provide an immediate overview of the sensitivity of each systematic parameter, assuming the current parameterization is in use, this study only partially represents the potential of the \ndup. The enhanced angular acceptance and reduced hadron threshold extend phase-space coverage and enable tailored observables, improving model discrimination in CC0$\pi$, CC1$\pi$, and $\nu_e$ channels. This will be clearly accompanied by a further development of the systematic parameterization with respect to the model described here, potentially also dealing with unexpected systematic effects. \ndup new data and samples will guide this development and its consequent tuning. 

\begin{table*}[htbp]
    \centering
    \begin{tabular}{ c | c | c}
    \hline\hline
    \textbf{Configuration} & $\nu_e$ CCQE-like & $\nu_\mu$ CCQE-like\\ 
    \hline
    Current        &  4.1\% &    3.8\%  \\ 
    \hline
    $1.08\times 10^{22}$ POT, FGD1+2      &  3.0\% &   2.8\% \\ 
    \hline
    $1.08\times 10^{22}$ POT, Super-FGD+FGD1+2 & 2.7\% &  2.5\% \\
    \hline
    $1.08\times 10^{22}$ POT, Super-FGD+FGD1+2, O/C 100\% corr. & 1.8\% & 1.7\% \\
    \hline\hline
    \end{tabular}
    \caption{Systematic errors on the CCQE-like appearance and disappearance samples at FD, after the ND280 fit constraints are applied. After reporting the current uncertainty (first line), several configurations are considered for an expected accumulated statistics at the far detector of $1.08\times 10^{22}$ POT (last line of \autoref{tab:t2k2POT}): the case where only FGD1 and FGD2 samples are used at the ND280 fit (2$^{nd}$ line), the case where also \sfgd samples are included (3$^{rd}$ line) and finally the case where an arbitrary 100\% prior correlation is applied before the ND280 fit between oxygen and carbon SF shell parameters.}
    \label{tab:upgradeSK}
\end{table*}

\section{Summary and conclusions}
\label{sec:summary-conclusions}
This work describes the methodology and results of the near-detector analysis performed in the context of the T2K neutrino oscillation measurement reported in Ref.~\cite{T2K:2025yoy}. 

We have introduced a refined flux prediction, a new neutrino cross-section model, new event selections for the off-axis ND280 detector with proton and photon tagging, and updated uncertainty estimates. The neutrino flux prediction has been updated using 2010 NA61/SHINE hadron-scattering data from a replica T2K target, together with improved modeling of the horn cooling water flow and non-hadronic uncertainties. Overall, these updates nearly halve the flux uncertainties in the 2-7 GeV range, slightly reduce them below 0.5 GeV, and slightly increase them between 0.5-2 GeV. The neutrino interaction model and associated uncertainties were significantly expanded to accommodate the new event selections and to include improved insights from nuclear theory and neutrino cross-section measurements. In particular, new theory-driven uncertainties were developed for the CCQE model, including effects from Pauli blocking, nuclear optical potentials, shell structure, and proton final-state interactions. Updated uncertainty treatments for other interaction channels include a new parameterization of the shape of the 2p2h cross section, new uncertainties on nuclear binding energy and resonance decay kinematics in pion production, and new parameters to modify multi-pion production in soft and deep inelastic scattering. Proton and photon tagging were incorporated into the ND280 selection of the $\nu_\mu$ $\text{CC}0\pi$ topology, increasing the number of selected event samples in $\nu$-mode beam from three to five. As a result, additional uncertainties related to proton secondary interactions and detector response were introduced in the analysis.

The ND280 data were analyzed using both Bayesian and hybrid-frequentist frameworks that minimize a negative log-likelihood. The complementarity of these frameworks allows a systematic check of the robustness of the results with respect to analysis choices. Within the hybrid-frequentist framework, a goodness-of-fit was performed, demonstrating that the model provides a good description of the data. New ways of visualizing and scrutinizing the results were introduced in this analysis for the first time, investigating the non-Gaussian behavior of the likelihood. Further robustness studies were conducted, including fits with alternative interaction models and consistency checks against publicly available cross-section measurements. These demonstrate that the use of our analysis as an input to neutrino oscillation measurements is robust against a wide range of modeling assumptions for the current level of statistics available at the far detector. These results also provide clear evidence that further reduction and improved modeling of systematic uncertainties will be essential for next-generation LBL experiments. In this work, we have identified some specific areas of neutrino interaction modeling that require further refinement. These include the treatment of the cross section for pion-less interactions at very forward scattering angles and the modeling of outgoing hadron kinematics in CCQE, 2p2h, and resonant pion production interactions. Bespoke improvements will be necessary to meet the precision demands of future oscillation analyses.

The Collaboration recently completed the upgrade of ND280, which introduced enhanced detector modules designed to improve acceptance and lower hadron thresholds. In this work, we have presented sensitivity studies using the new suite of detectors. They illustrate how the upgraded ND280 may be able to confront outstanding challenges in neutrino interaction modeling to further enhance the precision of future neutrino oscillation measurements.

\section*{Acknowledgements}
The T2K collaboration would like to thank the J-PARC staff for superb accelerator performance. We thank the CERN NA61/SHINE Collaboration for providing valuable particle production data. We acknowledge the support of MEXT, JSPS KAKENHI (JP16H06288, JP18K03682, JP18H03701, JP18H05537, JP19J01119, JP19J22440, JP19J22258, JP20H00162, JP20H00149, JP20J20304, JP24K17065) and bilateral programs (JPJSBP120204806, JPJSBP120209601),  Japan; UGent-BOF and FWO-Flanders, Belgium; NSERC, the NRC, and CFI, Canada; the CEA and CNRS/IN2P3, France; the Deutsche Forschungsgemeinschaft (DFG, German Research Foundation) 397763730, 517206441, Germany; the NKFIH (NKFIH 137812 and TKP2021-NKTA-64), Hungary; the INFN, Italy; the Ministry of Science and Higher Education (2023/WK/04) and the National Science Centre (UMO-2018/30/E/ST2/00441, UMO-2022/46/E/ST2/00336 and UMO-2021/43/D/ST2/01504), Poland;  the RSF (RSF 26-12-00495) and the Ministry of Science and Higher Education, Russia;  MICINN  (PID2022-136297NB-I00 /AEI/10.13039/501100011033/ FEDER, UE, PID2024-157541NB-I00 (UAM) and PID2023-146401NB-I00 (US), Severo Ochoa Centres of Excellence Programme 2025-2029 (CEX2024001441-S),  Government of Andalucia (FQM160) and the University of Tokyo ICRR's Inter-University Research Program FY2025 Ref. J1, and ERDF and European Union (UAM: H2020-MSCA-RISE-GA872549- SK2HK) and NextGenerationEU funds (PRTR-C17.I1) and  Generalitat de Catalunya (AGAUR 2021-SGR-01506, CERCA program) University of Seville grant (RYC2022-035203-I funded by MICIU/AEI/10.13039/501100011033, "ERDF a way of making Europe" and FSE+, Ayudas "Atracción de Investigadores con Alto Potencial". Ref. VIIPPIT-2025, and Secretariat for Universities and Research of the Ministry of Business and Knowledge of the Government of Catalonia and the European Social Fund (2022FI\_B 00336), Spain; the SNSF and SERI (200021\_185012, 200020\_188533, 20FL21\_186178I), Switzerland; the STFC and UKRI, UK; the DOE, USA; and NAFOSTED (103.99-2023.144, IZVSZ2.203433), Vietnam. We also thank CERN for the UA1/NOMAD magnet, DESY for the HERA-B magnet mover system, the BC DRI Group, Prairie DRI Group, ACENET, SciNet, and CalculQuebec consortia in the Digital Research Alliance of Canada, and GridPP in the United Kingdom, and the CNRS/IN2P3 Computing Center in France and NERSC (HEP-ERCAP0028625). In addition, the participation of individual researchers and institutions has been further supported by funds from the ERC (FP7), “la Caixa” Foundation  (ID 100010434, fellowship code LCF/BQ/IN17/11620050), the European Union’s Horizon 2020 Research and Innovation Programme under the Marie Sklodowska-Curie grant agreement numbers 713673 and 754496, and H2020 grant numbers  RISE-GA822070-JENNIFER2 2020 and RISE-GA872549-SK2HK, the Horizon Europe Marie Sklodowska-Curie Staff Exchange project JENNIFER3 grant 101183137; the JSPS, Japan; the Royal Society, UK; French ANR grant number ANR-19-CE31-0001 and ANR-21-CE31-0008; and  Sorbonne Université Emergences programmes; the SNF Eccellenza grant number PCEFP2\_203261;  the VAST-JSPS (No. QTJP01.02/20-22);  and the DOE Early Career programme, USA. For the purposes of open access, the authors have applied a Creative Commons Attribution license to any Author Accepted Manuscript version arising.

\bibliography{biblio}

\clearpage

\onecolumngrid
\begin{appendices}
    \section{Full comparisons to T2K cross-section data}
\label{app::more_plots}

This appendix shows the predictions of the neutrino interaction model before and after the near-detector fit against the full set of measurements discussed in the benchmarking section, \autoref{subsec:benchmark}.

\begin{figure*}[htbp]
    \centering 
        \includegraphics[ width=0.49\linewidth]{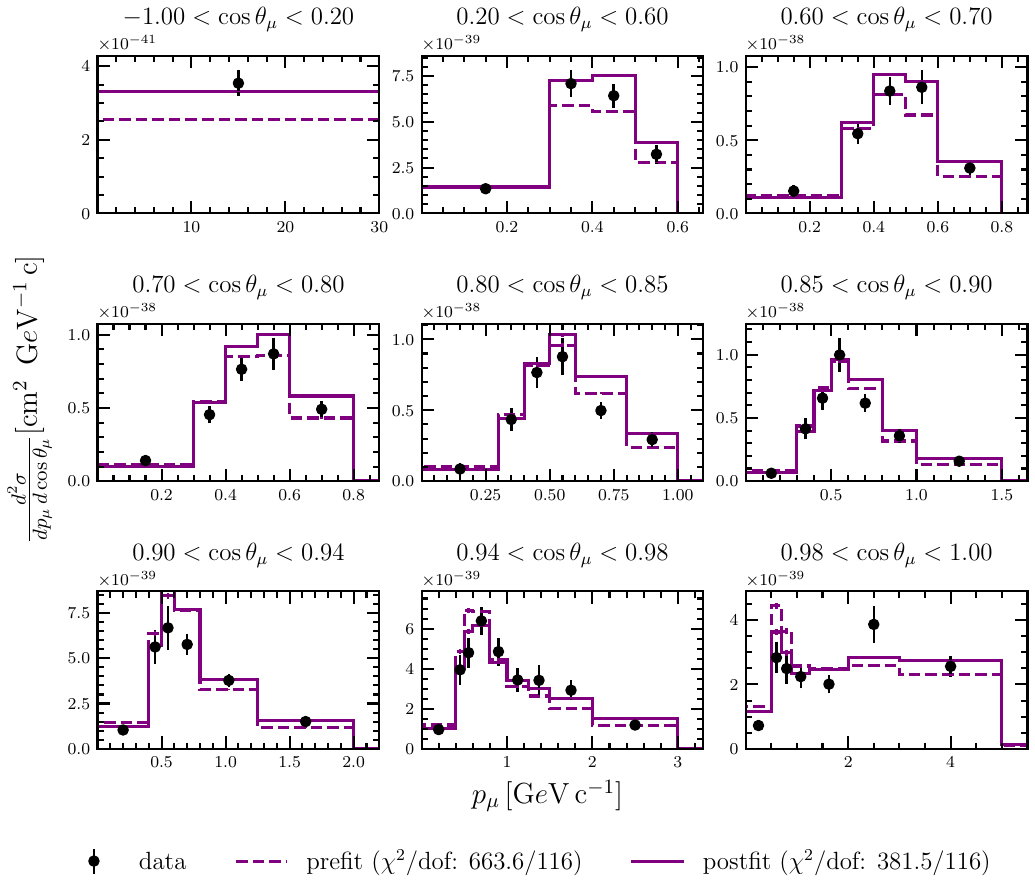}
        \includegraphics[ width=0.49\linewidth]{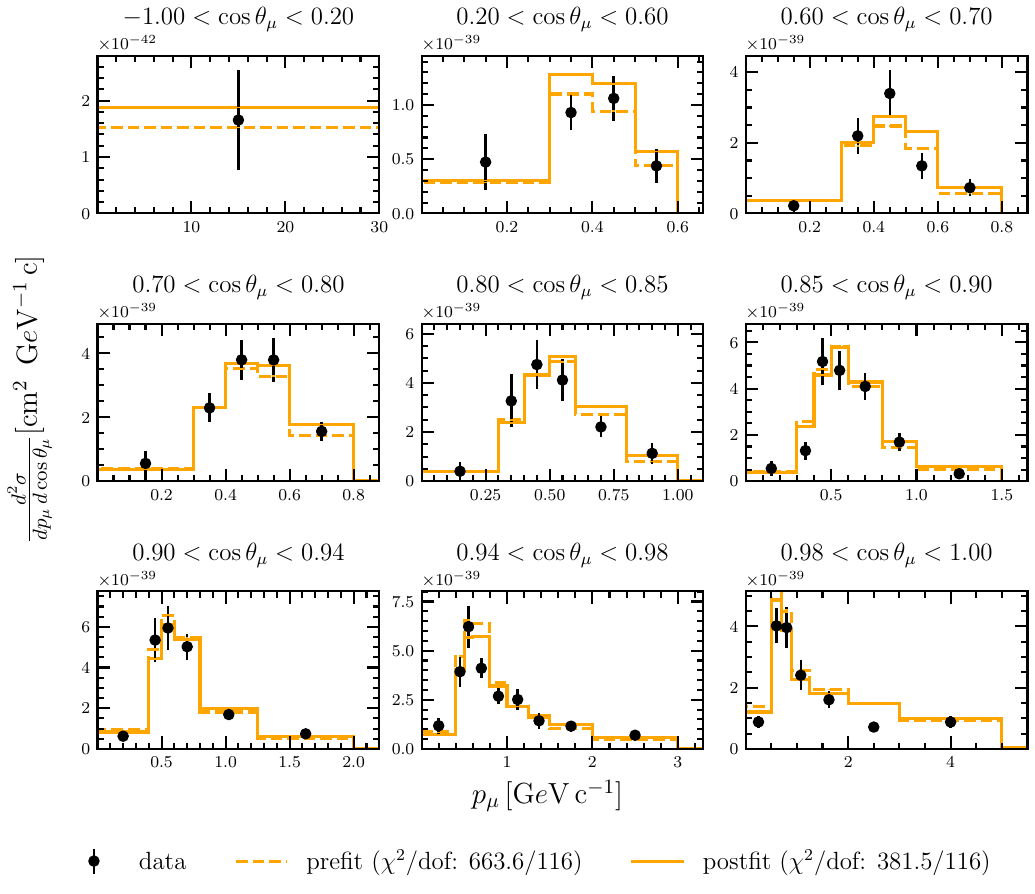}
    \caption{Predictions from the interaction model of the differential cross sections for T2K's $\num, \numb$ CC0$\pi$ measurement on a CH target in $p_\mu \cosmu$~\cite{T2K:2020sbd}, before and after the near-detector fit presented in this paper.
    The left panel in purple is muon neutrinos and the right panel in orange is muon antineutrinos.}
    \label{Fig::BANFF_NuAnu_2Dpcos}
\end{figure*}

\autoref{Fig::BANFF_NuAnu_2Dpcos} shows the simultaneous CC0$\pi$ neutrino and antineutrino measurement in $p_\mu \cosmu$, where the focus in the main body was around the improvements in the forward and backward angles in the peak muon momentum region.

\begin{figure*}[htb]
    \centering 
    \includegraphics[width=0.49\linewidth]{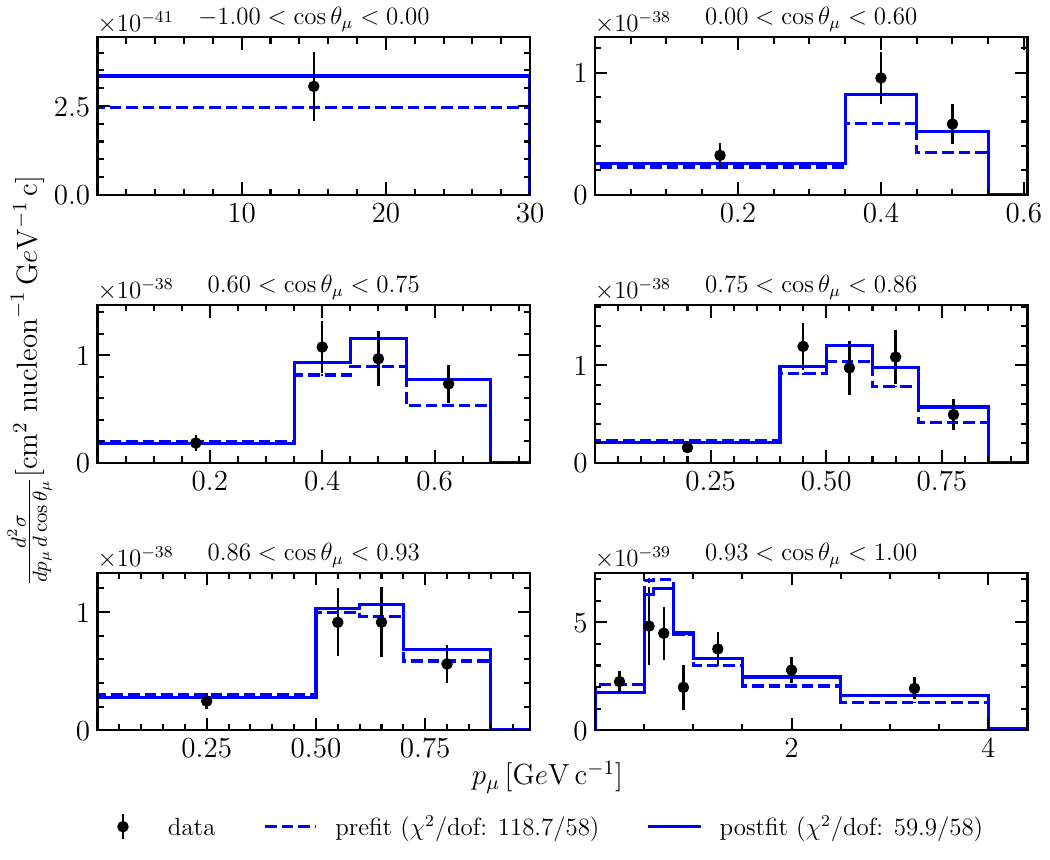}
    \includegraphics[width=0.49\linewidth]{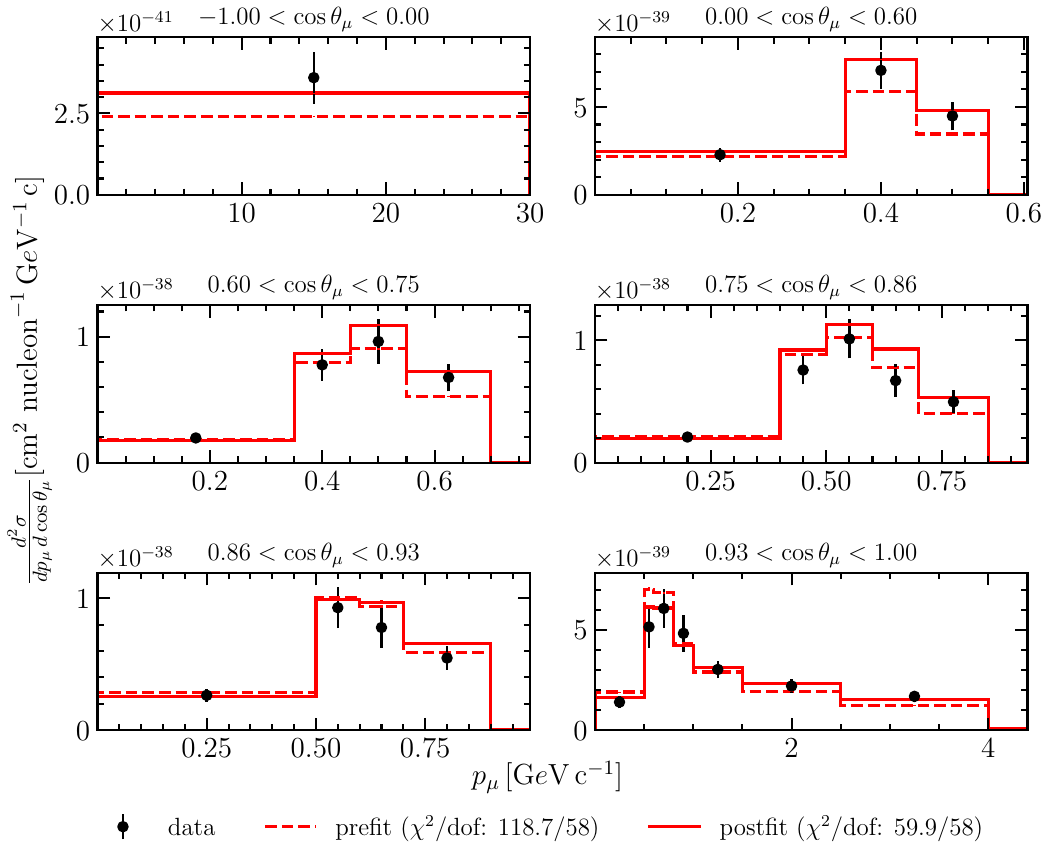}
    \caption{Predictions from the interaction model of the differential cross sections for T2K's CC0$\pi$ joint O-C measurement in $p_\mu \cosmu$~\cite{T2K:2020jav}, before and after the near-detector fit presented in this paper.
    The left panel in blue is oxygen and the right panel in red is carbon.}
    \label{Fig::BANFF_OC_2Dpcos}
\end{figure*}

\autoref{Fig::BANFF_OC_2Dpcos} shows the simultaneous measurement of the CC0$\pi$ cross section for muon neutrinos on carbon and oxygen. Both measurements see similar behavior before and after the output of the ND280 analysis is applied, with a $\chi^2$ indicating reasonable agreement between the postfit model and the measurement.

\begin{figure*}[htb]
    \centering 
    \subfloat{
        \includegraphics[ width=0.7\linewidth]{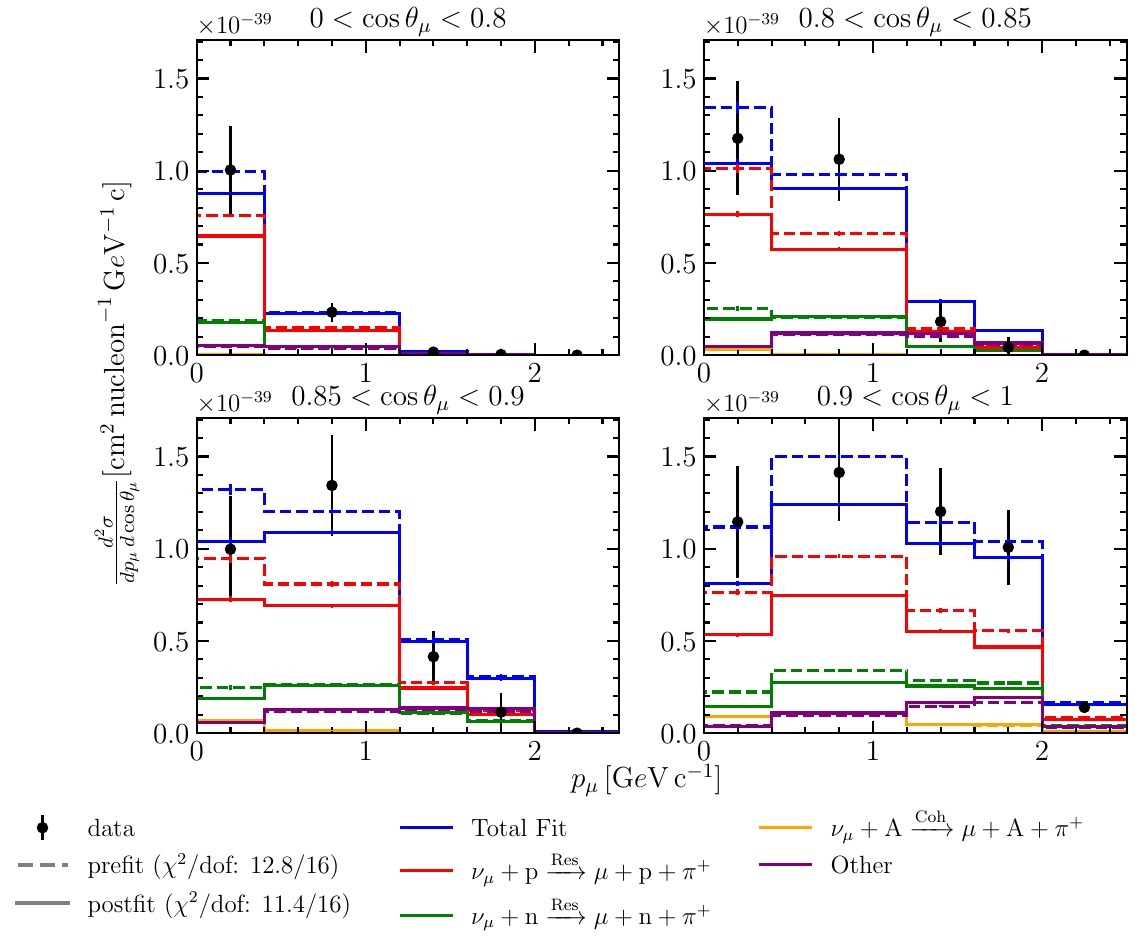}
    }
    \caption{Predictions from the interaction model of the differential cross sections for T2K's measurement of the CC1$\pi^+$ final state on a CH target in $p_\mu \cosmu$~\cite{T2K:2019yqu}, before and after the near-detector fit presented in this paper.}
    \label{Fig::BANFF_2Dpmucosmu}
\end{figure*}

\autoref{Fig::BANFF_2Dpmucosmu} shows the predictions for T2K's CC$1\pi^+$ measurement for the whole muon kinematics range reported in the data ($p_\mu>200~\text{MeV}/\mathrm{c}, \cosmu > 0$).
The overall $\chi^2$ marginally decreases for the postfit model, but is already agreeable before the ND280 analysis. Notably, the near-detector analysis does not affect the prediction for muons above 1 GeV/$c$ unless they are forward-going, which is not well described by the model. The non-resonant contributions (SIS, DIS) become increasingly dominant here, as seen in the channel breakdown shown in the figure.

\section{All Cross-section Parameters}

\cref{tab:xsecparameters} reports the full list of cross-section parameters used in the analysis, including the names, prior values, and errors, as well as a short explanation of the parameter meaning. All details are given in the \cref{subsec:xsecuncert}.

\begin{table*}[!ht]
\centering
\vspace{-0.5cm}
\small 
\scalebox{0.7}{ 
 \begin{tabular}{  l | c | l  }
 \hline\hline
Parameter      & Value   & Comment \\ 
\hline
\multicolumn{3}{l}{\textbf{1p1h}~(\autoref{sec:1p1huncert}) } \\
$M_A^{QE} (\text{GeV}/c^2)$    & $1.03\pm0.06$  & Axial mass for CCQE \\
High $Q^2$ norm 1  & $1\pm0.11$ & Normalisation for CCQE events with $0.25 < Q^2 < 0.5$ \\
High $Q^2$ norm 2  & $1\pm0.18$ & Normalisation for CCQE events with $0.5 < Q^2 < 1.0$ \\
High $Q^2$ norm 3  & $1\pm0.40$ & Normalisation for CCQE events with $Q^2>1.0$ \\
s-shell norm $^{12}$C   & $0\pm0.45$  & Normalisation of $s$-shell for carbon \\
p-shell norm $^{12}$C   & $0\pm0.20$  & Normalisation of $p$-shell for carbon \\
s-shell norm $^{16}$O   & $0\pm0.75$  & Normalisation of $s$-shell for oxygen \\
p$_{1/2}$-shell norm $^{16}$O & $0\pm0.20$  & Normalisation of $p_{1/2}$-shell for oxygen \\
p$_{3/2}$-shell norm $^{16}$O & $0\pm0.45$  & Normalisation of $p_{3/2}$-shell for oxygen \\
SRC norm $^{12}$C         & $1\pm2$  & Normalisation of short-range correlation component in $^{12}$C SF \\
SRC norm $^{16}$O         & $1\pm2$  & Normalisation of short-range correlation component in $^{16}$O SF \\
p-shell $p_{N}$ shape $^{12}$C & $1\pm1$ & Fixed at ND280 \\
p-shell $p_{N}$ shape $^{12}$C & $0\pm1$ & Fixed at ND280 \\
s-shell $p_{N}$ shape $^{16}$O & $0\pm1$  & Fixed at ND280 \\
p$_{1/2}$-shell $p_{N}$ shape $^{16}$O & $0\pm1$ & Fixed at ND280 \\
p$_{3/2}$-shell $p_{N}$ shape $^{16}$O & $0\pm1$ & Fixed at ND280 \\
Pauli blocking $^{12}$C $\nu$ & $0\pm1$         & Fermi surface momentum cut-off for nucleon in carbon for neutrinos \\
Pauli blocking $^{16}$O $\nu$ & $0\pm1$         & Fermi surface momentum cut-off for nucleon in oxygen for neutrinos \\
Pauli blocking $^{12}$C \nub & $0\pm1$   & Fermi surface momentum cut-off for nucleon in carbon for antineutrinos \\
Pauli blocking $^{16}$O \nub & $0\pm1$   & Fermi surface momentum cut-off for nucleon in oxygen for antineutrinos  \\
Optical potential $^{12}$C & 0 & Final-state interactions correction affecting lepton \\
Optical potential $^{16}$O & 0 & Final-state interactions correction affecting lepton \\
$\Delta E_\mathrm{rmv}$ $^{12}$C $\nu$       (MeV) &  $2\pm6$  & Removal energy \\
$\Delta E_\mathrm{rmv}$ $^{12}$C \nub (MeV) &  $0\pm6$  & Removal energy \\
$\Delta E_\mathrm{rmv}$ $^{16}$O $\nu$       (MeV) &  $4\pm6$  & Removal energy \\
$\Delta E_\mathrm{rmv}$ $^{16}$O \nub (MeV) &  $0\pm6$  & Removal energy \\
$\alpha$ correction ($q_3$)     & 0  & $q_3$ dependent correction to removal energy \\ 
\hline

\multicolumn{3}{l}{\textbf{2p2h}~(\autoref{sec:2p2huncert})} \\
2p2h norm $\nu$                & 1  & Normalisation of 2p2h for $\nu$ \\
2p2h norm \nub          & 1  & Normalisation of 2p2h for \nub \\
2p2h norm $^{12}$C to $^{16}$O & $1\pm0.2$ & Normalisation of 2p2h for interactions on oxygen  \\
PNNN shape                   & $0\pm0.33$ & Changes ratio of pn to nn/pp pairs in the Valencia 2p2h model \\
2p2h shape $^{12}$C NN       &  $0\pm3$   & Shifts 2p2h events or be more $\Delta$-like or non $\Delta$-like in the Valencia model \\
2p2h shape $^{12}$C np       &  $0\pm3$   & Shifts 2p2h events or be more $\Delta$-like or non $\Delta$-like in the Valencia model \\
2p2h shape $^{16}$O NN       &  $0\pm3$   & Shifts 2p2h events in the Valencia model to be more $\Delta$-like or non $\Delta$-like \\
2p2h shape $^{16}$O np       &  $0\pm3$   & Shifts 2p2h events or be more $\Delta$-like or non $\Delta$-like in the Valencia model \\
2p2h Edep low $E_{\nu}$         & 1  & Fixed at ND280 \\
2p2h Edep high $E_{\nu}$        & 1  & Fixed at ND280 \\
2p2h Edep low $E_{\nub}$   & 1  & Fixed at ND280 \\
2p2h Edep high $E_{\nub}$  & 1  & Fixed at ND280 \\
\hline

\multicolumn{3}{l}{\textbf{SPP}~(\autoref{sec:SPPuncert})} \\
$M_A^{RES} (\text{GeV}/c^2$)        & $0.91\pm0.1$ & Axial mass for single pion production via a resonance \\
$C^A_{5}$                           & $1.06\pm0.1$  & Axial form factor for single pion production via a resonance \\
Non-resonant $I_{1/2}$              & $1.21\pm0.27$ & Normalisation of $I_{1/2}$ non-resonant background in the Rein--Sehgal model \\
Non-resonant $I_{1/2}$, low p$_{\pi}$ & $1.3\pm1.3$  & Same as above, but FD only, for antineutrinos producing a $p_{\pi} < 200~\text{MeV}/c$ \\
RS $\Delta$ decay                     & 1  & Changes type of $\Delta$ decay between ``Delta''-like and isotropic in the Rein--Sehgal (RS) model \\
RES $\pi^0$ norm $\nu_{\mu}$          & $1\pm0.3$   & Normalisation of CC$1\pi^0$ \\
RES $\pi^0$ norm \numb    & $1\pm0.3$   & Normalisation of $\nub$CC $1\pi^0$ \\
RES $E_\mathrm{rmv}$ $^{12}$C $\nu_{\mu}$       (MeV) & $25\pm25$ & Removal energy  \\
RES $E_\mathrm{rmv}$ $^{16}$O $\nu_{\mu}$       (MeV) & $25\pm25$ & Removal energy  \\
RES $E_\mathrm{rmv}$ $^{12}$C \numb (MeV) & $25\pm25$ & Removal energy  \\
RES $E_\mathrm{rmv}$ $^{16}$O \numb (MeV) & $25\pm25$ & Removal energy  \\
CC coh $^{12}$C  & $1\pm0.3$  & Normalisation for CC Coherent \\
CC coh $^{16}$O  & $1\pm0.3$  & Normalisation for CC Coherent \\
NC coh           & $1\pm0.3$ & Fixed at ND280 \\
\hline

\multicolumn{3}{l}{\textbf{SIS and DIS}~(\autoref{sec:DISuncert})} \\
M$\pi$ Multi TotXSec  & 0$\pm$1  & Impact of AGKY on total cross section\\
M$\pi$ Multi Shape    & 0$\pm$1  & Impact of AGKY on W and $\pi$ multiplicity \\
M$\pi$ BY Vector      & 0$\pm$1  & Vector part of Bodek-Yang correction for M$\pi$ \\
M$\pi$ BY Axial       & 0$\pm$1  & Axial part of Bodek-Yang correction for M$\pi$ \\

CC BY DIS             &  0$\pm$1 & Bodek-Yang correction for DIS \\
CC DIS Multi$\pi$ Norm $\nu$       & 1$\pm$0.035 & Normalisation of DIS/Multi$\pi$ \\
CC DIS Multi$\pi$ Norm \nub & 1$\pm$0.065 & Normalisation of DIS/Multi$\pi$ \\
\hline

\multicolumn{3}{l}{\textbf{FSI}~(\autoref{sec:FSIuncert})} \\
$\pi$ FSI QE low E  & 1.069$\pm$0.313  & Pion QE FSI, $p_{\pi} < 500$ MeV/c \\ 
$\pi$ FSI QE high E & 1.824$\pm$0.859  & Pion QE FSI, $p_{\pi} > 500$ MeV/c \\ 
$\pi$ FSI Prod.     & 1.002$\pm$1.101  & Pion Production FSI \\  
$\pi$ FSI Abs.      & 1.404$\pm$0.432  & Pion Absorption FSI \\ 
$\pi$ FSI Cex low E & 0.697$\pm$0.305  & Pion CEX FSI, $p_{\pi} < 500$ MeV/c \\ 
$\pi$ FSI Cex high E & 1.8$\pm$0.288   & Pion CEX FSI, $p_{\pi} > 500$ MeV/c \\ 
Nucleon FSI    & 0$\pm$0.3 & Modifies nucleon transparency \\
\hline

\multicolumn{3}{l}{\textbf{Other}~(\autoref{sec:miscuncert})} \\

CC Misc          & 1$\pm$1    & Normalisation for CC RES 1$\gamma$,$K$, $\eta$ and CC diffractive \\
NC 1$\gamma$     & 1$\pm$1   & Fixed at ND280 \\
NC other near    & 1$\pm$0.3  & Normalisation of NC diffractive and $\eta$ for ND280\\
NC other far     & 1$\pm$0.3  & Normalisation of NC diffractive and $\eta$ for FD \\
CC norm $\nu$        & 1$\pm$0.2 & Coulomb correction normalisation in $0.3<E_{\nu}<0.6$ GeV \\
CC norm \nub  & 1$\pm$0.1 & Coulomb correction normalisation in $0.3<E_{\nu}<0.6$ GeV\\
$\nu_{e}$/$\nu_{\mu}$            & 1$\pm$0.0282843  & $\nu_{e}$ normalisation\\
\nueb/\numb & 1$\pm$0.0282843 & \nueb normalisation \\
\hline\hline
\end{tabular}
}
\caption{Summary of all cross-section parameters used in the analysis. Parameters without quoted uncertainties mean they were left unconstrained. Units are included when applicable.
}
\label{tab:xsecparameters}
\end{table*}

\clearpage

\section{Summary of potential biases}

\autoref{tab:bias_table_fakedata_full} reports the potential biases on the main oscillation parameters for the simulated data set not reported in \autoref{Tab:FakeData}.

\begin{table*}[htbp]
\centering
\begin{tabular}{ll ccc}
\hline \hline
Simulated data set & Relative to & $\sin{\theta_{23}}$ & $\Delta m^2_{32}$ & $\delta_{CP}$\\
\hline

\multirow{2}{*}{CCQE z-exp nom. var.}    & Total       & 0.2\% & 0.1\% & -0.3\%  \\
                                    & Syst.       & 0.3\% & 0.2\% & -1.1\%   
                                    \vspace{2mm}\\
\hline

\multirow{2}{*}{CCQE z-exp lower var.}     & Total       & 0.8\% & -2.2\% & -0.1\%   \\
                                    & Syst.       & 1.7\% & -5.6\% & -0.4\%   
                                    \vspace{2mm}\\
\hline
\multirow{2}{*}{CCQE 3-comp nom. var.}   & Total       & 0.8\% & -1.7\% & -0.5\%   \\
                                    & Syst.       & 1.7\% & -4.3\% & -1.7\%  
                                    \vspace{2mm} \\
\hline
\multirow{2}{*}{CCQE 3-comp upper var.}   & Total       & 3.9\% & -2.2\% & 0.03\%  \\
                                    & Syst.       & 7.8\% & -5.6\% & 0.1\%  
                                    \vspace{2mm}\\ 
\hline
\multirow{2}{*}{CCQE 3-comp lower var.}    & Total       & -0.4\% & 0.8\% & -0.4\%   \\
                                    & Syst.       & -0.7\% & 1.9\% & -1.6\%   
                                    \vspace{2mm}\\                                  

\hline
\multirow{2}{*}{LFG}                & Total       & -0.2\% & -16.6\% & 0.9\%  \\
                                    & Syst.       & -0.5\% & -44.0\% &3.7\%
                                    \vspace{2mm}\\ 
\hline

\multirow{2}{*}{CCQE Removal energy} & Total       & -0.5\% & -0.4\% & 0.1\% \\
                                     & Syst.       & -0.9\% & -0.9\% & 0.4\%
                                    \vspace{2mm}\\
\hline
\multirow{2}{*}{CCQE Removal energy with cubic interpolation} 
                                    & Total       & 3.6\% & -6.6\% & 0.8\% \\
                                    & Syst.       & 6.4\% & -17.3\% & 2.7\%
                                    \vspace{2mm}\\  
\hline
\multirow{2}{*}{CCQE Removal energy with linear interpolation} 
                                    & Total       & 2.2\% & -1.8\% & 0\%  \\
                                    & Syst.       & 3.9\% & -4.8\% & 0\%
                                    \vspace{2mm}\\

\hline
\multirow{2}{*}{Low-$Q^2$ suppression in CC-1$\pi$ int.}     
                                    & Total       & 7.7\% & 8.2\% & -1.3\% \\
                                    & Syst.       & 16.9\% & 20.4\% & -5.2\%  
                                    \vspace{2mm}\\
\hline
\multirow{2}{*}{1$\pi$ Kinematics -3$\sigma$}     
                                    & Total       & -1.6\% & 7.3\% & -0.6\% \\
                                    & Syst.       & -3.5\% & 18.6\% & -2.5\%  
                                    \vspace{2mm}\\
\hline
\multirow{2}{*}{1$\pi$ Kinematics +3$\sigma$}     
                                    & Total       & 0.8\% & -3.1\% & -0.2\%  \\
                                    & Syst.       & 1.6\% & -7.9\% & -0.7\% 
                                    \vspace{2mm}\\
\hline

\multirow{2}{*}{Data-driven pion}   & Total       & -0.8\% & -3.0\% & 0.3\% \\
                                    & Syst.       & -1.7\% & -7.6\% & 1.0\% 
                                    \vspace{2mm}\\
\hline
\multirow{2}{*}{Radiative Corrections} 
                                    & Total       & -3.4\% & 6.8\% & -0.2\% \\
                                    & Syst.       & -7.4\% & 16.9\% & -0.9\% \\ 

\hline\hline
\end{tabular}
\caption{Potential biases on the main oscillation parameters for simulated data set, calculated as the shift in the middle of the $1\sigma$ confidence interval relative to the overall uncertainty from systematic sources (``Syst.'') and the total (``Total'') to one decimal place.}
\label{tab:bias_table_fakedata_full}
\end{table*}

\end{appendices}
\twocolumngrid

\end{document}